\documentclass[a4paper,11pt]{article}

\usepackage[T1]{fontenc}
\usepackage[utf8]{inputenc}
\usepackage{lmodern}

\usepackage{tikz} 
\usepackage{xcolor}
\usepackage{graphicx}
\graphicspath{{Figures/}}

\usepackage[left=2.5cm,right=2.5cm,bottom=2.5cm,top=2.5cm]{geometry}
	
\usepackage[english]{babel}

\usepackage{microtype}
	
\usepackage{amsmath,amssymb}

\usepackage{float}

\usepackage{bm}

\usepackage{enumitem}

\usepackage[hang,flushmargin]{footmisc}
	
\usepackage{relsize,exscale}

\usepackage{fancyhdr}

\usepackage[justification=RaggedRight, singlelinecheck=false]{caption}
\usepackage{subcaption}
	
\usepackage{changepage}
	
\usepackage[most]{tcolorbox}
\tcbset{colback=yellow!10!white, colframe=red!50!black, 
	highlight math style= {enhanced, 
		colframe=red,colback=red!10!white,boxsep=0pt}}
	
\usepackage{empheq}

\usepackage{gensymb}

\usepackage{dsfont}

\usepackage[only,llbracket,rrbracket]{stmaryrd}

\usepackage{mathtools}

\usepackage{slashed}

\DeclareMathAlphabet{\mathdutchcal}{U}{dutchcal}{m}{n}
\SetMathAlphabet{\mathdutchcal}{bold}{U}{dutchcal}{b}{n}
\DeclareMathAlphabet{\mathdutchbcal}{U}{dutchcal}{b}{n}

\DeclareFontFamily{OT1}{pzc}{}
\DeclareFontShape{OT1}{pzc}{m}{it}{<-> s * [1.10] pzcmi7t}{}
\DeclareMathAlphabet{\mathpzc}{OT1}{pzc}{m}{it}

\usepackage{cite}

\makeatletter
\let\oldthebibliography\thebibliography
\renewcommand\thebibliography[1]{%
	\oldthebibliography{#1}%
	\setlength{\itemsep}{5pt}%
	\setlength{\parskip}{0pt}%
}
\makeatother

\usepackage{hyperref}
\hypersetup{
	pagebackref=true,
	colorlinks=true,
	linktoc=section,   
	citecolor=blue, 
	filecolor=blue,
	linkcolor=blue,
	urlcolor=blue,
	pdfborder={0 0 0.5},
	pdfmenubar=false,
	pdftoolbar=false
}
\numberwithin{equation}{section}

\DeclarePairedDelimiterX{\bra}[1]{\langle}{\rvert}{#1}
\DeclarePairedDelimiterX{\ket}[1]{\lvert}{\rangle}{#1}
\DeclarePairedDelimiterX{\makebraket}[1]{\langle}{\rangle}{#1}

\usepackage{xparse}
\NewDocumentCommand{\braket}{som}{%
	\begingroup\activatebraketbar
	\IfBooleanTF{#1}
	{\makebraket*{#3}}
	{\IfNoValueTF{#2}{\makebraket{#3}}{\makebraket[#2]{#3}}}%
	\endgroup
}

\makeatletter
\newcommand{\braketbar}{%
	\,\delimsize\vert\@ifnextchar|{\!}{\,\!\!\:}%
}
\makeatother
\newcommand{\activatebraketbar}{%
	\begingroup\lccode`~=`|\lowercase{\endgroup\let~}\braketbar
	\mathcode`|="8000
}

\newcommand{\ols}[1]{\mskip.5\thinmuskip\overline{\mskip-.5\thinmuskip {#1} \mskip-.5\thinmuskip}\mskip.5\thinmuskip} 
\newcommand{\olsi}[1]{\,\overline{\!{#1}}} 
\makeatletter
\newcommand\closure[1]{
	\tctestifnum{\count@stringtoks{#1}>1} 
	{\ols{#1}} 
	{\olsi{#1}} 
}
\long\def\count@stringtoks#1{\tc@earg\count@toks{\string#1}}
\long\def\count@toks#1{\the\numexpr-1\count@@toks#1.\tc@endcnt}
\long\def\count@@toks#1#2\tc@endcnt{+1\tc@ifempty{#2}{\relax}{\count@@toks#2\tc@endcnt}}
\def\tc@ifempty#1{\tc@testxifx{\expandafter\relax\detokenize{#1}\relax}}
\long\def\tc@earg#1#2{\expandafter#1\expandafter{#2}}
\long\def\tctestifnum#1{\tctestifcon{\ifnum#1\relax}}
\long\def\tctestifcon#1{#1\expandafter\tc@exfirst\else\expandafter\tc@exsecond\fi}
\long\def\tc@testxifx{\tc@earg\tctestifx}
\long\def\tctestifx#1{\tctestifcon{\ifx#1}}
\long\def\tc@exfirst#1#2{#1}
\long\def\tc@exsecond#1#2{#2}
\makeatother

\makeatletter
\newlength\xvec@height%
\newlength\xvec@depth%
\newlength\xvec@width%
\newcommand{\xvec}[2][]{%
	\ifmmode%
	\settoheight{\xvec@height}{$#2$}%
	\settodepth{\xvec@depth}{$#2$}%
	\settowidth{\xvec@width}{$#2$}%
	\else%
	\settoheight{\xvec@height}{#2}%
	\settodepth{\xvec@depth}{#2}%
	\settowidth{\xvec@width}{#2}%
	\fi%
	\def\xvec@arg{#1}%
	\def\xvec@dd{:}%
	\def\xvec@d{.}%
	\raisebox{.2ex}{\raisebox{\xvec@height}{\rlap{%
				\kern.05em
				\begin{tikzpicture}[scale=1]
					\pgfsetroundcap
					\draw (.05em,0)--(\xvec@width-.05em,0);
					\draw (\xvec@width-.05em,0)--(\xvec@width-.15em, .1em);
					\draw (\xvec@width-.05em,0)--(\xvec@width-.15em,-.1em);
					\ifx\xvec@arg\xvec@d%
					\fill(\xvec@width*.45,.5ex) circle (.5pt);%
					\else\ifx\xvec@arg\xvec@dd%
					\fill(\xvec@width*.30,.5ex) circle (.5pt);%
					\fill(\xvec@width*.65,.5ex) circle (.5pt);%
					\fi\fi%
				\end{tikzpicture}%
	}}}%
	#2%
}
\makeatother

\usepackage{contour}
\usepackage[normalem]{ulem}

\contourlength{0.8pt}

\newcommand{\myuline}[1]{%
	\uline{\phantom{#1}}%
	\llap{\contour{white}{#1}}%
}

\makeatletter
\g@addto@macro\bfseries{\boldmath}
\makeatother

\newcommand{\Englob}{E\raisebox{-2pt}{\scalebox{0.7}{$n$}}\!\!\!\,\raisebox{4.5pt}{\scalebox{0.65}{$(\text{glob})$}}} 
\newcommand{\Enloc}{E\raisebox{-2pt}{\scalebox{0.7}{$n$}}\!\!\!\,\raisebox{4.5pt}{\scalebox{0.65}{$(\text{loc})$}}} 
 
\newcommand{\Eloc}{E\raisebox{-2.5pt}{\scalebox{0.65}{$0$}}\!\!\!\:\raisebox{4.5pt}{\scalebox{0.65}{$(\text{loc})$}}} 

\newcommand{\psinglob}{\psi\raisebox{-2pt}{\scalebox{0.7}{$n$}}\!\!\!\,\raisebox{4.5pt}{\scalebox{0.65}{$(\text{glob})$}}} 
\newcommand{\psinloc}{\psi\raisebox{-2pt}{\scalebox{0.7}{$n$}}\!\!\!\,\raisebox{4.5pt}{\scalebox{0.65}{$(\text{loc})$}}}  
\newcommand{\psinlocLO}{\psi\raisebox{-2pt}{\scalebox{0.7}{$n,\scalebox{0.85}{\text{LO}}$}}\hspace{-0.6cm}\raisebox{4.75pt}{\scalebox{0.65}{$(\text{loc})$}}\hspace{0.18cm}}  

\newcommand{\psilocexp}{\psi\raisebox{-2.5pt}{\scalebox{0.7}{$\text{exp}$}}\!\!\!\!\!\!\raisebox{4.5pt}{\scalebox{0.65}{$(\text{loc})$}}} 
\newcommand{\psilocnonexp}{\psi\raisebox{-2.5pt}{\scalebox{0.7}{$n,\text{non-exp}$}}\hspace{-1.22cm}\raisebox{4.5pt}{\scalebox{0.65}{$(\text{loc})$}}\hspace{0.74cm}} 

\newcommand\scaleddot{\scalebox{.89}{.}}

\makeatletter
\renewcommand{\dddot}[1]{%
	{\mathop{\kern\z@#1}\limits^{\makebox[0pt][c]{\vbox to-2.2\ex@{\kern-\tw@\ex@
					\hbox{\normalfont\scaleddot\kern-0.5pt\scaleddot\kern-0.5pt\scaleddot}\vss}}}}}
\renewcommand{\ddddot}[1]{%
	{\mathop{\kern\z@#1}\limits^{\makebox[0pt][c]{\vbox to-2.2\ex@{\kern-\tw@\ex@
					\hbox{\normalfont\scaleddot\kern-0.5pt\scaleddot\kern-0.5pt\scaleddot\kern-0.5pt\scaleddot}\vss}}}}}
\makeatother

\newcounter{daggerfootnote}
\newcommand*{\daggerfootnote}[1]{%
	\setcounter{daggerfootnote}{\value{footnote}}%
	\renewcommand*{\thefootnote}{\fnsymbol{footnote}}%
	\footnote[2]{#1}%
	\setcounter{footnote}{\value{daggerfootnote}}%
	\renewcommand*{\thefootnote}{\arabic{footnote}}%
}

\begin{document}
\renewcommand{\headrulewidth}{0pt}
\pagestyle{fancy}
\pagenumbering{Roman}
\fancyhead{}
\cfoot{$-\;\!$\thepage $\;\!-$}

\begin{flushright}
	{\small
		\textcolor{black}{TUM--HEP--1546/24}\\
		\textcolor{black}{MPP--2024--257}
	}
\end{flushright}
\vspace{0.5cm}

\begin{center}
{\LARGE\bf False vacuum decay of excited states \\[0.05cm] in finite-time instanton calculus} \\[0.75cm]
	
 	\textsc{Björn Garbrecht}$^{1}$ and \textsc{Nils Wagner}$^{1,2,\dagger}$

	\vspace{0.5cm}
	
	{\it ${}^1$ Physik--Department T70, Technische Universität München, James--Franck--Straße 1, \\ D--85748 Garching, Germany \\[0.15cm]}
	{\it ${}^2$ Max--Planck--Institut für Physik (Werner--Heisenberg--Institut), Boltzmannstraße 8, \\ D--85748 Garching, Germany\\}
	
	\vspace{0.5cm}
	\emph{E--mail:} \href{mailto:garbrecht@tum.de}{{\tt garbrecht@tum.de}}, 
	\href{mailto:nils.wagner@tum.de}{{\tt nils.wagner@tum.de}}\daggerfootnote{Corresponding author}

	\medskip
	
\end{center}

\vskip1cm

\begin{abstract}
	\vspace{0.2cm}
	
	\noindent
	Extracting information about a system's metastable ground state energy employing functional methods usually hinges on utilizing the late-time behavior of the Euclidean propagator, practically impeding the possibility of determining decay widths of excited states. We demonstrate that such obstacles can be surmounted by working with bounded time intervals, adapting the standard instanton formalism to compute a finite-time amplitude corresponding to excited state decay. This is achieved by projecting out the desired resonant energies utilizing carefully chosen approximations to the excited state wave functions in the false vacuum region. To carry out the calculation, we employ unconventional path integral techniques by considering the emerging amplitude as a single composite functional integral that includes fluctuations at the endpoints of the trajectories. This way, we explicitly compute the sought-after decay widths, including their leading quantum corrections, for arbitrary potentials, demonstrating accordance with traditional WKB results. While the initial starting point of weighting Euclidean propagator contributions according to their endpoints using false vacuum states has been proposed earlier, we find several flaws in the published evaluation of the relevant amplitudes. Although we show that the previous proposition of employing a sequential calculation scheme---where the functional integral is evaluated around extremal trajectories with fixed endpoints, weighted only at a subsequent stage---can lead to the desired goal, the novel composite approach is found to be more concise and transparent.
\end{abstract} 
\newpage

\tableofcontents

\newpage

\renewcommand{\headrulewidth}{0pt}
\pagestyle{fancy}
\pagenumbering{arabic}
\fancyhead{}
\cfoot{$-\;\!$\thepage $\;\!-$}

\section{Introduction}
\label{sec:1_Introduction}

At low temperatures, quantum tunneling constitutes the driving factor for quantum phase transitions, both in the early universe~\cite{LindeCosmicPhaseTrans,KibbleCosmicPhaseTrans,WittenCosmicSeparationOfPhases,HoganGravRadiationFromPhaseTransition,MazumdarReviewCosmicPhaseTrans} and in matter~\cite{LangerCondensationPoint,ReactionRateTheoryHanggi,QuantumPhaseTransReview,VacDecaySpinChains,ZenesiniTunnelinCM}. With the Standard Model vacuum found to be metastable~\cite{IsidoriMetaStabilitySM,ButtazzoSMLifetime,SchwartzScaleInvInstantons}, tunneling considerations provide an intricate tool constraining BSM theories, necessitating accurate and reliable theoretical techniques. Especially in the infinite-dimensional setting of QFT, investigating tunneling phenomena poses unique challenges, as extending the usual WKB methods meets formidable difficulties~\cite{BenderMultiDimensionalWKB,BitarWKBFieldTheory,DarmeGeneralizedEscapePaths}. Similarly, numerical computations are hindered by significant challenges~\cite{BergesLattice,AlexandruLattice,MouRealTime,BradenVacDecay,NishimuraRealTimeTunneling}. These issues are usually addressed by utilizing the functional formulation in the guise of Feynman path integrals, allowing for the simple extraction of non-perturbative quantities~\cite{GelfandFunctionalIntegration,DashenNonperturbativeMethods,GildenerInstantonMethod}, especially tunneling rates, via the instanton method~\cite{LangerCondensationPoint,LangerMetastability,KobzarevMetastableVacuum,BRSTInstantons,tHooftInstantons,ColemanFateOfFalseVac1,CallanColemanFateOfFalseVac2}. Recent advancements in the mathematical understanding of path integrals, employing the language of Picard--Lefschetz theory~\cite{PhamPicardLefschetz,HowlsPicardLefschetzTheory,WittenAnalyticContinuation,UnsalTowardsPicLefschetz}, have led to a surge of activity in the field. With the primary focus lying on understanding the relationship between instanton solutions in Euclidean time and the real-time dynamics of quantum particles~\cite{BenderTunnelingAsAnomaly,TurokRealTimeTunneling,UnsalRealTimeInstantons,TanizakiLefschetz,LehnersComplexTimePaths,SchwartzDirectMethod,SchwartzPrecisionDecayRate,GarbrechtFunctionalMethods,NishimuraRealTimeTunneling,BlumRealTimeTunneling,LawrenceRealTimeTunneling,SteingasserFiniteTemp,SteingasserRealTimeInstantons,FeldbruggeRealTimeTunneling}, we chime in with the motivation to better understand these functional methods.\\

\noindent
Approaches for computing both ground state level splittings and decay rates primarily rely on the late-time behavior of the Euclidean propagator, effectively projecting out the desired observable. While this late-time limit heavily simplifies subsequent calculations, it obstructs the extraction of analogous quantities for excited states. This problem was initially circumvented by employing functional methods to capture the poles of the resolvent operator given by the Fourier-transformed real-time partition function~\cite{FreedTunneling,LevitBarrierPenetrationTraceMethod,PatrascioiuComplexTime,WeissHaeffnerMetastableDecayFiniteTemp}. Due to this ansatz being plagued by an infinite number of relevant saddle points, a fully rigorous semiclassical evaluation employing those techniques has, to the best of our knowledge, only been performed for trivial potentials~\cite{McLaughlinComplexTimeMethod,CarlitzClassicalPaths}. An alternative approach consists of weighting Euclidean propagator contributions according to their endpoints. Utilizing a set of approximate resonant energy eigenfunctions to project out their associated eigenvalues explicitly, one can isolate the desired decay rate using finite-time amplitudes~\cite{LiangPeriodicInstantons,LiangNonVacuumBounces,LiangTunneling,LiangSineGordon,LiangFiniteTempCalculation,MullerKirstenQMBook,DeunffInstantons}. This starting point was originally put forward by Liang and Müller-Kirsten, however their computations contain discernible flaws, which will be resolved in the present work.\\

\noindent
Focusing on metastable one-dimensional systems described by an arbitrary (smooth) potential, we systematically compute the full leading-order excited state decay rate from the corresponding finite-time amplitude. While this ansatz agrees with the one proposed by Liang and Müller-Kirsten, we pursue a different way of evaluating the ensuing expression: Instead of applying the semiclassical Laplace's method sequentially to all integrals involved in the desired amplitude, we treat the entire expression as a single functional integral, encompassing both fluctuations of the trajectories as well as their endpoints, weighted by the appropriate wave functions within the false vacuum region. We find that this is the most lucid way to understand how contributions from the wave functions as well as the exponentiated action conspire to reproduce the well-known tunneling suppression factor, given by the exponential of the Euclidean single-instanton action, alongside the non-trivial leading-order prefactor.\\

\noindent
The present paper is structured as follows: We briefly introduce the conventional instanton method in section~\ref{sec:2_CommonInstantonMethod}, familiarizing the reader with any of our idiosyncratic notation along the way. Proceeding that renowned discussion, section~\ref{sec:3_ModifiedInstantonMethod} portrays the extension of the usual methodology in order to encompass the decay of excited states. Our exposition will be rather pedagogical, thoroughly explaining the relevant steps in attaining the final result~\eqref{eq:2_Full_Result_Decay_Width_InstantonCalc}. We will then briefly demonstrate how the result can be extended to partially Wick-rotated time, showing that all relevant steps generalize easily, reflecting prior results for the analytic continuation of the traditional instanton method~\cite{GarbrechtFunctionalMethods}. Lastly, subsection~\ref{sec:3_9_BreakDown} explains the intrinsic limitations of our procedure. In chapter~\ref{sec:A_SequentialComputation}, we repeat the modified instanton calculation presented in section~\ref{sec:3_ModifiedInstantonMethod}, utilizing the alternative computation scheme of first evaluating the Euclidean propagator for fixed endpoints, while only subsequently weighting the arising contributions by the pertinent wave functions. After briefly comparing both approaches in subsection~\ref{sec:Comparison_Methods}, we discuss the ground-laying work by Liang and Müller-Kirsten, who started with the same functional ansatz to study excited state energy splittings and tunneling rates~\cite{LiangPeriodicInstantons,LiangNonVacuumBounces,LiangTunneling,LiangSineGordon,LiangFiniteTempCalculation,MullerKirstenQMBook}. We finally conclude the paper in section~\ref{sec:5_Conclusion} while collecting additional supplementary material and auxiliary computations in the subsequent appendices. For the work to be self-contained, appendix~\ref{sec:B_DecayWidthsTraditionalWKB} contains a complete derivation of the full leading-order decay rate of excited states utilizing traditional WKB arguments. Appendices~\ref{sec:C_UniformWKB} to~\ref{sec:F_AspectsOfSLProblems} outsource less instructive computations, which would otherwise clutter the main body of text. Appendix~\ref{sec:F_AspectsOfSLProblems} additionally reviews certain notions of Sturm--Liouville problems that are important during our calculations. Throughout the paper, we shall retain powers of $\hbar$, constituting the principal perturbation parameter for all asymptotic expansions.\\[0.25cm]

\section{Standard instanton method for false ground state decay}
\label{sec:2_CommonInstantonMethod}
 
In the following section, we will shortly portray the traditional instanton method of obtaining the leading-order semiclassical approximation of the ground state decay rate using functional methods, first conceptualized by Langer~\cite{LangerCondensationPoint,LangerMetastability} and later popularized by Callan and Coleman~\cite{ColemanFateOfFalseVac1,CallanColemanFateOfFalseVac2}. The underlying procedure will serve as a basis for the subsequent modification encompassing excited states, portrayed in section~\ref{sec:3_ModifiedInstantonMethod}. The ensuing discussion is heavily influenced by the lucid review of Andreassen et al.~\cite{SchwartzPrecisionDecayRate}.\\
\begin{figure}[H]
	\centering
	\includegraphics[width=0.98\textwidth]{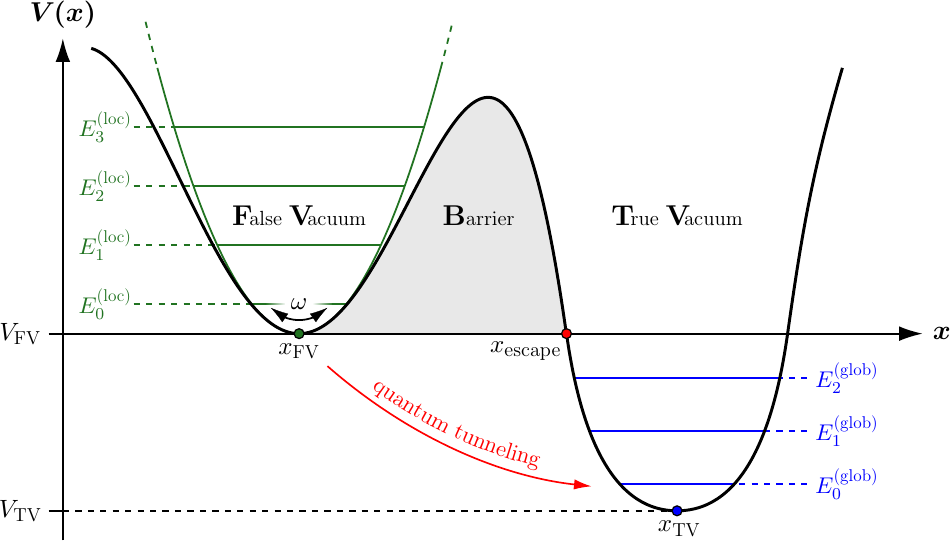}
	\caption{\hspace{0.001cm}}
	\vspace{-0.5cm}\hspace{1.3cm}
	\begin{tikzpicture}
		\node[anchor=center, inner sep=0pt, 
		minimum width=14.55cm, text width=14.55cm-0.4cm,align=justify,
		minimum height=2.5cm, 
		text depth=2.5cm-0.4cm] (0,0) {Illustration of a generic potential allowing for quantum tunneling from the metastable, false vacuum (FV) region toward the global minimum $x_\text{TV}$, dubbed the true vacuum (TV). Both regions are separated by a high barrier region (B). Additionally, the global and resonant (local) energies inside both wells are depicted, however their height has been greatly exaggerated for better visibility.};
	\end{tikzpicture}
	\label{fig:Generic_Tunneling_Potential}
\end{figure}

\noindent
We study the one-dimensional case of the probability decay of a quantum-mechanical point particle initially confined to a metastable region of a time-independent potential $V$. Figure~\ref{fig:Generic_Tunneling_Potential} depicts a typical potential under investigation, possessing a global (true) vacuum as well as an additional metastable (false) one, allowing for quantum tunneling through the shaded barrier region. The physical spectrum of the Hamiltonian is purely real, with the (global) energy eigenvalues denoted by $\Englob$---the lowest of them are illustrated qualitatively in figure~\ref{fig:Generic_Tunneling_Potential}. The FV region can separately be viewed as an open quantum system, allowing for resonances. These configurations satisfy the time-independent Schrödinger equation when nonphysical Gamow--Siegert boundary conditions are enforced~\cite{GamowAlphaDecay,SiegertRadiativeStates,IntroductionGamovVectors,ResonancesIntroduction,MizeraPolesSMatrix}. Instead of requiring the eigenfunctions to be normalizable, one seeks stationary solutions characterized by a purely outgoing flux at the escape point $x_\text{escape}$, directed toward the TV region. With these boundary conditions breaking Hermiticity, the corresponding eigenvalues $\Enloc$ to such steady-state solutions are found to be complex, possessing a small negative imaginary contribution related to the decay width of approximate eigenstates inside the FV region. Thus, the common goal of various methods for computing the decay width is to expose this relevant imaginary part attributed to the resonant energies residing inside the FV region. \\

\noindent
The starting point for the desired functional treatment is the path integral representation of the partially Wick-rotated propagator
\begin{subequations}
	\begin{align}
		\!K_\theta\scalebox{1.15}{$\big($}x_0,x_T\:\!;T\scalebox{1.15}{$\big)$}&\coloneqq K\!\!\;\scalebox{1.15}{$\big($}x_0,x_T\:\!;e^{-i\theta}T\scalebox{1.15}{$\big)$} = \mathlarger{\mathlarger{\sum}}_{n=0}^\infty \;\, \closure{\psi_n^{(\text{glob})}(x_0)} \: \psi_n^{(\text{glob})}(x_T) \,  \exp\!\left[\frac{-ie^{-i\theta} }{\hbar} \, E_n^{(\text{glob})} \:\!T\:\!\right] \label{eq:SpectralRepresentationPropagator}\\
		&= \mathlarger{\mathlarger{\int}}_{x(0)=x_0}^{x(0)=x_T} \mathcal{D}\llbracket x \rrbracket \, \exp\!\!\:\Bigg\{\frac{ie^{i\theta}}{\hbar} \mathlarger{\int}_{0}^{T} \frac{m\dot{x}(t)^2}{2} -e^{-2i\theta} V\!\!\:\big[x(t)\big] \,\text{d}t\Bigg\}\, ,
		\label{eq:PathIntegralRepresentationPropagator}
	\end{align}
	\label{eq:PropagatorRepresentations}
\end{subequations} 
where the wave functions $\psinglob$ are associated with the global energy eigenstates of the entire system to eigenvalue $\Englob$. The angular parameter $\theta\in \big(0,\pi/2\big]$ specifies by how much the time contour has been Wick-rotated, recovering the Euclidean propagator for $\theta=\pi/2$ and the Minkowski one in the limit $\theta\rightarrow 0^+$. The exponent appearing in the path integral is $i/\hbar$ times the analytically continued action $S_\theta\llbracket x\rrbracket$, where we choose to indicate functionals using double brackets.\\ 

\begin{figure}[H]
	\centering
	\includegraphics[width=0.95\textwidth]{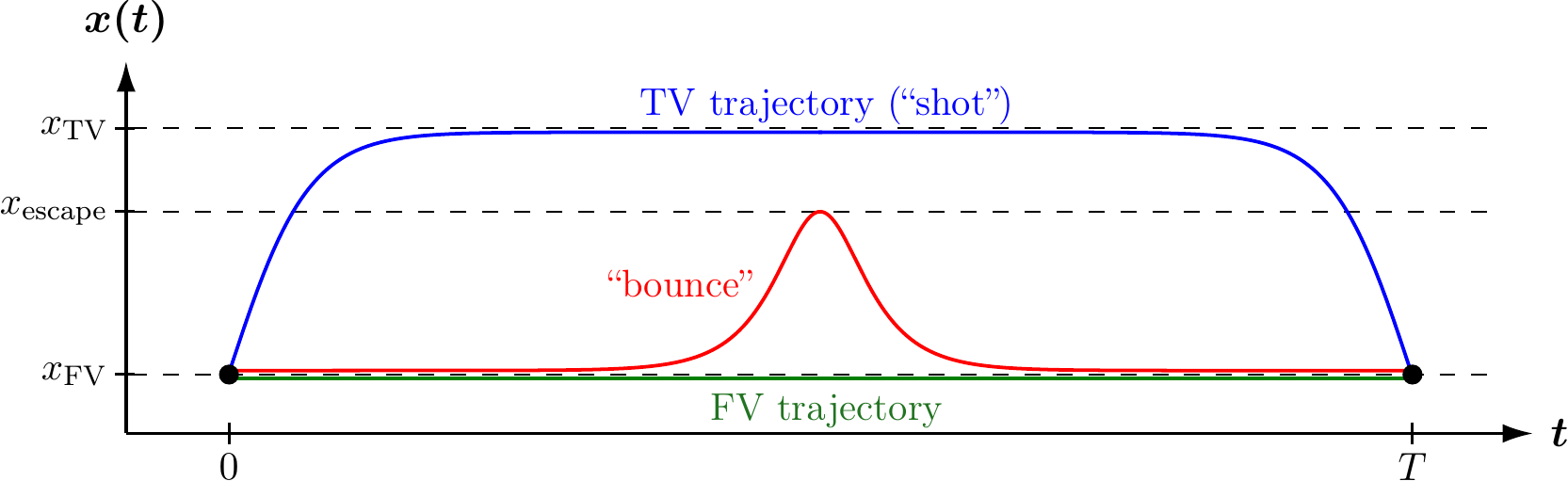}
	\caption{\hspace{0.001cm}}
	\vspace{-0.5cm}\hspace{1.3cm}
	\begin{tikzpicture}
		\node[anchor=center, inner sep=0pt, 
		minimum width=14.55cm, text width=14.55cm-0.4cm,align=justify,
		minimum height=0.5cm, 
		text depth=0.5cm-0.4cm] (0,0) {Pictorial comparison of the different classical trajectories encountered for large $T$, starting and ending at the FV~\cite{SchwartzPrecisionDecayRate}. };
	\end{tikzpicture}
	\label{fig:Comparison_Trajectories_FV_Bounce_Shot}
\end{figure}

\begin{figure}[h]
	\centering
	\includegraphics[width=0.7\textwidth]{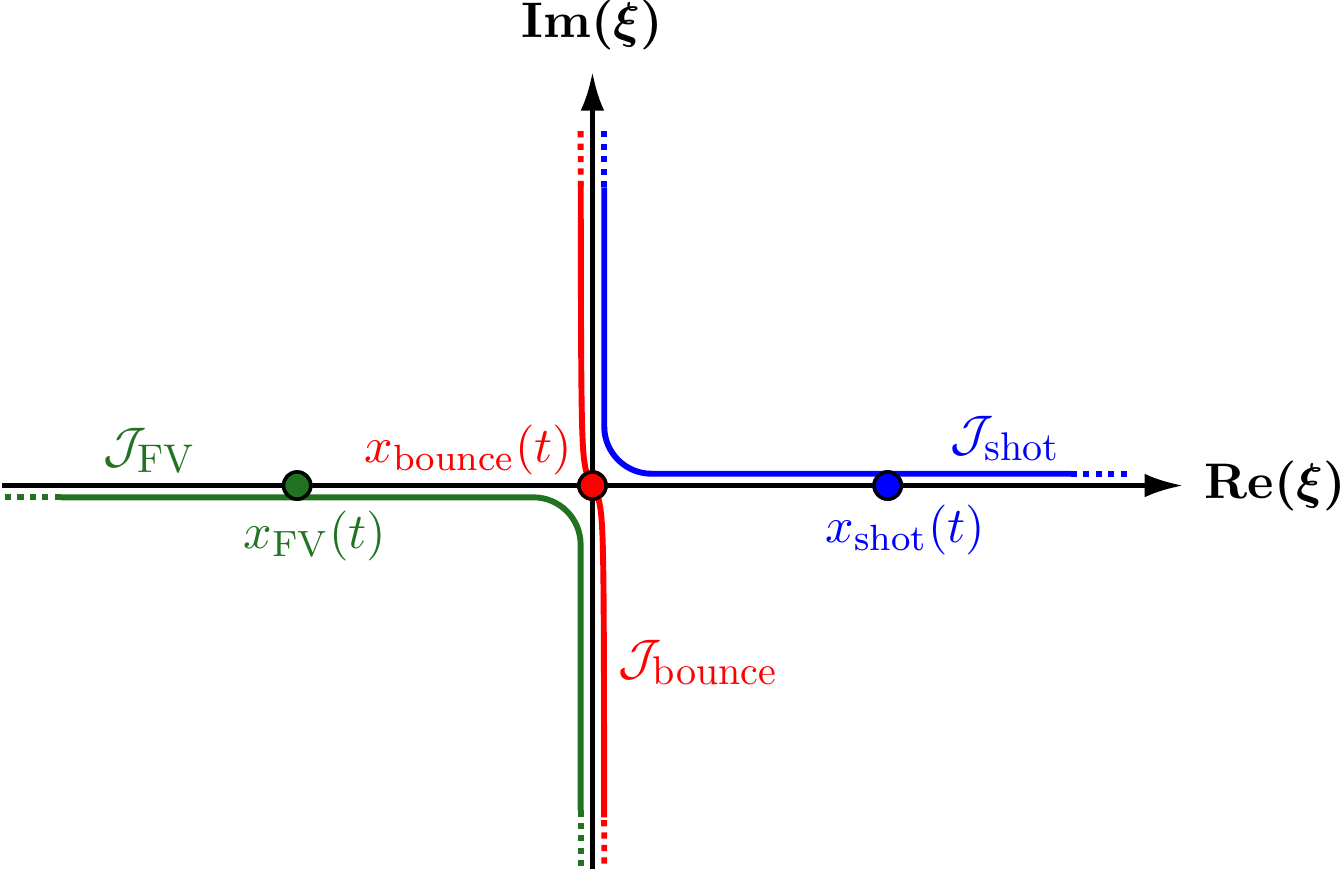}
	\caption{\hspace{0.001cm}}
	\vspace{-0.5cm}\hspace{1.5cm}
	\begin{tikzpicture}
		\node[anchor=center, inner sep=0pt, 
		minimum width=14.35cm, text width=14.35cm-0.4cm,align=justify,
		minimum height=2.5cm, 
		text depth=2.5cm-0.4cm] (0,0) {Rough sketch of the structure of the relevant steepest descent thimbles $\mathcal{J}$ encountered in the tunneling problem~\cite{SchwartzPrecisionDecayRate}. We only portray the important complexified field direction $\xi$, along which the different saddle points are connected by steepest descent/ascent thimbles due to the system being on a Stokes line. Note that the thimbles would in reality fully lay on top of one another for imaginary $\xi$, with the degeneracy only lifted for clarity.};
	\end{tikzpicture}
	\label{fig:Tunneling_Thimble_Structure}
\end{figure}

\noindent
In the formal limit $T\rightarrow \infty$, the spectral decomposition~\eqref{eq:SpectralRepresentationPropagator} can be truncated after the first term, thus projecting on the (global) ground state energy of the system, satisfying the equality \begin{equation}
	E_0^{(\text{glob})} = ie^{i\theta}\hbar \lim_{T \rightarrow \infty} \!\bigg\{T^{-1} \log\!\Big[K_\theta\scalebox{1.15}{$\big($}x_0,x_T\:\!;T\scalebox{1.15}{$\big)$}\Big] \!\bigg\} \, .
	\label{eq:4_Ground_State_Energy_ComplexifiedTime_Projection}
\end{equation}
The ad hoc arbitrary endpoints $x_0$ and $x_T$ can be chosen conveniently, enabling a more straightforward evaluation of the desired path integral. If one were purely interested in the global ground state energy, the clearest choice would amount to instating $x_0=x_T=x_\text{TV}$, for which the sole relevant saddle point of the path integral is given by the constant trajectory $x_\text{crit}(t)=x_\text{TV}$. However, due to our motivation to study the FV region, it will be more suitable to pick $x_0=x_T=x_\text{FV}$ instead. With this choice, one generally finds three relevant critical trajectories satisfying the classical Euler--Lagrange equation while pertaining to the enforced Dirichlet boundary conditions. Although the subsequent discussion can be generalized to arbitrary $\theta$ as shown by Ai et al.~\cite{GarbrechtFunctionalMethods}, let us focus on the simpler Euclidean case $\theta=\pi/2$. In that instance, the three classical trajectories in the (now inverted) potential, schematically depicted in figure~\ref{fig:Comparison_Trajectories_FV_Bounce_Shot}, are found to be:
\begin{enumerate}[label=\roman*)]
	\item \myuline{TV trajectory (``shot''):} \\
	\noindent Starting with a non-vanishing velocity, this trajectory rushes toward the TV peak, where it spends the majority of its time before slowing to a halt and reversing its motion. In the limit of large $T$, this trajectory is practically indistinguishable from the constant trajectory $x(t)=x_\text{TV}$, thus ensuring the exactness of relation~\eqref{eq:4_Ground_State_Energy_ComplexifiedTime_Projection}.
	\item \myuline{FV trajectory:}\\
	As we mandated both endpoints to be the FV $x_\text{FV}$, another admissible classical motion in the inverted potential is the trivial, constant trajectory $x_\text{crit}(t)=x_\text{FV}$. Similarly to the trivial trajectory sitting at the TV $x_\text{TV}$, the arising contribution would coincide with the ground state energy of the FV region if it were stabilized. Thus, this contribution is of considerable interest regarding the question of the resonant ground state energy $\Eloc$.
	\item \myuline{``Bounce'' trajectory:}\\
	Reaching a turning point at the halfway mark for large times $T$ can either be achieved if the motion considerably slows down before arriving at the turning point, as seen with the shot, or instead by initiating the motion with a minuscule velocity. While spending almost all time setting off the motion, staying close to the FV, the corresponding trajectory only takes a brief detour toward $x_\text{escape}$ before heading back. Due to its temporary nature, departing from a trivial behavior only for a short duration of time, such configurations are dubbed instantons. The important realization is that this trajectory is not a minimum of the Euclidean action but rather a saddle point---there exists a single direction in function space in which its Euclidean action decreases. This renders the contribution from the bounce purely imaginary, which again should spark interest as we expect $\Eloc$ to possess a non-vanishing imaginary part.
\end{enumerate}
When taking equation~\eqref{eq:4_Ground_State_Energy_ComplexifiedTime_Projection} at face value, only the minima of the Euclidean action (FV \& TV trajectory) would contribute. Meanwhile, all contributions arising from the bounce trajectory, constituting a genuine saddle point in the underlying real function space, are fully offset due to the overlapping structure of the corresponding steepest descent trajectories encountered in the complexified function space, as schematically illustrated in figure~\ref{fig:Tunneling_Thimble_Structure}.\footnote{This guarantees that the global ground state energy is a purely real quantity, which is bestowed upon us by the Hermiticity of the Hamiltonian when instating vanishing boundary conditions at spatial infinity.} However, when heuristically ignoring the effects of the shot trajectory, being attributed to the TV region, one exactly recovers the desired resonant energy $E\raisebox{-2.5pt}{\scalebox{0.65}{$0$}}\!\!\!\,\raisebox{4.75pt}{\scalebox{0.65}{$(\text{loc})$}}$. Technically, this aforementioned step is rather subtle and amounts to dropping the steepest descent thimble $\mathcal{J}_\text{shot}$ associated with the shot solution, introducing an important additional factor of $\frac{1}{2}$, arising from the enduring overlap of the relevant thimbles portrayed in figure~\ref{fig:Tunneling_Thimble_Structure}. Integrating over the remaining steepest descent contours provides a quantity possessing precisely the sought-after imaginary part corresponding to the ground state decay rate, stemming from the leftover non-overlapping part of the bounce thimble.\\

\noindent 
When computing the bounce contributions, one encounters another caveat in the form of a quasi-zero mode. Time-translation symmetry, weakly broken by the Dirichlet boundary conditions of the path integral and fully restored in the limit $T\rightarrow \infty$, yields a flat direction in function space that must be treated separately since a Gaussian integration proves insufficient. Introducing a collective coordinate~\cite{SakitaGervaisZeroModeFadeevPopov,BalitskyCollectiveCoordinates}, one obtains a linear factor $T$ arising from the volume of the underlying moduli space, which for sufficiently large times overcomes the exponential suppression factor of the bounce contribution. Consequently, one is required to also consider multi-bounce contributions, consisting of largely spaced single-bounce motions. These trajectories constitute parametrically flat directions on the scrutinized thimbles, identified with ``critical points at infinity''~\cite{BehtashSaddlesAtInfinity}, turning into exact saddle points only in the limit of infinite separation. Bypassing this ominous limit $T\rightarrow \infty$ is yet another reason justifying the interest in the modified method presented in section~\ref{sec:3_ModifiedInstantonMethod}. Incorporating the contributions arising from the previously mentioned multi-instanton solutions, the resonant ground state energy can be expressed as 
\begin{align}
	E_0^{(\text{loc})}= -\hbar \lim_{T \rightarrow \infty} \!\Bigg\{T^{-1} \log\!\bigg[K_{\text{E}}^{(\text{FV})}\scalebox{1.15}{$\big($}x_\text{FV},x_\text{FV}\:\!;T\scalebox{1.15}{$\big)$}+\mathlarger{\sum}_{n=1}^\infty\,\frac{1}{2^n} K_{\text{E}}^{(\text{$n$-bounce})}\scalebox{1.15}{$\big($}x_\text{FV},x_\text{FV}\:\!;T\scalebox{1.15}{$\big)$}\bigg] \!\Bigg\} \, .
	\label{eq:Resonant_Ground_State_Energy_Sum}
\end{align}
In the limit of large $T$, all multi-bounce contributions can be related to the single-bounce result, as for largely separated bounce motions, the action is additive, while the fluctuation determinant factorizes. Thus, the sum encountered in equation~\eqref{eq:Resonant_Ground_State_Energy_Sum} exponentiates, as was initially observed by Callan and Coleman~\cite{CallanColemanFateOfFalseVac2}, trivially neutralizing the overall logarithm. A semiclassical expansion of the path integrals in question then yields, up to subleading corrections in orders of $\hbar$, the ground state decay width $\Gamma_0$ as 
\begin{equation}
	\Gamma_0 = \underbrace{\sqrt{\frac{S_\text{E}\scalebox{1.3}{$\big\llbracket$}x_\text{bounce}^{(T=\infty)}\scalebox{1.3}{$\big\rrbracket$}}{2\pi\hbar}} \, \left\lvert \frac{\displaystyle{\text{det}_\zeta'\!\left\{-\frac{\text{d}^2}{\text{ d}t^2}+\!\!\:V''\!\!\:\scalebox{1.2}{$\big[$}x_\text{bounce}^{(T=\infty)}(t)\scalebox{1.2}{$\big]$}\!\right\}}}{\displaystyle{\text{det}_\zeta\!\left\{-\frac{\text{d}^2}{\text{d}t^2}+\!\!\:V''\!\!\:\scalebox{1.05}{$\big[$}x_\text{FV}(t)\scalebox{1.05}{$\big]$}\!\right\}}}\right\rvert^{\scalebox{0.95}{$-\frac{1}{2}$}}}_{\displaystyle{\begin{matrix}\text{``fluctuation factor''}\\ \text{(zero mode contribution $\times$ determinant ratio)}\end{matrix}}} \!\!\, \underbrace{\exp\!\!\:\bigg(\!\!\!\:-\!\!\:\frac{1}{\hbar}\, S_\text{E}\scalebox{1.3}{$\big\llbracket$}x_\text{bounce}^{(T=\infty)}\scalebox{1.3}{$\big\rrbracket$}\!\bigg)}_{\displaystyle\text{leading exponential}} \, ,
	\label{eq:GroundStateDecayWidth}
\end{equation} 
where the prime on the (subtracted) determinant indicates the omission of the zero mode and the subscript $\zeta$ discloses regularization, see appendix~\ref{sec:F1_Sturm_Liouville_Determinants}. We will abstain from covering the intricacies in attaining the above formula~\eqref{eq:GroundStateDecayWidth}, which can be found in the subject-relevant literature, see e.g.~\cite{CallanColemanFateOfFalseVac2,SchulmanPathIntegrals,SchwartzPrecisionDecayRate,DevotoTunnelingReview}. While withholding clarifying details in the present section, the upcoming parts portraying the modified instanton method will be much more explicit. Despite most steps in section~\ref{sec:3_2_GeneralEvaluationTactics} running parallel to the usual procedure, we have chosen not to waive necessary steps for the ensuing discussion to be fully self-contained.\\[-0.25cm]

\section{Modified instanton method encompassing excited state decay}
\label{sec:3_ModifiedInstantonMethod}

The traditional method, as portrayed in the previous section, entails certain limitations arising from the rather problematic limit $T\rightarrow \infty$. Despite simplifying the computation heavily, the limit restricts the procedure's applicability to considerations solely about the ground state while additionally requiring the incorporation of multi-bounce trajectories. This naturally leads us to the subsequently presented modifications to the standard computation scheme, encompassing the decay of excited states while also working with finite time intervals. We will first give a short motivation of the pursued ansatz before delving into the technical but yet instructive computation.\\ 

\noindent
As in the traditional calculation, the starting point should be the spectral representation~\eqref{eq:SpectralRepresentationPropagator} of the Wick-rotated propagator $K_\theta$. Projecting out the energy eigenvalues $\Englob$ by weighting the propagator contributions using the global eigenfunctions $\psinglob$, one obtains the exact relation
\begin{align}
	\begin{split}
		E_n^{(\text{glob})} &= \frac{i e^{i\theta}\hbar}{T}  \;\log\!\!\:\Bigg\{\mathlarger{\int}_{-\infty}^{\infty} \text{d}x_0 \mathlarger{\int}_{-\infty}^{\infty} \text{d}x_T\; \closure{\psi_n^{(\text{glob})}(x_T)} \:\psi_n^{(\text{glob})}(x_0)\, K_\theta\mathlarger{\big(}x_0,x_T\:\!;T\mathlarger{\big)}\Bigg\} \, .
	\end{split}
	\label{eq:Global_Energy_Exact_Projection}
\end{align} 
Notice that this identity meets the desired demands of extracting information about excited energies while bypassing the late-time limit, as equation~\eqref{eq:Global_Energy_Exact_Projection} holds for arbitrary lengths of the time interval $T$. The question arises of how one should adequately alter this relation to yield the sought-after resonant energy $\Enloc$ instead of the system's eigenenergy $\Englob$. It turns out that the most straightforward idea of replacing the global eigenfunctions $\psinglob$ with an appropriate local counterpart $\psinloc$ suffices. Even more so, we subsequently find that with the additional weight factors arising from the local wave functions, practically supported solely inside the FV region, shot-like paths, leaving the FV region for an elongated period of time, constitute no stationary trajectories anymore. Thus, the former requirement to manually drop such contributions is absent. To clarify why this replacement suffices, recall that a resonant state $\raisebox{0.5pt}{$\displaystyle{\smash{\ket[\big]{\raisebox{-0.5pt}{$\displaystyle{n\raisebox{4.25pt}{\scalebox{0.65}{$(\text{loc})$}}}$}}}}$}$ constitutes an eigenstate of the Hamiltonian $H$, albeit to radiative boundary conditions~\cite{IntroductionGamovVectors,ResonancesIntroduction}. Consequently, in case the state is normalized appropriately,\footnote{Let us ignore the subtle fact that resonant states are generally not normalizable on the real line due to approaching a plane wave solution past the escape point $x_{\scalebox{0.7}{\text{escape}}}$. Working locally around the FV, it will prove suitable to instead normalize these solutions on the constrained interval $\scalebox{1}{$\big(\!-\!\!\:\infty,x_{\scalebox{0.7}{\text{escape}}}\big]$}$.} it obeys the simple relation
\begin{align}
	\exp\!\left[\frac{-ie^{-i\theta} }{\hbar} \, E_n^{(\text{loc})} \:\!T\:\!\right] &= \braket*{n^{(\text{loc})}|\, \exp\!\left(\frac{-ie^{-i\theta} }{\hbar} \, H T\:\!\right)\:\!\!|\,n^{(\text{loc})}} \label{eq:Ansatz_MuellerKirsten_Liang} \\[0.1cm]
	& =\:\!\!\mathlarger{\mathlarger{\int}}_{-\infty}^\infty \:\!\!\text{d}x_0 \mathlarger{\mathlarger{\int}}_{-\infty}^\infty \:\!\!\text{d}x_T \, \braket*{n^{(\text{loc})}\:\!\!|x_T\:\!\!}\braket*{x_T|\, \exp\!\left(\frac{-ie^{-i\theta} }{\hbar} \, H T\:\!\right)\:\!\!|\;\! x_0\:\!\!}\braket*{x_0\:\!\!|\:\!n^{(\text{loc})}} \:\! ,\nonumber
\end{align}
demonstrating that the aforementioned replacement indeed has exactly the desired effect.\\ 

\noindent 
Let us again stress that this idea is not novel, as it had already been proposed by Liang and Müller-Kirsten, who successfully utilized the present ansatz~\eqref{eq:Ansatz_MuellerKirsten_Liang} to obtain expressions for both excited state level splittings~\cite{LiangPeriodicInstantons,LiangSineGordon} as well as decay widths~\cite{LiangNonVacuumBounces,LiangFiniteTempCalculation}. Even though their computations concluded with correct results for the special potentials under consideration, the employed methods are questionable, possessing severe shortcomings, which we elaborate on in section~\ref{sec:4_FormerAttempts}. The present study corrects and refines the given approach, systematically obtaining the leading-order WKB result for arbitrary time-independent potentials. \\

\noindent
Given expression~\eqref{eq:Global_Energy_Exact_Projection} after the replacement $\psinglob \mapsto \psinloc$, two sensible possibilities for a semiclassical evaluation present themselves:
\begin{enumerate}
	\item The most straightforward treatment is to sequentially compute the involved integrals, starting with the path integral expression for the propagator. After attaining the leading-order contribution to the propagator for freely varying $x_0$ and $x_T$, the remaining ``endpoint integrations'' are to be performed semiclassically. 
	\item The more lucid alternative is to merge the endpoint integrations with the ``inner'' path integral, thereby dropping the present Dirichlet boundary conditions. The arising composite functional integral with its open boundary conditions can be evaluated as usual, requiring only little technical generalizations. It will provide the advantage of time translation symmetry being manifest, greatly simplifying the treatment of the emerging zero mode.
\end{enumerate} 
While Müller-Kirsten and Liang opt for the first approach, we will initially adhere to the second method. For completeness, we will nonetheless include a sketch of the perhaps less instructive sequential calculation in section~\ref{sec:A_SequentialComputation}, also providing a short comparison between the two schemes, see subsection~\ref{sec:Comparison_Methods}. With the present discussion predominantly revolving around the composite path integral ansatz, we are interested in a semiclassical evaluation of the expression\footnote{We elaborate on the derivation of this formula from first principles in a subsequent work~\cite{WagnerComplexPI}. However, two key differences emerge compared to the naïve expression~\eqref{eq:Resonant_Energy_Conjecture} provided above: first, the functional integral should be taken over a complexified contour $\mathcal{C}\big([0,T], \Gamma\big)$ rather than the real function space $\mathcal{C}\big([0,T]\big)$; second, the complex conjugation of the wave function evaluated at $x_T$ is formally absent. Despite these differences, the calculation presented in this document remains unaffected, as briefly explained around equation (4.11) contained in~\cite{WagnerComplexPI}.}
\begin{align}
	E_n^{(\text{loc})} &= \frac{i e^{i\theta}\hbar}{T} \;\log\!\!\:\scalebox{1.3}{\Bigg\{}\!\!\:\underbrace{\mathlarger{\mathlarger{\int}}_{\vphantom{\xi}\smash{\scalebox{0.75}{$\mathcal{C}\big([0,T]\big)$}}} \mathcal{D}\llbracket x\rrbracket \;\, \closure{\psi_n^{(\text{loc})}\big[x(T)\big]} \:\psi_n^{(\text{loc})}\big[x(0)\big]\, \exp\!\!\:\bigg(\frac{iS_\theta \llbracket x\rrbracket}{\hbar}\bigg)\!}_{\displaystyle{\rule{0pt}{0.3cm}\text{composite path integral } \mathcal{P}_{\!\!\: n}^{(\theta)}\big(\hbar,T\big)}}\!\scalebox{1.3}{\Bigg\}} \, .
	\label{eq:Resonant_Energy_Conjecture}
\end{align}
As announced, we have absorbed the endpoint integrations into the path integral representation of $K_\theta$, extending the range of integration of the resulting functional integral to the unconstrained function space $\mathcal{C}\big([0,T]\big)$, denoting the set of all real-valued, continuous functions on the given time interval $[0,T]$.

\subsection{Preliminary remarks and auxiliary definitions}
\label{sec:3_1_AuxiliaryDefinitions}
 
In order to lighten the notation of all upcoming expressions, we shift the potential $V$ such that the FV point $x_\mathrm{FV}$ is located at the origin, where the potentials value is set to zero. As shown in figure~\ref{fig:Generic_Tunneling_Potential}, $\omega$ should denote the curvature of $V$ at the metastable minimum $x_\text{FV}=0$. Additionally, let us introduce the auxiliary potential $W$, implicitly defined through
\begin{align}
	V\!\!\:(x)=\frac{m\omega^2 x^2}{2} \Big[1+W\!\!\:(x)\Big] \, ,
	\label{eq:Auxiliary_Potential}
\end{align}
parametrizing the deviation from a purely harmonic potential in the FV region.\footnote{The given representation of $V$ allows us to more easily write down holomorpic expressions around the FV point. With subsequent formulas entailing square roots, one should be careful when dealing with $\sqrt{V(x)}$ due to additional, unnecessary branch cuts on the imaginary axis. We rather accept the light notational overhead by writing $x\sqrt{x^{-2}V(x)}$ in favor of the resulting expressions possessing a valid analytical continuation. While not mandatory for the inherently real Euclidean case, the generalization to arbitrary $\theta$ favors this subtle distinction.} For our purposes, taking $V$ as an analytic function, we infer $W\!\!\;(x)=\mathcal{O}(x)$. However, this requirement could be weakened, influencing ensuing validity bounds on relation~\eqref{eq:Resonant_Energy_Conjecture}. Lastly, let us introduce the two important quantities $\mathdutchbcal{A}$ and $\mathdutchbcal{B}$, allowing us to conveniently represent the arising tunneling rate, heavily lightening the notation. These desired quantities are defined through the relations 
\begin{subequations}
	\begin{align}
	\mathdutchbcal{A} &\coloneqq x_\text{escape}\; \exp\!\!\:\scalebox{1.15}{\Bigg\{} \mathlarger{\mathlarger{\int}}_{\scalebox{0.7}{$0$}}^{\scalebox{0.7}{$x_\text{escape}$}} \,\frac{1}{\xi} \Big[1+W(\xi)\Big]^{-\frac{1}{2}}- \frac{1}{\xi} \;\text{d}\xi\,\scalebox{1.15}{\Bigg\}} \, , \label{eq:2_Definition_A_Parameter_Tunneling}\\[0.1cm]
	\mathdutchbcal{B} &\coloneqq 2m\omega\:\! \mathlarger{\mathlarger{\int}}_{\scalebox{0.7}{$0$}}^{\scalebox{0.7}{$x_\text{escape}$}} \, \xi \Big[1+W(\xi)\Big]^{\frac{1}{2}} \:\text{d}\xi = S_\text{E}\scalebox{1.3}{$\big\llbracket$}x_\text{bounce}^{(T=\infty)}\scalebox{1.3}{$\big\rrbracket$}\, .
	\label{eq:2_Definition_B_Parameter_Tunneling}
	\end{align}
	\label{eq:2_Definition_AB_Parameter_Tunneling}%
\end{subequations}
With different definitions of the parameter $\mathdutchbcal{A}$ applied in various references, let us portray three commonly used alternatives, enabling an easier comparison of the later results. References~\cite{GargTunnelingRevisited,GarbrechtFunctionalMethods,WeissHaeffnerMetastableDecayFiniteTemp} all define a constant $A$ related to our parametrization via
\begin{equation}
	A = \left\{ \; \begin{matrix}
		\displaystyle{\!\!\sqrt{m\omega^3\mathdutchbcal{B}^{-1}} \,\mathdutchbcal{A}} &  \hspace{-1.13cm}\text{Ai et al.~\cite{GarbrechtFunctionalMethods}} \, , \\[0.2cm] 
		\displaystyle{\!\!2\sqrt{m\omega^3}\:\!\mathdutchbcal{A}} & \quad \text{Weiss \& Haeffner~\cite{WeissHaeffnerMetastableDecayFiniteTemp}} \, , \\[0.2cm]
		\displaystyle{\;\log\!\Big(x_\text{escape}^{-1} \:\!\mathdutchbcal{A} \Big)} & \hspace{-1.55cm}\text{Garg~\cite{GargTunnelingRevisited}} \, ,
	\end{matrix}  \right.
	\label{eq:Comparison_Parameters_References}
\end{equation}
which is then employed to express the non-exponential prefactor for the tunneling rate.\\

\noindent
Before diving into the computation, some additional preliminary remarks are in order. Striving for clarity, we will present the emerging computation of equation~\eqref{eq:Resonant_Energy_Conjecture} in its simplest form by working in Euclidean signature, i.e. setting $\theta=\pi/2$. This turns the path integral into a purely real, Laplace-type integral, initially avoiding the more thorough treatment via Picard--Lefschetz theory. Only later in section~\ref{sec:3_8_Extensions}, we briefly demonstrate that the treatment in Euclidean time indeed can be generalized to arbitrary $\theta$. Moreover, we will show that in all computationally accessible parameter regions, equation~\eqref{eq:Resonant_Energy_Conjecture} holds independently of the chosen time interval length $T$.\footnote{However, a na\"{\i}ve treatment will lead to a breakdown for large $T$, which is elaborated on in section~\ref{sec:3_9_BreakDown}.} As demonstrated in section~\ref{sec:4_FormerAttempts}, this was not achieved in previous treatments based on the given ansatz. \\

\noindent 
The calculation will proceed along the following line: We initially specify what expression should be put in place of the resonant wave function $\psinloc$ before thoroughly showing how an endpoint-weighted path integral is to be evaluated in the semiclassical limit $\hbar\to 0^+$. This leads us to investigate the critical paths of the composite functional integral before finally computing the contributions arising from the infinitesimal neighborhood of each stationary trajectory. With certain aspects of the computation being rather lengthy, we choose to outsource more uninstructive bits to appendices~\ref{sec:C_UniformWKB} to~\ref{sec:F_AspectsOfSLProblems}, intending to keep the main body of text easily accessible.

\subsection{Local wave functions}
\label{sec:3_1_WaveFunction}

Na\"{\i}vely, there exist two sensible options for $\psinloc$: Either one could think of approximating the resonant wave function as a harmonic oscillator state inside the FV region or by alternatively employing a traditional plane wave WKB ansatz. While both estimates would provide a sufficient approximation in certain narrow regions of the potential, they lack the sophistication to resolve the transition region between the FV region and the barrier to the required accuracy. As we will see later, the computation for short time scales $T$ directly probes the wave function inside this transition region, demanding the use of a uniform asymptotic approximation. Notice that if one aspired to compute the decay rate solely employing the wave function picture, illustrated in appendix~\ref{sec:B_DecayWidthsTraditionalWKB}, one would need to require global information about the resonant wave function $\psinloc$, demanding an elaborate matched asymptotic expansion. Contrasting, the pursued uniform WKB approximation relies more closely on local information around the FV point. \\[-0.075cm] 

\noindent
A detailed computation of the desired approximation can be found in appendix~\ref{sec:C_UniformWKB}. Given the previously introduced representation of the potential~\eqref{eq:Auxiliary_Potential} and the usual convention of denoting Hermite polynomials by $H_n$, the leading-order result for the local wave function is found to be\\[-0.55cm]
\begin{align}
	\psi_{n,\text{LO}}^{(\text{loc})}(x) &= \frac{1}{\sqrt{2^n n!}} \left(\frac{m\omega}{\pi\hbar}\right)^{\!\frac{1}{4}}\, \Big[1+W(x)\Big]^{\!\!\:-\frac{1}{4}} \scalebox{1.1}{\Bigg\{}\frac{2}{x^2}\mathlarger{\mathlarger{\int}}_0^x \,\xi \Big[1+W(\xi)\Big]^{\!\!\:\frac{1}{2}}\, \text{d}\xi\:\!\scalebox{1.1}{\Bigg\}}^{\!\!\!\:\scalebox{0.9}{$-\frac{n}{2}$}}  \nonumber \\[0.2cm] 
	&\;\;\;\;\times \exp\!\:\!\scalebox{1.1}{\Bigg\{}\frac{2n+1}{2}\mathlarger{\mathlarger{\int}}_{0}^x\, \frac{1}{\xi} \Big[1+W(\xi)\Big]^{-\frac{1}{2}}-\frac{1}{\xi} \; \text{d}\xi\scalebox{1.1}{\Bigg\}} \nonumber \\[0.2cm]
	&\;\;\;\;\times H_n\!\left(\!\!\:\scalebox{1.1}{\Bigg\{}\frac{2m\omega}{\hbar x^2}\mathlarger{\mathlarger{\int}}_0^x \,\xi \Big[1+W(\xi)\Big]^{\!\!\:\frac{1}{2}}\, \text{d}\xi\:\!\scalebox{1.1}{\Bigg\}}^{\!\!\!\:\frac{1}{2}} x\right) \, \exp\!\:\!\scalebox{1.1}{\Bigg\{}\!\!-\!\frac{m\omega}{\hbar}\mathlarger{\mathlarger{\int}}_0^x \,\xi \Big[1+W(\xi)\Big]^{\!\!\:\frac{1}{2}}\, \text{d}\xi\:\!\scalebox{1.1}{\Bigg\}} \nonumber \\[0.2cm]
	&\eqqcolon \underbrace{\vphantom{\scalebox{1.1}{\Bigg\{}}\frac{1}{\sqrt{2^n n!}} \left(\frac{m\omega}{\pi\hbar}\right)^{\!\frac{1}{4}}\,\psi_{n,\text{non-exp}}^{(\text{loc})}(x)\vphantom{\scalebox{1.1}{\Bigg\}}}}_{\displaystyle{\rule{0pt}{0.3cm}\text{polynomial $\hbar$-dependence}}} \;  \underbrace{\exp\!\:\!\scalebox{1.1}{\Bigg\{}\!\!-\!\frac{m\omega}{\hbar}\mathlarger{\mathlarger{\int}}_0^x \,\xi \Big[1+W(\xi)\Big]^{\!\!\:\frac{1}{2}}\, \text{d}\xi\:\!\scalebox{1.1}{\Bigg\}}}_{\displaystyle{\rule{0pt}{0.3cm}\text{singular $\hbar$-dependence}}}
	 \, . \label{eq:Resonant_Wave_Function}
\end{align}
In the last line we collected all terms possessing a purely polynomial $\hbar$-dependence in the newly defined function $\psilocnonexp$. The important singular term in the semiclassical limit $\hbar\rightarrow 0^+$ is found to be the usual exponential WKB suppression factor,\footnote{Note that in the Euclidean case, with all relevant paths being inherently real and $x>0$, we can utilize a simplified version of equation~\eqref{eq:Resonant_Wave_Function}, pulling factors of $\xi$ into the occurring square roots. This then manifests the exponential suppression factor under the integral to be given by the more commonly recognizable expression $\sqrt{2mV(\xi)}$.} which will have a non-trivial impact on the critical trajectories of the composite path integral~\eqref{eq:Resonant_Energy_Conjecture}. The $n$-independent exponent in relation~\eqref{eq:Resonant_Wave_Function} will subsequently be defined as $\psilocexp$, see equation~\eqref{eq:FullExponent}. \\[-0.075cm] 

\noindent
Two more technical details about the function $\psinloc$ are in order:
\begin{enumerate}
	\item The derivation of expression~\eqref{eq:Resonant_Wave_Function} utilizes that for $\hbar\ll 1$, the resonant energies deviate only slightly from a purely harmonic behavior, obeying
	\begin{equation}
		E_n^{\scalebox{0.8}{$(\text{loc})$}}=\hbar \omega\bigg(\!\!\:n+\frac{1}{2}\bigg) \Big[1+\mathcal{O}\scalebox{1.1}{\big(}\!\!\:\sqrt{\hbar}\,\scalebox{1.1}{\big)}\Big]\, ,
		\label{eq:2_Local_Energy_Ansatz_Definition_Epsn}
	\end{equation}
	 with both real and imaginary corrections being negligible to the desired order.\footnote{A crucial feature of the ensuing discussion is that a primitive estimate of the resonant energy eigenvalue utilized to construct the local wave function suffices to obtain the full leading-order decay rate. It turns out that recognizing the leading-order real part of the resonant energy $\Enloc$ is enough to reconstruct its associated imaginary part.} This requires $\Enloc=\mathcal{O}(\hbar)$ to be insignificant compared to the barrier height, allowing us to expand around the quadratic turning point $x_\text{FV}$. For larger energies approaching the top of the barrier, the given approximations~\eqref{eq:Resonant_Wave_Function} and~\eqref{eq:2_Local_Energy_Ansatz_Definition_Epsn} break down. Instead, one would have to expand around a linear turning point, yielding expressions involving Airy function.
	\item We constructed the wave function to be holomorphic,\footnote{More precisely, the function solely possesses a branch cut on the real axis beyond $x_{\scalebox{0.7}{\text{escape}}}$, a region not probed by the endpoints of all relevant critical trajectories at play. Crucially, this statement still holds true for the case of arbitrary Wick-rotation angles $\theta$.} so that the method of steepest descent can be applied to the desired path integral~\eqref{eq:Resonant_Energy_Conjecture}. Note that this argument requires the absence of complex conjugates, which is justified easily due to $\psinlocLO$ being purely real-valued on the real axis, thus the complex conjugation fully drops out even before the requirement of analytical continuation enters.
\end{enumerate}

\subsection{Evaluating an endpoint-weighted path integral}
\label{sec:3_2_GeneralEvaluationTactics}

Let us illustrate in detail how to derive an asymptotic expansion for the endpoint-weighted composite path integral~\eqref{eq:Resonant_Energy_Conjecture}, represented in the suggestive form
\begin{align}
	\label{eq:Pn:exp:nonexp}
	\mathcal{P}_{\!\!\: n}^{(\theta)}\big(\hbar,T\big)=\mathlarger{\mathlarger{\int}}_{\vphantom{\xi}\smash{\scalebox{0.75}{$\mathcal{C}\big([0,T]\big)$}}} \mathcal{D}\llbracket x\rrbracket \; f_\text{non-exp}\llbracket x\rrbracket  \, \exp\!\bigg(\!\!\!\:\!\!\:-\!\!\:\frac{f_\text{exp}\llbracket x\rrbracket}{\hbar}\bigg) \, .
\end{align}
Splitting the full functional integrand into its non-exponential piece $f_\text{non-exp}\llbracket x\rrbracket$ as well as the exponential term containing the relevant singular $\hbar$-dependence will be paramount when investigating the desired semiclassical limit $\hbar\rightarrow 0^+$. Initially focusing on the especially illuminating Euclidean case $\theta=\pi/2$, we have
\begin{align}
	f_{\text{exp}}\llbracket x\rrbracket&=\underbrace{\mathlarger{\mathlarger{\int}}_0^T \frac{m \dot{x}(t)^2}{2}+ V\!\!\:\mathlarger{\big[}x(t)\mathlarger{\big]}\,\text{d}t}_{\displaystyle{\vphantom{\psi_\text{exp}^{(\text{loc})}  \big[x(T)\big]}\text{Euclidean action $S_\text{E}\llbracket x\rrbracket$}}} \,+ \, \overbrace{\underbrace{m\omega \mathlarger{\mathlarger{\int}}_0^{x(0)} \xi \Big[1+W(\xi)\Big]^{\!\!\:\frac{1}{2}}\, \text{d}\xi}_{\displaystyle{\psi_\text{exp}^{(\text{loc})} \!\!\:\mathlarger{\big[}x(0)\mathlarger{\big]}}} \, +  \underbrace{\vphantom{\mathlarger{\mathlarger{\int}}_0^T}\Big[x(0) \leftrightarrow x(T)\Big]}_{\displaystyle{\psi_\text{exp}^{(\text{loc})}  \!\!\:\mathlarger{\big[}x(T)\mathlarger{\big]}}}}^{\displaystyle{\text{supplementary boundary terms}}} \, ,
	\label{eq:FullExponent}
\end{align}
where from now on we abbreviate the $n$-independent exponential piece of the resonant wave function~\eqref{eq:Resonant_Wave_Function} by $\psilocexp$.\footnote{The subscript exp on $\psi$ indicates that it is the exponent appearing in the exponential factor contributing to the wave function, not the exponential itself, see equation~\eqref{eq:Resonant_Wave_Function}.} \\ 

\noindent
As usual, due to the strong exponential suppression, dominant contributions to such Laplace-type integrals arise from the (infinitesimal) neighborhood around extremal paths $x_\text{crit}(t)$ of the functional $f_{\text{exp}}\llbracket x\rrbracket$, around which the path integral is subsequently expanded. With the integral possessing open boundary conditions, the variations $\Delta x(t)$ around any critical trajectory are fully unconstrained, requiring one to take great care of non-vanishing boundary contributions arising from integration by parts. Collecting all arising terms suitably, such an expansion yields the expression
\begin{align}
	\!\! f_\text{exp}\Big\llbracket \:\!x_\text{crit}+\Delta x\:\!\Big\rrbracket = f_\text{exp}\big\llbracket x_\text{crit}\big\rrbracket  &- \mathlarger{\mathlarger{\int}}_{0}^{T} \bigg\{\!\!\!\underbrace{m\ddot{x}_\text{crit}(t) - V'\!\!\:\mathlarger{\big[}x_\text{crit}(t)\mathlarger{\big]}}_{\displaystyle{\text{Euler--Lagrange equation}}} \!\!\! \!\,\bigg\} \,\Delta x(t) \,\text{d}t  \label{eq:Expanded_Exponent} \\[0.05cm]
	&+ \bigg\{\!\!\!\!\underbrace{m \dot{x}_\text{crit}(T) +\psi_\text{exp}^{(\text{loc})\prime}\!\!\:\mathlarger{\big[}x_\text{crit}(T)\mathlarger{\big]}}_{\displaystyle{\text{(right) transversality condition}}}\!\!\!\!\!\!\;\bigg\} \, \Delta x(T) \nonumber \\[0.1cm]
	&- \bigg\{\!\!\!\:\underbrace{m \dot{x}_\text{crit}(0) \!\;\!\;-\!\;\!\;\psi_\text{exp}^{(\text{loc})\prime}\!\!\:\mathlarger{\big[}x_\text{crit}(0)\mathlarger{\big]}}_{\displaystyle{\text{(left) transversality condition}}}\!\!\!\: \!\!\:\bigg\} \, \Delta x(0) \nonumber \\[0.1cm]
	&+ \frac{m\omega^2}{2} \mathlarger{\mathlarger{\int}}_{0}^{T} \Delta x(t)\:\!\Bigg\{\!\!\!\!\!\:\underbrace{\!-\, \frac{\text{d}^2}{\text{d}(\omega t)^2} + \frac{V'\!\!\;\big[x_\text{crit}(t)\big]}{m\omega^2} \!\!\:}_{\displaystyle{\text{fluctuation operator $O_\text{crit}$}}}\!\!\!\!\!\:\Bigg\} \,\Delta x(t) \,\text{d}t \nonumber \\[0.1cm]
	&+ \frac{1}{2}\,\bigg\{\!\!\!\!\:\underbrace{m\Delta \dot{x}(T)+\psi_\text{exp}^{(\text{loc})\prime\prime}\!\!\:\mathlarger{\big[}x_\text{crit}(T)\mathlarger{\big]} \Delta x(T)}_{\displaystyle{\text{(right) restriction on eigenfunctions}}}\!\!\!\!\:\bigg\} \, \Delta x(T) \nonumber \\[0.1cm]
	&- \frac{1}{2}\,\bigg\{\!\!\!\!\:\underbrace{m\Delta \dot{x}(0)-\psi_\text{exp}^{(\text{loc})\prime\prime}\!\!\:\mathlarger{\big[}x_\text{crit}(0)\mathlarger{\big]} \Delta x(0)}_{\displaystyle{\text{(left) restriction on eigenfunctions}}}\!\!\!\!\:\bigg\} \, \Delta x(0) + \mathcal{O}\big(\Delta x^3\big) \, . \nonumber 
\end{align}
Both for the potential and the exponential piece of the wave function, strokes denote a derivative with respect to the functions argument, while an overhead dot particularly emphasizes a temporal derivative. Demanding the first variation around any critical path $x_\text{crit}(t)$ to vanish for arbitrary fluctuations $\Delta x(t)$, we not only have to enforce the usual Euler--Lagrange equation for all $t\in (0,T)$ but additionally encounter supplementary constraints at both boundaries of the time interval, commonly referred to as transversality conditions~\cite{CalcOfVariations,OptimizationProblems}. We will study these constraints and the ensuing critical trajectories in detail in section~\ref{sec:CriticalPaths}. With all linear terms in $\Delta x(t)$ vanishing, the spotlight shifts to the quadratic term, which we need to diagonalize in order to proceed with a Gaussian integration. This is done by choosing an appropriate function basis $\big\{e_\mu(t)\!\!\:\big\}_{\scalebox{0.7}{$\mu\!\in\!\mathbb{N}^0$}}$, in which the fluctuations $\Delta x(t)$ are to be expanded as 
\begin{align}
	\Delta x(t)= \sqrt{\frac{\hbar}{m\omega}}\, \sum_{\mu=0}^\infty \, x_\mu e_\mu(t) \, ,
	\label{eq:ExpansionVariations}
\end{align}
with $x_\mu$ taking the role of real coordinates in the underlying function space $\mathcal{C}\big([0,T]\big)$. The additional prefactor ensures both the functions $e_\mu(t)$ as well as the coordinates $x_\mu$ to be dimensionless. Inspecting the expanded exponent~\eqref{eq:Expanded_Exponent}, it is most convenient to choose the set $\big\{e_\mu(t)\!\!\:\big\}_{\!\scalebox{0.7}{$\mu\!\in\!\mathbb{N}^0$}}$ as the eigenfunction basis associated to the regular Sturm--Liouville eigenvalue problem 
\begin{align}
	O_\text{crit} \,  e_\mu(t)=\scalebox{1.15}{$\bigg\{$}\!\!\!\:-\!\!\:\frac{\text{d}^2}{\text{d}(\omega t)^2} + \frac{V''\big[x_\text{crit}(t)\big]}{m\omega^2}\scalebox{1.15}{$\bigg\}$} \, e_\mu(t)=\lambda_\mu e_\mu(t)
	\label{eq:Fluctuation_Operator}
\end{align}
pertaining to Robin boundary conditions 
\begin{subequations}
	\begin{align}
		m\:\!\dot{e}_\mu(0)\,-\psi_\text{exp}^{(\text{loc})\prime\prime}\!\!\:\mathlarger{\big[}\;\!x_\text{crit}(0)\;\! \mathlarger{\big]}  \,e_\mu(0)\:&=0 \, , \label{eq:Robin_BoundaryConditions_A} \\[0.1cm]
		m\:\!\dot{e}_\mu(T)+\psi_\text{exp}^{(\text{loc})\prime\prime}\!\!\:\mathlarger{\big[}x_\text{crit}(T)\mathlarger{\big]} \,e_\mu(T)&=0 \, .
		\label{eq:Robin_BoundaryConditions_B}
	\end{align}
	\label{eq:Robin_BoundaryConditions}
\end{subequations}
Both relations~\eqref{eq:Robin_BoundaryConditions_A} and~\eqref{eq:Robin_BoundaryConditions_B} crucially ensure the leftover boundary contributions in equation~\eqref{eq:Expanded_Exponent} to drop out, allowing for the quadratic term to be purely diagonal in the coordinates $x_\mu$. It is generally known that the set of eigenfunctions constitute an orthogonal basis of $L^2\big([0,T]\big)$, normalized according to our needs as 
\begin{align}
	\braket[\big]{e_\mu,e_\nu} \coloneqq \int_{0}^{T} e_\mu(t) \:\! e_\nu(t) \,\text{d}(\omega t) = \delta_{\mu\nu} \, .
\end{align}
Again the supplementary factor of $\omega$ ensures the eigenfunctions to be dimensionless. Inserting the aforementioned decomposition~\eqref{eq:ExpansionVariations} together with the particular choice for the function basis $\big\{e_\mu(t)\!\!\;\big\}_{\!\!\:\scalebox{0.7}{$\mu\!\in\!\mathbb{N}^0$}}$ into equation~\eqref{eq:Expanded_Exponent}, the exponent is indeed cast into the desired quadratic form
\begin{align}
	f_\text{exp}\Big\llbracket \:\!x_\text{crit}+\Delta x\:\!\Big\rrbracket &=f_\text{exp}\big\llbracket x_\text{crit}\big\rrbracket + \frac{\hbar}{2} \sum_{\mu=0}^\infty \lambda_\mu x_\mu^2 + \mathcal{O}\scalebox{1.4}{(}\hbar^{\frac{3}{2}}\scalebox{1.4}{)} \, .
	\label{eq:Expanded_Exponent_Simplified}
\end{align}
With the (Euclidean) path integral measure expressed as
\begin{align}
	\mathcal{D}\big\llbracket \Delta x\big\rrbracket \coloneqq  \mathcal{N}_\text{E} \,\sqrt{\frac{\pi\hbar}{m\omega}}\: \prod_{\mu=0}^\infty \frac{\text{d}x_\mu}{\sqrt{2\pi}} \: ,
	\label{eq:EuclideanPathIntegralMeasure}
\end{align}
where we introduced the unspecified normalization constant $\mathcal{N}_\text{E}$, the integral decomposes into a product of purely Gaussian integrals, leaving only the non-exponential weight functional $f_\text{non-exp}\llbracket x\rrbracket$ left to be addressed. One can resort to the usual trick of introducing an external current $J(t)$ to cast the term into the more convenient exponential form
\begin{align}
	f_\text{non-exp}\Big\llbracket \:\!x_\text{crit}+\Delta x\:\!\Big\rrbracket &= \frac{1}{2^n n!} \sqrt{\frac{m\omega}{\pi\hbar}} \; \psi_{n,\text{non-exp}}^{(\text{loc})}\!\!\:\Big[x_\text{crit}(0)+\Delta x(0)\Big] \,  \psi_{n,\text{non-exp}}^{(\text{loc})}\!\!\:\Big[x_\text{crit}(T)+\Delta x(T)\Big] \nonumber \\[0.05cm]
	&=\frac{1}{2^n n!} \sqrt{\frac{m\omega}{\pi\hbar}} \; \scalebox{1.2}{$\Bigg\{$}\psi_{n,\text{non-exp}}^{(\text{loc})}\bigg[ \!\:\!\: x_\text{crit}(0)\!\:\!\:\!\:+\!\:\!\: \sqrt{\frac{\hbar}{m\omega}}\, \frac{\delta}{\delta J(0)}\!\:\!\:\bigg] \nonumber \\ 
	&\qquad\qquad\qquad\!\;\, \times  \psi_{n,\text{non-exp}}^{(\text{loc})}\bigg[ x_\text{crit}(T)+ \sqrt{\frac{\hbar}{m\omega}}\, \frac{\delta}{\delta J(T)}\bigg]  \nonumber \\[0.05cm] 
	&\qquad\qquad\qquad\!\;\, \times \exp\!\Bigg[\sum_{\mu=0}^\infty \, x_\mu \int_0^T \!e_\mu(t) \:\!J(t) \, \text{d}t\,\Bigg]\!\!\:\scalebox{1.2}{$\Bigg\}$}_{\!\scalebox{0.85}{$J(t)\!\!\:\!\!\: =\!\!\:\!\!\: 0$}} \, .
	\label{eq:5_Definition_fpoly_Via_Functional_Derivatives}
\end{align}
Pulling the functional derivatives out of the $\text{d}x_\mu$-integrals, one can straightforwardly perform the Gaussian integration, obtaining the leading-order contribution to the steepest descent thimble $\mathcal{J}_\text{crit}$ passing through the specific critical trajectory $x_\text{crit}(t)$ as
\begin{align} 
	\!\Big[\mathcal{P}_{\!\!\: n}^{(\pi/2)}\big(\hbar,T\big)\Big]_{\text{crit}}^{\text{(LO)}}&=\frac{\mathcal{N}_\text{E}}{2^n n!} \: \exp\!\!\,\scalebox{1.1}{$\bigg($} \!\!-\!\!\:\frac{f_\text{exp}\big\llbracket x_\text{crit}\big\rrbracket}{\hbar}\scalebox{1.1}{$\bigg)$} \, \Bigg(\prod_{\mu=0}^\infty \lambda_{\mu}\!\!\:\Bigg)^{\!\!\!\!\:-\frac{1}{2}} \scalebox{1.2}{$\Bigg\{$} \psi_{n,\text{non-exp}}^{(\text{loc})}\Big[ \ldots \Big] \, \psi_{n,\text{non-exp}}^{(\text{loc})}\Big[ \ldots \Big] \nonumber \\ 
	&\qquad\;\;\!\:\times \exp\!\Bigg[\,\frac{1}{2}\,\mathlarger{\int}_0^T \!\! \mathlarger{\int}_0^T J(t) \hspace{-0.65cm}\underbrace{\Bigg(\sum_{\mu=0}^\infty\frac{e_\mu(t)\, e_\mu(t')}{\lambda_\mu}\Bigg)}_{\displaystyle{\text{Green's function } G_\text{crit}\big(t,t'\big)}} \hspace{-0.65cm} J(t') \,\text{d}t\!\;\text{d}t'\Bigg] \!\!\: \scalebox{1.2}{$\Bigg\}$}_{\!\scalebox{0.85}{$J(t)\!\!\:\!\!\: =\!\!\:\!\!\: 0$}} .
	\label{eq:ResultNoZeroMode_Intermediate}
\end{align}
In this expression we can identify the product of all eigenvalues of the fluctuation operator $O_\text{crit}$ as its determinant, while the exponent contains the corresponding Green's function $G_\text{crit}$. The arbitrary normalization constant $\mathcal{N}_\text{E}$ is later offset by the implicit scaling of the $\zeta$-regularized fluctuation determinant $\text{det}_\zeta$, see appendix~\ref{sec:F1_Sturm_Liouville_Determinants}. Beware that both the determinant and the Green's function are to be computed using the Robin boundary conditions specified in equation~\eqref{eq:Robin_BoundaryConditions}. As usual, the above leading-order result inherits corrections of order $\sqrt{\hbar}$ due to higher-order terms arising in the expansion~\eqref{eq:Expanded_Exponent}, which we deliberately truncated. Retaining these corrections yields the common perturbative expansion in terms of Feynman diagrams. \\ 

\noindent 
The given result~\eqref{eq:ResultNoZeroMode_Intermediate} can still be simplified by noting that the non-exponential terms in our case are only evaluated at the endpoints of the time interval $[0,T]$, allowing the functional derivatives to be turned into ordinary ones. This then yields the final, streamlined expression
\begin{align}
	\!\!\Big[\mathcal{P}_{\!\!\: n}^{(\pi/2)}\big(\hbar,T\big)\Big]_{\text{crit}}^{\text{(LO)}}&=\frac{\mathcal{N}_\text{E}}{2^n n!}\; \text{det}_\zeta\big(O_\text{crit}\big)^{-\frac{1}{2}}\,  \exp\!\!\,\scalebox{1.1}{$\bigg($} \!\!-\!\!\:\frac{f_\text{exp}\big\llbracket x_\text{crit}\big\rrbracket}{\hbar}\scalebox{1.1}{$\bigg)$}  \label{eq:Final_Result_NoZeroMode} \\ 
	&\phantom{=}\times \Bigg\{ \psi_{n,\text{non-exp}}^{(\text{loc})}\bigg[ x_\text{crit}(0)+ \sqrt{\frac{\hbar}{m\omega}}\, \frac{\partial}{\partial J_0}\bigg] \, \psi_{n,\text{non-exp}}^{(\text{loc})}\bigg[ x_\text{crit}(T)+ \sqrt{\frac{\hbar}{m\omega}}\, \frac{\partial}{\partial J_T}\bigg] \nonumber \\
	&\phantom{=}\times \exp\!\scalebox{1.1}{\bigg[}\,\frac{1}{2}\,J_0\big.^{\!\!\!\: 2}\, G_\text{crit}\big(0,0\big)+J_0\:\! J_T\,  G_\text{crit}\big(0,T\big)+\frac{1}{2}\,J_T\big.^{\!\!\! 2}\, G_\text{crit}\big(T,T\big)\scalebox{1.1}{\bigg]}\!\!\: \Bigg\}_{\substack{\scalebox{0.85}{$J_0\!\, =\! 0$}\\[0.025cm] \scalebox{0.85}{$J_T\! =\! 0$}}}  \: , \nonumber 
\end{align}
where we have utilized the symmetry of $G_\text{crit}$ under a swap of both its arguments. As one would expect, the obtained expression is solely dependent on the Green's function evaluated at the boundary of the time interval.\\ 

\noindent
Before computing the different critical trajectories and the relevant quantities related to the fluctuation operator pertaining to these solutions, we should stress that the above procedure has to be adapted in case $O_\text{crit}$ possesses a (quasi-)zero mode. Given such a parametrically flat direction in function space due to an underlying symmetry, the Gaussian integration is rendered insufficient. Instead, one should expand not about a sole critical point, but the emerging one-parameter family $x\raisebox{-2.5pt}{\scalebox{0.7}{\text{crit}}}\!\!\!\!\!\!\!\,\raisebox{5pt}{\scalebox{0.65}{$(t_0)$}}$ of (approximate) solutions to the equation of motion, parametrized by a collective coordinate $t_0$~\cite{SakitaGervaisZeroModeFadeevPopov,BalitskyCollectiveCoordinates}. While the fluctuations perpendicular to the zero mode direction are treated in the same way as before, the remaining $t_0$-integration is to be performed exact. The modified expansion is given by
\begin{align}
	x(t)=x_\text{crit}^{(t_0)}(t)+\Delta x^{(t_0)}(t)=x_\text{crit}^{(t_0)}(t)+\sqrt{\frac{\hbar}{m\omega}}\, \sum_{\mu=1}^\infty \, x^\perp_\mu e_\mu^{(t_0)}(t) \, ,
	\label{eq:ExpansionVariationsCollectiveCoord}
\end{align}
where the quasi-zero mode $e_0^{(t_0)}(t)\propto\partial_{t_0} \:\! x_\text{crit}^{(t_0)}(t)$ has been traded for the collective coordinate $t_0$. Note that we adopt the prescription of denoting the zero mode by $e_0$ rather than utilizing Coleman's original convention of using $e_1$~\cite{CallanColemanFateOfFalseVac2}, instead ordering the eigenfunctions based on the magnitude of their respective eigenvalue. Given the parametrization~\eqref{eq:ExpansionVariationsCollectiveCoord}, all previous steps~\eqref{eq:Expanded_Exponent} to~\eqref{eq:Final_Result_NoZeroMode} are repeated with minimal adjustments. Assuming the linear term in $\Delta x^{(t_0)}(t)$ to be fully negligible for any value of the collective coordinate $t_0$, the final result is found to be 
\begin{align}
	\!\!\Big[\mathcal{P}_{\!\!\: n}^{(\pi/2)}\Big]_{\text{crit,ZM}}^{\text{(LO)}} &=\frac{\mathcal{N}_\text{E}}{2^n n!} \, \mathlarger{\mathlarger{\int}}_{t_{0,\text{min}}}^{t_{0,\text{max}}} \frac{\text{d}t_0}{\sqrt{2\pi}} \;\exp\!\bigg(\!\!\!\:-\!\!\:\frac{1}{\hbar} \, f_\text{exp}\raisebox{0.5pt}{\Big\llbracket} x_\text{crit}^{(t_0)}\raisebox{0.5pt}{\Big\rrbracket} \bigg) \, \text{det}'_\zeta\raisebox{1pt}{\Big[}O_\text{crit}^{(t_0)}\raisebox{1pt}{\Big]}^{-\frac{1}{2}} \, \text{J}_\text{det}\scalebox{1.15}{\big(}t_0,x_\mu^\perp\!\!\:=\!\!\:0\scalebox{1.15}{\big)} \nonumber \\[0.1cm] 
	&\phantom{=}\times \Bigg\{ \psi_{n,\text{non-exp}}^{(\text{loc})}\bigg[ x_\text{crit}^{(t_0)}(0)+ \sqrt{\frac{\hbar}{m\omega}}\, \frac{\partial}{\partial J_0}\bigg] \, \psi_{n,\text{non-exp}}^{(\text{loc})}\bigg[ x_\text{crit}^{(t_0)}(T)+ \sqrt{\frac{\hbar}{m\omega}}\, \frac{\partial}{\partial J_T}\bigg] \label{eq:Final_Result_ZeroMode} \\[0.05cm] 
	&\phantom{=}\times \exp\!\scalebox{1.1}{\bigg[}\:\!\frac{1}{2}\,J_0\big.^{\!\!\!\: 2}\, G_\text{crit}^{\perp,(t_0)}\big(0,0\big)+J_0\:\! J_T\,  G_\text{crit}^{\perp,(t_0)}\big(0,T\big)+\frac{1}{2}\,J_T\big.^{\!\!\! 2}\, G_\text{crit}^{\perp,(t_0)}\big(T,T\big)\scalebox{1.1}{\bigg]}\!\!\: \Bigg\}_{\substack{\scalebox{0.85}{$J_0\!\, =\! 0$}\\[0.025cm] \scalebox{0.85}{$J_T\! =\! 0$}}} \: . \nonumber
\end{align}
The striking difference to the result~\eqref{eq:Final_Result_NoZeroMode} is the remaining integration over the collective coordinate $t_0$, ranging from $t_{0,\text{min}}$ to $t_{0,\text{max}}$. With the zero mode treated separately, it has been subtracted from both the appearing determinant and the Green's function, leading to the primed determinant $\text{det}'_\zeta$ as well as the subtracted Green's function $G^\perp$, as explained in appendices~\ref{sec:F1_Sturm_Liouville_Determinants} and~\ref{sec:F3_Subtracted_Greens_Function} respectively. Additionally, the path integral measure accommodates an additional Jacobian determinant $\smash{\text{J}_\text{det}\big(t_0,x^\perp_\mu\big)}$ arising from the non-trivial coordinate transformation $\smash{\big\{x_\mu\big\}\raisebox{-3.5pt}{$\!\!\:\scalebox{0.7}{$\mu\!\in\!\mathbb{N}^0$}$}}\mapsto \smash{\big\{t_0,x^\perp_\mu\big\}\raisebox{-3.5pt}{$\!\!\:\scalebox{0.7}{$\mu\!\in\!\mathbb{N}$}$}\,}$. To leading-order in $\hbar$, the Jacobian is evaluated at the critical point, thus allowing us to fully neglect the subleading $x_\mu^\perp$-dependence. The computation is outsourced to appendix~\ref{sec:D_JacobianDeterminant}, being based upon the excellent discussion given by Andreassen et al.~\cite{SchwartzPrecisionDecayRate}, adapted to our normalization convention.

\subsection{Discussion of critical trajectories}
\label{sec:CriticalPaths}

Now that we established how to compute the individual contributions arising from each saddle point of the path integral, we should investigate which paths actually constitute critical trajectories. Looking at the expansion~\eqref{eq:Expanded_Exponent} and employing the specific form of the exponential wave function terms $\psilocexp$ given in equation~\eqref{eq:FullExponent}, each extremal path has to satisfy the three conditions 
\begin{subequations}
	\begin{align}
		\dot{x}_\text{crit}(t)\big.^2 &=\:\!\omega^2 \, x_\text{crit}(t)\big.^2 \;\!\!\:\bigg\{\!\:1 +\!\; W\!\!\:\mathlarger{\big[}x_\text{crit}(t)\mathlarger{\big]}\!\!\:\bigg\} + \frac{2E_\text{crit}}{m}\, , & & \begin{matrix}
			\text{(integrated)} \\
			\text{Euler--Lagrange equation}
		\end{matrix} \label{eq:Euler_Lagrange_Equation} \\
		\dot{x}_\text{crit}(0) &= \phantom{+}\omega \:\! x_\text{crit}(0) \!\:\:\bigg\{1+\!\: W\!\!\:\mathlarger{\big[} x_\text{crit}(0)\mathlarger{\big]}\!\!\:\bigg\}^{\!\!\:\frac{1}{2}}\, ,& &\; \begin{matrix}
			\text{transversality condition} \\
			\text{at the left boundary}
		\end{matrix} \label{eq:Left_Transversality_Condition} \\
		\dot{x}_\text{crit}(T) &= -\omega \:\! x_\text{crit}(T) \!\: \bigg\{1+W\!\!\:\mathlarger{\big[} x_\text{crit}(T)\mathlarger{\big]}\!\!\:\bigg\}^{\!\!\:\frac{1}{2}}\, . & &\; \begin{matrix}
			\text{transversality condition} \\
			\text{at the right boundary}
		\end{matrix} \label{eq:Right_Transversality_Condition}
	\end{align}
	\label{eq:All_Extremal_Conditions}%
\end{subequations}
Note that the assumption of the potential $V$ being time-independent enters relation~\eqref{eq:Euler_Lagrange_Equation}, allowing the original Euler--Lagrange equation to be integrated, obtaining the given first order ordinary differential equation (ODE) with the conserved classical energy $E_\text{crit}$ as its integration constant. Squaring both transversality conditions, it is immanent that
\begin{align}
E_\text{crit}=0
\end{align}
is enforced on critical trajectories. Besides the trivial FV trajectory $x_\text{crit}(t)=x_\text{FV}=0$, the only other classical solution satisfying this condition is, up to time-translations, the usual (infinite-time) bounce motion $x\raisebox{-2.5pt}{\scalebox{0.7}{\text{bounce}}}\!\!\!\!\!\!\!\!\!\!\!\!\!\!\;\raisebox{5pt}{\scalebox{0.65}{$(T\!=\!\infty)$}}(t)$, possessing turning points at $x_\text{FV}$ and $x_\text{escape}$. An important consequence arising from the requirement of vanishing classical energy is that shot-like trajectories are never critical due to their positive energy. Compared to the conventional treatment, this offers the significant advantage that we do not have to discard these solutions manually, as weighting the contributions appropriately takes care of this unsatisfactory step of the standard approach. 
\begin{figure}[H]
	\centering
	\includegraphics[width=\textwidth]{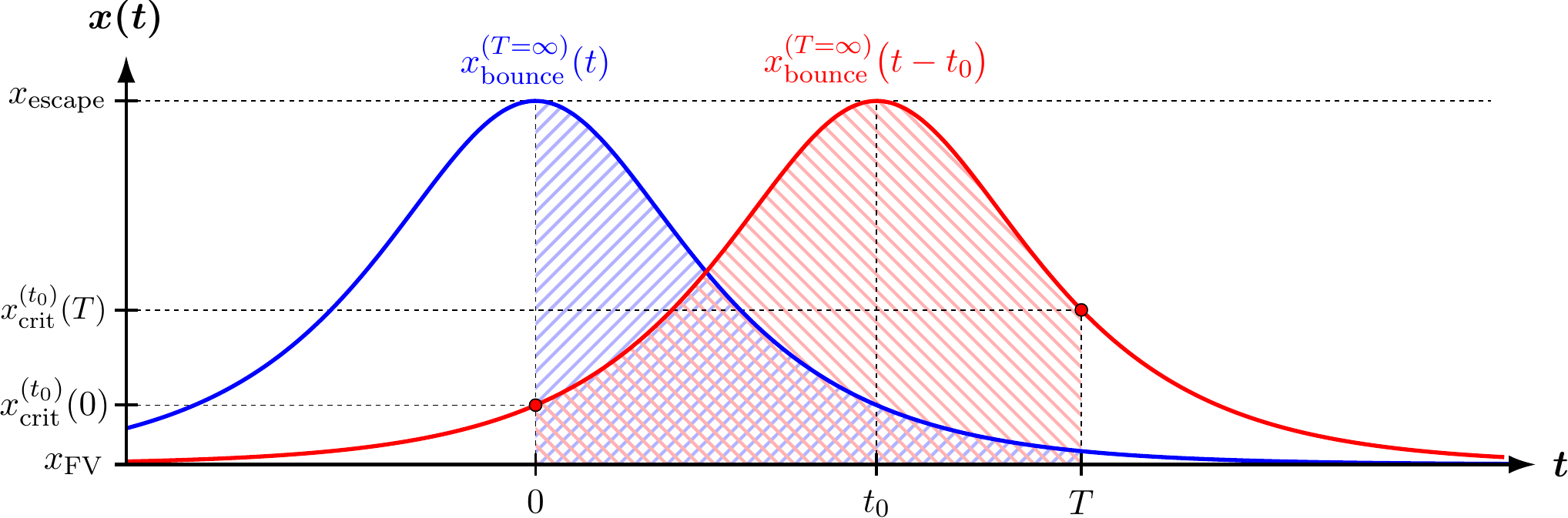}
	\caption{\hspace{0.001cm}}
	\vspace{-0.5cm}\hspace{1.5cm}
	\begin{tikzpicture}
		\node[anchor=center, inner sep=0pt, 
		minimum width=14.35cm, text width=14.35cm-0.4cm,align=justify,
		minimum height=2.0cm, 
		text depth=2.0cm-0.4cm] (0,0) {Depiction of two admissible finite-time bounce motions, seen as cutouts of the (time-translated) infinite-time bounce $x\raisebox{-2.5pt}{\scalebox{0.7}{\text{bounce}}}\!\!\!\!\!\!\!\!\!\!\!\!\!\!\:\raisebox{5pt}{\scalebox{0.65}{$(T\!=\!\infty)$}}(t)$, defined as the representative that turns at exactly $t=0$. The time translation parameter $t_0$ is required to lie inside the interval $[0,T]$ for the motion to satisfy both transversality conditions~\eqref{eq:Left_Transversality_Condition} and~\eqref{eq:Right_Transversality_Condition}.};
	\end{tikzpicture}
	\label{fig:Bounce_Motion_Collective_Coordinate}
\end{figure}
\vspace{-0.5cm}

\noindent
Since the path integral ranges over functions living on the time interval $[0,T]$, the so-found critical bounce motions constitute only a portion of one full bounce period, spanning an infinite time, see figure~\ref{fig:Bounce_Motion_Collective_Coordinate}. While time-translated bounces trivially satisfy the Euler--Lagrange equation~\eqref{eq:Euler_Lagrange_Equation}, they only obey both transversality conditions~\eqref{eq:Left_Transversality_Condition} and~\eqref{eq:Right_Transversality_Condition} in case the motion traverses the turning point $x_\text{escape}$ during the given time interval $[0,T]$, as the supplementary boundary conditions impose a sign flip of the velocity. Insisting that $x\raisebox{-2.5pt}{\scalebox{0.7}{\text{bounce}}}\!\!\!\!\!\!\!\!\!\!\!\!\!\!\:\raisebox{5pt}{\scalebox{0.65}{$(T\!=\!\infty)$}}(t)$ reaches $x_\text{escape}$ at exactly $t=0$ as shown in figure~\ref{fig:Bounce_Motion_Collective_Coordinate}, we find the one-parameter family of critical paths $x\raisebox{-2.5pt}{\scalebox{0.7}{\text{crit}}}\!\!\!\!\!\!\!\,\raisebox{5pt}{\scalebox{0.65}{$(t_0)$}}\:\!(t)=x\raisebox{-2.5pt}{\scalebox{0.7}{\text{bounce}}}\!\!\!\!\!\!\!\!\!\!\!\!\!\!\:\raisebox{5pt}{\scalebox{0.65}{$(T\!=\!\infty)$}}\big(t-t_0\big)$, parameterized by the collective coordinate $t_0\in [0,T]$. Notably, the constraint on the collective coordinate $t_0$ imposed from the requirement to hit the turning point during the motion fixes the volume of the underlying moduli space to the desired linear factor $T$.\\

\noindent
Having determined the relevant trajectories to expand around, we are now fit to compute their contributions to the overall path integral. As previously derived in equations~\eqref{eq:Final_Result_NoZeroMode} and~\eqref{eq:Final_Result_ZeroMode}, the leading-order result only requires knowledge about the (primed) determinant of the fluctuation operator and its corresponding (subtracted) Green's function at the boundary of the time interval. Section~\ref{sec:3_4_FVContribution} deals with the computation for the FV trajectory, while section~\ref{sec:3_5_BounceContribution} faces the more difficult one-parameter family of bounce saddles.

\subsection{FV contribution}
\label{sec:3_4_FVContribution}

Let us initially study the rather simple case of the constant FV trajectory $x_\text{crit}(t)=x_\text{FV}=0$. We begin by noting that for the given trajectory, one trivially finds $f_\text{exp}\llbracket x_\text{crit}\rrbracket=0$, since both the Euclidean action as well as the exponential weight factors $\psilocexp$ vanish identically. With this out of the way, we can directly delve into the study of the corresponding fluctuation operator $O_\text{FV}$, which is given by
\begin{align}
	O_\text{FV}=-\:\!\frac{\text{d}^2}{\text{d}(\omega t)^2} +1 \, .
\end{align}
The eigenfunctions $e_\mu(t)$ are required to satisfy the Robin boundary conditions
\begin{align}
	\begin{pmatrix} 1 & -1 \\ 0 & 0 \end{pmatrix} \begin{pmatrix} \;\;\;\;\;\;\, e_\mu(0) \\ \omega^{-1} \,\dot{e}_\mu(0) \end{pmatrix} + \begin{pmatrix} 0 & 0 \\ 1 & 1 \end{pmatrix} \begin{pmatrix} \;\;\;\;\;\;\, e_\mu(T) \\ \omega^{-1} \,\dot{e}_\mu(T) \end{pmatrix} = 0 \, ,
	\label{eq:FVBoundaryConditions}
\end{align}
where we employed equation~\eqref{eq:Robin_BoundaryConditions} together with expression~\eqref{eq:FullExponent} for $\psilocexp$. With the usual Gel'fand-Yaglom method of computing functional determinants readily generalizing to Robin boundary conditions as portrayed in appendix~\ref{sec:F1_Sturm_Liouville_Determinants}, the determinant of an operator $O$ can be computed utilizing the boundary behavior of two particular linearly independent solutions $\smash{\mathpzc{y}\raisebox{-3.25pt}{\scalebox{0.75}{$\lambda\!\!\:\!\!\:=\!\!\:\!\!\:0$}}\!\!\!\!\!\!\!\!\!\:\!\:\!\raisebox{4.75pt}{\scalebox{0.65}{$(1,2)$}}(t)}$ of the ODE $O \mathpzc{y}=0$~\cite{GelfandFunctionalIntegration,KirstenMcKaneFuncDets1,KirstenMcKaneFuncDets2}. With this in mind, we already collected the boundary conditions~\eqref{eq:FVBoundaryConditions} appropriately to compare them to the general expression~\eqref{eq:A3_Bouncary_Conditions_Eigenvalue_Problem}, allowing us to read off the important matrices $M_\text{FV}$ and $N_\text{FV}$ required for the determinant computation. Given the requirements~\eqref{eq:A3_Boundary_Conditions_y12} and the above operator $O_\text{FV}$, one obtains the relevant solutions 
\begin{subequations}
	\begin{align}
		\mathpzc{y}_{\scalebox{0.8}{$\lambda\!\!\:\!\!\:=\!\!\:\!\!\:0$}}^{\scalebox{0.8}{$(1)$}}(t)&=\cosh\!\big(\omega t\big) \, , \label{eq:SpecialFunctionsFVSaddle_A} \\[0.1cm] 
		\mathpzc{y}_{\scalebox{0.8}{$\lambda\!\!\:\!\!\:=\!\!\:\!\!\:0$}}^{\scalebox{0.8}{$(2)$}}(t)&=\sinh\!\big(\omega t\big) \, , 
		\label{eq:SpecialFunctionsFVSaddle_B}
	\end{align}
	\label{eq:SpecialFunctionsFVSaddle}%
\end{subequations}
such that employing equation~\eqref{eq:A3_Definition_Y_Matrix} together with the results~\eqref{eq:A3_Gelfand_Yaglom_Determinant_Ratio} and~\eqref{eq:ProportionalityConstantDeterminants}, the desired regularized determinant is found to be 
\begin{align}
	\text{det}_\zeta\big(O_\text{FV}\big) = \frac{\mathcal{N}_\text{E}\big.^{\!\!\!\!\: 2}}{2} \: \text{det}\scalebox{1.1}{\Bigg[}\!\begin{pmatrix} 1 & -1 \\0 & 0 \end{pmatrix}+\begin{pmatrix} 0 & 0 \\ 1 & 1 \end{pmatrix} \begin{pmatrix} \cosh\!\big(\omega T\big)  & \sinh\!\big(\omega T\big)  \\[0.15cm] \sinh\!\big(\omega T\big)  & \cosh\!\big(\omega T\big)  \end{pmatrix} \!\scalebox{1.1}{\Bigg]} =  \mathcal{N}_\text{E}\big.^{\!\!\!\!\: 2} \, e^{\omega T} \, .
	\label{eq:6_False_Vacuum_Determinant_Full}
\end{align}
At this point, the sole remaining quantity to be computed is the FV Green's function $G_\text{FV}$. For a detailed derivation consult appendix~\ref{sec:F2_GreensFunction}, where, given generic Robin boundary conditions, we derived the complete expression for the Green's function in terms of the auxiliary functions $\mathpzc{y}\raisebox{-3.25pt}{\scalebox{0.75}{$\lambda\!\!\:\!\!\:=\!\!\:\!\!\:0$}}\!\!\!\!\!\!\!\!\!\!\:\;\!\raisebox{4.75pt}{\scalebox{0.65}{$(1,2)$}}(t)$. Plugging the previously presented solutions~\eqref{eq:SpecialFunctionsFVSaddle_A} and~\eqref{eq:SpecialFunctionsFVSaddle_B} into the so-obtained general result~\eqref{eq:A3_Free_Greens_Function_Result} yields the desired Green's function
\begin{align}
	G_\text{FV}\big(t,t'\big) = \frac{1}{2} \exp\!\Big(\!\!\!\:-\!\!\:\omega \big\lvert t-t'\big\rvert\Big) \, .
	\label{eq:6_Greens_Function_FV}
\end{align}
Thereupon inserting expressions~\eqref{eq:6_False_Vacuum_Determinant_Full} and~\eqref{eq:6_Greens_Function_FV} into the previously derived relation~\eqref{eq:Final_Result_NoZeroMode}, we find, up to subleading corrections, the intermediate representation
\begin{align}
	\begin{split}
	\Big[\mathcal{P}_{\!\!\: n}^{(\pi/2)}\big(\hbar,T\big)\Big]_{\text{FV}}^{\text{(LO)}} &=\frac{e^{-\frac{\omega T}{2}}}{2^n n!}\; \scalebox{1.1}{\Bigg\{} \:\! \psi_{n,\text{non-exp}}^{(\text{loc})}\scalebox{1.15}{\bigg(}\!\!\:\sqrt{\frac{\hbar}{m\omega}}\, \frac{\partial}{\partial J_0}\scalebox{1.15}{\bigg)} \: \psi_{n,\text{non-exp}}^{(\text{loc})}\scalebox{1.15}{\bigg(}\!\!\:\sqrt{\frac{\hbar}{m\omega}}\, \frac{\partial}{\partial J_T}\scalebox{1.1}{\bigg)} \\[0.1cm]
	& \hspace{4.65cm}\times \exp\!\scalebox{1.15}{\bigg(}\frac{J_0\big.^{\!\!\!\: 2}+J_T\big.^{\!\!\! 2}}{4} + \frac{e^{-\omega T}}{2} \,J_0 \:\!J_T\scalebox{1.1}{\bigg)} \!\!\:\scalebox{1.1}{\Bigg\}}_{\substack{\scalebox{0.85}{$J_0\!\, =\! 0$}\\[0.025cm] \scalebox{0.85}{$J_T\! =\! 0$}}}  \, . 
	\end{split}
\end{align}
Due to the Green's function being of order unity, the arguments inside $\psilocnonexp$ are strictly of order $\smash{\sqrt{\hbar}}\ll 1$. This allows us to expand the wave function $\smash{\psilocnonexp}$ for small arguments, whose leading-order behavior~\eqref{eq:4_Asymptotics_Non_Exponential_Wave_Function_Small_x} is purely harmonic. Together with the auxiliary relation~\eqref{eq:6_Special_Result_Auxiliary_Function_HermitePolys} for Hermite polynomials derived in appendix~\ref{sec:E_HermiteIdentity}, we finally obtain the leading-order result
\begin{align}
		\Big[\mathcal{P}_{\!\!\: n}^{(\pi/2)}\big(\hbar,T\big)\Big]_{\text{FV}}^{\text{(LO)}} &=\frac{e^{-\frac{\omega T}{2}}}{2^n n!}\;\Bigg\{ H_{n}\bigg(\frac{\partial}{\partial J_0}\bigg)\, H_{n}\bigg(\frac{\partial}{\partial J_T}\bigg) \,\exp\!\scalebox{1.15}{\bigg(}\frac{J_0\big.^{\!\!\!\: 2}+J_T\big.^{\!\!\! 2}}{4} + \frac{e^{-\omega T}}{2} \,J_0 \:\!J_T\scalebox{1.1}{\bigg)} \!\!\:\Bigg\}_{\substack{\scalebox{0.85}{$J_0\!\, =\! 0$}\\[0.025cm] \scalebox{0.85}{$J_T\! =\! 0$}}} \nonumber \\ 
		&= \frac{e^{-\frac{\omega T}{2}}}{2^n n!} \:\, \mathcal{I}_n\bigg(0,0;\frac{1}{4},\frac{1}{4},\frac{e^{-\omega T}}{2}\bigg) = \exp\!\!\!\;\scalebox{1.1}{$\bigg[$}\!-\!\!\:\omega T \:\!\bigg(\!\:\!n+\frac{1}{2}\bigg)\:\!\!\scalebox{1.1}{$\bigg]$} \, .
		\label{eq:FV_Contribution}
\end{align}
This result is expected---more yet, it must turn out this way---as for the construction of the wave function $\psinloc$ we explicitly assumed $\Enloc$ to be only slightly perturbed around the usual harmonic oscillator eigenenergies, see equation~\eqref{eq:2_Local_Energy_Ansatz_Definition_Epsn}. With this compulsory part under control, we can investigate the more interesting contributions arising from the family of bounce trajectories.

\subsection{Bounce contributions}
\label{sec:3_5_BounceContribution}

Let us now finally turn toward the set of critical bounce trajectories $x\raisebox{-2.5pt}{\scalebox{0.7}{\text{crit}}}\!\!\!\!\!\!\!\,\raisebox{5pt}{\scalebox{0.65}{$(t_0)$}}\:\!(t)=x\raisebox{-2.5pt}{\scalebox{0.7}{\text{bounce}}}\!\!\!\!\!\!\!\!\!\!\!\!\!\!\:\raisebox{5pt}{\scalebox{0.65}{$(T\!=\!\infty)$}}\big(t-t_0\big)$, where we begin our discussion by briefly demonstrating the presence of an exact zero mode for any value of the collective coordinate $t_0$. This turns out to be beneficial, as the found zero mode heavily simplifies the computation of both the primed determinant and the Jacobian factor encountered in relation~\eqref{eq:Final_Result_ZeroMode}. As previously stated in section~\ref{sec:CriticalPaths}, all representatives of the family of critical paths $x\raisebox{-2.5pt}{\scalebox{0.7}{\text{crit}}}\!\!\!\!\!\!\!\,\raisebox{5pt}{\scalebox{0.65}{$(t_0)$}}\:\!(t)$ satisfy the Euler--Lagrange equation~\eqref{eq:Euler_Lagrange_Equation} as well as both transversality conditions~\eqref{eq:Left_Transversality_Condition} and~\eqref{eq:Right_Transversality_Condition} due to the infinite-time bounce motion obeying
\begin{align}
	\frac{\text{d} \:\!x^{(T=\infty)}_\text{bounce}(t)}{\text{d}(\omega t)}  
	&= -\,\text{sign}(t)\:x^{(T=\infty)}_\text{bounce}(t) \, \bigg\{1+W\!\!\:\Big[x^{(T=\infty)}_\text{bounce}(t)\Big]\!\!\:\bigg\}^{\frac{1}{2}} \, ,
	\label{eq:BounceODE}
\end{align}
ensuring all linear terms arising in the expansion~\eqref{eq:Expanded_Exponent} to fully drop out. Utilizing the non-integrated equation of motion~\eqref{eq:Expanded_Exponent} as well as its first time derivative, we find the bounce fluctuation operator~\eqref{eq:Fluctuation_Operator} as
\begin{align}
	O_\text{bounce}^{(t_0)}=-\frac{\text{d}^2}{\text{d}(\omega t)^2} + \frac{V''\scalebox{1.2}{\big[}x^{(t_0)}_\text{crit}(t)\scalebox{1.2}{\big]}}{m\omega^2} = \omega^{-2} \:\! \scalebox{1.15}{\bigg\{} \!\!\!\:-\!\!\:\frac{\text{d}^2}{\text{d}t^2} +\raisebox{0.5pt}{\Big[} \dot{x}_\text{crit}^{(t_0)}(t)\raisebox{0.5pt}{\Big]}^{-1} \dddot{x}^{(t_0)}_\text{crit}(t) \scalebox{1.15}{\bigg\}} \, ,
	\label{eq:Bounce_FluctOperator}
\end{align}
with the corresponding Robin boundary conditions~\eqref{eq:Robin_BoundaryConditions} given by
\begin{subequations}
\begin{align}
	\dot{e}^{(t_0)}_\mu(0)\,&=\phantom{+}\frac{1}{m}\: \psi_\text{exp}^{(\text{loc})\prime\prime}\raisebox{0.5pt}{\Big[} \;\!x_\text{crit}^{(t_0)}(0)\;\!\raisebox{0.5pt}{\Big]} \,e^{(t_0)}_\mu(0)\;\!\;\!=\ddot{x}^{(t_0)}_\text{crit}(0)\: \raisebox{0.5pt}{\Big[} \;\! \dot{x}_\text{crit}^{(t_0)}(0)\;\!\raisebox{0.5pt}{\Big]}^{-1} e^{(t_0)}_\mu(0)  \, , \label{eq:BoundaryConditionsEigenfunctionsBounce_Left} \\[0.1cm]
	\dot{e}^{(t_0)}_\mu(T)&=-\frac{1}{m}\: \psi_\text{exp}^{(\text{loc})\prime\prime}\raisebox{0.5pt}{\Big[} x_\text{crit}^{(t_0)}(T)\raisebox{0.5pt}{\Big]} \,e^{(t_0)}_\mu(T)=\ddot{x}^{(t_0)}_\text{crit}(T) \raisebox{0.5pt}{\Big[} \dot{x}_\text{crit}^{(t_0)}(T)\raisebox{0.5pt}{\Big]}^{-1} e^{(t_0)}_\mu(T) \, . \label{eq:BoundaryConditionsEigenfunctionsBounce_Right}
\end{align} 
\label{eq:BoundaryConditionsEigenfunctionsBounce}%
\end{subequations}
These derivations again crucially took advantage of the specific form of the exponential WKB suppression factor~\eqref{eq:FullExponent}. Notice that we additionally employed $\dot{x}\raisebox{-2.5pt}{\scalebox{0.7}{\text{crit}}}\!\!\!\!\!\!\!\,\raisebox{5pt}{\scalebox{0.65}{$(t_0)$}}\:\!(0)>0$ and $\dot{x}\raisebox{-2.5pt}{\scalebox{0.7}{\text{crit}}}\!\!\!\!\!\!\!\,\raisebox{5pt}{\scalebox{0.65}{$(t_0)$}}\:\!(T)<0$ when deriving both relations~\eqref{eq:BoundaryConditionsEigenfunctionsBounce_Left} and~\eqref{eq:BoundaryConditionsEigenfunctionsBounce_Right}, implicitly using $t_0\in [0,T]$. It becomes immanent that $e\raisebox{-2.75pt}{\scalebox{0.65}{0}}\!\!\raisebox{5pt}{\scalebox{0.65}{$(t_0)$}}(t)\propto\dot{x}\raisebox{-2.5pt}{\scalebox{0.7}{\text{crit}}}\!\!\!\!\!\!\!\,\raisebox{5pt}{\scalebox{0.65}{$(t_0)$}}\:\!(t)$ is the underlying exact zero mode, being annihilated by the fluctuation operator~\eqref{eq:Bounce_FluctOperator} while satisfying the required boundary conditions~\eqref{eq:BoundaryConditionsEigenfunctionsBounce}, regardless of the chosen $t_0$. While this might sound utterly trivial, notice that for finite time intervals one usually does not find an exact zero mode. Instead, fixing Dirichlet boundary conditions explicitly breaks time translation invariance, with the symmetry only being restored in the infinite-time limit. It is thus rather remarkable that the Robin boundary conditions involving the WKB suppression factor arising from the wave function weights manifestly preserve time-translation symmetry for any value of the parameter $T$. The preceding considerations justify the use of equation~\eqref{eq:Final_Result_ZeroMode}, which we will now dissect piece by piece. \\ 

\noindent
Let us start by investigating the full exponent $f_\text{exp}$ as given in equation~\eqref{eq:FullExponent}, being undeniably the dominant factor in equation~\eqref{eq:Pn:exp:nonexp} due to accommodating the singular $\hbar$-dependence. Given that the bounce motions fulfill the ODE~\eqref{eq:BounceODE}, the Euclidean bounce action is found to be 
\begin{align}
	S_\text{E}\raisebox{1pt}{\Big\llbracket} x_\text{crit}^{(t_0)}\raisebox{1pt}{\Big\rrbracket} &= \mathlarger{\int}_0^T \frac{m}{2} \:\! \raisebox{0.5pt}{\Big[} \dot{x}_\text{crit}^{(t_0)}(t)\raisebox{0.5pt}{\Big]}^2  + V\!\!\:\raisebox{0.5pt}{\Big[} x_\text{crit}^{(t_0)}(t)\raisebox{0.5pt}{\Big]} \, \text{d}t = 2\:\!\mathlarger{\int}_0^T V\!\!\:\raisebox{0.5pt}{\Big[} x_\text{crit}^{(t_0)}(t)\raisebox{0.5pt}{\Big]} \, \text{d}t \nonumber\\[0.2cm]
	&= 2\:\!\mathlarger{\int}_{\scalebox{0.8}{$x_\text{crit}^{(t_0)}(0)$}}^{\scalebox{0.8}{$x_\text{crit}^{(t_0)}(t_0)$}} V(\xi) \, \raisebox{0.5pt}{\Big[} \dot{x}_\text{crit}^{(t_0)}(t)\raisebox{0.5pt}{\Big]}^{-1}_{\scalebox{0.8}{$t\!\!\:\!\!\:=\!\!\:\!\!\:t(\xi)$}} \text{d}\xi \,+ 2\:\!\mathlarger{\int}^{\scalebox{0.8}{$x_\text{crit}^{(t_0)}(T)$}}_{\scalebox{0.8}{$x_\text{crit}^{(t_0)}(t_0)$}} V(\xi) \, \raisebox{0.5pt}{\Big[} \dot{x}_\text{crit}^{(t_0)}(t)\raisebox{0.5pt}{\Big]}^{-1}_{\scalebox{0.8}{$t\!\!\:\!\!\:=\!\!\:\!\!\:t(\xi)$}} \text{d}\xi \nonumber\\[0.25cm]
	&= m\omega\:\! \mathlarger{\int}_{\scalebox{0.8}{$x_\text{crit}^{(t_0)}(0)$}}^{\scalebox{0.8}{$x_\text{escape}$}} \xi \Big[1+W(\xi)\Big]^{\!\!\:\frac{1}{2}} \,\text{d}\xi + m \omega\:\! \mathlarger{\int}_{\scalebox{0.8}{$x_\text{crit}^{(t_0)}(T)$}}^{\scalebox{0.8}{$x_\text{escape}$}} \xi \Big[1+W(\xi)\Big]^{\!\!\:\frac{1}{2}} \,\text{d}\xi\, ,
	\label{eq:6_Shifted_Bounce_Action}
\end{align}
where we employed $x\raisebox{-2.5pt}{\scalebox{0.7}{\text{crit}}}\!\!\!\!\!\!\!\,\raisebox{5pt}{\scalebox{0.65}{$(t_0)$}}\:\!(t_0)=x_\text{escape}$ as well as invertibility of the bounce velocity for both halves $t\in [0,t_0)$ and $t\in (t_0,T]$. We see that the contributions arising from the local wave functions~\eqref{eq:FullExponent} exactly complement the action over the time interval $[0,T]$, meticulously constituting the missing terms for the correct exponential suppression factor to emerge. We thus obtain the full exponent
\begin{align}
	\begin{split}
		f_\text{exp}\raisebox{1pt}{\Big\llbracket}\:\! x_\text{crit}^{(t_0)}\:\!\raisebox{1pt}{\Big\rrbracket} &= 2m\omega\:\! \mathlarger{\int}_{\scalebox{0.8}{$0$}}^{\scalebox{0.8}{$x_\text{escape}$}} \xi \Big[1+W(\xi)\Big]^{\!\!\:\frac{1}{2}} \,\text{d}\xi =S_\text{E}\raisebox{1pt}{\Big\llbracket} x^{(T=\infty)}_\text{bounce}\raisebox{1pt}{\Big\rrbracket} = \mathdutchbcal{B} 
	\end{split}\,,
	\label{eq:6_Full_Exponent_Bounce}%
\end{align}
matching the well-known result from both instanton calculus over infinite time intervals as well as ususal WKB calculations.\footnote{A similar observation of the exponential piece of the WKB wave function complementing the Euclidean action of a classical tunneling trajectory has been previously reported in reference~\cite{BlumRealTimeTunneling}.} As expected, the exponent is fully $t_0$-independent due to the presence of the underlying zero mode. More yet, we indeed find that the obtained exponential suppression factor is independent of $T$, constituting the usual Euclidean action in case the finite-time bounce is extended to an infinite time interval. While certainly being a nice intermediate result, it is arguably bound to emerge per construction of the local wave functions $\psinloc$, entailing the typical WKB suppression factor. Thus, let us proceed with the investigation of the less trivial prefactor.\\

\noindent
Taking advantage of the fact that the uncovered exact zero mode $e\raisebox{-2.75pt}{\scalebox{0.65}{0}}\!\!\!\,\raisebox{5pt}{\scalebox{0.65}{$(t_0)$}}\:\!(t)$ is proportional to $\dot{x}\raisebox{-2.5pt}{\scalebox{0.7}{\text{crit}}}\!\!\!\!\!\!\!\,\raisebox{5pt}{\scalebox{0.65}{$(t_0)$}}\:\!(t)$, both the primed determinant and the Jacobian factor are easily computed using formulas~\eqref{eq:StreamlinedPrimedDeterminant} and~\eqref{eq:JacobianFactor}. Note the convenient cancellation of the zero mode norm, yielding 
\begin{align}
	\!\!\mathcal{N}_\text{E} \: \text{det}'_\zeta\raisebox{1pt}{\Big[}O_\text{crit}^{(t_0)}\raisebox{1pt}{\Big]}^{-\frac{1}{2}} \:\! \text{J}_\text{det}\scalebox{1.15}{\big(}t_0,x_\mu'\!\!\!\: =\! 0\scalebox{1.15}{\big)} &= \sqrt{\frac{2m\omega}{\hbar}} \; \Big[\dot{x}_\text{crit}^{(t_0)}(0) \, \dot{x}_\text{crit}^{(t_0)}(T)\Big]^{\frac{1}{2}} \nonumber \\
	&= \pm i\,\sqrt{\frac{2m\omega^3}{\hbar}} \, \Big[x_\text{crit}^{(t_0)}(0)\, x_\text{crit}^{(t_0)}(T)\Big]^{\frac{1}{2}} \!\:\bigg\{1+W\!\!\;\raisebox{0.5pt}{\Big[} \;\!x_\text{crit}^{(t_0)}(0)\;\!\raisebox{0.5pt}{\Big]}\!\!\;\!\!\:\bigg\}^{\!\!\:\frac{1}{4}} \nonumber \\ 
	&\phantom{\pm i\;\;\:}\qquad\qquad\qquad\qquad\qquad\quad\;\:\, \times \bigg\{1+W\!\!\;\raisebox{0.5pt}{\Big[} x_\text{crit}^{(t_0)}(T)\raisebox{0.5pt}{\Big]}\!\!\;\!\!\:\bigg\}^{\!\!\:\frac{1}{4}} \!.
	\label{eq:DeterminantJacobianFactorBounce}
\end{align}
Enforcing a sign flip of the bounce velocity renders the determinant factor negative, leading to the usual result of bounce contributions being purely imaginary. The overall sign has to be fixed by physical considerations, i.e. requesting the decay rate to be positive, which in the present case amounts to picking the plus sign. \\ 

\noindent 
The remaining unknown quantities entering expression~\eqref{eq:Final_Result_ZeroMode} arise from the endpoints of the shifted bounce trajectory  $x\raisebox{-2.5pt}{\scalebox{0.7}{\text{crit}}}\!\!\!\!\!\!\!\,\raisebox{5pt}{\scalebox{0.65}{$(t_0)$}}\:\!(0,T)$ as well as the subtracted Green's function at the boundary of the time interval $[0,T]$. With the Green's function being of order unity, see the discussion in section~\ref{sec:F3_Subtracted_Greens_Function}, we may potentially drop the derivative terms due to the factor $\sqrt{\hbar}\ll 1$ multiplying them, i.e. 
\begin{align}
	\psi_{n,\text{non-exp}}^{(\text{loc})}\scalebox{1.1}{\bigg[} x_\text{crit}^{(t_0)}\big(0,T\big)+ \sqrt{\frac{\hbar}{m\omega}}\, \hspace{-1.9cm}\underbrace{\frac{\partial}{\partial J_{0,T}}}_{\displaystyle{\mathcal{O}(1) \text{ due to } G_\text{bounce}^{\perp,(t_0)}=\mathcal{O}(1)}}\hspace{-1.85cm}\scalebox{1.1}{\bigg]} \;\xrightarrow{\,\scalebox{0.8}{$x_\text{crit}^{(t_0)}\big(0,T\big)\gg \sqrt{\hbar}$}\;} \; \psi_{n,\text{non-exp}}^{(\text{loc})}\Big[ x_\text{crit}^{(t_0)}\big(0,T\big)\Big] \, .
	\label{eq:Simplification_Local_Wave_Functions}
\end{align}
 This however requires the endpoint contributions to not be parametrically suppressed themselves, which strongly depends on the scale of $T$. To investigate the magnitude of the endpoints, we can again utilize the invertibility of the bounce velocity in appropriate domains, relating the endpoints to the time it takes for the trajectory to reach the turning point $x_\text{escape}$. For $x\raisebox{-2.5pt}{\scalebox{0.7}{\text{crit}}}\!\!\!\!\!\!\!\,\raisebox{5pt}{\scalebox{0.65}{$(t_0)$}}\:\!(0)$, one finds the expression
\begin{align}
	\omega t_0=\mathlarger{\mathlarger{\int}}_0^{t_0} \text{d}(\omega t) = \mathlarger{\mathlarger{\int}}_{\scalebox{0.75}{$x_\text{crit}^{(t_0)}(0)$}}^{\scalebox{0.75}{$x_\text{crit}^{(t_0)}(t_0)$}}  \omega \raisebox{0.5pt}{\Big[} \dot{x}_\text{crit}^{(t_0)}(t)\raisebox{0.5pt}{\Big]}^{-1}_{\scalebox{0.8}{$t\!\!\:\!\!\:=\!\!\:\!\!\:t(\xi)$}} \, \text{d}\xi = \mathlarger{\mathlarger{\int}}_{\scalebox{0.75}{$x_\text{crit}^{(t_0)}(0)$}}^{\scalebox{0.75}{$x_\text{escape}$}} \: \frac{1}{\xi} \Big[1+W(\xi)\Big]^{-\frac{1}{2}} \, \text{d}\xi \, .
\end{align}
Subtracting the dominant logarithmic dependence from the integral and employing definition~\eqref{eq:2_Definition_A_Parameter_Tunneling}, we find the important exact relations
\begin{subequations}
\begin{align}
	\omega t_0&=\log\!\!\:\Bigg[\;\!\raisebox{2pt}{$\displaystyle{\frac{\mathdutchbcal{A}}{x_\text{crit}^{(t_0)}(0)}}$}\;\!\Bigg] \, -  \mathlarger{\mathlarger{\int}}^{\scalebox{0.75}{$x_\text{crit}^{(t_0)}(0)\;\;\!\!$}}_{\scalebox{0.75}{$0$}} \frac{1}{\xi} \Big[1+W(\xi)\Big]^{-\frac{1}{2}} \; -\frac{1}{\xi}\; \text{d}\xi \, , \label{eq:ExactTimeRelationx0}   \\
	\omega \big(T-t_0\big)&= \log\!\!\:\Bigg[\raisebox{2pt}{$\displaystyle{\frac{\mathdutchbcal{A}}{x_\text{crit}^{(t_0)}(T)}}$}\Bigg] \, \:\!\underbrace{-\mathlarger{\mathlarger{\int}}^{\scalebox{0.75}{$x_\text{crit}^{(t_0)}(T)$}}_{\scalebox{0.75}{$0$}} \: \frac{1}{\xi} \Big[1+W(\xi)\Big]^{-\frac{1}{2}} \; -\frac{1}{\xi}\; \text{d}\xi}_{\displaystyle{\text{negligible in case of $x_\text{crit}^{(t_0)}\big(0,T\big)\ll 1$}}} \, .
	\label{eq:ExactTimeRelationxT} 
\end{align}
\label{eq:ExactTimeRelations}%
\end{subequations}
For increasing $T$, these relations constitute a lower bound $\mathdutchbcal{A} e^{-\omega T}$ on the magnitude of both endpoints, scaling according to  $x\raisebox{-2.5pt}{\scalebox{0.7}{\text{crit}}}\!\!\!\!\!\!\!\,\raisebox{5pt}{\scalebox{0.65}{$(t_0)$}}\:\!(0)\sim \mathdutchbcal{A}\:\! e^{-\omega t_0}$ and $x\raisebox{-2.5pt}{\scalebox{0.7}{\text{crit}}}\!\!\!\!\!\!\!\,\raisebox{5pt}{\scalebox{0.65}{$(t_0)$}}\:\!(T)\sim \mathdutchbcal{A}\:\! e^{\omega (t_0-T)}$ respectively. Thus, in case of moderately large $T$, satisfying $e^{-\omega T}\gg \sqrt{\hbar}$, we can safely ignore the Green's function contributions inside $\psilocnonexp$ as the terms arising from the endpoints dominate the argument of the wave function~\eqref{eq:Simplification_Local_Wave_Functions}. This allows us to split our discussion into two distinct cases, for which a separate set of simplifications can be inferred, being valid for different ranges of the free parameter $T$.

\subsubsection{Small and moderately large $T$}
\label{sec:Small_Moderate_T}

Following the above assessment, for $T\ll \omega^{-1} \log\!\big(\hbar^{-\frac{1}{2}}\big)$ we are allowed to drop all terms involving the subtracted Green's function from equation~\eqref{eq:Final_Result_ZeroMode}. Utilizing this simplification as well as the previous result~\eqref{eq:DeterminantJacobianFactorBounce}, we find the leading contribution arising from bounce-like motions to equation~\eqref{eq:Final_Result_ZeroMode} given by 
\begin{align}
	\!\!\Big[\mathcal{P}_{\!\!\: n}^{(\pi/2)}\big(\hbar,T\big)\Big]_{\text{bounce}}^{\text{(LO)}} &=\frac{i}{2^n n!} \, \sqrt{\frac{2m\omega^3}{\hbar}}\: \exp\!\bigg(\!\!-\!\!\:\frac{\mathdutchbcal{B}}{\hbar}\bigg)  \mathlarger{\mathlarger{\int}}_{0}^{T} \frac{\text{d}t_0}{\sqrt{2\pi}}\;  \Big[x_\text{crit}^{(t_0)}(0)\, x_\text{crit}^{(t_0)}(T)\Big]^{\frac{1}{2}} \nonumber \\[-0.05cm]  
	&\qquad\qquad\qquad\qquad\qquad\quad\:\: \times \bigg\{1+W\!\!\;\raisebox{0.5pt}{\Big[} x_\text{crit}^{(t_0)}(0)\raisebox{0.5pt}{\Big]}\!\!\:\bigg\}^{\!\!\:\frac{1}{4}} \bigg\{1+W\!\!\;\raisebox{0.5pt}{\Big[} x_\text{crit}^{(t_0)}(T)\raisebox{0.5pt}{\Big]}\!\!\:\bigg\}^{\!\!\:\frac{1}{4}} \nonumber \\[0.1cm] 
	&\qquad\qquad\qquad\qquad\qquad\quad\:\: \times  \psi_{n,\text{non-exp}}^{(\text{loc})}\!\!\:\raisebox{0.5pt}{\Big[} x_\text{crit}^{(t_0)}(0)\raisebox{0.5pt}{\Big]} \, \psi_{n,\text{non-exp}}^{(\text{loc})}\!\!\:\raisebox{0.5pt}{\Big[} x_\text{crit}^{(t_0)}(T)\raisebox{0.5pt}{\Big]} \, .
	\label{eq:BounceContributionIntermediate}
\end{align}
With the above expression~\eqref{eq:Resonant_Wave_Function} in mind, notice that the argument of the Hermite polynomial contained in $\psilocnonexp(x)$ entails a factor $\smash{\hbar^{-\frac{1}{2}}}$, rendering the combined argument much larger than unity. This again hinges upon the assumption of moderately large $T$, putting the lower bound $\mathdutchbcal{A} e^{-\omega T}$ on the size of $x\raisebox{-2.5pt}{\scalebox{0.7}{\text{crit}}}\!\!\!\!\!\!\!\,\raisebox{5pt}{\scalebox{0.65}{$(t_0)$}}\:\!(0,T)$, entering the argument linearly.  Thus, only the highest monomial influences the leading-order result, allowing us to infer the simplification
\begin{align}
	H_n\!\!\:\scalebox{1.3}{\Bigg(}\!\!\:\scalebox{1.1}{\Bigg\{}\frac{2m\omega}{\hbar x^2}\mathlarger{\mathlarger{\int}}_0^x \,\xi \Big[1+W(\xi)\Big]^{\!\!\:\frac{1}{2}}\, \text{d}\xi\:\!\scalebox{1.1}{\Bigg\}}^{\!\frac{1}{2}} x \scalebox{1.3}{\Bigg)} = 2^n \, \scalebox{1.1}{\Bigg\{}\frac{2m\omega}{\hbar x^2}\mathlarger{\mathlarger{\int}}_0^x \,\xi \Big[1+W(\xi)\Big]^{\!\!\:\frac{1}{2}}\, \text{d}\xi\:\!\scalebox{1.1}{\Bigg\}}^{\!\frac{n}{2}} \!x^n +\ldots \, ,
	\label{eq:HermiteSimplificytionLargeArgument}
\end{align}
indicating subleading corrections with dots. Inserting the given relation~\eqref{eq:HermiteSimplificytionLargeArgument} into the definition~\eqref{eq:Resonant_Wave_Function} for $\psilocnonexp(x)$, we find an intricate cancellation of several terms. With the additional factors present in equation~\eqref{eq:BounceContributionIntermediate}, we deduce 
\begin{align}
	&\,\Big[x_\text{crit}^{(t_0)}(0)\Big]^{\frac{1}{2}} \bigg\{1+W\!\!\;\raisebox{0.5pt}{\Big[} x_\text{crit}^{(t_0)}(0)\raisebox{0.5pt}{\Big]}\!\!\:\bigg\}^{\!\!\:\frac{1}{4}} \psi_{n,\text{non-exp}}^{(\text{loc})}\!\!\:\raisebox{0.5pt}{\Big[} x_\text{crit}^{(t_0)}(0)\raisebox{0.5pt}{\Big]}  \nonumber \\ 
	= &\, \bigg(\frac{4m\omega}{\hbar}\bigg)^{\!\!\frac{n}{2}} \scalebox{1.3}{\Bigg(} x_\text{crit}^{(t_0)}(0) \: \exp\!\:\!\scalebox{1.2}{\Bigg\{}\mathlarger{\mathlarger{\int}}_{0}^{\scalebox{0.7}{$\displaystyle{x_\text{crit}^{(t_0)}(0)}$}}\, \frac{1}{\xi} \Big[1+W(\xi)\Big]^{-\frac{1}{2}}-\frac{1}{\xi} \; \text{d}\xi\scalebox{1.2}{\Bigg\}}\!\!\: \scalebox{1.3}{\Bigg)}^{\!\!\!\scalebox{0.85}{$\frac{2n+1}{2}$}} \! \bigg\{1+\mathcal{O}\Big[\sqrt{\hbar} \: x_\text{crit}^{(t_0)}(0)\big.^{-1}\Big]\!\!\:\bigg\} \nonumber \\[0.1cm] 
	= &\, \bigg(\frac{4m\omega}{\hbar}\bigg)^{\!\!\frac{n}{2}} \Big(\mathdutchbcal{A} e^{-\omega t_0}\Big)^{\!\scalebox{0.85}{$\frac{2n+1}{2}$}}  \bigg\{1+\mathcal{O}\Big[\sqrt{\hbar} \: e^{\omega T}\Big]\!\!\:\bigg\} \, ,
	\label{eq:Simplified_Integrand}
\end{align}
where in the last line we employed the exact relation~\eqref{eq:ExactTimeRelationx0}. The above deduction can be repeated identically for all terms entailing the remaining endpoint $x\raisebox{-2.5pt}{\scalebox{0.7}{\text{crit}}}\!\!\!\!\!\!\!\,\raisebox{5pt}{\scalebox{0.65}{$(t_0)$}}\:\!(T)$; using equation~\eqref{eq:ExactTimeRelationxT} instead, the sole difference is the replacement $t_0\mapsto T-t_0$ in the final result~\eqref{eq:Simplified_Integrand}. Notice that due to utilizing~\eqref{eq:ExactTimeRelations}, we at no point required to assume the endpoints to be small, which, for short times $T$, would not be true. Multiplying both expressions, the $t_0$-dependence fully drops out, such that the collective coordinate integral over $t_0$ yields the desired linear factor $T$. We have thus seen that it is less trivial to spot that the integrand to leading-order really is independent of the collective coordinate $t_0$, even though this must be the case in the end. Dropping all subleading corrections, we finally find the bounce contribution given by 
\begin{align}
	\Big[\mathcal{P}_{\!\!\: n}^{(\pi/2)}\big(\hbar,T\big)\Big]_{\text{bounce}}^{\text{(LO)}} &=\frac{i}{n!} \, \frac{\omega T}{\sqrt{2\pi}}\:  \scalebox{1.1}{$\bigg($}\frac{2m\omega \mathdutchbcal{A}^2 e^{-\omega T}}{\hbar}\scalebox{1.1}{$\bigg)$}^{\!\!\!\!\: \scalebox{0.85}{$\frac{2n+1}{2}$}}  \exp\!\bigg(\!\!-\!\!\:\frac{\mathdutchbcal{B}}{\hbar}\bigg)  \, .
	\label{eq:BounceContributionFinal}
\end{align}
Remarkably, this result is valid for a wide range of parameters $T$, adhering to the previously stated bound, allowing us to safely neglect corrections of order $\sqrt{\hbar} \: e^{\omega T}\ll 1$. Moreover, it should be noted that the previously seen cancellation~\eqref{eq:Simplified_Integrand} of factors inside the integrand is solely possible due to the speficic form of $\psilocnonexp(x)$. For small times $T$ and consequently endpoints close to $x_\text{escape}$, the wave function is probed deep inside the barrier region, making it inevitable to employ the provided uniform WKB approximation~\eqref{eq:Resonant_Wave_Function}. The demand for resolving the correct asymptotic behavior both in the harmonic region of the potential as well as deep inside the barrier, while still possessing a smooth transition between the two, retrospectively explains this necessary accuracy of the approximation that may formerly have appeared arbitrary.

\subsubsection{Large $T$}
\label{sec:Large_T}

For increasing $T$, the approximations~\eqref{eq:Simplification_Local_Wave_Functions} and~\eqref{eq:HermiteSimplificytionLargeArgument} eventually break down. Instead of requiring $e^{-\omega T} \gg \sqrt{\hbar}$ as in the previous section, let us resume under the condition  $e^{-\omega T}\ll 1$.\footnote{However, to subsequently reproduce equation~\eqref{eq:BounceContributionFinal}, we will eventually find that $T$ cannot be arbitrarily large. This leads to the roadblock that a further simplification based on $e^{-\omega T} \gg \hbar$ becomes mandatory for the case $n \neq 0$. The necessity of this initially superfluous constraint will be explained in section~\ref{sec:3_9_BreakDown}.} Notice that due to working in the semiclassical regime $\hbar \ll 1$, both cases significantly overlap, guaranteeing a seamless transition. Under the present assumption of large $T$, the magnitude of both endpoints $x\raisebox{-2.5pt}{\scalebox{0.7}{\text{crit}}}\!\!\!\!\!\!\!\,\raisebox{5pt}{\scalebox{0.65}{$(t_0)$}}\:\!\,(0,T)$ is generally found to be strongly suppressed.\footnote{Beware of the collective-coordinate integration regions for which $t_0$ is either very small or close to $T$, such that one of the two endpoints is still unsuppressed. In these regions all terms containing the unsuppressed endpoint can be treated using the procedure illustrated in section~\ref{sec:Small_Moderate_T}, such that we may proceed under the premise of both endpoints satisfying $x\raisebox{-2.5pt}{\scalebox{0.7}{\text{crit}}}\!\!\!\!\!\!\!\,\raisebox{5pt}{\scalebox{0.65}{$(t_0)$}}\:\!\,(0,T)\ll 1$.} Due to the aforementioned assumption we can expand the local wave function $\psilocnonexp(x)$ as well as the remaining terms entailed in relation~\eqref{eq:BounceContributionIntermediate} for small arguments. This allows us to once more employ the simplified relation~\eqref{eq:4_Asymptotics_Non_Exponential_Wave_Function_Small_x} for the non-exponential wave function $\psilocnonexp$, from which we conclude that we can approximate parts of equation~\eqref{eq:Final_Result_NoZeroMode} utilizing
\begin{align}
	\psi_{n,\text{non-exp}}^{(\text{loc})}\scalebox{1.1}{$\bigg[$} \,\underbrace{\!x_\text{crit}^{(t_0)}\big(0,T\big)+ \sqrt{\frac{\hbar}{m\omega}}\, \frac{\partial}{\partial J_{0,T}}\!}_{\rule{0pt}{0.32cm}\displaystyle{\text{both addends $\ll 1$}}}\,\scalebox{1.1}{$\bigg]$} = H_n\scalebox{1.1}{$\bigg[$} \sqrt{\frac{m\omega}{\hbar}}\; x_\text{crit}^{(t_0)}\big(0,T\big)+ \frac{\partial}{\partial J_{0,T}}\scalebox{1.1}{$\bigg]$} + \text{subleading} \, .
	\label{eq:Simplified_Wave_Function}
\end{align}
Similarly, the exact result~\eqref{eq:DeterminantJacobianFactorBounce} can be drastically simplified by dropping terms containing the auxiliary potential $W$, which for small endpoints yield negligible contributions. Utilizing the mentioned approximations together with relation~\eqref{eq:ExactTimeRelations}, truncated to leading-order in both endpoints, one obtains the simple $t_0$-independent expression
\begin{align}
	\!\!\mathcal{N}_\text{E} \: \text{det}'_\zeta\raisebox{1pt}{\Big[}O_\text{crit}^{(t_0)}\raisebox{1pt}{\Big]}^{-\frac{1}{2}} \:\! \text{J}_\text{det}\scalebox{1.15}{\big(}t_0,x_\mu'\!\!\!\: =\! 0\scalebox{1.15}{\big)} &= i\,\sqrt{\frac{2m\omega^3}{\hbar}} \, \mathdutchbcal{A} \, \exp\!\bigg(\!\!\!\!\:-\!\frac{\omega T}{2}\bigg) \, \bigg\{1+\mathcal{O}\Big[x_\text{crit}^{(t_0)}\big(0,T\big)\Big]\!\bigg\} \, .
	\label{eq:DeterminantJacobianFactorBounce_Simplified}
\end{align}
Inserting the above relations~\eqref{eq:Simplified_Wave_Function} and~\eqref{eq:DeterminantJacobianFactorBounce_Simplified} into the full expression~\eqref{eq:Final_Result_ZeroMode} for the endpoint-weighted path integral, we obtain the intermediate leading-order result
\begin{align}
	\!\!\Big[\mathcal{P}_{\!\!\: n}^{(\pi/2)}\Big]_{\text{bounce}}^{\text{(LO)}} &=\frac{i}{2^n n!}\,\sqrt{\frac{2m\omega^3}{\hbar}} \, \mathdutchbcal{A} \, \exp\!\bigg(\!\!\!\!\:-\!\frac{\omega T}{2}\bigg) \, \exp\!\bigg(\!\!-\!\!\:\frac{\mathdutchbcal{B}}{\hbar}\bigg)   \nonumber \\[0.1cm] 
	&\phantom{=}\times \mathlarger{\mathlarger{\int}}_{0}^{T} \frac{\text{d}t_0}{\sqrt{2\pi}} \; \Bigg\{ H_n\scalebox{1.1}{$\bigg[$} \sqrt{\frac{m\omega}{\hbar}}\; x_\text{crit}^{(t_0)}(0)+ \frac{\partial}{\partial J_{0}}\scalebox{1.1}{$\bigg]$} \: H_n\scalebox{1.1}{$\bigg[$} \sqrt{\frac{m\omega}{\hbar}}\; x_\text{crit}^{(t_0)}(T)+ \frac{\partial}{\partial J_{T}}\scalebox{1.1}{$\bigg]$} \label{eq:IntermediateResultBounceLargeT} \\[0.05cm] 
	&\phantom{=}\times \exp\!\scalebox{1.1}{\bigg[}\:\!\frac{1}{2}\,J_0\big.^{\!\!\!\: 2}\, G_\text{bounce}^{\perp,(t_0)}\big(0,0\big)+J_0\:\! J_T\,  G_\text{bounce}^{\perp,(t_0)}\big(0,T\big)+\frac{1}{2}\,J_T\big.^{\!\!\! 2}\, G_\text{bounce}^{\perp,(t_0)}\big(T,T\big)\scalebox{1.1}{\bigg]}\!\!\: \Bigg\}_{\substack{\scalebox{0.85}{$J_0\!\, =\! 0$}\\[0.025cm] \scalebox{0.85}{$J_T\! =\! 0$}}} \: . \nonumber
	\label{eq:Bounce_contribution_Intermediate_LargeT}
\end{align}
The integrand found this way is of the specific form~\eqref{eq:6_Definition_Auxiliary_Function_HermitePolys}, which we investigate in appendix~\ref{sec:E_HermiteIdentity}. Employing the simplified parametrization
\eqref{eq:Simplified_Parametrization_I_n}, we obtain the relevant auxiliary constants $\sigma_{1,2}$ and $\kappa_{1,2,3}$ in the schematic form 
\begin{align}
	\begin{matrix}
		\displaystyle{\sigma_1 =\sqrt{\frac{m\omega}{\hbar}}\:  \frac{2 x_T G_{0T}-x_0 \big(2G_{TT}-1\big) }{\big(2G_{00}-1\big) \big(2G_{TT}-1\big)-4 G_{0T}^2} \, ,} \\[0.55cm] 
		\displaystyle{\sigma_2 =\sqrt{\frac{m\omega}{\hbar}}\:  \frac{2 x_0 G_{0T}-x_T \big(2G_{00}-1\big) }{\big(2G_{00}-1\big) \big(2G_{TT}-1\big)-4 G_{0T}^2} \, ,} 
	\end{matrix}
	\qquad \qquad 
	\begin{matrix}
		\kappa_1 = 2G_{00}-1\, , \\[0.25cm]
		\;\,\kappa_2 = 2G_{TT}-1\, , \\[0.25cm]
		\!\!\!\!\!\!\!\!\kappa_3 = 4G_{0T}\, .
	\end{matrix} \quad
\end{align}
The study of the subtracted Green's function was outsourced to appendices~\ref{sec:F3_Subtracted_Greens_Function} and~\ref{sec:Homogeneous_Solutions}, yielding the final result~\eqref{eq:Subtracted_Greens_Function_Bounce_Final}. Assuming $W'(0)\neq 0$, all terms involving $G_{0T}$ are strongly suppressed and can be rightfully dropped, yielding (up to subleading corrections)\footnote{While we assume all subsequent conclusions including the result~\eqref{eq:Result_In_LargeT_Approx} to still hold in the case of $W'(0)$ being parametrically suppressed (or zero), the leading-order approximation of the Green's function~\eqref{eq:Subtracted_Greens_Function_Bounce_Final} would not suffice for such an analysis. For brevity, we thus restrict our discussion to the simpler case $W'(0)=\mathcal{O}(1)$.}
\begin{align}
	\sigma\coloneqq \sigma_1&=\sigma_2= \sqrt{\frac{m\omega}{\hbar}}\: \Big[2W'(0)\Big]^{-1} \, , & \kappa_1 &= -2W'(0)\, x_\text{crit}^{(t_0)}(0)\, , & \kappa_2 &= -2W'(0)\, x_\text{crit}^{(t_0)}(T)\, .
\end{align} 
With this, the full integrand $\mathcal{I}_n$ encountered in relation~\eqref{eq:IntermediateResultBounceLargeT} takes the simple form 
\begin{align}
	\!\!\mathcal{I}_n&=e^{-(\kappa_1+\kappa_2)\:\!\sigma^2} \; \frac{\partial^{2n}}{\partial v^n\partial w^n}\, \bigg\{ \! \exp\!\Big[\kappa_1 v^2+\kappa_2 w^2\Big]\!\!\:\bigg\}_{\! \scalebox{0.75}{$\,v\!\!\:=\!\!\: w \!\!\: =\!\!\: \sigma$}} \Bigg(1+ \mathcal{O}\:\!\scalebox{1.3}{\Big\{}\!\!\;x_\text{crit}^{(t_0)}\big(0,T\big)\:\! \log\!\!\!\:\raisebox{-0.25pt}{$\scalebox{1.35}{\big[}$} x_\text{crit}^{(t_0)}\big(0,T\big)\raisebox{-0.25pt}{$\scalebox{1.35}{\big]}$}\!\!\!\: \scalebox{1.3}{\Big\}}\!\Bigg) \nonumber \\[0.1cm]
	&= \bigg(\frac{2}{\sigma}\bigg)^{\!\!\!\: 2n} \Bigg[\sqrt{\pi}\: e^{-\kappa_1\:\!\sigma^2} \, _2\widetilde{F}_2\!\left(\frac{1}{2},1;\frac{1-n}{2},\frac{2-n}{2};\kappa_1 \sigma^2\right)\!\!\:\Bigg] \!\!\:\times\!\!\: \Big[\kappa_1\leftrightarrow \kappa_2\Big] \!\!\:\times\!\!\: \Big[1+\mathcal{O}\big(T e^{-\omega T}\big)\Big] \!\: ,
	\label{eq:Result_In_LargeT}
\end{align}	
where $_2\widetilde{F}_2$ denotes a regularized hypergeometric function. Note that the term in square brackets is simply a polynomial in the parameter $\kappa_1\sigma^2$. With $\sigma$ constituting an overall constant solely dependent on the shape of the potential, it is immanent that the behavior of the integrand $\mathcal{I}_n$ is fully controlled by the two parameters 
\begin{align}
	\kappa_{1,2} \:\!\sigma^2 = -\frac{m\omega}{2\hbar \:\! W'(0)}\: x_\text{crit}^{(t_0)}\big(0,T\big) \, .
\end{align}
Once more, notice that the magnitude of the bounce endpoints is bounded from below, being larger than $\mathdutchbcal{A} e^{-\omega T}$. Thus, in the case $e^{-\omega T} \gg \hbar$ we conclude $\kappa_{1,2} \:\!\sigma^2 \gg 1$, with which expression~\eqref{eq:Result_In_LargeT} simplifies substantially since only the highest monomial power contributes, yielding 
\begin{align}
	\mathcal{I}_n&= \bigg(\frac{2}{\sigma}\bigg)^{\!\!\!\: 2n} \big(\kappa_1\sigma^2\big)^n \big(\kappa_2\sigma^2\big)^n \, \bigg\{1+\mathcal{O}\Big[\big(\kappa_{1,2}\sigma^2\big)^{-1},T e^{-\omega T}\Big]\!\!\;\bigg\} \nonumber \\[0.05cm]
	&= \bigg(\frac{2m\omega \mathdutchbcal{A}^2 e^{-\omega T}}{\hbar}\bigg)^{\!\! n} \,\Big[1+\mathcal{O}\scalebox{1.1}{\big(}\hbar e^{\omega T},T e^{-\omega T}\scalebox{1.1}{\big)}\Big]\: .
	\label{eq:Result_In_LargeT_Approx}
\end{align}	
Similar to the previous discussion, the collective coordinate integral is left trivial, such that after collecting all results, we fully recover the above expression~\eqref{eq:BounceContributionFinal} as derived for small and moderately large $T$. As the presented relations hold for times $T$ exceeding the previous bound demanded in section~\ref{sec:Small_Moderate_T}, we were partly successful in slightly extending the range of applicability of the computation. However, once $e^{-\omega T}$ and $\hbar$ become of comparable size, there is no hope of maintaining the found relation~\eqref{eq:BounceContributionFinal} and the provided ansatz eventually breaks down.\footnote{Note that for $n=0$, this problematic behavior does not materialize as one can directly evaluate expression~\eqref{eq:Bounce_contribution_Intermediate_LargeT}, being fully independent of the employed $T$ (still provided $e^{-\omega T} \ll 1$). Thus, $n=0$ constitutes the sole case for which this breakdown does not occur naturally, which can be easily understood given the reasoning in section~\ref{sec:3_9_BreakDown}.} This is most easily seen when consulting the full result~\eqref{eq:Result_In_LargeT}, increasingly deviating from the required expression~\eqref{eq:Result_In_LargeT_Approx} for growing $T$. As we will argue in section~\ref{sec:3_9_BreakDown}, this behavior is expected, showing that there is nothing wrong with our computation per se, but instead we treated the provided ansatz~\eqref{eq:Resonant_Energy_Conjecture} without sufficient care. Let us refrain from reasoning this breakdown at this point and let us first collect the obtained results in order to see that indeed, for reasonably large $T$, we obtain the known results.

\subsection{Extracting the decay rate}
\label{sec:3_7_Results}

To conclude the calculation, it remains to insert the results~\eqref{eq:FV_Contribution} and~\eqref{eq:BounceContributionFinal} into relation~\eqref{eq:Resonant_Energy_Conjecture}, given $\theta=\pi/2$. It is important to stress that, even though there is no shot trajectory present in the given expansion, the previously seen overlap of FV and bounce thimbles as schematically depicted in figure~\ref{fig:Tunneling_Thimble_Structure} appears to remain valid, i.e. we have to instate an additional factor $\frac{1}{2}$ for the bounce solution. Up to subleading corrections neglected in the derivation of relations~\eqref{eq:FV_Contribution} and~\eqref{eq:BounceContributionFinal}, we obtain 
\begin{align}
	E_n^{(\text{loc})} &= -\frac{\hbar}{T} \;\log\!\!\:\scalebox{1.3}{\Bigg\{}\!\!\!\: \exp\!\!\!\;\scalebox{1.1}{$\bigg[$}\!-\!\!\:\omega T \:\!\bigg(\!\:\!n+\frac{1}{2}\bigg)\:\!\!\scalebox{1.1}{$\bigg]$} \scalebox{1.1}{\Bigg[}1+\frac{1}{2} \frac{i}{n!} \, \frac{\omega T}{\sqrt{2\pi}}\:  \scalebox{1.1}{$\bigg($}\frac{2m\omega \mathdutchbcal{A}^2}{\hbar}\scalebox{1.1}{$\bigg)$}^{\!\!\!\!\: \scalebox{0.85}{$\frac{2n+1}{2}$}}  \exp\!\bigg(\!\!-\!\!\:\frac{\mathdutchbcal{B}}{\hbar}\bigg)\scalebox{1.1}{\Bigg]}\!\scalebox{1.3}{\Bigg\}} \nonumber \\ 
	&= \hbar\omega \bigg(\!\:\!n+\frac{1}{2}\bigg) - \frac{i}{n!} \, \frac{\hbar \omega}{\sqrt{8\pi}}\:  \scalebox{1.1}{$\bigg($}\frac{2m\omega \mathdutchbcal{A}^2}{\hbar}\scalebox{1.1}{$\bigg)$}^{\!\!\!\!\: \scalebox{0.85}{$\frac{2n+1}{2}$}}  \exp\!\bigg(\!\!-\!\!\:\frac{\mathdutchbcal{B}}{\hbar}\bigg) \, \bigg[1+\mathcal{O}\Big(T e^{-\mathdutchbcal{B}/\hbar}\Big)\bigg] \, .
	\label{eq:Resonant_Energy_Conjecture_Inserted}
\end{align}
Separating both factors by utilizing the additive properties of the logarithm, the remaining term has been expanded using $\log(1+x)=x+\mathcal{O}(x^2)$, once more employing that $T$ cannot be arbitrarily large, otherwise invalidating several steps in the derivation.\footnote{For very large times $T\sim e^{-\mathdutchbcal{B}/\hbar}$, multi-instantons cannot be ignored. However, it is found that the computation is always bound to fail for such large time scales, explicitly excluding multi-instanton contributions to ever be important when utilizing the pursued ansatz~\eqref{eq:Resonant_Energy_Conjecture}, see section~\ref{sec:3_9_BreakDown}.} This directly leads us to the decay width $\Gamma\raisebox{-1.5pt}{$\scalebox{0.75}{\!$n$}$}\:\!$, arriving at the full leading-order result 
\begin{equation}
	\Gamma\raisebox{-1.5pt}{$\scalebox{0.75}{\!$n$}$}=-\frac{2}{\hbar} \,\text{Im}\Big[E_n^{\scalebox{0.8}{$(\text{loc})$}}\Big]=\frac{1}{n!}\left(\frac{2m\omega \mathdutchbcal{A}^2}{\hbar}\right)^{\!\!\scalebox{0.8}{$n$}} \hspace{-0.58cm}\underbrace{\sqrt{\frac{m\omega^3 \mathdutchbcal{A}^2}{\pi \hbar}} \, \exp\!\left(\!-\frac{\mathdutchbcal{B}}{\hbar}\right)}_{\rule{0pt}{0.35cm}\displaystyle{\text{ground state decay width }\Gamma\raisebox{-2pt}{$\scalebox{0.7}{\!0}$}}}\hspace{-0.58cm}\, .
	\label{eq:2_Full_Result_Decay_Width_InstantonCalc}
\end{equation}
Notice that the initial assumption $\Enloc=\mathcal{O}(\hbar)$ has the effect of all non-trivial behavior being encapsulated inside the non-exponential prefactor, whereas the exponential suppression factor $\mathdutchbcal{B}$ is left invariant. The result~\eqref{eq:2_Full_Result_Decay_Width_InstantonCalc} coincides with the result~\eqref{eq:2_Full_Result_Decay_Width_WKB} obtained from a straightforward WKB analysis, independently derived in appendix~\ref{sec:B_DecayWidthsTraditionalWKB}. Moreover, employing the previously mentioned substitutions~\eqref{eq:Comparison_Parameters_References}, we find agreement with the general result obtained by Weiss and Haeffner~\cite{WeissHaeffnerMetastableDecayFiniteTemp}, which had been previously derived in references~\cite{PatrascioiuComplexTime,LevitBarrierPenetrationTraceMethod} without representing it in the condensed form~\eqref{eq:2_Full_Result_Decay_Width_InstantonCalc}. We also find accordance with exemplary results for specific potentials given in e.g. references~\cite{BenderAnharmonicOscillator,LiangPeriodicInstantons,MarinoAdvancedQM} as well as the usual WKB expressions found in discussions of tunnel splittings~\cite{GargTunnelingRevisited}. We crucially kept the parameter $T$ variable during the complete derivation of the result~\eqref{eq:2_Full_Result_Decay_Width_InstantonCalc}, such that we find the ansatz~\eqref{eq:Resonant_Energy_Conjecture} to indeed be viable for practically arbitrary $T$. As we have seen in section~\ref{sec:Large_T}, only for very large $T$ obeying $e^{-\omega T} \slashed{\gg} \hbar$, the derivation eventually is rendered invalid, which we will further elaborate on in section~\ref{sec:3_9_BreakDown}. As we will uncover, this breakdown is actually expected, resulting from a na\"{\i}ve treatment of the provided ansatz~\eqref{eq:Resonant_Energy_Conjecture}.

\subsection{Extension to arbitrary $\theta$}
\label{sec:3_8_Extensions} 

Before delivering a heuristic explanation for the found breakdown of the approximation~\eqref{eq:Resonant_Energy_Conjecture} to the resonant energy at large times $T$, let us briefly sketch how the presented method can be generalized to incorporate arbitrary Wick-rotation angles $\theta$. As we will see, the derivation runs perfectly parallel to the previous discussion, requiring no additional groundbreaking ideas. Despite this, it will provide us with some additional insights on the method’s failure for sizable $T$ while illuminating the relation to former works regarding real-time tunneling
calculations. With a full real-time picture for quantum tunneling utilizing the path integral approach being of considerable interest, the present subsection may provide useful material toward this goal. However, we note that the remainder of the paper can be understood independently of the present subsection, which is rather technical. In case of the traditional instanton method for computing the ground state decay width $\Gamma\raisebox{-2pt}{$\scalebox{0.7}{\!0}$}\,$, the extension to arbitrary $\theta$ has been carried out by Ai et al.~\cite{GarbrechtFunctionalMethods}. In the general case $\theta\neq \pi/2$, the formerly employed Laplace's method presented in section~\ref{sec:3_2_GeneralEvaluationTactics} is exchanged in favor of the method of steepest descent,\footnote{Beware that even the real Euclidean case is to be treated using this more general framework, however for the sake of simplicity we held back from this rather technical discussion.} which, in case of multi-dimensional integrals, exploits ideas from Picard--Lefschetz theory~\cite{PhamPicardLefschetz,HowlsPicardLefschetzTheory,WittenAnalyticContinuation,TanizakiLefschetz,UnsalTowardsPicLefschetz}.\\ 

\begin{figure}[h]
	\centering
	\includegraphics[width=0.78\textwidth]{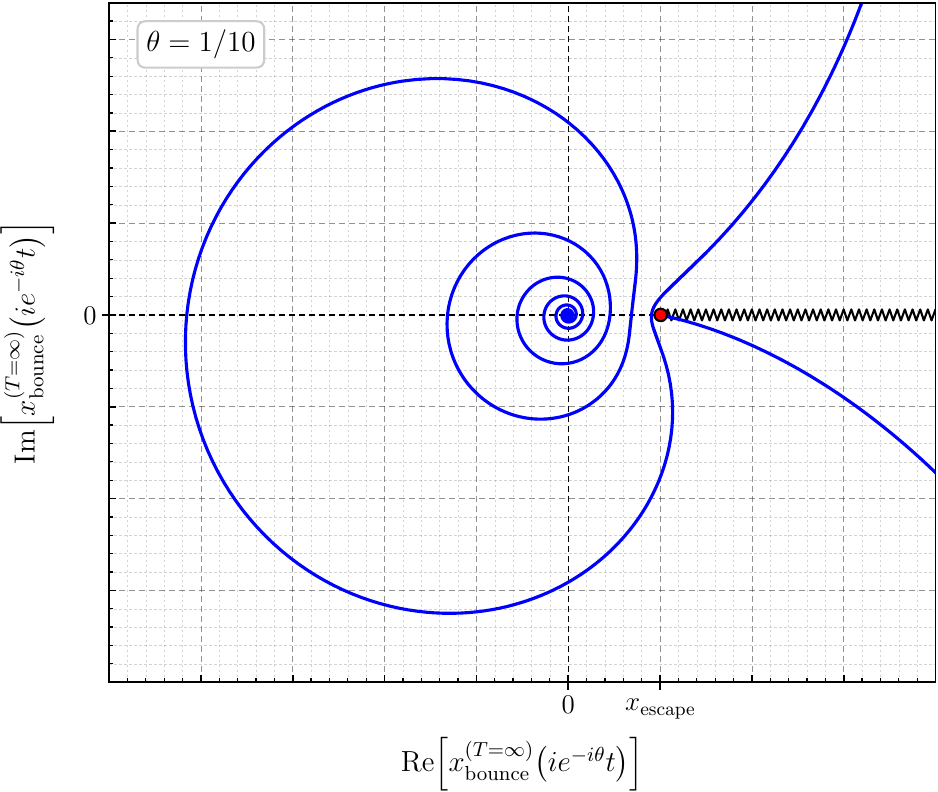}
	\caption{\hspace{0.001cm}}
	\vspace{-0.54cm}\hspace{1.5cm}
	\begin{tikzpicture}
		\node[anchor=center, inner sep=0pt, 
		minimum width=14.35cm, text width=14.35cm-0.4cm,align=justify,
		minimum height=3.0cm, 
		text depth=3.0cm-0.4cm] (0,0) {Schematic depiction of an exemplary complexified bounce trajectory $x\raisebox{-2.5pt}{\scalebox{0.7}{\text{bounce}}}\!\!\!\!\!\!\!\!\!\!\!\!\!\!\:\raisebox{5pt}{\scalebox{0.65}{$(T\!=\!\infty)$}}\big(ie^{-i\theta}t\big)$ for $\theta=1/10$. The (full) motion starts for $t\to -\infty$ at the origin with infinitesimal velocity, spiraling outwards before approaching $x_\text{escape}$, where the motion ultimately reverses at $t=0$. Note that parts of the outermost revolution before reaching $x_\mathrm{escape}$ are not fitted within the cutout. The zig-zag line indicates the branch cut of the formerly constructed resonant wave function $\psinlocLO$. While coming arbitrarily close, the trajectory never crosses this strip, no matter the chosen value for $\theta$.};
	\end{tikzpicture}
	\label{fig:Complexified_Bounce}
\end{figure}

\noindent
With the exponential functional $f_\text{exp}\llbracket x\rrbracket$ taking complex values for $\theta \neq \pi/2$, we are required to seek critical solutions in the full complexified function space $\mathcal{C}^{\:\!\mathbb{C}}\big([0,T]\big)$. The emerging discussion is identical to the Euclidean case presented in section~\ref{sec:CriticalPaths}, yielding the three conditions 
\begin{subequations}
	\begin{align}
		\dot{x}_\text{crit}(t)\big.^2 &=\:\!\big(ie^{-i\theta}\omega\big)^2 \, x_\text{crit}(t)\big.^2 \;\!\!\:\bigg\{\!\:1 +\!\; W\!\!\:\mathlarger{\big[}x_\text{crit}(t)\mathlarger{\big]}\!\!\:\bigg\} + \frac{2E_\text{crit}}{m}\, , & & \begin{matrix}
			\text{(integrated)} \\
			\text{Euler--Lagrange equation}
		\end{matrix} \label{eq:Euler_Lagrange_Equation_ArbitraryTheta} \\
		\dot{x}_\text{crit}(0) &= \;\;\; \phantom{+}ie^{-i\theta}\omega \:\!\:\! x_\text{crit}(0) \!\:\:\bigg\{1+\!\: W\!\!\:\mathlarger{\big[} x_\text{crit}(0)\mathlarger{\big]}\!\!\:\bigg\}^{\!\!\:\frac{1}{2}}\, ,& &\; \begin{matrix}
			\text{transversality condition} \\
			\text{at the left boundary}
		\end{matrix} \label{eq:Left_Transversality_Condition_ArbitraryTheta} \\
		\dot{x}_\text{crit}(T) &= \;\;\; -ie^{-i\theta}\omega \:\!\:\! x_\text{crit}(T) \!\: \bigg\{1+W\!\!\:\mathlarger{\big[} x_\text{crit}(T)\mathlarger{\big]}\!\!\:\bigg\}^{\!\!\:\frac{1}{2}}\, . & &\; \begin{matrix}
			\text{transversality condition} \\
			\text{at the right boundary}
		\end{matrix} \label{eq:Right_Transversality_Condition_ArbitraryTheta}
	\end{align}
	\label{eq:All_Extremal_Conditions_ArbitraryTheta}%
\end{subequations}
Once more, the critical energy $E_\text{crit}$ is required to vanish, constraining the set of admissible critical trajectories to the trivial FV solution as well as a family of bounce-related trajectories. Closely inspecting all constraints~\eqref{eq:All_Extremal_Conditions_ArbitraryTheta}, the resulting bounce-like trajectories are related to the Euclidean bounce solution by the simple analytic continuation $t\mapsto ie^{-i\theta}t$, basically reverting the effects of Euclidean time. The solution obtained this way, schematically illustrated in figure~\ref{fig:Complexified_Bounce}, possesses precisely the same form as the highly oscillatory trajectories usually encountered when engaged with problems concerning real-time tunneling~\cite{BenderTunnelingAsAnomaly,TurokRealTimeTunneling,UnsalRealTimeInstantons,GarbrechtFunctionalMethods,NishimuraRealTimeTunneling,BlumRealTimeTunneling,SteingasserRealTimeInstantons}. As these solutions arise by analytic continuation, it is straightforward to confirm that their partially Wick-rotated action $S_\theta$ coincides with the Euclidean pendant~\eqref{eq:6_Shifted_Bounce_Action}. Due to constructing the local wave functions $\psinloc$ to be holomorphic inside the domain of interest,\footnote{Note that while the analytically continued bounce motion probes the wave function at arbitrarily large values deep inside the FV region, it never crosses the branch cut of $\psinloc$ on the real axis, see figure~\ref{fig:Complexified_Bounce}. This guarantees that we are allowed to utilize expression~\eqref{eq:Resonant_Wave_Function}.} the full exponent in equation~\eqref{eq:Pn:exp:nonexp} can be acquired by adding the previously found integral representations of the finite-time action and the exponential wave function contribution $\psilocexp$, despite the endpoints now generally taking complex values. Once more, both parts intricately supplement each other in order for the real, infinite-time bounce action $\mathdutchbcal{B}$ to emerge. Similarly to before, all $t_0$-dependence in the exponent is fully canceled. \\ 

\noindent 
The fluctuation factors are less trivial to come by, as the desired eigenfunction basis $\big\{e_\mu(t)\!\!\:\big\}_{\scalebox{0.7}{$\mu\!\in\!\mathbb{N}^0$}}$ is required to satisfy the modified eigenvalue problem
\begin{align}
	O_\text{crit}^{(\theta)} \,  e_\mu(t)=-ie^{i\theta}\scalebox{1.15}{$\bigg\{$}\!\!\!\:-\!\!\:\frac{\text{d}^2}{\text{d}(\omega t)^2} - e^{-2i\theta} \frac{V''\big[x_\text{crit}(t)\big]}{m\omega^2}\scalebox{1.15}{$\bigg\}$} \, e_\mu(t)=\lambda_\mu \closure{e_\mu(t)}
	\label{eq:Fluctuation_Operator_ArbitraryTheta}
\end{align}
to real, non-negative eigenvalue $\lambda_\mu \geq 0$, obeying the formerly faced Robin boundary conditions~\eqref{eq:Robin_BoundaryConditions}, now also containing an additional factor $ie^{-i\theta}$. The worrisome difference to the Euclidean case~\eqref{eq:Fluctuation_Operator} is the newly entailed complex conjugation in equation~\eqref{eq:Fluctuation_Operator_ArbitraryTheta}, arising from solving the linearized gradient flow equation in order to parameterize the (linearized) steepest descent thimbles. For more details, consult the references covering Picard--Lefschetz theory in greater depth, e.g. references~\cite{WittenAnalyticContinuation,GarbrechtFunctionalMethods,TanizakiLefschetz}. Constituting a further caveat, the integration measure entails a supplementary phase factor arising from the angle of intersection between the original integration contour and the linearized Lefschetz thimble. However, it was shown that instead of computing the arising determinant of the modified eigenvalue problem~\eqref{eq:Fluctuation_Operator_ArbitraryTheta}, one can similarly study the determinant resulting from the proper eigenvalue equation $O\;\!\!\raisebox{-2.5pt}{\scalebox{0.7}{\text{crit}}}\!\!\!\!\!\!\raisebox{5pt}{\scalebox{0.65}{$(\theta)$}}\; \hat{e}_\mu(t)=\hat{\lambda}_\mu \hat{e}_\mu(t)$, readily incorporating the complete aforementioned phase information~\cite{GarbrechtFunctionalMethods}. The proper eigenfunctions $\hat{e}_\mu(t)$ are constrained to be purely real, whereas the corresponding eigenvalues $\smash{\hat{\lambda}_\mu}$ are generally complex, allowing us to employ the usual Gel'fand-Yaglom method introduced in appendix~\ref{sec:F1_Sturm_Liouville_Determinants}. Again, the relevant solutions $\mathpzc{y}\raisebox{-3.25pt}{\scalebox{0.75}{$\lambda\!\!\:\!\!\:=\!\!\:\!\!\:0$}}\!\!\!\!\!\!\!\!\,\!\raisebox{4.75pt}{\scalebox{0.65}{$(1,2)$}}(t)$ to eigenvalue zero required to compute the determinant are trivially related to their previously encountered Euclidean pendants by the same analytical continuation, $t\mapsto ie^{-i\theta}t$. The resulting change in the FV determinant~\eqref{eq:6_False_Vacuum_Determinant_Full} is given by the simple replacement $\smash{e^{\omega T} \mapsto e^{ie^{-i\theta}\omega T}}$, whereas for bounce-like motions one attains a similar expression to~\eqref{eq:DeterminantJacobianFactorBounce}, with the former critical trajectory replaced by the $\theta$-dependent one found above. Note that relation~\eqref{eq:DeterminantJacobianFactorBounce} will also acquire an additional overall factor $ie^{-i\theta}$ arising from the Jacobian~\eqref{eq:JacobianFactor} due to expressing $e\raisebox{-2.75pt}{\scalebox{0.65}{0}}\!\!\!\,\raisebox{5pt}{\scalebox{0.65}{$(t_0)$}}\:\!(t)$ in terms of the $t_0$-derivative of $x\raisebox{-2.5pt}{\scalebox{0.7}{\text{crit}}}\!\!\!\!\!\!\!\,\raisebox{5pt}{\scalebox{0.65}{$(t_0)$}}\:\!\,(t)$. \\

\noindent 
In case of the trivial FV trajectory, an expression for the Green's function is found similarly to the procedure employed to arrive at relation~\eqref{eq:6_Greens_Function_FV}, as only the analytically continued solutions $\mathpzc{y}\raisebox{-3.25pt}{\scalebox{0.75}{$\lambda\!\!\:\!\!\:=\!\!\:\!\!\:0$}}\!\!\!\!\!\!\!\!\!\!\:\:\!\raisebox{4.75pt}{\scalebox{0.65}{$(1,2)$}}(t)$ enter the discussion. Without repeating the previous steps, the final result is indeed the expected expression
\begin{align}
	\Big[\mathcal{P}_{\!\!\: n}^{(\theta)}\big(\hbar,T\big)\Big]_{\text{FV}}^{\text{(LO)}} &=\exp\!\!\!\;\scalebox{1.1}{$\bigg[$}\!\!\!\: -\!\!\: ie^{-i\theta}\omega T \:\!\bigg(\!\:\!n+\frac{1}{2}\bigg)\:\!\!\scalebox{1.1}{$\bigg]$} \, .
	\label{eq:FV_Contribution_ArbotraryTheta}
\end{align}
For the remaining discussion concerned with all bounce-like contributions, let us focus on the case of moderately large $T$ presented in section~\ref{sec:Small_Moderate_T}. The reasoning of the exact relations~\eqref{eq:ExactTimeRelations} proceeds as before, again with the sole difference that the time coordinate $t$ entering all involved quantities is replaced by $ie^{-i\theta}t$. This once more leads to the previously seen cancellation~\eqref{eq:Simplified_Integrand}, such that the collective coordinate integration is again left trivial, yielding the volume factor $T$. With the additional Jacobian factor of $ie^{-i\theta}$, one finds the bounce contributions given by 
\begin{align}
	\!\Big[\mathcal{P}_{\!\!\: n}^{(\theta)}\big(\hbar,T\big)\Big]_{\text{bounce}}^{\text{(LO)}} &=-\frac{i}{n!} \, \frac{ie^{-i\theta}\omega T}{\sqrt{2\pi}}\:  \scalebox{1.1}{$\bigg($}\frac{2m\omega \mathdutchbcal{A}^2}{\hbar}\scalebox{1.1}{$\bigg)$}^{\!\!\!\!\: \scalebox{0.85}{$\frac{2n+1}{2}$}}  \exp\!\bigg(\!\!-\!\!\:\frac{\mathdutchbcal{B}}{\hbar}\bigg) \:\! \exp\!\!\!\;\scalebox{1.1}{$\bigg[$}\!\!\!\: -\!\!\: ie^{-i\theta}\omega T \:\!\bigg(\!\:\!n+\frac{1}{2}\bigg)\:\!\!\scalebox{1.1}{$\bigg]$} \, .
	\label{eq:BounceContributionFinal_ArbitraryTheta}
\end{align}
Notice that both relations~\eqref{eq:FV_Contribution_ArbotraryTheta} and~\eqref{eq:BounceContributionFinal_ArbitraryTheta} show that the relevant contributions transform trivially under the analytical continuation $t\mapsto ie^{-i\theta}t$ from Euclidean time back to ``less Wick-rotated'' time, as was already observed for the traditional instanton method~\cite{GarbrechtFunctionalMethods}. This demonstrates that the ansatz~\eqref{eq:Resonant_Energy_Conjecture} is valid for a generic choice of $\theta$, with the calculation leading up to that result being more technical, but similar in nature to the Euclidean case.\\

\noindent
One striking insight from the present discussion is that the simplifications inferred in section~\ref{sec:Small_Moderate_T} are now valid over a larger range of the parameter $T$, since exponential enhancements due to factors of $e^{ie^{-i\theta}\omega T}$ are weakened in case of $\theta <\pi/2$. Thus, $\hbar^{-1}$ stays the dominant quantity for larger time scales---instead of $\smash{\omega T\ll \log\!\big(\hbar^{-\frac{1}{2}}\big)}$ we rather only require $\smash{\omega T\ll \sin(\theta)^{-1} \log\!\big(\hbar^{-\frac{1}{2}}\big)}$. Repeating the analysis provided in section~\ref{sec:Large_T} for the complexified case, one once more finds that the range of admissible $T$ can be slightly extended to $\omega T\ll \sin(\theta)^{-1} \log\!\big(\hbar^{-1}\big)$, at which point the aforementioned breakdown would occur. Note that in the real-time limit $\theta\rightarrow 0^+$, the above assessments allow us to choose $T$ fully arbitrary. However, beware that our computation is still capped, as the last step in the derivation of relation~\eqref{eq:Resonant_Energy_Conjecture_Inserted} requires $T e^{-\mathdutchbcal{B}/\hbar} \ll 1$. For even larger $T$, multi-instanton contributions become increasingly relevant, spoiling the given approximation.

\subsection{On the inevitable breakdown of the computation}
\label{sec:3_9_BreakDown}

As we have seen in the previous section~\ref{sec:Large_T}, the ansatz~\eqref{eq:Resonant_Energy_Conjecture} for $n\neq 0$ appears to reliably break down for time scales $T$ that do not satisfy the bound $\omega T\ll \sin(\theta)^{-1} \log\!\big(\hbar^{-1}\big)$.\footnote{Note that we also gathered circumstantial evidence to argue that the same breakdown is to be expected in the (Euclidean) sequential calculation presented in section~\ref{sec:A_SequentialComputation}, see especially subsection~\ref{sec:Small_Endpoints}.} The present section attempts to provide the reason why this breakdown occurs. However, beware that even though the given explanation accounts for all observed intricacies, there may be further concealed effects that went unnoticed. Looking back at relation~\eqref{eq:Resonant_Energy_Conjecture}, we note that it relies on the subliminal assumption that inside the FV region, the propagator can be well-approximated by
\begin{align}
 	\!K_\theta\scalebox{1.15}{$\big($}x_0,x_T\:\!;T\scalebox{1.15}{$\big)$} &\cong \mathlarger{\mathlarger{\sum}}_{n=0}^\infty \;\, \closure{\psi_{n}^{(\text{loc})}(x_0)} \: \psi_{n}^{(\text{loc})}(x_T) \,  \exp\!\scalebox{1.15}{\bigg\{}\!\!\;\frac{-ie^{-i\theta} }{\hbar} \, E_n^{(\text{loc})} \:\!T\scalebox{1.15}{\bigg\}} \;\;\;\, \Big(\text{for } x_0,x_T \in \text{FV}\Big) \, , 
 	\label{eq:SpectralRep_FV_Propagator}
\end{align}
asserting the set of resonant states $\raisebox{0.5pt}{$\displaystyle{\smash{\ket[\big]{\raisebox{-0.5pt}{$\displaystyle{n\raisebox{4.25pt}{\scalebox{0.65}{$(\text{loc})$}}}$}}}}$}$ to form a genuine basis. It is clear that in case the FV region is stabilized, this relation is rendered exact. In case of a sizable barrier region separating both FV and TV region, only the modes involving high $n$ are significantly modified, such that this assertion is still reasonable. In similar spirit, the low-energy eigenfunctions $\psinloc$ inside the FV region are well-approximated assuming there to be only the single turning point $x_\text{FV}$, which is exactly captured by the uniform WKB approximation portrayed in section~\ref{sec:C_UniformWKB}. This, at least heuristically, yields an alternative justification for the presented method. Simultaneously,  relation~\eqref{eq:SpectralRep_FV_Propagator} more clearly reveals where the computation leads us astray for large $T$. To see this, observe that for any practical evaluation of equation~\eqref{eq:Resonant_Energy_Conjecture}, instead of using an exact representation for $\psinloc$, one employs the leading-order expression $\smash{\psinlocLO}$ portrayed in equation~\eqref{eq:Resonant_Wave_Function}. However, this approximation deviates from the proper solution for the wave function $\psinloc$ at order $\sqrt{\hbar}$, see the derivation in appendix~\ref{sec:C_UniformWKB}. Thus, when projecting out the $n^\text{th}$ excited resonant energy $\Enloc$, the projection is generally contaminated by terms of similar order, i.e. we have 
\begin{align}
	\braket[\Big]{\psi_{n}^{(\text{loc})} , \psi_{m,\text{LO}}^{(\text{loc})}} = \delta_{nm} + \sqrt{\hbar} \: \Delta_{nm}\, ,
	\label{eq:ErrorProjection}
\end{align}
with $\Delta_{nm}=\mathcal{O}(1)$. Thus, when computing the desired leading-order result, due to the incomplete projection, we are expected to encounter additional terms of the form
\begin{align}
	\Big[\mathcal{P}_{\!\!\: n}^{(\theta)}\big(\hbar,T\big)\Big]^{(\text{LO})}&=\mathlarger{\int}_{-\infty}^{\infty} \text{d}x_0 \mathlarger{\int}_{-\infty}^{\infty} \text{d}x_T\; \closure{\psi_{n,\text{LO}}^{(\text{loc})}(x_T)} \:\psi_{n,\text{LO}}^{(\text{loc})}(x_0)\, K_\theta\mathlarger{\big(}x_0,x_T\:\!;T\mathlarger{\big)} \label{eq:LeadingOrderExpression}\\[0.05cm]  
	&= \exp\!\left[\frac{-ie^{-i\theta} }{\hbar} \, E_n^{(\text{loc})} \:\!T\:\!\right] \Bigg\{ 1+2\sqrt{\hbar} \; \text{Re}\big(\Delta_{nn}\big)  \nonumber \\[-0.25cm] 
	&\qquad\qquad\qquad\qquad\!\!\;\: \qquad\qquad \underbrace{\! +\, \hbar\;  \mathlarger{\sum}_{m=0}^\infty \; \big\lvert \Delta_{nm}\big\rvert^2\, \exp\!\scalebox{1.1}{\bigg[} \Big(E_m^{(\text{loc})}-E_n^{(\text{loc})}\Big) \:\! \frac{ie^{-i\theta}T}{\hbar}\,\scalebox{1.1}{\bigg]}\!}_{\displaystyle{\rule{0pt}{0.35cm}\text{terms not accounted for previously}}} \!\!\: \Bigg\} \, . \nonumber 
\end{align}
Besides negligible $\mathcal{O}\big(\!\!\:\sqrt{\hbar}\,\big)$ corrections to the sought-after contribution, there remain leftover off-diagonal terms of order $\hbar$. The arising problem is quickly noticed, as for increasing $T$ all terms adhering to $m<n$ bestow an exponentially growing error upon us. With the lowest energy eigenvalues only slightly deviating from a purely harmonic behavior, we obtain the approximate relation
\begin{align}
	\Big[\mathcal{P}_{\!\!\: n}^{(\theta)}\big(\hbar,T\big)\Big]^{(\text{LO})}&\approx \Big[\mathcal{P}_{\!\!\: n}^{(\theta)}\big(\hbar,T\big)\Big]^{(\text{exact})} \Bigg\{1+\underbrace{\hbar\;  \mathlarger{\sum}_{m=0}^\infty \; \big\lvert \Delta_{nm}\big\rvert^2\, \exp\!\Big[ ie^{-i\theta} \big(m-n\big)\:\! \omega T\,\Big]}_{\displaystyle{\begin{matrix}
				\text{worrisome error, increasing} \\
				\text{with growing $T$ for $m<n$}
	\end{matrix}}} \!\!\!\:\Bigg\} \, . 
\label{eq:ErrorAnsatz}
\end{align}
In case of $n\neq 0$, there is always one addend $m<n$ dominating the expression for large $T$, thus spoiling the computation. This is exactly the result found in section~\ref{sec:3_5_BounceContribution}, as only the ground state decay width $n=0$ is exempt from the artificial restriction on $T$. Moreover, the reasoning given here directly encapsulates the fact that for $\theta<\pi/2$, the breakdown is substantially delayed, up to the point that in the limit $\theta\rightarrow 0^+$ the time interval $T$ can be chosen arbitrary.\footnote{In case of small $\theta$, the imaginary part of $\Enloc$ becomes increasingly important, constituting the dominant real part of the exponent found in relation~\eqref{eq:LeadingOrderExpression}. This yields the bound $T e^{-\mathdutchbcal{B}/\hbar}\ll 1$ on $T$, even in case of minuscule $\theta$, constraining the calculation to the single-instanton sector, in line with the discussion in section~\ref{sec:3_8_Extensions}. If one heuristically approximates multi-instanton contributions, one indeed would find no exponentiation of the sum due to the additional endpoint weighting. Thus, multi-instanton contributions are naturally prevented from entering the discussion.} What is still missing is a transparent reason justifying the attained bound $\omega T\ll \sin(\theta)^{-1} \log\!\big(\hbar^{-1}\big)$, as na\"{\i}vely we would anticipate the $m=0$ addend to dominate, such that the expected bound would be lowered to $\omega T\ll n^{-1} \sin(\theta)^{-1} \log\!\big(\hbar^{-1}\big)$. The only reasonable explanation is that for larger differences between $n$ and $m$, the higher off-diagonal terms $\Delta_{nm}$ in the projection~\eqref{eq:ErrorProjection} are more strongly suppressed, i.e. $\sqrt{\hbar}\,\Delta_{nm}\lesssim \hbar^{\lvert n-m\rvert/2}$. While certainly plausible, it is not fully clear if this suppression is strong enough to restore the observed bound or if other intricacies enter the discussion.\\[-0.25cm]

\section{Modified instanton method: Customary sequential derivation}
\label{sec:A_SequentialComputation}

As discussed at the onset of section~\ref{sec:3_ModifiedInstantonMethod}, one can choose from two admissible schemes when evaluating the desired finite-time amplitude~\eqref{eq:Global_Energy_Exact_Projection} after instating the replacement $\psinglob \mapsto \psinloc$: Either one employs the previously presented composite approach or instead opts for the more na\"{\i}ve sequential one, which will be sketched in the following section. This does not amount to an entirely unnecessary effort, as it reveals the (dis-)advantages of the composite approach while constituting the basis on which we will later discuss the original work by Liang and Müller-Kirsten, who exclusively utilized this secondary computation scheme. Once more, let us focus on the Euclidean case $\theta=\pi/2$.\\

\noindent
As a starting point, we desire to obtain the complete leading-order expression for the Euclidean propagator to fixed endpoints in the semiclassical limit $\hbar\to 0^+$, which should then be inserted into the initial amplitude~\eqref{eq:Global_Energy_Exact_Projection}. While such a feat is infeasible in full generality, we can make meaningful progress by assuming the propagator not to possess a soft mode in the domain of interest,\footnote{While one can explicitly check that this indeed holds as long as we again satisfy the bound $\smash{\omega T\ll \log\!\big(\hbar^{-1}\big)}$, the required computation is laborious and hardly illustrative, thus we choose to refrain from presenting it explicitly. One would indeed find that for large $T$ and $x_0\sim x_T$, the propagator entails a quasi-zero mode $\lambda_0\sim e^{-\omega T}$ around bounce-like saddle points. For $T$ exceeding the formerly mentioned bound, this soft mode has to be treated separately, as the Gaussian integration required to arrive at the Van Vleck formula~\eqref{eq:VanVleck_Propagator} is rendered invalid.} allowing us to employ the common semiclassical Van Vleck approximation~\cite{VanVleckDeterminant}
\begin{equation}
	K_\text{E}\scalebox{1.15}{$\big($}x_0,x_T\:\!;T\scalebox{1.15}{$\big)$} \sim \mathlarger{\mathlarger{\sum}}_{\raisebox{1pt}{$\scalebox{0.8}{$x_\text{crit}$}$}}\;  \Bigg\{\!\!\!\:-\!\!\:\frac{1}{2\pi\hbar}\,\frac{\partial^2 S_\text{E}^{(\text{crit})}\big(x_0,x_T\:\!;T\big)}{\partial x_0\partial x_T}\Bigg\}^{\!\!\:\frac{1}{2}} \, \exp\!\!\:\Bigg\{\!\!\!\:-\!\!\:\frac{S_\text{E}^{(\text{crit})}\big(x_0,x_T\:\!;T\big)}{\hbar}\Bigg\}\, .
	\label{eq:VanVleck_Propagator}
\end{equation}
Here $\smash{S_\text{E}\big(x_0,x_T\:\!;T\big)}$ denotes the classical on-shell action function, assigning, to given endpoints $x_0$ and $x_T$, the Euclidean action of the classical trajectory $\smash{x\raisebox{-2.75pt}{\scalebox{0.75}{$\text{crit}$}}\!\!\!\!\!\!\!\!\:\raisebox{5.5pt}{\scalebox{0.65}{$(x_0,x_T\:\!;T)$}}(t)}$ satisfying the Dirichlet boundary conditions $\smash{x\raisebox{-2.75pt}{\scalebox{0.75}{$\text{crit}$}}\!\!\!\!\!\!\!\!\:\raisebox{5.5pt}{\scalebox{0.65}{$(x_0,x_T\:\!;T)$}}}(0)=x_0$ and $\smash{x\raisebox{-2.75pt}{\scalebox{0.75}{$\text{crit}$}}\!\!\!\!\!\!\!\!\:\raisebox{5.5pt}{\scalebox{0.65}{$(x_0,x_T\:\!;T)$}}}(T)=x_T$. Due to there generally existing multiple solutions to this boundary value problem, the so-defined function is usually multi-valued, with $S\raisebox{-3pt}{\scalebox{0.7}{\text{E}}}\!\!\!\;\!\raisebox{5.25pt}{\scalebox{0.65}{$(\text{crit})$}}\big(x_0,x_T\:\!;T\big)$ distinguishing a single branch. The Van Vleck propagator~\eqref{eq:VanVleck_Propagator} allows us to obtain the fluctuation determinant from solely computing the action of (families of) classical paths satisfying the traditional equation of motion, enabling the subsequent discussion. Utilizing the formerly computed representation~\eqref{eq:Resonant_Wave_Function} for $\psinlocLO$, we once more desire to evaluate 
\begin{align}
	\!\Big[\mathcal{P}_{\!\!\: n}^{(\pi/2)}\big(\hbar,T\big)\Big]^{(\text{LO})}&=\mathlarger{\int}_{-\infty}^{\infty} \text{d}x_0 \mathlarger{\int}_{-\infty}^{\infty} \text{d}x_T\; \closure{\psi_{n,\text{LO}}^{(\text{loc})}(x_T)} \:\psi_{n,\text{LO}}^{(\text{loc})}(x_0)\, K_\text{E}\mathlarger{\big(}x_0,x_T\:\!;T\mathlarger{\big)} \nonumber \\[0.05cm] 
	&= \frac{1}{2^n n!} \left(\frac{m\omega}{\pi\hbar}\right)^{\!\frac{1}{2}} \, \mathlarger{\mathlarger{\sum}}_{\raisebox{1pt}{$\scalebox{0.8}{$x_\text{crit}$}$}}\; \mathlarger{\mathlarger{\int}}_{-\infty}^{\infty} \text{d}x_0 \mathlarger{\mathlarger{\int}}_{-\infty}^{\infty} \text{d}x_T\; \psi_{n,\text{non-exp}}^{(\text{loc})}(x_0) \, \psi_{n,\text{non-exp}}^{(\text{loc})}(x_T) \nonumber \\[-0.1cm]  
	&\qquad\qquad \times \Bigg\{\!\!\!\:-\!\!\:\frac{1}{2\pi\hbar}\,\frac{\partial^2 S_\text{E}^{(\text{crit})}\big(x_0,x_T\:\!;T\big)}{\partial x_0\partial x_T}\Bigg\}^{\!\!\:\frac{1}{2}} \: \exp\!\!\:\Bigg\{\!\!\!\:-\!\!\:\frac{f_\text{exp}^{(\text{crit})}\big(x_0,x_T\big)}{\hbar}\Bigg\} \, , 
	\label{eq:P_Expression_Sequential}
\end{align}
with the full exponential function given by
\begin{align}
	f_\text{exp}^{(\text{crit})}\big(x_0,x_T\big)\coloneqq S_\text{E}^{(\text{crit})}\big(x_0,x_T\:\!;T\big) + \psi_{n,\text{exp}}^{(\text{loc})}(x_0) + \psi_{n,\text{exp}}^{(\text{loc})}(x_T) \, ,
	\label{eq:Full_Exponent_Sequential}
\end{align}
leaving the fixed time interval $T$ implicit. Whereas the sum encountered in equation~\eqref{eq:VanVleck_Propagator} runs over the critical trajectories $\smash{x\raisebox{-2.75pt}{\scalebox{0.75}{$\text{crit}$}}\!\!\!\!\!\!\!\!\:\raisebox{5.5pt}{\scalebox{0.65}{$(x_0,x_T\:\!;T)$}}(t)}$ obtained for fixed $x_0$ and $x_T$, the sum in relation~\eqref{eq:P_Expression_Sequential} encompasses all ``overall'' critical paths, which are also required to constitute stationary points for the endpoint integrations. The interpretation is simple: One should first perform a semiclassical analysis of the path integral for varying endpoints, yielding different families of critical trajectories $\smash{x\raisebox{-2.75pt}{\scalebox{0.75}{$\text{crit}$}}\!\!\!\!\!\!\!\raisebox{5.5pt}{\scalebox{0.65}{$(x_0,x_T\:\!;T)$}}(t)}$ that vary smoothly under perturbations of $x_0$ and $x_T$. From those distinct families of critical paths, only certain representatives are again critical points of both remaining Laplace-type endpoint integrals---exactly these $x_\text{crit}(t)$ are to be included in the sum seen in equation~\eqref{eq:P_Expression_Sequential}.

\subsection{Discussion of critical trajectories}
\label{sec:Critical_Paths_Sequential}

As usual, we start by determining the critical paths of the full exponent~\eqref{eq:Full_Exponent_Sequential}. The Van Vleck propagator~\eqref{eq:VanVleck_Propagator} already requires us to only consider classical trajectories satisfying the usual Euler--Lagrange equation~\eqref{eq:Euler_Lagrange_Equation}, the endpoint integrations then yield the additional two transversality conditions 
\begin{subequations}
	\begin{align}
		\Bigg[\frac{\partial S_\text{E}^{(\text{crit})}\big(x_0,x_T\:\!;T\big)}{\partial x_0} \;\! + \;\! \frac{\text{d} \psi_{n,\text{exp}}^{(\text{loc})}(x_0)}{\text{d} x_0} \;\! \Bigg]_{\substack{\scalebox{0.85}{$x_0\!\, =\! x\raisebox{-2.75pt}{\scalebox{0.65}{$0$}}\!\!\raisebox{4.75pt}{\scalebox{0.6}{$(\text{crit})$}}$}\\[0.025cm] \scalebox{0.85}{$x_T\! =\! x\raisebox{-2.75pt}{\scalebox{0.65}{$T$}}\!\!\!\!\:\raisebox{4.75pt}{\scalebox{0.6}{$(\text{crit})$}}$}}} &\overset{!}{=} 0 \, , & &\; \begin{matrix}
			\text{transversality condition} \\
			\text{for the endpoint $x\raisebox{-2.75pt}{\scalebox{0.65}{$0$}}\!\!\raisebox{4.75pt}{\scalebox{0.6}{$(\text{crit})$}}$}
		\end{matrix} \label{eq:Left_Transversality_Condition_Sequential} \\[0.2cm] 
		\Bigg[\frac{\partial S_\text{E}^{(\text{crit})}\big(x_0,x_T\:\!;T\big)}{\partial x_T} + \frac{\text{d} \psi_{n,\text{exp}}^{(\text{loc})}(x_T)}{\text{d} x_T}\Bigg]_{\substack{\scalebox{0.85}{$x_0\!\, =\! x\raisebox{-2.75pt}{\scalebox{0.65}{$0$}}\!\!\raisebox{4.75pt}{\scalebox{0.6}{$(\text{crit})$}}$}\\[0.025cm] \scalebox{0.85}{$x_T\! =\! x\raisebox{-2.75pt}{\scalebox{0.65}{$T$}}\!\!\!\!\:\raisebox{4.75pt}{\scalebox{0.6}{$(\text{crit})$}}$}}} &\overset{!}{=} 0 \, . & &\; \begin{matrix}
			\text{transversality condition} \\
			\text{for the endpoint $x\raisebox{-2.75pt}{\scalebox{0.65}{$T$}}\!\!\!\!\:\raisebox{4.75pt}{\scalebox{0.6}{$(\text{crit})$}}$}
		\end{matrix} \label{eq:Right_Transversality_Condition_Sequential}
	\end{align}
	\label{eq:All_Extremal_Conditions_Sequential}%
\end{subequations}
These conditions constrain the set of stationary endpoints $x\raisebox{-2.75pt}{\scalebox{0.65}{$0$}}\!\!\raisebox{4.75pt}{\scalebox{0.6}{$(\text{crit})$}}$ and $x\raisebox{-2.75pt}{\scalebox{0.65}{$T$}}\!\!\!\!\:\raisebox{4.75pt}{\scalebox{0.6}{$(\text{crit})$}}$ for each family of critical trajectories $\smash{x\raisebox{-2.75pt}{\scalebox{0.75}{$\text{crit}$}}\!\!\!\!\!\!\!\raisebox{5.5pt}{\scalebox{0.65}{$(x_0,x_T\:\!;T)$}}(t)}$, leaving us with the overall critical paths $x_\text{crit}(t)$.  Employing the well-known Hamilton-Jacobi equations
\begin{subequations}
	\begin{align}
		\frac{\partial S_\text{E}^{(\text{crit})}\big(x_0,x_T\:\!;T\big)}{\partial x_0} &=\quad\:\, - \: \sqrt{2m\Big[V(x_0)\;\!+\:\!E_\text{crit}\big(x_0,x_T\big)\Big]} = - m \dot{x}_\text{crit}^{\scalebox{0.8}{$\big(x_0,x_T\:\!;T\big)$}}(0)\, , \label{eq:Partial_Derivative_Euclidean_Action_wrt_Endpoints_A} \\[0.1cm]
		\frac{\partial S_\text{E}^{(\text{crit})}\big(x_0,x_T\:\!;T\big)}{\partial x_T} &=  (-1)^\ell  \sqrt{2m\Big[V(x_T)+E_\text{crit}\big(x_0,x_T\big)\Big]} = \phantom{-}m  \dot{x}_\text{crit}^{\scalebox{0.8}{$\big(x_0,x_T\:\!;T\big)$}}(T)\, ,
	\end{align}
	\label{eq:Partial_Derivative_Euclidean_Action_wrt_Endpoints}%
\end{subequations}
one can relate the partial derivatives of the on-shell action $\smash{S\raisebox{-3pt}{\scalebox{0.7}{\text{E}}}\!\!\!\;\!\raisebox{5.25pt}{\scalebox{0.65}{$(\text{crit})$}}\big(x_0,x_T\:\!;T\big)}$ to the canonical momentum of the classical motion at its endpoints. This is a rather common procedure, already employed by various authors, for recent examples see e.g.~\cite{SchiffWeightedPathIntegrals,TurokRealTimeTunneling,DarmeGeneralizedEscapePaths,DraperVacuumDecayTimdeDepBackground}. Notably, we implicitly assumed the considered motion to initially have a non-negative starting velocity, while the sign of the velocity at $t=T$ is determined by the number of turning points $\ell$ traversed during the motion. Together with the particular form of $\psilocexp$ given in equation~\eqref{eq:FullExponent}, the transversality conditions~\eqref{eq:All_Extremal_Conditions_Sequential} yield the two relations
\begin{subequations}
	\begin{align}
		\sqrt{2mV\!\!\:\scalebox{1.25}{$\big[$} x_0^{(\text{crit})} \scalebox{1.25}{$\big]$} } &\overset{!}{=} \phantom{(-1)^{\ell+1}} \sqrt{2m \:\!\scalebox{1.15}{$\Big\{$} V\!\!\:\scalebox{1.25}{$\big[$}x_0^{(\text{crit})}\scalebox{1.25}{$\big]$}+E_\text{crit}\scalebox{1.25}{$\big[$}x_0^{(\text{crit})},x_T^{(\text{crit})}\scalebox{1.25}{$\big]$}\!\!\:\scalebox{1.15}{$\Big\}$} } \; , \\
		\sqrt{2mV\!\!\:\raisebox{0.5pt}{\scalebox{1.25}{$\big[$}} x_T^{(\text{crit})} \raisebox{0.5pt}{\scalebox{1.25}{$\big]$}} }  &\overset{!}{=} (-1)^{\ell+1} \sqrt{2m \:\!\scalebox{1.15}{$\Big\{$} V\!\!\:\scalebox{1.25}{$\big[$}x_T^{(\text{crit})}\scalebox{1.25}{$\big]$}+E_\text{crit}\scalebox{1.25}{$\big[$}x_0^{(\text{crit})},x_T^{(\text{crit})}\scalebox{1.25}{$\big]$}\!\!\:\scalebox{1.15}{$\Big\}$} } \; .
		\label{eq:Necessary_Transversality_Condition_Critical_Paths_Sequential_A}
	\end{align}
	\label{eq:Necessary_Transversality_Condition_Critical_Paths_Sequential}%
\end{subequations}
Identically to the discussion in section~\ref{sec:CriticalPaths}, one again finds that the classical energy $E_\text{crit}$ is required to vanish. Moreover, the second condition~\eqref{eq:Necessary_Transversality_Condition_Critical_Paths_Sequential_A} either demands the overall critical path $x_\text{crit}(t)$ to possess an odd number of turning points, or the potential is mandated to vanish at $x\raisebox{-2.75pt}{\scalebox{0.65}{$T$}}\!\!\!\!\:\raisebox{4.75pt}{\scalebox{0.6}{$(\text{crit})$}}$, leaving only the options $x\raisebox{-2.75pt}{\scalebox{0.65}{$T$}}\!\!\!\!\:\raisebox{4.75pt}{\scalebox{0.6}{$(\text{crit})$}} \in \big\{x_\text{FV},x_\text{escape}\big\}$. Just as in the former evaluation, this solely leaves us with either the trivial FV trajectory or segments of a time-translated, infinite-time bounce, provided the turning point $x_\text{escape}$ is traversed during the motion. This condition again constrains the volume of the moduli space, which in the sequential picture will amount to a restriction on the starting position $x\raisebox{-2.75pt}{\scalebox{0.65}{$0$}}\!\!\raisebox{4.75pt}{\scalebox{0.6}{$(\text{crit})$}}$.

\subsection{FV contribution}

In case of the trivial FV trajectory $x_\text{crit}(t)=x_\text{FV}=0$, we can directly determine an approximate expression for the relevant family of critical paths $\smash{x\raisebox{-2.75pt}{\scalebox{0.75}{$\text{crit}$}}\!\!\!\!\!\!\!\raisebox{5.5pt}{\scalebox{0.65}{$(x_0,x_T\:\!;T)$}}(t)}$ given arbitrary endpoints. This is feasible as only small fluctuations around $x\raisebox{-2.75pt}{\scalebox{0.65}{$0$}}\!\!\raisebox{4.75pt}{\scalebox{0.6}{$(\text{crit})$}}=x\raisebox{-2.75pt}{\scalebox{0.65}{$T$}}\!\!\!\!\:\raisebox{4.75pt}{\scalebox{0.6}{$(\text{crit})$}}=0$ contribute toward the expression, thus the prevalent boundary value problem is practically identical to the case of a purely harmonic potential, yielding
\begin{align}
	x_\text{FV}^{(x_0,x_T\:\!;T)}(t) = x_0 \cosh(\omega t)+\frac{x_T-x_0 \cosh(\omega T)}{\sinh(\omega T)} \, \sinh(\omega t) + \mathcal{O}\scalebox{1.2}{$\big($}x_0^{2},x_T^{2},x_0x_T\scalebox{1.2}{$\big)$}\, .
\end{align} 
With the Euclidean action of that particular family of classical paths trivially given by 
\begin{align}
	S_\text{E}^{(\text{FV})}\big(x_0,x_T\:\!;T\big) = \frac{m\omega}{2 \sinh(\omega T)} \:\!  \bigg[\big(x_0^2+x_T^2\big) \cosh(\omega T)&- 2 x_0 x_T\bigg] \Big[1+ \mathcal{O}\big(x_0,x_T\big)\Big]\, ,
	\label{eq:EuclideanAction_FV_Trajec_Sequential}
\end{align}
the propagator for small $x_0,x_T$ is equivalent to that of a harmonic oscillator. Inserting this result into the desired expression~\eqref{eq:P_Expression_Sequential} while expanding the wave function $\psinlocLO$ for small endpoints yields the intermediate result
\begin{align}
	\!\!\!\Big[\mathcal{P}_{\!\!\: n}^{(\pi/2)}\big(\hbar,T\big)\Big]_\text{FV}&= \frac{1}{2^n n!} \left(\frac{m\omega}{\pi\hbar}\right)^{\!\frac{1}{2}} \, \left[\frac{m\omega}{2\pi \hbar \sinh(\omega T)} \right]^{\!\frac{1}{2}} \; \mathlarger{\mathlarger{\int}}_{-\infty}^{\infty} \text{d}x_0 \mathlarger{\mathlarger{\int}}_{-\infty}^{\infty} \text{d}x_T \;\,  H_n\!\left(\sqrt{\frac{m\omega}{\hbar}}\,x_0\right) \label{eq:P_Expression_Sequential_FV} \\
	&\quad \times \!\!\: H_n\!\left(\sqrt{\frac{m\omega}{\hbar}}\,x_T\right) \! \Big[1+\mathcal{O}\big(x_0,x_T\big)\Big] \exp\!\scalebox{1.2}{\bigg\{}\!\!\!\:\!\!\:-\!\!\:\frac{m\omega \big(x_0^2+x_T^2\big)}{2\hbar}\, \Big[1+\mathcal{O}\big(x_0,x_T\big)\Big]\!\!\,\scalebox{1.2}{\bigg\}} \nonumber \\ 
	&\quad\times \exp\!\!\:\Bigg\{\!\!\!\:-\!\!\:\frac{m\omega}{2\hbar  \sinh(\omega T)} \:\! \bigg[\big(x_0^2+x_T^2\big) \cosh(\omega T)- 2 x_0 x_T\bigg] \Big[1+ \mathcal{O}\big(x_0,x_T\big)\Big]\!\!\,\Bigg\} \, . \nonumber
\end{align}
Rescaling $x_{0,T} \mapsto \sqrt{\hbar}\, x_{0,T}$ shows that the formerly neglected terms are truly subleading, at which point one solely has to compute the remaining integral. Performing standard algebra, the leading-order term can be brought into the final form
\begin{align}
	\!\!\Big[\mathcal{P}_{\!\!\: n}^{(\pi/2)}\big(\hbar,T\big)\Big]^{(\text{LO})}_\text{FV}&= \frac{\sinh(\omega T)^{-\frac{1}{2}}}{2^{n+\frac{1}{2}} \pi \:\!n!} \mathlarger{\mathlarger{\int}}_{-\infty}^{\infty} \text{d}x  \mathlarger{\mathlarger{\int}}_{-\infty}^{\infty} \text{d}y \; H_n(x)\,H_n(y) \!\; \exp\!\;\!\Bigg\{\frac{2 x y-\big(x^2+y^2\big) \:\!e^{\omega T}}{2 \sinh(\omega T)} \!\:\!\Bigg\} \nonumber \\[0.15cm] 
	&= \exp\!\!\!\;\scalebox{1.1}{$\bigg[$}\!-\!\!\:\omega T \:\!\bigg(\!\:\!n+\frac{1}{2}\bigg)\:\!\!\scalebox{1.1}{$\bigg]$} \, , \label{eq:Result_FV_Sequential}
\end{align}
where one first diagonalizes the exponent by the change of variables $x, y \mapsto \frac{1}{2}\:\! \big(x\pm y\big)$ before utilizing the identity
\begin{align}
	\mathlarger{\mathlarger{\int}}_{-\infty}^{\infty} \text{d}x  \mathlarger{\mathlarger{\int}}_{-\infty}^{\infty} \text{d}y \; H_n\big(x+y\big) H_n\big(x-y\big) \exp\!\bigg[\!\!\!\;-\!\!\:\frac{2x^2}{1+\alpha} -\frac{2y^2}{1-\alpha}\bigg]&= 2^{n-1} \pi \:\! n!  \;\! \alpha^{n} \sqrt{1-\alpha^2}
\end{align}
for $\alpha= e^{-\omega T}$. As stated before, the found result~\eqref{eq:Result_FV_Sequential} contains basically no new information, being practically fully redundant due to the employed ansatz~\eqref{eq:2_Local_Energy_Ansatz_Definition_Epsn} when constructing the wave function $\psinloc$. Arguably, this derivation is much simpler than the one presented in section~\eqref{eq:FV_Contribution}, which instead required a lot of preparatory technical work to arrive at this rather trivial result. However, as we will see for the bounce contribution, the former method is conceptually much clearer. Additionally, with the technical details worked out once, the composite path integral method is highly versatile and can be employed to a plethora of different cases.

\subsection{Bounce contributions}

\begin{figure}[H]
	\begin{subfigure}[b]{0.545\textwidth}
		\centering
		\includegraphics[width=\textwidth]{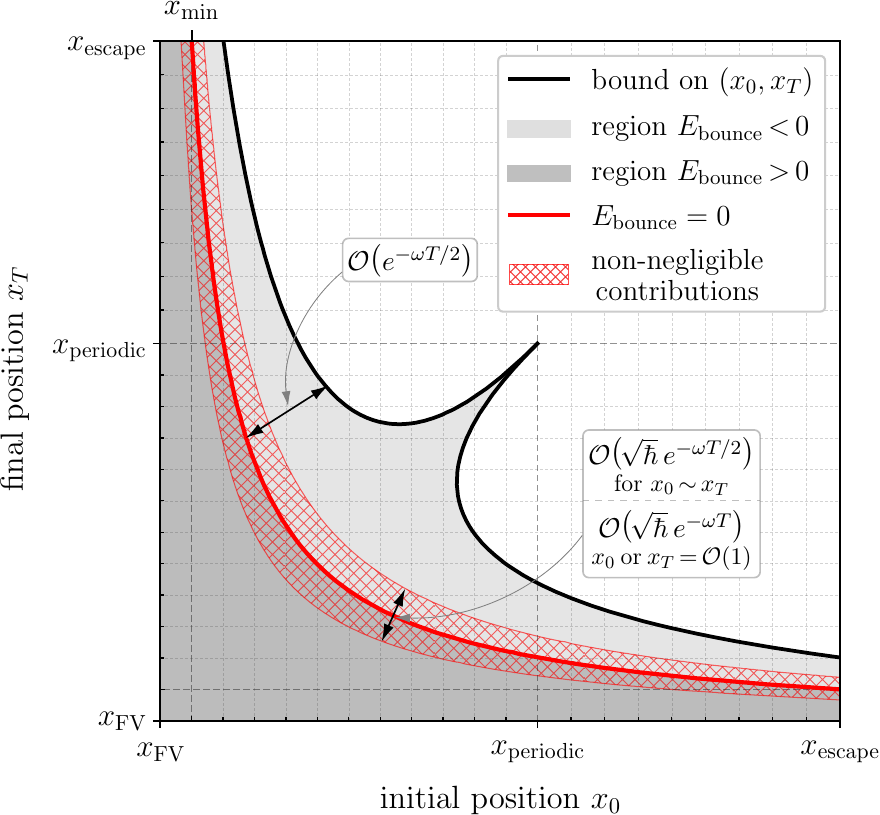}
		\vspace*{-3mm}
		\caption{Different (single-)bounce regions for fixed $T$.}
		\label{fig:BounceRegions}
	\end{subfigure}
	\hfill
	\begin{subfigure}[b]{0.435\textwidth}
		\centering
		\includegraphics[width=\textwidth]{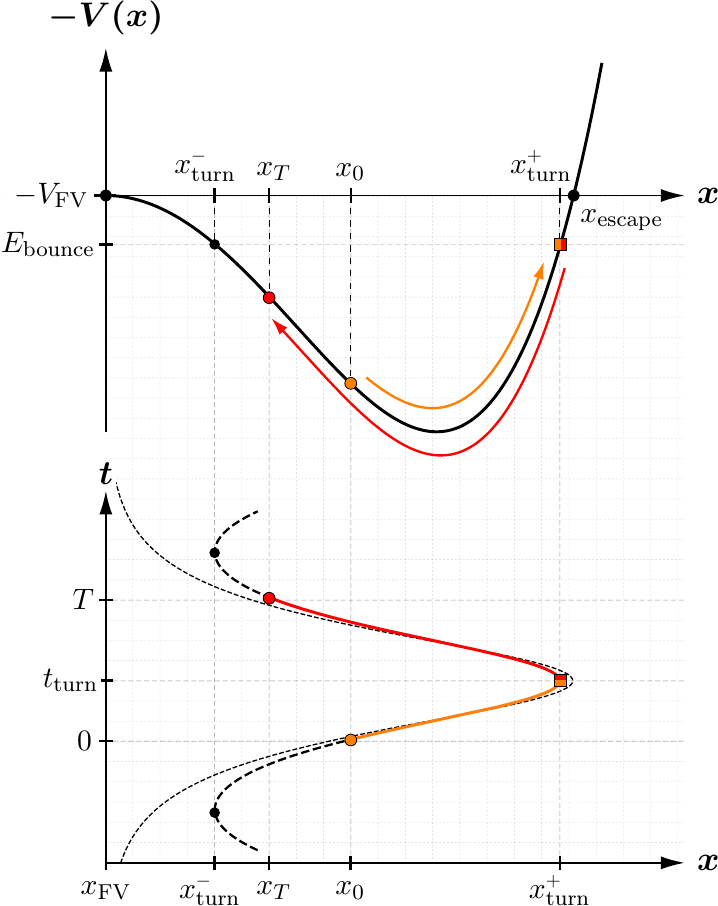}
		\caption{Exemplary bounce with negative energy.}
		\label{fig:BounceMotion}
	\end{subfigure}
	\caption{\hspace{0.001cm}}
	\label{fig:Bounces_InvertedQuarticPotential}
	\vspace{-0.44cm}\hspace{1.3cm}
	\begin{tikzpicture}
		\node[anchor=center, inner sep=0pt, 
		minimum width=14.55cm, text width=14.55cm-0.4cm,align=justify,
		minimum height=3.5cm, 
		text depth=3.5cm-0.2cm] (0,0) {Illustration of the different regions emerging when viewing bounce motions in the $(x_0,x_T)$-plane, illustrating the critical strip around $E_\text{bounce}=0$ from which the dominant contributions will arise (a), as well as an exemplary bounce trajectory with negative energy (b). Both figures show the situation for an inverted quartic potential, which is however representative for more general situations. Plot (b) hereby only shows the part of the inverted potential which is essential to the depicted bounce motion and its continuation beyond the time interval $[0,T]$ (loosely dashed line), being illustrated together with the infinite-time bounce (densely dashed line) for reference.};
	\end{tikzpicture}
\end{figure}

With the FV trajectory covered, let us turn toward all bounce-like contributions. Again, we desire to obtain a suitable approximation for the action of a general bounce motion $\smash{x\raisebox{-2.75pt}{\scalebox{0.75}{$\text{bounce}$}}\!\!\!\!\!\!\!\!\!\!\!\!\!\!\!\:\raisebox{5.5pt}{\scalebox{0.65}{$(x_0,x_T\:\!;T)$}}(t)}$ in order to apply the Van Vleck formula~\eqref{eq:VanVleck_Propagator}. Note that while all critical bounce motions possess $E_\text{bounce}=0$, the determinant factor requires us to also study solutions in their vicinity, for which the classical energy is small but non-zero. Thus, let us at first give some basic remarks on general single-bounce motions, possessing an arbitrary energy $E_\text{bounce}$---figure~\ref{fig:BounceMotion} depicts such an exemplary bounce. The sole turning point $x_\text{turn}^+$ of the assessed family of motions is always given by the closest point to $x_\text{escape}$ satisfying $V\big(x_\text{turn}^+\big)=-E_\text{bounce}$. Figure~\ref{fig:BounceRegions} depicts the landscape of so-found bounce-like motions encountered for varying endpoints $x_0$ and $x_T$, given a fixed time interval length $T$. The highlighted hatched strip surrounding the line of critical bounce motions, possessing $E_\text{bounce}=0$, constitutes the region from which non-negligible contributions to the overall endpoint integrals will eventually arise from.\footnote{We will not cover the rather cumbersome approximation for the widths of the different regions provided in figure~\ref{fig:Bounces_InvertedQuarticPotential}. For large $T$, the red line of critical bounce motions satisfies $x_0 x_T \approx \mathdutchbcal{A}^2 e^{-\omega T}$, whereas the black bounding line would obey $x_0 x_T \approx 2\mathdutchbcal{A}^2 e^{-\omega T}$. Also note that while only bounce trajectories with energy $E_{\scalebox{0.7}{\text{bounce}}}=\mathcal{O}\big(\:\!\!\sqrt{\hbar}\,\big)$ will contribute significantly, when translating this condition into the $(x_0,x_T)$-plane one would find the additional enhancement factor $e^{-\omega T/2}$ or $e^{-\omega T}$ depending on the magnitude of $x_{0}$ and $x_T$. That way, one always avoids the inclusion of trajectories lying beyond the black bounding line, residing well outside the relevant strip which yields significant contributions.} Let us furthermore emphasize that when constraining our view to motions possessing a single turning point $x_\text{turn}^+$, the exact relation
\begin{align}
	T&= \mathlarger{\mathlarger{\int}}_{\scalebox{0.8}{$x_0$}}^{\scalebox{0.8}{$x_{\text{turn}}^{\scalebox{0.8}{$+$}}$}} \bigg\{\frac{2}{m}\scalebox{1.25}{$\big[$}V(x)+E_\text{bounce}\scalebox{1.25}{$\big]$}\!\!\:\bigg\}^{\!-\frac{1}{2}} \,\text{d}x +  \mathlarger{\mathlarger{\int}}_{\scalebox{0.8}{$x_T$}}^{\scalebox{0.8}{$x_{\text{turn}}^{\scalebox{0.8}{$+$}}$}} \bigg\{\frac{2}{m}\scalebox{1.25}{$\big[$}V(x)+E_\text{bounce}\scalebox{1.25}{$\big]$}\!\!\:\bigg\}^{\!-\frac{1}{2}} \,\text{d}x \, ,
	\label{eq:4_Total_Time_Bounce_Motion}
\end{align}
restricts the region of admissible bounces in the $(x_0,x_T)$-plane, as certain choices of the endpoints admit no real solution for the bounce energy. Again note that $x_\text{turn}^+$ is a function of $E_\text{bounce}$, thus fixing the pair $(x_0,x_T)$ in equation~\eqref{eq:4_Total_Time_Bounce_Motion} leaves only $E_\text{bounce}$ undetermined. Past the bounding line encountered in figure~\ref{fig:BounceRegions}, there only lie multi-bounce trajectories, lending subleading contributions.

\noindent
As depicted in figure~\ref{fig:BounceMotion}, for negative energy $E_\text{bounce}<0$ the relevant bounce trajectories arise as cutouts of periodic motions inside the flipped barrier region between the two turning points $x_\text{turn}^\pm$. The black bounding line illustrated in figure~\ref{fig:BounceRegions} thereby comprises all negative energy solutions starting or ending with exactly vanishing velocity. The special point $(x_0,x_T)=(x_\text{periodic},x_\text{periodic})$ encountered in figure~\ref{fig:BounceRegions} corresponds to a periodic bounce motion with period $T$ inside the inverted potential, starting and ending at exactly $x_\text{turn}^-=x_\text{periodic}$. Notice that if it were only for the Euclidean action, this periodic single-bounce motion would constitute its global minimum (among the family of bounces), however the additional weight contributions from the local wave functions biases the endpoints of the overall critical paths toward the origin.\\

\noindent
Let us now come back to studying the Euclidean action of these configurations. Note that each given pair $(x_0,x_T)$ implicitly determines $E_\text{bounce}$ though relation~\eqref{eq:4_Total_Time_Bounce_Motion}, allowing the computation of the action given by 
\begin{align}
	\begin{split}
		S_\text{E}^{(\text{bounce})}\big(x_0,x_T\:\!;T\big)\!\!\: &=  \mathlarger{\mathlarger{\int}}_{\scalebox{0.8}{$x_0$}}^{\scalebox{0.8}{$x_{\text{turn}}^{\scalebox{0.8}{$+$}}$}} \! \sqrt{2m\scalebox{1.15}{$\big[$}V(x)+E_\text{bounce}\scalebox{1.15}{$\big]$}} \,\text{d}x + \big(x_0\leftrightarrow x_T\big) -\!\!\:E_\text{bounce} \;\! T \, ,
	\end{split} \label{eq:4_Euclidean_Action_Bounce_Motion}
\end{align}
As before, we employed the classical EOM as well as invertibility of the bounce velocity in respective domains to arrive at equation~\eqref{eq:4_Euclidean_Action_Bounce_Motion}. It is immanent that for vanishing classical energy, the previous relation~\eqref{eq:6_Shifted_Bounce_Action} is recovered, which, added to the weight functions $\psilocexp$, once more constitutes the usual infinite-time bounce action $\mathdutchbcal{B}$. To conveniently parameterize the hatched strip containing the relevant fluctuations around the critical bounces, let us invoke the coordinate change
\begin{align}
	\big\{x_0,x_T\big\} \mapsto \Big\{x_0,E_\text{bounce}(x_0,x_T)\Big\} \, ,
	\label{eq:Coordinate_Transformation}
\end{align}
which inside the narrow tubular region is well-defined. This transformation allows us to conveniently bring the exponent $f\raisebox{-2.5pt}{\scalebox{0.75}{$\text{exp}$}}\!\!\!\!\!\!\!\:\!\raisebox{5pt}{\scalebox{0.65}{$(\text{bounce})$}}\big[x_0,E_\text{bounce}(x_0,x_T)\big]$ into a purely quadratic form in the coordinate $E_\text{bounce}$, leaving us only with the integration over $x_0$, acting as a collective coordinate. While the integration domain for the $E_\text{bounce}$-integral can be na\"{\i}vely taken to be the entire real axis, the collective coordinate $x_0$ only runs over the region $x_0 \in\big(x_\text{min},x_\text{escape}\big)$. The newly introduced constant $x_\text{min}$ denotes the point closest to $x_\text{FV}=0$ for which a trajectory with vanishing classical energy can still reach its turning point $x_\text{escape}$ in the given time $T$,\footnote{Or vice versa, the position reached after time $T$ when starting a motion at $x_{\scalebox{0.7}{\text{escape}}}$ with vanishing velocity.} see figure~\ref{fig:BounceRegions}. Additionally, beware of the Jacobian
\begin{align}
	\text{J}_\text{det}\big(x_0,E_\text{bounce}\big) = \det\!\bigg[\frac{\partial(x_0,x_T)}{\partial(x_0,E_\text{bounce})}\bigg] = \frac{\partial x_T\big(x_0,E_\text{bounce}\big)}{\partial E_\text{bounce}} \, ,
\end{align}
arising from the given coordinate transformation~\eqref{eq:Coordinate_Transformation}. With these preliminary considerations out of the way, let us proceed by expanding the exponent in orders of the bounce energy. The zeroth order result $f\raisebox{-2.5pt}{\scalebox{0.75}{$\text{exp}$}}\!\!\!\!\!\!\!\:\!\raisebox{5pt}{\scalebox{0.65}{$(\text{bounce})$}}\big(x_0,E_\text{bounce}\!\!\:=\!\!\:0\big)=\mathdutchbcal{B}$ is clear from the previous discussion around equation~\eqref{eq:4_Euclidean_Action_Bounce_Motion}, such that we can turn our attention to the higher order terms. Utilizing the relations\footnote{In case the derivative acts on $x^+_{\scalebox{0.7}{\text{turn}}}$, the arising term drops out as the integrand vanishes when evaluated at the upper boundary.} 
\begin{subequations}
	\begin{align}
		\frac{\partial}{\partial E_\text{bounce}} \mathlarger{\mathlarger{\int}}_{\scalebox{0.8}{$x_0$}}^{\scalebox{0.8}{$x_{\text{turn}}^{\scalebox{0.8}{$+$}}$}} \! \sqrt{2m\scalebox{1.15}{$\big[$}V(x)+E_\text{bounce}\scalebox{1.15}{$\big]$}} \,\text{d}x   &=  \!\mathlarger{\mathlarger{\int}}_{\scalebox{0.8}{$x_0$}}^{\scalebox{0.8}{$x_{\text{turn}}^{\scalebox{0.8}{$+$}}$}} \bigg\{\frac{2}{m}\scalebox{1.15}{$\big[$}V(x)+E_\text{bounce}\scalebox{1.15}{$\big]$}\!\!\:\bigg\}^{\!-\frac{1}{2}} \,\text{d}x \, ,\\[0.1cm]
		\frac{\partial}{\partial E_\text{bounce}} \mathlarger{\mathlarger{\int}}_{\scalebox{0.8}{$x_T$}}^{\scalebox{0.8}{$x_{\text{turn}}^{\scalebox{0.8}{$+$}}$}} \! \sqrt{2m\scalebox{1.15}{$\big[$}V(x)+E_\text{bounce}\scalebox{1.15}{$\big]$}} \,\text{d}x &= \!\mathlarger{\mathlarger{\int}}_{\scalebox{0.8}{$x_T$}}^{\scalebox{0.8}{$x_{\text{turn}}^{\scalebox{0.8}{$+$}}$}} \bigg\{\frac{2}{m}\scalebox{1.15}{$\big[$}V(x)+E_\text{bounce}\scalebox{1.15}{$\big]$}\!\!\:\bigg\}^{\!-\frac{1}{2}} \,\text{d}x \\
		&\qquad\!   -\frac{\partial x_T\big(x_0,E_\text{bounce}\big)}{\partial E_\text{bounce}} \, \sqrt{2m\scalebox{1.15}{$\big[$}V(x_T)+E_\text{bounce}\scalebox{1.15}{$\big]$}} \, , \nonumber 
		\\[0.15cm]
		\frac{\partial}{\partial E_\text{bounce}} \mathlarger{\mathlarger{\int}}^{\scalebox{0.8}{$x_T$}}_{\scalebox{0.8}{$0$}}\! \sqrt{2mV(x)} \,\text{d}x &= \frac{\partial x_T\big(x_0,E_\text{bounce}\big)}{\partial E_\text{bounce}} \,\sqrt{2mV(x_T)} \, ,
	\end{align}
	\label{eq:Partial_Derivatives_Ebounce}%
\end{subequations}
together with equations~\eqref{eq:4_Total_Time_Bounce_Motion} and~\eqref{eq:4_Euclidean_Action_Bounce_Motion}, we find the simple expression
\begin{align}
	\!\frac{\partial f_\text{exp}^{(\text{bounce})}\big(x_0,E_\text{bounce}\big)}{\partial E_\text{bounce}}&= \frac{\partial x_T\big(x_0,E_\text{bounce}\big)}{\partial E_\text{bounce}} \,  \bigg\{\!\!\:\sqrt{2mV(x_T)}-\sqrt{2m\scalebox{1.15}{$\big[$}V(x_T)+E_\text{bounce}\scalebox{1.15}{$\big]$}}\, \bigg\} \, .
	\label{eq:Partial_Derivative_Exponent_Energy_Bounce}
\end{align}
As demanded by the previous discussion in section~\ref{sec:Critical_Paths_Sequential}, the first derivative necessarily vanishes for critical paths obeying $E_\text{bounce}=0$, independent of the value of the collective coordinate $x_0$. Differentiating equation~\eqref{eq:Partial_Derivative_Exponent_Energy_Bounce} again and evaluating it at vanishing bounce energy, we see that only the term involving the derivative acting on the bracketed expression survives, yielding the intermediate result 
\begin{align}
	f_\text{exp}^{(\text{bounce})}\big(x_0,E_\text{bounce}\big)&=\mathdutchbcal{B} - \frac{E_\text{bounce}^2}{2} \, \frac{\partial x_T\big(x_0,E_\text{bounce}\big)}{\partial E_\text{bounce}}\scalebox{1.15}{$\bigg\vert$}_{\scalebox{0.8}{$E_\text{bounce}\!\!\:\!\!\:=\!\!\:\!\!\:0$}} \nonumber \\ 
	&\qquad\qquad\qquad \times  \bigg\{\frac{2}{m}\,V\!\!\:\Big[x_T\big(x_0,E_\text{bounce}\!\!\:=\!\!\:0\big)\Big]\!\!\:\bigg\}^{\!-\frac{1}{2}} + \mathcal{O}\scalebox{1.1}{$\big($}E_\text{bounce}^3\scalebox{1.1}{$\big)$} \, .
\end{align}
Given this expansion, we can directly perform the Laplace-type $E_\text{bounce}$-integration, yielding 
\begin{align}
	\!\Big[\mathcal{P}_{\!\!\: n}^{(\pi/2)}\big(\hbar,T\big)\Big]^{(\text{LO})}_\text{bounce} &= \frac{1}{2^n n!} \left(\frac{m\omega}{\pi\hbar}\right)^{\!\frac{1}{2}}\, \exp\!\bigg(\!\!-\!\!\:\frac{\mathdutchbcal{B}}{\hbar}\bigg)  \mathlarger{\mathlarger{\int}}_{\scalebox{0.8}{$x_\text{min}$}}^{\scalebox{0.8}{$x_\text{escape}$}} \text{d}x_0 \mathlarger{\mathlarger{\int}}_{-\infty}^{\infty} \!\text{d}E \; \frac{\partial x_T\big(x_0,E\big)}{\partial E} \:  \psi_{n,\text{non-exp}}^{(\text{loc})}(x_0)\nonumber \\[-0.1cm]  
	& \quad \times  \psi_{n,\text{non-exp}}^{(\text{loc})}\Big[x_T\big(x_0,E\big)\Big] \,   \Bigg\{\!\!\!\:-\!\!\:\frac{1}{2\pi\hbar}\,\frac{\partial^2 S_\text{E}^{(\text{bounce})}\big(x_0,E\:\!;T\big)}{\partial x_0\,\partial  E} \scalebox{1.1}{\bigg[}\frac{\partial x_T\big(x_0,E\big)}{\partial E}\scalebox{1.1}{\bigg]}^{-1 }\Bigg\}^{\!\!\:\frac{1}{2}} \nonumber \\ 
	&\quad \times \exp\!\!\:\Bigg(\frac{E^2}{2\hbar} \, \frac{\partial x_T\big(x_0,E\big)}{\partial E}\scalebox{1.15}{$\bigg\vert$}_{\scalebox{0.8}{$E\!\!\:\!\!\:=\!\!\:\!\!\:0$}} \bigg\{\frac{2}{m}\,V\!\!\:\Big[x_T\big(x_0,E\!\!\:=\!\!\:0\big)\Big]\!\!\:\bigg\}^{\!-\frac{1}{2}} \Bigg) \, , \label{eq:P_Expression_Sequential_Bounce} \\[0.2cm] 
	&= \frac{1}{2^n n!} \left(\frac{m\omega}{\pi\hbar}\right)^{\!\frac{1}{2}}\, \exp\!\bigg(\!\!-\!\!\:\frac{\mathdutchbcal{B}}{\hbar}\bigg)  \mathlarger{\mathlarger{\int}}_{\scalebox{0.8}{$x_\text{min}$}}^{\scalebox{0.8}{$x_\text{escape}$}} \text{d}x_0 \; \Bigg\{\frac{\partial^2 S_\text{E}^{(\text{bounce})}\big(x_0,E\:\!;T\big)}{\partial x_0\,\partial  E}\scalebox{1.15}{$\bigg\vert$}_{\scalebox{0.8}{$E\!\!\:\!\!\:=\!\!\:\!\!\:0$}}\Bigg\}^{\!\!\:\frac{1}{2}} \nonumber \\
	&\quad \times \psi_{n,\text{non-exp}}^{(\text{loc})}(x_0) \, \psi_{n,\text{non-exp}}^{(\text{loc})}\Big[x_T\big(x_0,E\!\!\:=\!\!\:0\big)\Big] \, \bigg\{\frac{2}{m}\,V\!\!\:\Big[x_T\big(x_0,E\!\!\:=\!\!\:0\big)\Big]\!\!\:\bigg\}^{\!\frac{1}{4}} \, .  \nonumber
\end{align}
Note that in the last line, all non-exponential terms are evaluated at $E_\text{bounce}=0$, as each higher order in $E_\text{bounce}$ arises with an additional factor of $\sqrt{\hbar}$.\footnote{Unfortunately, as we will argue in section~\ref{sec:Small_Endpoints}, this simplification is invalid for large $T$, as the higher order terms in $E_{\scalebox{0.7}{\text{bounce}}}$ are found to be parametrically enhanced.} This allows us to get rid of all terms involving $\partial x_T/\partial E$, leaving only the partial derivative of the Euclidean action undetermined. Employing the Hamilton-Jacobi formula~\eqref{eq:Partial_Derivative_Euclidean_Action_wrt_Endpoints_A} again, we find 
\begin{align}
	\frac{\partial^2 S_\text{E}^{(\text{bounce})}\big(x_0,E_\text{bounce}\:\!;T\big)}{\partial x_0\,\partial  E_\text{bounce}}\scalebox{1.2}{$\bigg\vert$}_{\scalebox{0.8}{$E_\text{bounce}\!\!\:\!\!\:=\!\!\:\!\!\:0$}} = -\bigg[\frac{m}{2V(x_0)}\bigg]^{\frac{1}{2}} \, , 
\end{align}
thus providing us with the intermediate result
\begin{align}
	\Big[\mathcal{P}_{\!\!\: n}^{(\pi/2)}\big(\hbar,T\big)\Big]^{(\text{LO})}_\text{bounce} &= \frac{\pm i}{2^{n} n!} \left(\frac{m\omega}{\pi\hbar}\right)^{\!\frac{1}{2}}\, \exp\!\bigg(\!\!-\!\!\:\frac{\mathdutchbcal{B}}{\hbar}\bigg) \label{eq:IntermediateResult_Bounce_Sequential} \\[0.1cm] 
	&\quad \times \mathlarger{\mathlarger{\int}}_{\scalebox{0.8}{$x_\text{min}$}}^{\scalebox{0.8}{$x_\text{escape}$}} \text{d}x_0 \, \Bigg\{\!\!\:\bigg[\frac{V(x_T)}{V(x_0)}\bigg]^{\frac{1}{4}} \psi_{n,\text{non-exp}}^{(\text{loc})}(x_0) \, \psi_{n,\text{non-exp}}^{(\text{loc})}(x_T)\Bigg\}_{\scalebox{0.8}{\!$x_T\big(x_0,E\!\!\:=\!\!\:0\big)$}} \, . \nonumber 
\end{align}
In order to evaluate the remaining collective coordinate integral, we require to instate a simplification for $\psilocnonexp$. As before, the ensuing discussion hinges on the magnitude of $x_{0,T}$,\footnote{Whereas section~\ref{sec:3_5_BounceContribution} was split purely based on the magnitude of $T$, let us now rather distinguish the cases based on the magnitude of their endpoints. In the former discussion, this more lucid distinction could not be utilized to its full extent due to the additional Green's function terms.} with both to-be-distinguished cases heavily overlapping. Recall that we initially shifted the potential such that the FV point $x_\mathrm{FV}$ is located at the origin, thus all subsequent statements about the magnitude of the endpoints $x_{0,T}$ concern their deviation from this reference point $x_\mathrm{FV}=0$.

\subsubsection{Moderately small $x_{0,T}$}
\label{sec:ModeratelySmall_Endpoints}

Let us at first view the situation for $x_{0,T}\gg \sqrt{\hbar}$, which in case of moderately large $T$ satisfying $e^{-\omega T}\gg \sqrt{\hbar}$ is always guaranteed due to relation~\eqref{eq:ExactTimeRelations}. In that case, the previous expression~\eqref{eq:IntermediateResult_Bounce_Sequential} is reminiscent of relation~\eqref{eq:BounceContributionIntermediate} derived in section~\ref{sec:Small_Moderate_T}. Utilizing the identical reasoning that lead to equation~\eqref{eq:Simplified_Integrand} and noticing that in the above expression~\eqref{eq:IntermediateResult_Bounce_Sequential} the point $(x_0,x_T)$ always lies on the critical strip $E_\text{bounce}=0$ since the perpendicular coordinate direction has already been integrated out, for $\smash{T\ll \omega^{-1} \log\!\big(\hbar^{-\frac{1}{2}}\big)}$ we can again invoke the relations
\begin{subequations}
	\begin{align}
		V(x_0)^{\frac{1}{4}} \, \psi_{n,\text{non-exp}}^{(\text{loc})}(x_0) &= \bigg(\frac{m\omega^2}{2}\bigg)^{\!\!\!\:\frac{1}{4}} \bigg(\frac{4m\omega}{\hbar}\bigg)^{\!\!\frac{n}{2}} \Big(\mathdutchbcal{A} e^{\scalebox{0.75}{$-\omega t_{\text{turn}}$}}\Big)^{\!\scalebox{0.85}{$\frac{2n+1}{2}$}}  \bigg\{1+\mathcal{O}\Big[\sqrt{\hbar} \: e^{\omega T}\Big]\!\!\:\bigg\} \, , \\ 
		V(x_T)^{\frac{1}{4}} \, \psi_{n,\text{non-exp}}^{(\text{loc})}(x_T) &= \bigg(\frac{m\omega^2}{2}\bigg)^{\!\!\!\:\frac{1}{4}} \bigg(\frac{4m\omega}{\hbar}\bigg)^{\!\!\frac{n}{2}} \Big(\mathdutchbcal{A} e^{\scalebox{0.75}{$\omega t_{\text{turn}}-\omega T$}}\Big)^{\!\scalebox{0.85}{$\frac{2n+1}{2}$}}  \bigg\{1+\mathcal{O}\Big[\sqrt{\hbar} \: e^{\omega T}\Big]\!\!\:\bigg\} \, ,
	\end{align}
	\label{eq:Simplified_NonExp_WaveFunction}%
\end{subequations}
see the previous equation~\eqref{eq:Simplified_Integrand}. Again notice that these relations hold for arbitrarily large endpoints and only fail once the magnitude of $x_0,x_T$ becomes strongly suppressed. This greatly simplifies the $x_0$-integral, arriving at the leading-order expression
\begin{align}
	\!\!\!\Big[\mathcal{P}_{\!\!\: n}^{(\pi/2)}\big(\hbar,T\big)\Big]^{(\text{LO})}_\text{bounce} \! =\!\!\: \frac{\pm i}{n!} \left(\frac{m^2\omega^3}{2\pi\hbar}\right)^{\!\!\!\:\frac{1}{2}}\, \!\!\!\:\exp\!\bigg(\!\!-\!\!\:\frac{\mathdutchbcal{B}}{\hbar}\bigg) \bigg(\frac{2m\omega}{\hbar}\bigg)^{\!\!n} \Big(\mathdutchbcal{A} e^{-\omega T}\Big)^{\!\scalebox{0.85}{$\frac{2n+1}{2}$}} \!\mathlarger{\mathlarger{\int}}_{\scalebox{0.8}{$x_\text{min}$}}^{\scalebox{0.8}{$x_\text{escape}$}} \!\frac{\text{d}x_0}{\sqrt{V(x_0)}} \,.
	\label{eq:Bounce_Contribution_Sequential}
\end{align}
Let us remind ourselves on how we obtained the two bounds on the remaining collective coordinate integral. As can be seen schematically in figure~\ref{fig:BounceRegions}, the largest $x_0$ lying on the critical strip is just the escape point $x_\text{escape}$, while the minimal possible one is $x_\text{min}$, belonging to a critical bounce trajectory ending at $x_T=x_\text{escape}$ after precisely time $T$. This allows us to utilize relation~\eqref{eq:4_Total_Time_Bounce_Motion}, such that the arising integral can be computed exactly, yielding $\smash{2^{-\frac{1}{2}}m^{\frac{1}{2}} T}$. In the given approach, this collective coordinate factor due to time translation invariance is hidden until the very end, since the Dirichlet boundary conditions of the initial Euclidean propagator break the underlying symmetry explicitly. Inserting the value for the last integral into the above expression~\eqref{eq:Bounce_Contribution_Sequential} fully reproduces the former result~\eqref{eq:BounceContributionFinal} after a suitable choice of the overall sign.

\subsubsection{Strongly suppressed $x_{0,T}$}
\label{sec:Small_Endpoints}

As we have seen in section~\ref{sec:Large_T}, the former result should be generalizable to incorporate $T$ satisfying $e^{-\omega T/2} \lesssim \sqrt{\hbar} \ll e^{-\omega T}$, however in the sequential computation scheme this extension seems to be unjustifiably hard to prove. Therefore, let us limit this discussion to highlighting the challenges that arise when attempting to extend the result without proper care. Aside from the subsequent discussion for $\omega T\ll \log\!\big(\hbar^{-1}\big)$, the present approach would ultimately fail for $e^{-\omega T}$ approaching $\hbar$, as in that case the Euclidean propagator evolves a quasi-zero mode $\lambda_0\sim e^{-\omega T}$, which renders the use of the Van Vleck propagator~\eqref{eq:VanVleck_Propagator} ill-advised. Thus, we expect an even deeper breakdown to occur in that range of $T$, as was found in section~\ref{sec:Large_T}. \\

\noindent
In case the endpoints are strongly suppressed, the replacements~\eqref{eq:Simplified_NonExp_WaveFunction} would cease to work as the lower order monomial terms inside the Hermite polynomial are non-negligible. Instead of assuming $x_{0,T}\lesssim \sqrt{\hbar}$, let us work with the more convenient assumption $x_{0,T}\ll 1$, for which we already know the behavior~\eqref{eq:4_Asymptotics_Non_Exponential_Wave_Function_Small_x} of the wave function $\psilocnonexp$. Additionally, note that for $x_0$ and $x_T$ adhering to a bounce motion with vanishing energy, equations~\eqref{eq:ExactTimeRelationx0} and~\eqref{eq:ExactTimeRelationxT} yield the relation $x_T\big(x_0,E\!\!\:=\!\!\:0\big)=\mathdutchbcal{A}^2 e^{-\omega T}/x_0+\text{subleading}$. Utilizing these simplifications and employing that the potential is approximately harmonic for small $x_{0,T}$, the prevalent case yields\footnote{Note that the inferred simplifications on the integrand are only applicable if both $x_0$ and $x_T$ are small in magnitude, thus at the boundary of the integral we pick up an error. However, one will find this to be an acceptable error to make due to both boundary regions of the critical line only yielding subleading $\mathcal{O}(1)$ contributions, whereas the intermediate region $x_0\sim x_T$ yields the desired moduli space volume factor $T\gg 1$. However, due to the integration region for $x_0$ being heavily contracted near $x_{\scalebox{0.7}{\text{min}}}$, this is hardly spotted in how the integral is written down. Rather than choosing the lower bound $x_{\scalebox{0.7}{\text{min}}}$, one could also split the $(x_0,x_T)$-plane diagonally into two parts, instating a symmetry factor of two while integrating only over the triangular region $x_0\geq x_T$. Consequently, the lower bound of the integral would be the point on the critical line for which $x_0=x_T\sim \mathcal{A}e^{-\omega T/2}\ll 1$ holds, thus the relevant contributions of the integral are seen to arise from a region where the instated simplifications are manifestly valid.}
\begin{align}
	\Big[\mathcal{P}_{\!\!\: n}^{(\pi/2)}\big(\hbar,T\big)\Big]^{(\text{LO})}_\text{bounce} &\approx \frac{\pm i}{2^{n} n!} \left(\frac{m\omega}{\pi\hbar}\right)^{\!\frac{1}{2}}\, \exp\!\bigg(\!\!-\!\!\:\frac{\mathdutchbcal{B}}{\hbar}\bigg) \: \mathdutchbcal{A}\, \exp\!\!\:\bigg(\!\!-\!\!\:\frac{\omega T}{2}\bigg) \nonumber \\[0.1cm] 
	&\qquad\qquad \times  \mathlarger{\mathlarger{\int}}_{\scalebox{0.8}{$x_\text{min}$}}^{\scalebox{0.8}{$x_\text{escape}$}} \frac{\text{d}x_0}{x_0} \,  H_n\scalebox{1.1}{\bigg(}\!\!\;\sqrt{\frac{m\omega}{\hbar}}\,x_0\!\!\:\scalebox{1.1}{\bigg)} \, H_n\scalebox{1.1}{\bigg(}\!\!\;\sqrt{\frac{m\omega}{\hbar}}\,\frac{\mathdutchbcal{A}^2 e^{-\omega T}}{x_0}\scalebox{1.1}{\bigg)} \, .  
	\label{eq:BounceContribution_LargeT_Sequential}
\end{align}
As one would find, the above expression~\eqref{eq:BounceContribution_LargeT_Sequential} only gives rise to the correct result in case of $e^{-\omega T/2} \gg \hbar$, which was already entailed in the discussion portrayed in section~\ref{sec:ModeratelySmall_Endpoints}. A deeper investigation reveals that the second line of equation~\eqref{eq:P_Expression_Sequential_Bounce} breaks down already for $T$ exceeding the former time scale, which can be attributed to the term $\partial x_T/\partial E$ being greatly enhanced, such that formerly neglected terms are actually relevant. More specifically, for small $\smash{x_0,x_T=\mathcal{O}\big(e^{-\omega T/2}\big)}$, one finds
\begin{align}
	\frac{\partial x_T\big(x_0,E_\text{bounce}\big)}{\partial E_\text{bounce}}\scalebox{1.15}{$\bigg\vert$}_{\scalebox{0.8}{$E_\text{bounce}\!\!\:=\!\!\:0$}} \approx \frac{x_T\big(x_0,E_\text{bounce}\!\!\:=\!\!\:0\big)}{m\omega^2} \bigg[x_0^{-2}+x_T\big(x_0,E_\text{bounce}\!\!\:=\!\!\:0\big)^{-2}\,\bigg]\, ,
\end{align}
thus there is a conceivable error when dropping higher order terms of the relation
\begin{align}
	x_T\big(x_0,E\big)& =x_T\big(x_0,E\!\!\:=\!\!\:0\big)+\frac{\partial x_T\big(x_0,E\big)}{\partial E}\scalebox{1.15}{$\bigg\vert$}_{\scalebox{0.8}{$E\!\!\:\!\!\:=\!\!\:\!\!\:0$}}\: E + \mathcal{O}\big(E^2\big) \nonumber \\
	&\approx x_T\big(x_0,E\!\!\:=\!\!\:0\big) \raisebox{0.5pt}{\Bigg[}1+\underbrace{\frac{x_0^{-2}+x_T\big(x_0,E\!\!\:=\!\!\: 0\big)^{-2}}{m\omega^2} \, E}_{\displaystyle{=\mathcal{O}\big(e^{\omega T} \sqrt{\hbar}\,\big)}}\raisebox{0.5pt}{\Bigg]} + \mathcal{O}\big(E^2\big) \, .
\end{align}
While usually $\smash{E_\text{bounce}=\mathcal{O}\big(\:\!\!\sqrt{\hbar}\,\big)}$ has the effect that for any appearance of $E_\text{bounce}$, the energy is simply set to zero, this is no longer valid in case the next-to-leading-order term attains a conflicting enhancement. Thus, the previously instated replacement $x_T\big(x_0,E\big)\mapsto x_T\big(x_0,E\!\!\:=\!\!\:0\big)$ in equation~\eqref{eq:P_Expression_Sequential_Bounce} is invalid for $T$ surpassing the former bound. A similar statement can be made for the replacement of $E_\text{bounce}=0$ inside the $\partial x_T/\partial E$ term encountered in relation~\eqref{eq:P_Expression_Sequential_Bounce}. As we have seen, even though for $\omega T\ll \log\!\big(\hbar^{-1}\big)$ the provided ansatz can theoretically still be employed, we would have to retain higher-order corrections in $E_\text{bounce}$ when performing the Gaussian $E_\text{bounce}$-integration~\eqref{eq:P_Expression_Sequential_Bounce}. Such an endeavor seems unreasonable, compelling us to conclude this discussion prematurely.

\subsection{Comparison between sequential and composite evaluation}
\label{sec:Comparison_Methods}

With both evaluation schemes presented, let us briefly compare them to highlight their respective advantages and disadvantages. As we have seen, the sequential ansatz is rather straightforward, providing only minor technical roadblocks. Due to rightfully being able to employ the Van Vleck propagator~\eqref{eq:VanVleck_Propagator} in the relevant case $\omega T \ll \log\big(\hbar^{-1}\big)$, the sequential evaluation boils down to manipulating expressions involving classical trajectories, e.g. Hamilton-Jacobi equations~\eqref{eq:Partial_Derivative_Euclidean_Action_wrt_Endpoints}, which have been long-studied and are thus well-understood. However, one still faces certain difficulties within this approach. From the above treatment, it remains unclear from which integral the negative mode providing the imaginary factor arises.\footnote{If one inspects this problem more closely, one would find that, for small $T$, the endpoint integration (more precisely the $E_{\scalebox{0.7}{$\text{bounce}$}}$-integral) would be divergent due to $\partial x_T/\partial E_{\scalebox{0.7}{$\text{bounce}$}}>0$, whereas for large $T$ the Van Vleck formula, i.e. the path integral, encapsulates the negative mode. In intermediate time ranges, the origin of the negative mode depends on the value of the collective coordinate $x_0$.} One finds that for different values of $T$, the endpoint integrals and the inner path integral trade the negative mode, making its correct interpretation ambiguous. In principle, the same problem could potentially arise in the presence of quasi-zero modes, being split between the different integrals in intricate ways. However, even in case the symmetry is fully contained in one of the integrals, as was the case for our computation, the origin of the zero mode and the emerging moduli space volume factor are much harder to spot.\\ 

\noindent
In comparison, the composite path integral approach required considerable initial thought as well as several rather technical computations extending the conventional formulas to the case of Robin boundary conditions. However, once the technical foundations are established, the method can be employed to more general situations while being conceptually much clearer. As we have seen, the approach manifestly preserves time translation symmetry, strongly elucidating the origin of the underlying zero mode. Additionally, the composite path integral computation is more simply extended to larger $T$, explicitly showing the inevitable breakdown of the provided ansatz. This feat seems practically unachievable using the sequential ansatz, as it proves exceptionally difficult to obtain good bounds on the error of many involved quantities. This can be traced back to performing several Gaussian integrals, each of which can develop soft modes while adding supplemental caveats due to underlying symmetries being explicitly broken at intermediate stages of the calculation. Thus, although the customary sequential approach offers an orthogonal perspective on solving the problem at hand, we find that it partly obscures the agreement both with WKB results for excited states and standard instanton methods applied to the false ground state.

\subsection{Comments on previous works}
\label{sec:4_FormerAttempts}

While our techniques utilized to compute the endpoint-weighted expression~\eqref{eq:Resonant_Energy_Conjecture} are partly novel, the initial ansatz~\eqref{eq:Ansatz_MuellerKirsten_Liang} is not, as it has formerly been suggested by Liang and Müller-Kirsten (LiMK). Initially investigating excited state level splittings~\cite{LiangPeriodicInstantons}, LiMK also employed the presented idea to obtain tunneling rates~\cite{LiangNonVacuumBounces}, with numerous similar papers by them portraying the method applied to various specific cases~\cite{LiangTunneling,LiangSineGordon,LiangFiniteTempCalculation,MullerKirstenQMBook}. Even though their expositions conclude with correct results, it has not been possible to reproduce several key steps in their derivations, as we will report on in the present section. We explicitly demonstrate how their computations in spite of these flaws were able to conclude with the correct exponential factor. Contrary to this purely classical bit, comprehending their calculation of the fluctuation factor proves to be less straightforward. Even though we are able to pinpoint several steps that are in clear conflict with a proper treatment, it is still obscure how the correct answer emerges from these computations. With the subsequent discussion hinging on a sequential evaluation of expression~\eqref{eq:Ansatz_MuellerKirsten_Liang}, the previous considerations presented in chapter~\ref{sec:A_SequentialComputation} gain importance.\\

\begin{figure}[H]
	\centering
	\includegraphics[width=0.8\textwidth]{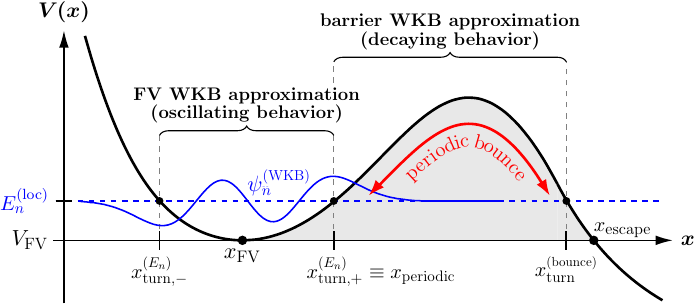}
	\caption{\hspace{0.001cm}}
	\vspace{-0.5cm}\hspace{1.3cm}
	\begin{tikzpicture}
		\node[anchor=center, inner sep=0pt, 
		minimum width=14.55cm, text width=14.55cm-0.4cm,align=justify,
		minimum height=2.5cm, 
		text depth=2.5cm-0.4cm] (0,0) {Illustration of the periodic finite-time bounce motion investigated by LiMK, depicted in the original potential $V(x)$. The classical turning point $x\raisebox{-2.5pt}{\scalebox{0.7}{$\text{turn}$}}\hspace{-0.5cm}\raisebox{4.5pt}{\scalebox{0.6}{$(\text{bounce})$}}$ of this periodic instanton as well as the two additional turning points $x\raisebox{-2.5pt}{\scalebox{0.7}{$\text{turn},\pm$}}\hspace{-0.82cm}\raisebox{4.5pt}{\scalebox{0.6}{$(E_n)$}}\hspace{0.45cm}$ to energy $E\raisebox{-2.5pt}{\scalebox{0.7}{$n$}}\!\!\raisebox{4.5pt}{\scalebox{0.6}{$(\text{loc})$}}$, around which the utilized plane wave WKB approximation changes from an oscillatory to a decaying behavior, are portrayed.};
	\end{tikzpicture}
	\label{fig:5_Liang_Kirsten_Bounce}
\end{figure}

\noindent
LiMK start their calculation with the observation that amongst all admissible bounce motions with fixed time $T$ and arbitrary endpoints $x_{0,T}$, the Euclidean action $S_\text{E}\llbracket x\rrbracket$ is minimized by periodic motions inside the inverted barrier region. These trajectories, dubbed ``periodic instantons'', possess a period length of exactly $T$, starting and ending their motion with vanishing velocity. Locating the so-found trajectory in figure~\ref{fig:BounceRegions}, depicting the different admissible bounce regions for varying endpoints $x_{0,T}$, the periodic instanton solution is precisely the tip of the black bounding line, amounting to coinciding endpoints $x_0=x_T=x_\text{periodic}$. From minimizing the Euclidean action, LiMK conclude that these solutions constitute the dominant tunneling trajectory around which both path and exterior integrals are to be expanded. This assessment however initially ignores the non-negligible exponential weight contributions arising from the wave functions $\psilocexp$, decisively influencing the endpoints of the critical bounce trajectory. In our approach of taking a uniform WKB ansatz, we saw that the critical endpoints get biased toward the FV point, resulting in the critical strip shown in figure~\ref{fig:BounceRegions}. However, LiMK opted to utilize a usual plane wave WKB approximation of the form
\begin{align}
	\psi_{n}^{(\text{loc})}(x)& = C_n \, \bigg\{2m\Big[V(x)-E_n^{(\text{loc})}\Big]\!\bigg\}^{\!\!\:-\frac{1}{4}}\,\exp\!\left\{\!\!\:-\:\!\frac{1}{\hbar}\mathlarger{\mathlarger{\int}}_{\scalebox{0.7}{$x_{\text{turn},+}^{(E_n)}$}}^{\scalebox{0.7}{$x$}}\sqrt{2m\Big[V(\xi)-E_n^{(\text{loc})}\Big]}\,\text{d}\xi\right\}\, ,
	\label{eq:4_WKB_Wave_Function_Inside_Barrier_Liang}
\end{align}
oscillating inside the FV region while being purely decaying inside the barrier region as schematically illustrated in figure~\ref{fig:5_Liang_Kirsten_Bounce}. Before inspecting the consequences of this choice of $\psinloc$ regarding the computation of the fluctuation factor, let us investigate its implications on the overall critical bounce trajectory. Similar to the derivation of equation~\eqref{eq:Necessary_Transversality_Condition_Critical_Paths_Sequential}, employing the Hamilton-Jacobi equations at both endpoints yields the modified transversality conditions
\begin{subequations}
	\begin{align}
		\sqrt{2m \:\!\scalebox{1.15}{$\Big\{$} V\!\!\:\scalebox{1.2}{$\big[$}x_0^{(\text{bounce})}\scalebox{1.2}{$\big]$}-E_n^{(\text{loc})}\scalebox{1.15}{$\Big\}$} } &\overset{!}{=} \phantom{(-1)^{\ell+1}} \sqrt{2m \:\!\scalebox{1.15}{$\Big\{$} V\!\!\:\scalebox{1.2}{$\big[$}x_0^{(\text{bounce})}\scalebox{1.2}{$\big]$}+E_\text{bounce}\scalebox{1.15}{$\Big\}$} } \, , \\
		\sqrt{2m \:\!\scalebox{1.15}{$\Big\{$} V\!\!\:\scalebox{1.2}{$\big[$}x_T^{(\text{bounce})}\scalebox{1.2}{$\big]$}-E_n^{(\text{loc})}\scalebox{1.15}{$\Big\}$} } &\overset{!}{=} (-1)^{\ell+1} \sqrt{2m \:\!\scalebox{1.15}{$\Big\{$} V\!\!\:\scalebox{1.2}{$\big[$}x_T^{(\text{bounce})}\scalebox{1.2}{$\big]$}+E_\text{bounce}\scalebox{1.15}{$\Big\}$} } \, .
	\end{align}
	\label{eq:4_Restrictions_Critical_Path}%
\end{subequations}
As in equation~\eqref{eq:Necessary_Transversality_Condition_Critical_Paths_Sequential}, $\ell$ denotes the number of turning points traversed during the motion. Notice that because we only considered the case $\Enloc=\mathcal{O}(\hbar)$, the additional $\Enloc$-dependence was negligible in determining the critical trajectory as it could always be fully absorbed into the non-exponential prefactor $\psilocnonexp$, solely entering the discussion of the fluctuation factor. While this simplified our treatment of the critical paths, shifting ensuing complications into the computation of the overall fluctuation factor, the method by LiMK of retaining the resonant energy inside the exponent is equally admissible. In this more generalized setting by LiMK, the magnitude of the classical bounce energy has to match the energy of the excited state one decays from, essentially demanding $E_\text{bounce}=-\Enloc$. The authors never explicitly state this relation clearly and in fact at no point even distinguish between these two ad hoc unrelated energies entering the discussion, implicitly setting them equal from the start.\footnote{Beware of the fact that LiMK define the classical bounce energy with a negative sign, such that both quantities are indeed identical.} While this can be considered careless, it rescues their computation, as it ensures that they indeed expand around the dominant bounce trajectory found by proper weighting arguments, fortunately coinciding with the Euclidean action minimum. It is immanent that the correct exponential factor emerges from the discussion. However, one caveat remains, as the time interval $T$ cannot be chosen freely, but is instead fully fixed after specifying the energy $\Enloc$ of the decaying initial state $\psinloc$. Demanding the bounce trajectory to be periodic in the first place intricately ties the formerly arbitrary constant $T$ to the resonant energy in question. This renders our portrayed calculation more faithful to the setup of the problem, since we manifestly treat $T$ as the unconstrained parameter it is. 
\vspace{-0.5cm}

\noindent
In order to obtain the non-exponential fluctuation factor, one now faces several challenges. The first problem arises when considering the determinant pertaining to the Euclidean propagator. Because the initial and final velocity of the periodic bounce motion vanish identically, the spectrum of the fluctuation operator contains the exact zero mode $e_0(t)=\dot{x}_\text{bounce}(t)$, leading to a divergent determinant prefactor.\footnote{Note that in the sequential computation presented in appendix~\ref{sec:A_SequentialComputation}, this problem does not arise since the critical bounce velocities at the endpoints never vanish. Instead, one finds the soft mode to be unproblematic in the considered limits, mandating no special treatment.} While such phenomena can usually be treated utilizing collective coordinates, the zero mode in the present case does not correspond to a symmetry of the path integral, as time translation invariance is explicitly broken by the Dirichlet boundary conditions of the propagator. Thus, the correct way to proceed would amount to retaining quartic order fluctuations in the zero mode $e_0(t)$ when expanding the path integral, which is a rather daunting task.\footnote{The same holds for fluctuations around the critical bounce involving shifts in the endpoint $x_0$ and/or $x_T$, for which the na\"{\i}ve determinant factor is not divergent, but rather parametrically enhanced. This regime still requires special care, as na\"{\i}ve semiclassical formulas generally cease to yield appropriate results.} A second problem becomes immanent when inspecting equation~\eqref{eq:4_WKB_Wave_Function_Inside_Barrier_Liang}, displaying the chosen wave function $\psinloc$. With the critical bounce motion starting and ending at the linear turning point $x\raisebox{-2.5pt}{\scalebox{0.7}{$\text{turn},-$}}\hspace{-0.82cm}\raisebox{4.5pt}{\scalebox{0.6}{$(E_n)$}}\hspace{0.35cm}=x_\text{periodic}$ satisfying $V(x_\text{periodic})=\Enloc$, the na\"{\i}ve WKB estimate diverges. This is expected, since the utilized plane wave WKB approximation breaks down near classical turning points. A rigorous analysis would once more require us to construct a uniform WKB approximation around the linear turning point $x_\text{periodic}$ in order to resolve the proper behavior of the wave function close to that point. One must conclude that the employed wave function is inadequate for the desired task, as one is required to fully resolve the neighborhood of the relevant turning point correctly, which the above ansatz~\eqref{eq:4_WKB_Wave_Function_Inside_Barrier_Liang} is not able to achieve. Once more, this analysis shows explicitly why it was crucial to choose a uniform WKB approximation for our analysis. \\ 

\noindent
Despite acknowledging being faced with both previously discussed divergences arising from the wave function $\psinloc$ and the propagator determinant, LiMK practically ignore the inherited issues completely. They state that since the bare singularity is only present when evaluating the questionable expressions exactly at the critical bounce solution, the divergence gets ``smoothed out'' by the additional endpoint integrations. However, this procedure is clearly not admissible as the origin of the present singularity is completely unphysical, stemming solely from an improper treatment of both the propagator and the wave function in the relevant region of interest. Exacerbating the situation, they utilize the common shifting method~\cite{GelfandFunctionalIntegration,DashenNonperturbativeMethods}, supplying the functional determinant as
\begin{equation}
	\text{det}_\zeta\!\left\{-\frac{\text{d}^2}{\text{d}(\omega t)^2}+\frac{V''\big[x_\text{bounce}(t)\big]}{m\omega^2}\right\} = 2\:\! N\!\!\;(0)\:\! N\!\!\;(T) \mathlarger{\int}_{0}^T \:\!\frac{1}{N\!\!\;(t)^2}\: \text{d}t \: ,
	\label{eq:3_Shifting_Method_Determinant}
\end{equation}
where $N\!\!\;(t)$ is an arbitrary zero mode of the portrayed differential operator $O_\mathrm{bounce}$, provided the function $N\!\!\;(t)$ entails no zero crossings in the desired interval $[0,T]$.\footnote{The additional requirement of $N\!\!\;(t)$ possessing no zero crossings enters through demanding the coordinate transformation utilized in the derivation of the result~\eqref{eq:3_Shifting_Method_Determinant} to be well-defined, see e.g. reference~\cite{GelfandFunctionalIntegration}.} LiMK na\"{\i}vely employ this formula by setting $N\!\!\;(t)=\dot{x}_\text{bounce}(t)$, possessing a zero crossing due to the bounce motion traversing the turning point $x\raisebox{-2.5pt}{\scalebox{0.7}{$\text{turn}$}}\hspace{-0.5cm}\raisebox{4.5pt}{\scalebox{0.6}{$(\text{bounce})$}}$, thus rendering the integral in equation~\eqref{eq:3_Shifting_Method_Determinant} ill-defined even in a principal value sense. After stoic manipulation of all these inherently flawed expressions, the derivations by LiMK strikingly conclude with correct results. Despite providing numerous steps in their derivation, it remains unclear to us how and why these errors offset, enabling them to ultimately unveil the desired expression.\\

\noindent
Lastly, LiMK also consider multi-instanton solutions, na\"{\i}vely exponentiating the single-bounce result. While this is essentially unnecessary due to the constraints put on the magnitude of $T$, it is also incorrect in the given instance because the weight functions $\psinloc$ only enter the discussion once, spoiling the usual property of multi-bounce contributions exponentiating easily. Although this issue ultimately does not affect the result, it highlights that LiMK's work fails to accurately account for the limitations of their own methods. With these computations lacking reliable approximations, and the correct result hinging on an effective cancellation of initially flawed arguments, the present paper provides the first systematic approach to evaluating finite-time amplitudes for the decay of excited states through tunneling using functional methods.\\

\noindent 
Before concluding, let us briefly highlight two works by Schiff et al.~\cite{SchiffWeightedPathIntegrals} and by Turok~\cite{TurokRealTimeTunneling}, where the idea of weighting propagator contributions to amplitudes involving wave functions has been utilized to study the real-time evolution of the quantum-mechanical wave function. Although these discussions only entail a single endpoint integral for the initial time, the utilized methods parallel our previous considerations. Turok studies the exponential factor for the decay rate by employing a sequential approach to extract the action of the relevant critical path associated to a real-time tunneling event for an initial Gaussian wave function centered inside the FV region. While yielding the desired exponential suppression factor of the wave function far outside the FV region for the special potentials under consideration, the relation to the decay rate has not been explained directly. Additionally, an extraction of the non-exponential prefactor employing his method appears yet infeasible. Meanwhile, the work of Schiff et al. is noteworthy as, to the best of our knowledge, it is the first to adopt a composite approach for computing the resulting expression to derive a general asymptotic expansion of the wave function after some time has evolved. They however opt not to split the initial wave function into an exponential and a non-exponential piece, which in practice leads to a non-trivial mixing between orders in the desired $\hbar$-expansion. Consequentially, several crucial simplifications as for example the vanishing of $E_\mathrm{crit}$ for critical trajectories would not be easily achievable given this approach, complicating an explicit systematic expansion as shown in the present work.

\section{Conclusion}
\label{sec:5_Conclusion}

In this work, we have systematically shown how to obtain the decay width of excited states employing predominantly functional methods. As we found, one does not gain much ground without supplementing the pursued approach with an appropriate guess for the excited state wave functions $\psinloc$ in the guise of a uniform WKB approximation. However, this sophisticated demand renders the given ansatz inconvenient for almost all practical purposes. The sole advantage over traditional WKB methods is that we spare ourselves the tedious matching procedure, which could be particularly convenient in multi-dimensional settings, where a purely local analysis of resonant wave functions may be preferential compared to studying their global features. Despite these shortcomings, the presented approach reveals some additional structure underlying the standard instanton method, which will be explored in an upcoming paper.\\ 

\noindent 
In addition to yielding yet another alternative way to extract tunneling rates, the present work introduces some novel methods facilitating a composite approach to endpoint-weighted path integrals, bypassing the usual inefficient sequential computation of the involved integrals. While these techniques require supplementary preparatory setup, once all auxiliary computations have been provided, the portrayed methods can easily be generalized to encompass more involved problems. Perhaps the most valuable feature of such a combined analysis is the manifestation of symmetries, granting an exceptionally clear treatment of encapsulated zero modes. The suggestive composite computation scheme especially shines when calculating functional determinants, as the commonly employed Gel'fand-Yaglom method readily encompasses the case of generalized Robin boundary conditions. \\

\noindent
The demonstrated techniques could possibly provide further clarification on a purely real-time picture of quantum tunneling since the time evolution of any given state can be similarly represented as an endpoint-weighted path integral of the form
\begin{align}
	\!\!\psi\big(x_T,T\big)=\!\!\:\mathlarger{\mathlarger{\int}}_{-\infty}^{\infty} \psi_{T=0}(x_0)\, K\!\!\;\scalebox{1.15}{$\big($}x_0,x_T\:\!;T\scalebox{1.15}{$\big)$}\:\mathrm{d}x_0 = \!\!\:\mathlarger{\mathlarger{\int}}_{x(0) \mathrm{\;free}}^{x(T)=x_T} \!\!\:\mathcal{D}\llbracket x\rrbracket\;  \psi_{T=0}\big[x(0)\big]\:\! \exp\!\!\:\bigg(\frac{iS \llbracket x\rrbracket}{\hbar}\bigg)\:\! ,
\end{align} 
with only the starting point being unconstrained. While Schiff et al.~\cite{SchiffWeightedPathIntegrals} and Turok~\cite{TurokRealTimeTunneling} have made strides in this direction, a plethora of questions still remain unanswered. We leave this intriguing direction for exploration to future work.

\vfill
\subsection*{Acknowledgments}

N.W. is supported by the German Academic Scholarship Foundation as well as the International Max Planck Research School on Elementary Particle Physics (IMPRS EPP) run by the Max Planck Institute for Physics. During earlier stages of the work, N.W. was partly supported by a scholarship from the Max Weber Program of the state of Bavaria.

\newpage

\appendix 

\section{Decay widths using the traditional WKB formalism}
\label{sec:B_DecayWidthsTraditionalWKB}

\begin{figure}[H]
	\centering
	\includegraphics[width=0.85\textwidth]{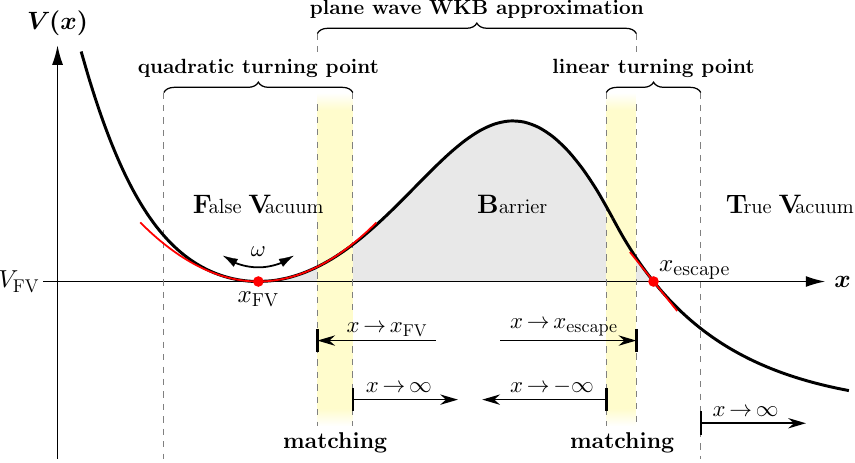}
	\caption{\hspace{0.001cm}}
	\vspace{-0.5cm}\hspace{1.3cm}
	\begin{tikzpicture}
		\node[anchor=center, inner sep=0pt, 
		minimum width=14.55cm, text width=14.55cm-0.4cm,align=justify,
		minimum height=2.5cm, 
		text depth=2.5cm-0.4cm] (0,0) {Portrayal of the WKB matching procedure employed for the method of matched asymptotic expansions. The plane wave WKB approximation obtained inside the barrier region is matched to the exact solutions around the turning points where the original WKB expansion is not applicable. The potential past the linear turning point $x_\text{escape}$ will be found to have no influence on the tunneling rate, as tunneling events are seen to occur toward $x\rightarrow \infty$.};
	\end{tikzpicture}
	\label{fig:2_WKB_Matching_Procedure}
\end{figure}

For the work to be self-contained, let us briefly present a more traditional way of obtaining the desired excited-state decay widths, exclusively employing the wave function picture. We will closely follow reference~\cite{GarbrechtFunctionalMethods}, providing the subsequent reasoning for the ground state decay rate. As briefly touched upon in section~\ref{sec:2_CommonInstantonMethod}, we desire to solve the time-independent Schrödinger equation 
\begin{equation}
	\frac{\text{d}^2}{\text{d} x^2} \,  \psi_n^{(\text{loc})}(x)= \frac{2m}{\hbar^2} \Big[V(x)-E_n^{(\text{loc})}\Big] \psi_n^{(\text{loc})}(x)
\end{equation}
to outgoing Gamow--Siegert boundary conditions, revealing a non-perturbatively small imaginary contribution to the emerging resonant energy eigenvalue $\Enloc$. In order to tackle this problem, one commonly employs the method of matched asymptotic expansions~\cite{BenderAdvancedMathematicalMethods,GarbrechtFunctionalMethods}, schematically illustrated in figure~\ref{fig:2_WKB_Matching_Procedure}. The central idea consists in solving for the leading-order behavior of the wave function both inside the barrier region as well as near the classical turning points $x_\text{FV}$ and $x_\text{escape}$, where a na\"{\i}ve plane wave WKB approximation fails. These individual approximations, valid in constrained parts of the potential, are then asymptotically matched in their respective overlap regions. Mandating the so-obtained solution to be a purely outgoing wave past the escape point $x_\mathrm{escape}$ quantizes the desired resonant energies, with their imaginary part corresponding to the sought-after decay rate. We stress that this procedure relies on global information about the to-be-constructed solution $\psinloc(x)$, while for the functional treatment $\psinloc(x)$ is only required in the vicinity of the FV region, thus a different ansatz is employed, see appendix~\ref{sec:C_UniformWKB}.
\noindent
Note that we utilize the appropriate prescription of matching the WKB solution to parabolic cylinder functions around the quadratic turning point $x_\text{FV}$. This will yield the correct numerical factors compared to frequently encountered derivations only employing linear turning point connection formulas, sometimes even leading to debates over the reliability of WKB methods, see the illuminating review by Garg~\cite{GargTunnelingRevisited}.\\

\noindent
As before, without loss of generality we shift the potential such that $x_\text{FV}=0$ and $V_\text{FV}=0$ holds. Let us remind ourselves of the important definitions~\eqref{eq:2_Definition_AB_Parameter_Tunneling} and the fact that we are perturbing around approximate harmonic oscillator states inside the FV region. Thus, the resonant eigenenergies are given by
\begin{equation}
	E_n^{\scalebox{0.8}{$(\text{loc})$}}=\hbar \omega\bigg(n+\frac{1}{2}+\epsilon_n\bigg) \, ,
\end{equation} 
with $n\in \mathbb{N}^0$ and $\epsilon_n\ll 1$ being a small correction term later found to be the desired imaginary contribution.\footnote{NLO corrections to the real part of the resonant eigenenergies are fully ignored as they are insignificant to our discussion.} With these preliminary considerations out of the way, we are set to compute the wave functions in the different regions, starting with the behavior inside the barrier region.

\vspace{0.2cm}

\subsection{WKB solution inside the barrier region}
\label{sec:2_3_2_WKB_Solution_Inside_Barrier}
Inside the barrier region, far away from the turning points $x_\text{FV}$ and $x_\text{escape}$, we can use the familiar plane wave approximation, yielding the well-known leading-order WKB result    
\begin{align}
	\psi_n^{(\text{WKB})}(x)\sim \frac{c_\pm}{\sqrt{p(x)}}\,\exp\!\left[\pm\:\!\frac{1}{\hbar}\int_{\scalebox{0.8}{$x_\text{escape}$}}^{\scalebox{0.8}{$x$}} p(\xi)\,\text{d}\xi\right] \quad \text{with } \; p(x)=\sqrt{2m\Big[V(x)-E_n^{\scalebox{0.8}{$(\text{loc})$}}\Big]} \, .
	\label{eq:2_WKB_Solution_Barrier_Ansatz}
\end{align}
Note that for convenience, we choose $x_\text{escape}$ instead of $x_\text{FV}$ as our reference point, simplifying the subsequent discussion. Deep inside the barrier region we can furthermore employ the approximation $V(x)\gg E\raisebox{-2.5pt}{\scalebox{0.75}{$n$}}\!\!\!\!\:\raisebox{4pt}{\scalebox{0.65}{$(\text{loc})$}}$ to expand $p(x)$ to second order, obtaining
\begin{align}
	p(x)=\sqrt{2m\Big[V(x)-E_n^{\scalebox{0.8}{$(\text{loc})$}}\Big]} = \sqrt{2mV(x)} \left\{1-\frac{E_n^{\scalebox{0.8}{$(\text{loc})$}}}{2V(x)}+ \mathcal{O}\!\left[\frac{(n\hbar \omega)^2}{V(x)^2}\right]\!\!\:\right\} .
\end{align}
While the zeroth order in $\hbar$ suffices to determine the overall prefactor, the first order needs to be retained as the exponential term comes with an additional factor $\hbar^{-1}$, see equation~\eqref{eq:2_WKB_Solution_Barrier_Ansatz}. With this, the exponent is given by
\begin{align}
	\mathlarger{\int}_{\scalebox{0.8}{$x_\text{escape}$}}^{\scalebox{0.8}{$x$}} p(\xi)\,\text{d}\xi = -\mathlarger{\int}_{\scalebox{0.8}{$x$}}^{\scalebox{0.8}{$x_\text{escape}$}} \sqrt{2mV(\xi)} \,\text{d}\xi + mE_n^{\scalebox{0.8}{$(\text{loc})$}}\mathlarger{\int}_{\scalebox{0.8}{$x$}}^{\scalebox{0.8}{$x_\text{escape}$}} \frac{1}{\sqrt{2mV(\xi)}} \,\text{d}\xi  + \mathcal{O}\big(\hbar^2\big).
	\label{eq:2_Expansion_Exponential_Function}
\end{align}
As shown pictorially in figure~\ref{fig:2_WKB_Matching_Procedure}, the method of matched asymptotic approximations demands the behavior of the WKB solution near the two turning points, so let us begin by studying the solution in the limit $x\rightarrow 0^+$. The first integral in equation~\eqref{eq:2_Expansion_Exponential_Function}, expanded around the origin, yields
\begin{align}
	\mathlarger{\int}_{\scalebox{0.8}{$x$}}^{\scalebox{0.8}{$x_\text{escape}$}} \sqrt{2mV(\xi)} \,\text{d}\xi &= \mathlarger{\int}_{\scalebox{0.8}{$0$}}^{\scalebox{0.8}{$x_\text{escape}$}} \sqrt{2mV(\xi)} \,\text{d}\xi - \mathlarger{\int}_{\scalebox{0.8}{$0$}}^{\scalebox{0.8}{$x$}} \sqrt{2mV(\xi)} \,\text{d}\xi \nonumber\\
	&= \frac{\mathdutchbcal{B}}{2} - m\omega \mathlarger{\int}_{\scalebox{0.8}{$0$}}^{\scalebox{0.8}{$x$}} \xi\:\!\Big[1+W(\xi)\Big]^{\frac{1}{2}}  \,\text{d}\xi = \frac{\mathdutchbcal{B}}{2} - \frac{m\omega x^2}{2} \Big[1+\mathcal{O}(x)\Big] \, .
	\label{eq:2_Approximation_First_Integral_Towards_FV}
\end{align}
Note that we used $x>0$, assuming the particle to escape toward $x\rightarrow \infty$, as portrayed in figure~\ref{fig:2_WKB_Matching_Procedure}. For the second integral, one has to be careful as the integrand possesses a singularity at $x_\text{FV}=0$. We extract the singular behavior by isolating the diverging part, finding
\begin{align}
	\mathlarger{\int}_{\scalebox{0.8}{$x$}}^{\scalebox{0.8}{$x_\text{escape}$}} \frac{1}{\sqrt{2mV(\xi)}} \,\text{d}\xi &=\frac{1}{m\omega} \mathlarger{\int}_{\scalebox{0.8}{$x$}}^{\scalebox{0.8}{$x_\text{escape}$}} \,\frac{1}{\xi}\:\!\Big[1+W(\xi)\Big]^{\!-\frac{1}{2}}  - \frac{1}{\xi} \,\text{d}\xi + \frac{1}{m\omega} \Big[\log(\xi)\Big]_{\scalebox{0.8}{$x$}}^{\scalebox{0.8}{$x_\text{escape}$}} \nonumber \\
	&=  \frac{1}{m\omega} \log\!\bigg(\frac{\mathdutchbcal{A}}{x}\:\!\bigg) + \mathcal{O}(x) \, .
	\label{eq:2_Approximation_Second_Integral_Towards_FV}
\end{align}
Note that we utilized definition~\eqref{eq:2_Definition_A_Parameter_Tunneling} to represent the $\mathcal{O}(1)$ contribution in simple form. Using the WKB ansatz~\eqref{eq:2_WKB_Solution_Barrier_Ansatz} together with the expansion~\eqref{eq:2_Expansion_Exponential_Function} and the intermediate results~\eqref{eq:2_Approximation_First_Integral_Towards_FV} and~\eqref{eq:2_Approximation_Second_Integral_Towards_FV}, we find the full expression in the limit $x\rightarrow 0^+$ to be given by
\begin{align}
	\scalebox{0.98}{$\psi_n^{(\text{WKB})}(x)$} &\,\scalebox{0.98}{$\displaystyle{\xrightarrow{\,x\rightarrow 0^+\!\:}   \frac{c_\pm}{\sqrt{m\omega x}}\, \exp\!\Bigg\{\!\mp\frac{1}{\hbar}\!\left[\frac{\mathdutchbcal{B}}{2} - \frac{m\omega x^2}{2}-\omega^{-1} E_n^{\scalebox{0.75}{$(\text{loc})$}} \log\!\bigg(\frac{\mathdutchbcal{A}}{x}\:\!\bigg) + \mathcal{O}\big(\hbar x\big) \right]\!\!\!\:\Bigg\} \Big[1+\mathcal{O}\big(\hbar, x\big)\Big]}$} \nonumber \\
	&\,\scalebox{0.98}{$\displaystyle{\xrightarrow{\,x\rightarrow 0^+\!\:}  \frac{c_\pm}{\sqrt{m\omega \mathdutchbcal{A}}} \,  \exp\!\left(\:\!\mp\;\!\frac{\mathdutchbcal{B}-m\omega x^2}{2\hbar}\right) \Big(\mathdutchbcal{A}^{-1} x \Big)^{\scalebox{0.8}{$\mp \frac{2n+1}{2}\!\!\:\!\!\:-\!\!\:\!\!\:\frac{1}{2}$}} \Big[1+\mathcal{O}\big(\hbar, x,\epsilon_n\big)\Big] .}$}
	\label{eq:2_WKB_Result_near_Quadratic_Turning_Point}
\end{align}
Here we approximated the prefactor as $p(x)=\sqrt{2mV(x)}+\mathcal{O}(\hbar)=m\omega x+\mathcal{O}\big(\hbar,x^{2}\big)$ and neglected the small energy correction $\lvert \epsilon_n\rvert\ll 1$ inside the exponent. This concludes the discussion of the left limit, such that we now turn to the asymptotic behavior of the WKB approximation~\eqref{eq:2_WKB_Solution_Barrier_Ansatz} in the limit $x\rightarrow x_\text{escape}^-$. Linearizing the potential around $x_\text{escape}$, we obtain 
\begin{align}
	\mathlarger{\int}_{\scalebox{0.8}{$x$}}^{\scalebox{0.8}{$x_\text{escape}$}} \sqrt{2mV(\xi)} \,\text{d}\xi =& \, \sqrt{-2m V'(x_\text{escape})} \mathlarger{\int}_{\scalebox{0.8}{$x$}}^{\scalebox{0.8}{$x_\text{escape}$}}  \sqrt{x_\text{escape}-\xi} \: \Big[1+\mathcal{O}\big(x_\text{escape}-\xi\big)\Big] \,\text{d}\xi \nonumber \\
	=& \, \frac{2}{3}\sqrt{2m \big\lvert V'(x_\text{escape})\big\rvert \big(x_\text{escape}-x\big)^3} \: \Big[1+\mathcal{O}\big(x_\text{escape}-x\big)\Big] \, .
	\label{eq:2_Approximation_First_Integral_Towards_xturn}
\end{align}
We will subsequently abbreviate $V'_\text{escape}\coloneqq V'(x_\text{escape})$, which in the case of the particle escaping toward $x\rightarrow \infty$ is negative. With the choice of our reference point taken to be $x_\text{escape}$, we can ignore the next-to-leading-order correction in relation~\eqref{eq:2_Expansion_Exponential_Function} as it vanishes linearly, thus being fully negligible. With the former considerations, we find the leading-order behavior near the turning point $x_\text{escape}$ to be given by
\begin{align}
	\psi_n^{(\text{WKB})}(x) \xrightarrow{\scalebox{0.8}{$x\!\!\:\!\!\:\rightarrow\!\!\:\!\!\: x_\text{escape}^-\,$}} \frac{c_\pm}{\sqrt[4]{2m\big\lvert V'_\text{escape}\big\rvert\big(x_\text{escape}-x\big)}}\:\! \exp\!\!\!\:\left[ \mp\frac{\sqrt{8m\big\lvert V'_\text{escape}\big\rvert \big(x_\text{escape}-x\big)^3}}{3\hbar} \right]  .
	\label{eq:2_WKB_Result_near_Linear_Turning_Point}
\end{align}

\subsection{Solution around the quadratic turning point}
\label{sec:2_3_3_Solution_near_Quadratic_Turning_Point}

In the FV region, the potential is dominated by its harmonic part, for which the solutions to the Schrödinger equation
\begin{equation}
	-\frac{\hbar^2}{2m} \frac{\text{d}^2}{\text{d} x^2}\, \psi_n^{(\text{FV})}(x)=\Bigg[\hbar\omega\bigg(n+\frac{1}{2}+\epsilon_n\bigg)-\frac{m\omega^2 x^2}{2}\Bigg] \psi_n^{(\text{FV})}(x) 
	\label{eq:2_Schrödinger_Equation_Quadratic_Turning_Point}
\end{equation} 
can be given exactly in terms of parabolic cylinder functions $D_\nu(x)$. The two fundamental solutions to the above ODE are found to be
\begin{align}
	u_1^{\scalebox{0.8}{$(n\!\!\:\!\!\:+\!\!\:\!\!\:\epsilon_n)$}}(x)&= D_{\scalebox{0.8}{$n\!\!\:\!\!\:+\!\!\:\!\!\:\epsilon_n$}}\!\left(\sqrt{\frac{2m\omega}{\hbar}}\, x\right) \quad \text{and} \quad  u_2^{\scalebox{0.8}{$(n\!\!\:\!\!\:+\!\!\:\!\!\:\epsilon_n)$}}(x)= i D_{\scalebox{0.8}{$-n\!\!\:\!\!\:-\!\!\:\!\!\:1\!\!\:\!\!\:-\!\!\:\!\!\:\epsilon_n$}}\!\left(i\sqrt{\frac{2m\omega}{\hbar}}\, x\right) \, .
	\label{eq:2_Exact_Solution_Quadratic_Turning_Point}
\end{align}
However, in the end we only require the ratio of the coefficients multiplying both particular solutions in order to compute $c_+/c_-$. To obtain this information, we observe that for the unperturbed problem $\epsilon_n=0$, for which the potential has been rendered stable at $x_\text{FV}$, equation \eqref{eq:2_Schrödinger_Equation_Quadratic_Turning_Point} reduces to that of a pure harmonic oscillator (HO), with the normalizable
solution given by $\psi\raisebox{-2.5pt}{\scalebox{0.7}{$n$}}\!\!\!\!\:\raisebox{4.5pt}{\scalebox{0.65}{$(\text{HO})$}}(x)=u\raisebox{-3pt}{\scalebox{0.7}{$1$}}\!\!\raisebox{5pt}{\scalebox{0.7}{$(n)$}}(x)$. Notice that one can safely ignore any supplementary
normalization factors, as in the end we will be content with solely computing ratios. The aforementioned solution also shows that, to leading-order, we require $n\in \mathbb{N}^0$ to recover the usual harmonic oscillator wave functions given in terms of Hermite polynomials. Setting $\psi\raisebox{-2.5pt}{\scalebox{0.7}{$n$}}\!\!\!\!\:\raisebox{4.5pt}{\scalebox{0.65}{$(\text{FV})$}}=\psi\raisebox{-2.5pt}{\scalebox{0.7}{$n$}}\!\!\!\!\:\raisebox{4.5pt}{\scalebox{0.65}{$(\text{HO})$}}+\epsilon_n\:\!\delta\psi\raisebox{-2.5pt}{\scalebox{0.7}{$n$}}\!\!\!\!\:\raisebox{4.5pt}{\scalebox{0.65}{$(\text{FV})$}}$, we obtain the inhomogeneous ODE 
\begin{equation}
	\frac{\text{d}^2}{\text{d} x^2}\, \delta\psi_n^{(\text{FV})}(x)+\left[\frac{m\omega}{\hbar}\big(2n+1\big)-\frac{m^2\omega^2 x^2}{\hbar^2}\right] \delta\psi_n^{(\text{FV})}(x)=-\frac{2m\omega}{\hbar}\, u_1^{\scalebox{0.8}{$(n)$}}(x) \, , 
	\label{eq:2_ODE_Perturbed_Solution_Around_HO}
\end{equation} 
which we have already provided in canonical form. The particular solution is then simply given by
\begin{align*}
	\scalebox{0.97}{$\displaystyle{\delta\psi_n^{(\text{FV})}(x)=-\frac{2m\omega}{\hbar} \:\! W\!\!\:\Big[u_1^{\scalebox{0.8}{$(n)$}},u_2^{\scalebox{0.8}{$(n)$}}\Big]^{-1}\Bigg\{\!\!\!\:-u_1^{\scalebox{0.8}{$(n)$}}(x)\!\!\:\mathlarger{\int}_{-\infty}^{\scalebox{0.8}{$x$}} u_1^{\scalebox{0.8}{$(n)$}}(\xi)\:\! u_2^{\scalebox{0.8}{$(n)$}}(\xi)\,\text{d}\xi +u_2^{\scalebox{0.8}{$(n)$}}(x)\!\!\:\mathlarger{\int}_{-\infty}^{\scalebox{0.8}{$x$}} \Big[u_1^{\scalebox{0.8}{$(n)$}}(\xi)\Big]^2 \,\text{d}\xi\Bigg\} \, ,}$}
\end{align*}
where we already used the fact that the Wronskian of the linearly independent solutions $u\raisebox{-3pt}{\scalebox{0.7}{$1$}}\!\!\raisebox{5pt}{\scalebox{0.65}{$(n)$}}(x)$ and $u\raisebox{-3pt}{\scalebox{0.7}{$2$}}\!\!\raisebox{5pt}{\scalebox{0.65}{$(n)$}}(x)$ is a constant, found to be given by
\begin{align}
	W\!\!\:\Big[u_1^{\scalebox{0.8}{$(n)$}},u_2^{\scalebox{0.8}{$(n)$}}\Big] \coloneqq u_1^{\scalebox{0.8}{$(n)$}}(x) \:\frac{\text{d}u_2^{\scalebox{0.8}{$(n)$}}(x)}{\text{d}x} - \frac{\text{d}u_1^{\scalebox{0.8}{$(n)$}}(x)}{\text{d}x}\: u_2^{\scalebox{0.8}{$(n)$}}(x) = \sqrt{\frac{2m\omega}{\hbar}}\,(-i)^n \, .
	\label{eq:2_Wronskian_Parabolic_Cylinder_Functions}
\end{align}
This shows that when perturbing around the harmonic oscillator wave function, we obtain an additional outward growing contribution $u\raisebox{-3pt}{\scalebox{0.7}{$2$}}\!\!\raisebox{5pt}{\scalebox{0.65}{$(n)$}}(x)$. The asymptotic behavior of the parabolic cylinder functions is given by 
\begin{subequations}
	\begin{align}
		D_{\scalebox{0.8}{$n$}}(x) &\xrightarrow{x\rightarrow \infty} \exp\!\left(\!-\frac{x^2}{4}\right) x^{n} \Big[1+\mathcal{O}\big(x^{-2}\big)\Big] \,, \\
		iD_{\scalebox{0.8}{$-n\!\!\:\!\!\:-\!\!\:\!\!\:1$}}(i x) &\xrightarrow{x \rightarrow \infty} \exp\!\left(\!+\frac{x^2}{4}\right) x^{-n} \left[\frac{i^{3n}}{x}+\mathcal{O}\big(x^{-3}\big)\right] \, ,
	\end{align}
	\label{eq:2_Asymptotics_Parabolic_Cylinder_Functions}%
\end{subequations}
such that we can compute the leading-order correction $\delta\psi\raisebox{-2pt}{\scalebox{0.7}{$n$}}\!\!\!\!\:\raisebox{4.5pt}{\scalebox{0.65}{$(\text{FV})$}}$ easily. While the $\epsilon_n$ corrections received by the  $u\raisebox{-3pt}{\scalebox{0.7}{$2$}}\!\!\raisebox{5pt}{\scalebox{0.65}{$(n)$}}$-component constitute its leading-order, the corrections to the $u\raisebox{-3pt}{\scalebox{0.7}{$1$}}\!\!\raisebox{5pt}{\scalebox{0.65}{$(n)$}}$ factor are already next-to-leading-order, with the dominant contribution arising from the unperturbed wave function. Thus, when considering the leading-order expression of the ratio between the corresponding coefficients, we can safely ignore the $u\raisebox{-3pt}{\scalebox{0.7}{$1$}}\!\!\raisebox{5pt}{\scalebox{0.65}{$(n)$}}$ corrections. With this, we arrive at
\begin{align}
	\delta\psi_n^{(\text{FV})}(x)&=-\sqrt{\frac{2m\omega}{\hbar}} \: i^n \, u_2^{\scalebox{0.8}{$(n)$}}(x)\!\!\:\mathlarger{\int}_{-\infty}^{\scalebox{0.8}{$x$}} \Big[u_1^{\scalebox{0.8}{$(n)$}}(\xi)\Big]^2 \,\text{d}\xi + \Big[u_1^{\scalebox{0.8}{$(n)$}}\text{ contributions}\Big] \, ,
	\label{eq:2_Particular_Solution_Around_Quadratic_Turning_Point_Preliminary}
\end{align}
such that in the limit $x\rightarrow \infty$ the integration ranges over the entire real line, providing a well-known integral. Ignoring the $u\raisebox{-3pt}{\scalebox{0.7}{$1$}}\!\!\raisebox{5pt}{\scalebox{0.65}{$(n)$}}(x)$ contributions, we obtain the final result
\begin{align}
	\delta\psi_n^{(\text{FV})}(x) \xrightarrow{x\rightarrow \infty\,}  -\sqrt{\frac{2m\omega}{\hbar}} \: i^n \, u_2^{\scalebox{0.8}{$(n)$}}(x)\!\!\:\underbrace{\int_{-\infty}^\infty \Big[u_1^{\scalebox{0.8}{$(n)$}}(\xi)\Big]^2 \,\text{d}\xi}_{\displaystyle{=\sqrt{\frac{\pi\hbar}{m\omega}} \, n!}} = -\sqrt{2\pi}\, i^{n} n! \, u_2^{\scalebox{0.8}{$(n)$}}(x)  .
	\label{eq:2_Particular_Solution_Around_Quadratic_Turning_Point_Final}
\end{align}
With the previously given asymptotic behavior of the parabolic cylinder functions~\eqref{eq:2_Asymptotics_Parabolic_Cylinder_Functions} we deduce
\begin{align}
	\begin{split}
		\psi_n^{(\text{FV})}(x) \xrightarrow{x\rightarrow \infty\,}& \; \bigg(\frac{2m\omega}{\hbar}\bigg)^{\!\!\!\:\frac{n}{2}} \exp\!\left(\!-\frac{m\omega x^2}{2\hbar}\right) x^{n} \Big[1+\mathcal{O}\big(\epsilon_n,x^{-2}\big)\Big] \\ 
		&\qquad\quad -\sqrt{2\pi} \, \epsilon_n n! \bigg(\frac{2m\omega}{\hbar}\bigg)^{\!\!\!\:-\frac{n+1}{2}} \exp\!\left(\!\!\:\frac{m\omega x^2}{2\hbar}\right) x^{-n-1} \Big[1+\mathcal{O}\big(\epsilon_n,x^{-2}\big)\Big] \, .
	\end{split} 
	\label{eq:2_Particular_Solution_Around_Quadratic_Turning_Point_Asymptotics}
\end{align}
Comparing this to equation~\eqref{eq:2_WKB_Result_near_Quadratic_Turning_Point}, one can directly read off the desired ratio $c_+/c_-$, obtaining the important intermediate relation
\begin{align}
	\frac{c_+}{c_- } = -\sqrt{2\pi}\;\! \epsilon_n n!\left(\frac{2m\omega \mathdutchbcal{A}^2}{\hbar}\right)^{\!\!\scalebox{0.8}{$-\frac{2n+1}{2}$}} \exp\!\left(\frac{\mathdutchbcal{B}}{\hbar}\right) .
	\label{eq:2_WKB_Ratio_c+_c-_From_Quadratic_Turning_Point}
\end{align}

\vspace{0.3cm}

\subsection{Solution around the linear turning point}
\label{sec:2_3_5_Solution_near_Linear_Turning_Point}

The remaining aspect is to find the solution around $x_\text{escape}$, obtaining a second expression for the ratio $c_+/c_-$, which then allows us to extract $\epsilon_n$. Near the classical turning point $x_\text{escape}$, the linearized potential is of the form $V(x)= \big\lvert V'(x_\text{escape})\big\rvert  \big(x_\text{escape}-x\big) +\mathcal{O}\big(x_\text{escape}-x\big)$ such that after substituting $y=x_\text{escape}-x$ into the Schrödinger equation, we obtain  
\begin{align}
	\frac{\text{d}^2}{\text{d} y^2} \,\psi_n^{(\text{Airy})}(y) &= -\bigg[\underbrace{\frac{m \omega}{\hbar}\big(2n+1+2\epsilon_n\big)}_{\displaystyle{\eqqcolon \rule{0pt}{0.26cm}\alpha}}-\underbrace{\frac{2m \big\lvert V'(x_\text{escape})\big\rvert }{\hbar^2}}_{\displaystyle{\eqqcolon \beta}} \!\: y\:\!\bigg] \psi_n^{(\text{Airy})}(y) \, .
	\label{eq:2_Schrödinger_Equation_Linear_Turning_Point}
\end{align}
For notational convenience, we defined the auxiliary, positive parameters $\alpha$ and $\beta$. Again, the exact solutions are known and can be given in terms of the Airy functions $\text{Ai}(x)$ and $\text{Bi}(x)$, with the general solution found to be
\begin{align}
	\psi_n^{(\text{Airy})}(y) &= c_1 \, \text{Ai}\!\left(\frac{-\alpha+\beta y}{\beta^{2/3}}\right) + c_2 \, \text{Bi}\!\left(\frac{-\alpha+\beta y}{\beta^{2/3}}\right) \, .
	\label{eq:2_Exact_Solution_Linear_Turning_Point}
\end{align}
With our interest solely confined to the behavior for $y\rightarrow \pm \infty$, one can neglect the contribution arising from $\alpha$. The leading asymptotic behavior of the Airy functions in the limit $x \rightarrow \infty$ can be given by
\begin{subequations}
	\begin{align}
		\text{Ai}(x) &\sim \frac{x^{-\frac{1}{4}}}{2\sqrt{\pi}}\, \exp\!\left(\!\!\:-\frac{2}{3}\,x^{\frac{3}{2}}\right)\, , \qquad\quad  \text{Ai}(-x) \sim \frac{x^{-\frac{1}{4}}}{\sqrt{\pi}}\, \sin\!\left(\frac{2}{3}\,x^{\frac{3}{2}}+\frac{\pi}{4}\right)\, ,\\[0.2cm] 
		\text{Bi}(x) &\sim \frac{x^{-\frac{1}{4}}}{\sqrt{\pi}}\, \exp\!\left(\!\!\:+\frac{2}{3}\,x^{\frac{3}{2}}\right) \, , \qquad\quad \,  \text{Bi}(-x) \sim \frac{x^{-\frac{1}{4}}}{\sqrt{\pi}}\, \cos\!\left(\frac{2}{3}\,x^{\frac{3}{2}}+\frac{\pi}{4}\right) \, ,
	\end{align}
	\label{eq:2_Asymptotics_Airy_Functions}%
\end{subequations}
where corrections to the above relations are of order $x^{-\frac{3}{2}}$. In the limit $x\rightarrow \infty$, i.e. $y\rightarrow -\infty$, we demand a purely outgoing plane wave to satisfy the formerly introduced Gamow--Siegert boundary conditions, see section~\ref{sec:2_CommonInstantonMethod}. Ignoring contributions from $\alpha$, we find the asymptotic behavior
\begin{align}
	\begin{split}
		\psi_n^{(\text{Airy})}(x) &\xrightarrow{x\rightarrow \infty\,}\,  \frac{\beta^{-\frac{1}{12}} \big(x-x_\text{escape}\big)^{-\frac{1}{4}}}{\sqrt{\pi}} \Bigg\{\:\! c_1\;\!\sin\!\left[\frac{2}{3}\:\!\beta^{\frac{1}{2}}\big(x-x_\text{escape}\big)^{\frac{3}{2}}+\frac{\pi}{4}\right] \\[-0.15cm] 
		&\qquad\qquad\qquad\qquad\qquad\qquad\qquad\qquad +c_2\,\cos\!\left[\frac{2}{3}\:\!\beta^{\frac{1}{2}}\big(x-x_\text{escape}\big)^{\frac{3}{2}}+\frac{\pi}{4}\right] \!\Bigg\} \, .
	\end{split}
	\label{eq:2_Solution_Linear_Turning_Point_Asymptotics_Infinity}
\end{align}
In order for this to be a purely outgoing wave into the positive $x$-direction, we require the ratio $c_1/c_2=i$. At last, we have to match the Airy solution around the linear turning point to the WKB solution inside the barrier, which relates both ratios $c_1/c_2$ and $c_+/c_-$. Using the asymptotic relations~\eqref{eq:2_Asymptotics_Airy_Functions}, now for $x\rightarrow -\infty$, i.e. $y\rightarrow \infty$, we find the solution given by 
\begin{align}
	\begin{split}
		\psi_n^{(\text{Airy})}(x) &\xrightarrow{x\rightarrow -\infty\,}\,  \frac{\beta^{-\frac{1}{12}} \big(x_\text{escape}-x\big)^{-\frac{1}{4}}}{\sqrt{\pi}} \Bigg\{\:\! \frac{c_1}{2}\;\!\exp\!\left[-\frac{2}{3}\:\!\beta^{\frac{1}{2}}\big(x_\text{escape}-x\big)^{\frac{3}{2}}\right] \\[-0.15cm]  
		&\qquad\qquad\qquad\qquad\qquad\qquad\qquad\qquad\qquad  +c_2\,\exp\!\left[\frac{2}{3}\:\!\beta^{\frac{1}{2}}\big(x_\text{escape}-x\big)^{\frac{3}{2}}\right] \!\Bigg\} \, ,
	\end{split}
	\label{eq:2_Solution_Linear_Turning_Point_Asymptotics_Minus_Infinity}
\end{align}
with $\beta^{\frac{1}{2}}=\hbar^{-1}\sqrt{2m\big\lvert V'(x_\text{escape})\big\rvert}$. Comparing equation~\eqref{eq:2_Solution_Linear_Turning_Point_Asymptotics_Minus_Infinity} to the previous result~\eqref{eq:2_WKB_Result_near_Linear_Turning_Point}, we find the exponents to match perfectly, with the coefficients $c_{1,2}$ necessarily satisfying the relations
\begin{align}
	c_1= \frac{2\sqrt{\pi} \:\! \beta^{-\frac{1}{6}}}{\sqrt{\hbar}} \,c_+  \qquad \text{and} \qquad c_2= \frac{\sqrt{\pi} \:\! \beta^{-\frac{1}{6}}}{\sqrt{\hbar}} \,c_- \, ,
	\label{eq:2_Result_c+_c-_Linear_Turning_Point}
\end{align}
providing the sought-after ratio $c_1/c_2=2\:\! c_+/c_-=i$.

\vspace{0.5cm}
\subsection{Extracting the decay rate}
\label{sec:2_3_5_Extracting_Decay_Rate_WKB}

Equating relation~\eqref{eq:2_WKB_Ratio_c+_c-_From_Quadratic_Turning_Point} with the obtained ratio $c_+/c_- = i/2$, we can directly infer
\begin{align}
	\epsilon_n &= -\frac{i}{2\sqrt{2\pi} \:\! n!} \left(\frac{2m\omega \mathdutchbcal{A}^2}{\hbar}\right)^{\!\!\!\!\:\scalebox{0.8}{$\frac{2n+1}{2}$}} \exp\!\left(\!-\frac{\mathdutchbcal{B}}{\hbar}\right) \, ,
	\label{eq:2_Full_Result_Imaginary_Energy_WKB}
\end{align}
which, as anticipated, is a negative imaginary contribution. This finally leads us to the expression for the decay rate given by
\begin{equation}
	\Gamma_n=-\frac{2}{\hbar} \,\text{Im}\Big[E_n^{\scalebox{0.8}{$(\text{loc})$}}\Big]=\frac{1}{n!}\left(\frac{2m\omega \mathdutchbcal{A}^2}{\hbar}\right)^{\!\!\scalebox{0.8}{$n$}}\sqrt{\frac{m\omega^3 \mathdutchbcal{A}^2}{\pi \hbar}} \,  \exp\!\left(\!-\frac{\mathdutchbcal{B}}{\hbar}\right)\, ,
	\label{eq:2_Full_Result_Decay_Width_WKB}
\end{equation}
which agrees with given literature results, compare e.g. to references~\cite{WeissHaeffnerMetastableDecayFiniteTemp,PatrascioiuComplexTime,LevitBarrierPenetrationTraceMethod,GargTunnelingRevisited}.\\[-0.25cm]

\section{Uniform WKB wave function}
\label{sec:C_UniformWKB}

Just as in the previous appendix, we require the leading-order result for the wave function $\psinloc(x)$ corresponding to the appropriate resonant energy $\Enloc$. The difference for the functional computation is that we require the wave function solely in the vicinity of the FV region, where bounce trajectories start and end. The former treatment is found to be insufficient, as the transition region between the FV and the barrier region is not captured to the required accuracy. This problem can be evaded by utilizing a so-called uniform WKB approximation, properly capturing the behavior past the turning points~\cite{BerryWKBinWaveMechanics,DingleAsymptoticExpansions}. Introducing the small parameter $\varepsilon\coloneqq  (2m\omega)^{-1}\hbar\ll 1$ and employing parametrization~\eqref{eq:Auxiliary_Potential}, the initial Schrödinger equation reads
\begin{equation}
	\varepsilon^2 \!\: \frac{\text{d}^2}{\text{d} x^2} \,  \psi_{\nu}^{(\text{loc})}(x)=  \Bigg\{\frac{x^2}{4} \Big[1+W(x)\Big]-\varepsilon\,\bigg(\nu+\frac{1}{2}\bigg)\!\!\:\Bigg\} \:\psi_{\nu}^{(\text{loc})}(x) \, ,
\end{equation}
which can be solved perturbatively in the parameter $\varepsilon$ by making a parabolic cylinder function ansatz of the form
\begin{align}	
	\psi_{\nu}^{(\text{loc})}(x)=\frac{C_{\nu}}{\sqrt{\phi'(x)}} \, D_{\scalebox{0.75}{$\nu$}}\!\!\left[\frac{\phi(x)}{\sqrt{\varepsilon\,}}\right] \, .
	\label{eq:AnsatzUniformWKB}
\end{align}
Inserting the previous ansatz into the former ODE grants the new differential equation
\begin{align}
	\phi'(x)^2 \Big[\phi(x)^2-2\:\!\varepsilon\!\; \big(2\nu+1\big)\Big]+3\:\! \varepsilon^2 \;\!\frac{\phi''(x)^2}{\phi'(x)^2}-2\:\! \varepsilon^2 \;\! \frac{\phi'''(x)}{\phi'(x)} = x^2 \Big[1+W(x)\Big] - 2\:\!\varepsilon\!\; \big(2\nu+1\big) \, .
\end{align}
Inserting power series ansätze
\begin{align}
	\phi(x)=\sum_{m=0}^\infty \frac{\varepsilon^m}{m!} \, \phi^{\scalebox{0.75}{$(m)$}}(x) \, , \qquad \qquad \nu=\sum_{m=0}^\infty \frac{\varepsilon^m}{m!} \, \nu^{\scalebox{0.75}{$(m)$}}+\:\begin{matrix}\text{non-perturbative} \\[-0.1cm] \text{contributions}\end{matrix}\, ,
\end{align}
the zeroth order result is found to be $\nu$-independent, satisfying the simple first order ODE
\begin{align}
	\phi^{\scalebox{0.75}{$(0)$}}(x)\big.^2 \left[\frac{\text{d}}{\text{d}x}\;\!\phi^{\scalebox{0.75}{$(0)$}}(x)\right]^2 = \left\{\frac{1}{2}\frac{\text{d}}{\text{d}x} \Big[\phi^{\scalebox{0.75}{$(0)$}}(x)\Big]^2\right\}^2 = \frac{2V(x)}{m \omega^2} = x^2 \Big[1+W(x)\Big] \, .
	\label{eq:A1_Phi0_Differential_Equation}
\end{align}
This equation allows for four different solutions, falling into identical pairs when inserted into the ansatz~\eqref{eq:AnsatzUniformWKB}, leaving us with either an exponentially decaying or growing solution inside the barrier region. While the full wave function will be given by a superposition of both solutions, notice that, as seen in section~\ref{sec:B_DecayWidthsTraditionalWKB}, the exponentially increasing part only becomes relevant past $x_\text{escape}$, which is outside the region of interest to us. Retaining only the exponentially decreasing contribution, we find the holomorphic solution
\begin{align}
	\phi^{\scalebox{0.75}{$(0)$}}(x)=  \Bigg\{\frac{2}{x^2}\mathlarger{\mathlarger{\int}}_0^x \,\xi \Big[1+W(\xi)\Big]^{\!\!\:\frac{1}{2}}\, \text{d}\xi\:\!\Bigg\}^{\!\!\:\frac{1}{2}} \: x \, .
	\label{eq:A1_Phi0_Full_Solution}
\end{align}
The integration bound is fixed by the requirement to obtain the correct behavior for the unperturbed problem $W(x)=0$, demanding $\phi^{\scalebox{0.75}{$(0)$}}(x)/x \:\!\raisebox{-1pt}{$\xrightarrow{x\rightarrow \:\!0}$}\:\! 1$. This also bestows the usual result $\nu^{\scalebox{0.75}{$(0)$}}=n\in \mathbb{N}^0$ upon us\footnote{With our interest only lying in the lowest-order approximation of $\psinloc(x)$, we effectively utilize the much more na\"{\i}ve Rayleigh--Schrödinger perturbation theory of setting $\nu=n$, neglecting any corrections to the energy level entering the expansion~\cite{DunneUnsalUniformWKB}.}, as well as fixing the normalization constant (up to order $\varepsilon$ terms) as
\begin{align}
	C_n=\frac{1}{\sqrt{n!}} \left(\frac{m\omega}{\pi\hbar}\right)^{\!\frac{1}{4}} \, .
\end{align}
Hereby we utilized the fact that for $n\in \mathbb{N}^0$, the parabolic cylinder functions are given by
\begin{align}
	D_n(x) = 2^{-\frac{n}{2}} \, H_n\!\!\!\:\left(\frac{x}{\sqrt{2}}\right)\:\! \exp\!\left(\!-\frac{x^2}{4}\right) \, .
\end{align}
Due to the factor $\varepsilon^{-\frac{1}{2}}$ entailed in the ansatz~\eqref{eq:AnsatzUniformWKB}, we also require retaining $\phi^{\scalebox{0.75}{$(1)$}}(x)$, still yielding $\mathcal{O}(1)$ contributions. Only from the second order onward are the overall effects on $\psinloc(x)$ negligible, with further terms constituting subleading contributions involving positive powers of $\varepsilon$. The ODE obtained for the first non-trivial order is 
\begin{align}
	2\:\!\phi^{\scalebox{0.75}{$(0)$}}(x)^2\;\! \phi'^{\scalebox{0.75}{$(0)$}}(x) \;\!\phi'^{\scalebox{0.75}{$(1)$}}(x) + \Big[2\:\!\phi^{\scalebox{0.75}{$(0)$}}(x)\;\! \phi^{\scalebox{0.75}{$(1)$}}(x)- 2\big(2n+1\big)\Big] \phi'^{\scalebox{0.75}{$(0)$}}(x)^2 &= -2\big(2n+1\big) \, ,
\end{align}
which can be cast into the simple form
\begin{align}
	\frac{\text{d}}{\text{d}x} \Big[\phi^{\scalebox{0.75}{$(0)$}}(x) \;\! \phi^{\scalebox{0.75}{$(1)$}}(x)\Big]  &= \frac{2n+1}{\phi^{\scalebox{0.75}{$(0)$}}(x)\;\! \phi'^{\scalebox{0.75}{$(0)$}}(x)} \Big[\phi'^{\scalebox{0.75}{$(0)$}}(x)^2-1\Big] \, .
\end{align}
Again, this first-order ODE can be straightforwardly integrated. We require choosing the constant of integration for $\phi^{\scalebox{0.75}{$(0)$}}(x) \;\! \phi^{\scalebox{0.75}{$(1)$}}(x)$ such that the product vanishes linearly near $0$, otherwise $\phi^{\scalebox{0.75}{$(1)$}}(0)$ would be singular due to $\phi^{\scalebox{0.75}{$(0)$}}(x)/x \,\raisebox{-1pt}{$\xrightarrow{x\rightarrow \:\!0}$}\, 1$. Thus, we find
\begin{align}
	\phi^{\scalebox{0.75}{$(0)$}}(x) \;\! \phi^{\scalebox{0.75}{$(1)$}}(x) &= \big(2n+1\big) \mathlarger{\mathlarger{\int}}_{0}^x \frac{\phi'^{\scalebox{0.75}{$(0)$}}(\xi)^2-1}{\phi^{\scalebox{0.75}{$(0)$}}(\xi)\;\! \phi'^{\scalebox{0.75}{$(0)$}}(\xi)} \, \text{d}\xi \, ,
	\label{eq:A1_Product_Phi0_Phi1}
\end{align}
which one can easily check satisfies the condition of vanishing linearly at the origin. Massaging~\eqref{eq:A1_Product_Phi0_Phi1} by splitting the fraction and subtracting the singular behavior, we find
\begin{align}
	\frac{1}{2n+1}\:\,\!\phi^{\scalebox{0.75}{$(0)$}}(x) \;\! \phi^{\scalebox{0.75}{$(1)$}}(x) &= \mathlarger{\mathlarger{\int}}_{0}^x \frac{\phi'^{\scalebox{0.75}{$(0)$}}(\xi)}{ \phi^{\scalebox{0.75}{$(0)$}}(\xi)} - \frac{1}{ \phi^{\scalebox{0.75}{$(0)$}}(\xi)\;\!\phi'^{\scalebox{0.75}{$(0)$}}(\xi)}  \, \text{d}\xi \nonumber\\[0.1cm]
	&= \mathlarger{\mathlarger{\int}}_{0}^x \frac{\text{d}}{\text{d}\xi}\!\;\bigg\{ \!\log\!\Big[\xi^{-1}\phi^{\scalebox{0.75}{$(0)$}}(\xi)\Big]\bigg\} - \Bigg\{\frac{1}{\xi} \Big[1+W(\xi)\Big]^{-\frac{1}{2}}-\frac{1}{\xi}\Bigg\} \; \text{d}\xi \nonumber\\[0.05cm]
	&= \log\!\Big[x^{-1}\phi^{\scalebox{0.75}{$(0)$}}(x)\Big] - \mathlarger{\mathlarger{\int}}_{0}^x \frac{1}{\xi} \Big[1+W(\xi)\Big]^{-\frac{1}{2}}-\frac{1}{\xi} \; \text{d}\xi \, ,
\end{align}
where we utilized relation~\eqref{eq:A1_Phi0_Differential_Equation} as well as the limit $\phi^{\scalebox{0.75}{$(0)$}}(x)/x \,\raisebox{-1pt}{$\xrightarrow{x\rightarrow \:\!0}$}\, 1$. This results in the convenient expression 
\begin{align}
	\scalebox{0.98}{$\displaystyle{\exp\!\bigg[\!-\!\!\:\frac{1}{2}\;\!\phi^{\scalebox{0.75}{$(0)$}}(x) \;\! \phi^{\scalebox{0.75}{$(1)$}}(x) \bigg] = \Big[x\;\! \phi^{\scalebox{0.75}{$(0)$}}(x)^{-1}\Big]^{\scalebox{0.8}{$\frac{2n+1}{2}$}} \exp\left\{\rule{0pt}{0.75cm}\frac{2n+1}{2}\mathlarger{\mathlarger{\int}}_{0}^x \frac{1}{\xi} \Big[1+W(\xi)\Big]^{-\frac{1}{2}}-\frac{1}{\xi} \; \text{d}\xi\right\} ,}$}
	\label{eq:Exponential_Term_Uniform_WKB}
\end{align}
for which the connection to the important tunneling parameter $\mathdutchbcal{A}$ introduced in equation~\eqref{eq:2_Definition_A_Parameter_Tunneling} becomes manifest. Collecting the relevant terms then leads to the previously portrayed result~\eqref{eq:Resonant_Wave_Function}. \\

\noindent 
With the deviation from the harmonic potential $W(x)=\mathcal{O}(x)$ being small for $x\rightarrow 0$, we recover the usual harmonic oscillator behavior in that limit. Employing $\phi^{\scalebox{0.75}{$(0)$}}(x)=x+\mathcal{O}(x^2)$ and expanding the former result, we obtain
\begin{align}
	\psi_{n,\text{non-exp}}^{(\text{loc})}(x) &=  H_n\scalebox{1.2}{\bigg\{}\!\!\;\sqrt{\frac{m\omega}{\hbar}}\:x \;\!\Big[1+\mathcal{O}(x)\Big]\!\!\:\scalebox{1.2}{\bigg\}} \Big[1+\mathcal{O}(x)\Big] =  H_n\scalebox{1.1}{\bigg(}\!\!\;\sqrt{\frac{m\omega}{\hbar}}\,x\scalebox{1.1}{\bigg)} \Big[1+\mathcal{O}(x)\Big] \, .
	\label{eq:4_Asymptotics_Non_Exponential_Wave_Function_Small_x}
\end{align}
Moreover, the former approximation yields the correct exponential suppression inside the barrier, matching the traditional plane wave WKB result
\begin{align}
	\psi_n^{(\text{loc})}(x) \sim \text{non-exponential terms} \cdot \exp\!\left[-\frac{1}{\hbar}\int_0^x \sqrt{2mV(\xi)} \; \text{d}\xi\right] \, ,
\end{align}
thus constituting a uniformly valid approximation.\\[-0.25cm]

\section{(Functional) Jacobian determinant}
\label{sec:D_JacobianDeterminant}

The important Jacobian factor $\text{J}_\text{det}(t_0,x_\mu^\perp)$ arising from the non-trivial coordinate transformation to collective coordinates is mostly computed heuristically without proper care. This section closely follows Andreassen et al.~\cite{SchwartzPrecisionDecayRate}, constituting, together with reference~\cite{SchwartzCollectiveCoordinate}, the sole rigorous derivations known to the authors. We adapt their reasoning to our idiosyncratic normalization choice~\eqref{eq:ExpansionVariations}, while marginally generalizing the procedure, allowing for general zero modes $e_0(t)$ not related to time translation symmetry. Given an arbitrary quasi-zero mode $e_0^{(t_0)}(t)\propto\partial_{t_0} \:\! x_\text{crit}^{(t_0)}(t)$, both previously encountered decompositions~\eqref{eq:ExpansionVariations} and~\eqref{eq:ExpansionVariationsCollectiveCoord} of the fluctuations around a (family of) critical paths read
\begin{align}
	x(t) =  x_\text{crit}(t) + \sqrt{\frac{\hbar}{m\omega}}\: \mathlarger{\sum}_{\mu=0}^\infty \;x_\mu \:\! e_\mu(t) = x_\text{crit}^{\scalebox{0.8}{$(t_0)$}}(t) + \sqrt{\frac{\hbar}{m\omega}}\:\sum_{\mu=1}^\infty \, x^\perp_\mu e_\mu^{\scalebox{0.8}{$(t_0)$}}(t) \, ,
\end{align}
where it is crucial to differentiate between the inequivalent coordinates $x_\mu \neq x_\mu^\perp$. With both sets $\smash{\big\{e_\mu(t)\big\}\raisebox{-3pt}{\scalebox{0.8}{$\mu\!\!\:\!\!\:\in\!\!\:\!\!\: \mathbb{N}^0$}}}$ and $\smash{\big\{e\raisebox{-2.25pt}{\scalebox{0.75}{$\mu$}}\!\!\!\raisebox{4.5pt}{\scalebox{0.65}{$(t_0)$}}(t)\big\}\raisebox{-3pt}{\scalebox{0.8}{$\mu\!\!\:\!\!\:\in\!\!\:\!\!\: \mathbb{N}^0$}}}$ (for any fixed $t_0$) constituting an orthonormal basis of the underlying function space, we can project out $x_\mu$, inferring
\begin{align}
	\begin{split}
		x_\mu&=\sqrt{\frac{m\omega}{\hbar}}\mathlarger{\int}_0^T \Big[x(t)-x_\text{crit}(t)\Big] e_\mu(t) \, \text{d}(\omega t) \\
		&=\sqrt{\frac{m\omega}{\hbar}}\mathlarger{\int}_0^T\Big[\raisebox{-0.5pt}{$x_\text{crit}^{\scalebox{0.8}{$(t_0)$}}(t)-x_\text{crit}(t)$}\Big] e_\mu(t) \, \text{d}(\omega t) + \mathlarger{\sum}_{\nu=1}^\infty \; x^\perp_\nu \mathlarger{\int}_0^T  e_\nu^{\scalebox{0.8}{$(t_0)$}}(t) \,e_\mu(t) \, \text{d} (\omega t) \, . 
	\end{split}
\end{align}
It is then straightforward to obtain the partial derivatives required to compute the Jacobian determinant, yielding
\begin{subequations}
	\begin{align}
		\frac{\partial x_\mu}{\partial x^\perp_\nu}&= \int_0^T  e_\mu(t)\, e_\nu^{\scalebox{0.8}{$(t_0)$}}(t) \, \text{d}(\omega t) \, ,\\[0.15cm]
		\frac{\partial x_\mu}{\partial t_0}&=\sqrt{\frac{m\omega}{\hbar}} \mathlarger{\int}_0^T \frac{\partial x_\text{crit}^{\scalebox{0.8}{$(t_0)$}}(t)}{\partial t_0} \, e_\mu(t) \, \text{d}(\omega t) + \mathlarger{\sum}_{\nu= 1}^\infty \; x^\perp_\nu \mathlarger{\int}_0^T  \frac{\partial e_\nu^{\scalebox{0.8}{$(t_0)$}}(t)}{\partial t_0} \,e_\mu(t) \, \text{d}(\omega t) \, .
		\label{eq:3_Partial_Derivatives_Coordinate_Change}
	\end{align}
\end{subequations}
To lighten the notation, it proves useful to introduce the auxiliary quantity
\begin{align}
	U_{\mu\nu}^{\scalebox{0.8}{$(t_0)$}}= \int_0^T  e_\mu(t) \,e_\nu^{\scalebox{0.8}{$(t_0)$}}(t)  \, \text{d}(\omega t)
	\label{eq:3_Definition_Auxiliary_Orthogonal_Matrix}
\end{align}
with $\mu,\nu\in \mathbb{N}^0$, which is an orthogonal matrix as can be checked by a direct computation
\begin{align}
	\sum_{\nu=0}^\infty\, U_{\mu\nu}^{\scalebox{0.8}{$(t_0)$}} \Big[U_{\nu\rho}^{\scalebox{0.8}{$(t_0)$}}\Big]^T 
	&= \int_0^T\!\!\!\int_0^T e_\mu(t) \,e_\rho(t') \hspace{-0.85cm}\underbrace{\Bigg[\sum_{\nu=0}^\infty  e_\nu^{\scalebox{0.8}{$(t_0)$}}(t) \, e_\nu^{\scalebox{0.8}{$(t_0)$}}(t') \Bigg]}_{\displaystyle{=\delta\big[\omega(t-t')\big] \; \text{ (completeness)}}} \hspace{-0.9cm}\, \text{d}(\omega t)\!\;\text{d}(\omega t') \nonumber\\ 
	&= \int_0^T e_\mu(t) \, e_\rho(t)\,\text{d}(\omega t) = \delta_{\mu\rho} \, ,
\end{align}
where again the completeness of both bases is used. With $U\!\raisebox{-2.5pt}{\scalebox{0.75}{$\mu\nu$}}\!\!\!\!\:\!\:\!\:\!\!\raisebox{4.5pt}{\scalebox{0.65}{$(t_0)$}}$ being orthogonal, we directly deduce $\det\!\big[U\:\raisebox{-2.5pt}{\scalebox{0.75}{$\!\!\!\mu\nu$}}\!\!\!\!\:\!\:\!\:\!\!\raisebox{4.5pt}{\scalebox{0.65}{$\;\;\;(t_0)$}}\big]=1$. Using this newly defined matrix, the Jacobian $\text{J}_\text{det}(t_0,x_\mu^\perp)$ of the coordinate change can be cast into the simple form
\begin{align}
	\text{J}_\text{det}\big(t_0,x_\mu^\perp\big)=\:\!\det\!\left(\!\begin{array}{c|c c c} 
		\vdots & \vdots & \vdots & \\[0.15cm]
		\displaystyle{\frac{\partial x_\mu}{\partial t_0}} & U_{\mu 1}^{\scalebox{0.8}{$(t_0)$}} & U_{\mu 2}^{\scalebox{0.8}{$(t_0)$}} & \ldots \\[0.15cm]
		\vdots & \vdots & \vdots &
	\end{array}\!\right) .
	\label{eq:3_Jacobian_Matrix_Form}
\end{align}
The final step consists in expanding $\partial_{t_0} x_\mu$ in terms of $U\!\raisebox{-2.5pt}{\scalebox{0.75}{$\mu\nu$}}\!\!\!\!\:\!\:\!\:\!\!\raisebox{5pt}{\scalebox{0.65}{$(t_0)$}}$, as the determinant vanishes in case two columns are linearly dependent. Inserting a completeness relation into equation~\eqref{eq:3_Partial_Derivatives_Coordinate_Change} and using definition~\eqref{eq:3_Definition_Auxiliary_Orthogonal_Matrix} then yields
\begin{align}
	\frac{\partial x_\mu}{\partial t_0}&=\sqrt{\frac{m\omega}{\hbar}} \, \mathlarger{\sum}_{\rho=0}^\infty \; \mathlarger{\int}_0^T\!\!\mathlarger{\int}_0^T \frac{\partial x_\text{crit}^{\scalebox{0.8}{$(t_0)$}}(t)}{\partial t_0} \, e_\rho^{\scalebox{0.8}{$(t_0)$}}(t) \, e_\rho^{\scalebox{0.8}{$(t_0)$}}(t') \, e_\mu(t') \, \text{d}(\omega t') \,\text{d}(\omega t) \nonumber\\[-0.15cm]
	&\qquad\qquad\qquad\qquad\quad + \mathlarger{\sum}_{\rho=0}^\infty\mathlarger{\sum}_{\nu=1}^\infty \; x^\perp_\nu \mathlarger{\int}_0^T\!\!\mathlarger{\int}_0^T \frac{\partial e_\nu^{\scalebox{0.8}{$(t_0)$}}(t)}{\partial t_0} \,e_\rho^{\scalebox{0.8}{$(t_0)$}}(t)\, e_\rho^{\scalebox{0.8}{$(t_0)$}}(t')\,e_\mu(t') \, \text{d}(\omega t')\,\text{d}(\omega t) \nonumber\\[0.25cm]
	&= \mathlarger{\sum}_{\rho=0}^\infty \left\{\sqrt{\frac{m\omega}{\hbar}}\mathlarger{\int}_0^T \frac{\partial x_\text{crit}^{\scalebox{0.8}{$(t_0)$}}(t)}{\partial t_0} \, e_\rho^{\scalebox{0.8}{$(t_0)$}}(t) \, \text{d}(\omega t)+\mathlarger{\sum}_{\nu= 1}^\infty \; x^\perp_\nu \mathlarger{\int}_0^T \frac{\partial e_\nu^{\scalebox{0.8}{$(t_0)$}}(t)}{\partial t_0} \,e_\rho^{\scalebox{0.8}{$(t_0)$}}(t) \, \text{d}(\omega t)\right\} U_{\mu\rho}^{\scalebox{0.8}{$(t_0)$}} \, .
\end{align}
Given the previous considerations around equality~\eqref{eq:3_Jacobian_Matrix_Form}, only the addend pertaining to $\rho=0$ survives, leaving us with the exact result
\begin{equation}
	\text{J}_\text{det}\big(t_0,x_\mu^\perp\big)=\sqrt{\frac{m\omega}{\hbar}} \mathlarger{\int}_0^T \frac{\partial x_\text{crit}^{\scalebox{0.8}{$(t_0)$}}(t)}{\partial t_0} \, e_0^{\scalebox{0.8}{$(t_0)$}}(t) \, \text{d}(\omega t)+\!\!\!\!\!\!\!\!\!\!\!\underbrace{\:\mathlarger{\sum}_{\mu= 1}^\infty \; x_\mu^\perp \mathlarger{\int}_0^T \frac{\partial e_\mu^{\scalebox{0.8}{$(t_0)$}}(t)}{\partial t_0} \,e_0^{\scalebox{0.8}{$(t_0)$}}(t) \, \text{d}(\omega t)}_{\rule{0pt}{0.3cm}\displaystyle{\text{subleading in powers of $\hbar$ for $x_\mu^\perp=\mathcal{O}(1)$}}} \!\!\!\!\!\!\!\!\!\! \, .
	\label{eq:3_Jacobian_Coordinate_Trafo_Full}
\end{equation}
Employing $e_0^{(t_0)}(t)\propto\partial_{t_0} \:\! x_\text{crit}^{(t_0)}(t)$, we can further simplify the leading-order contribution, yielding the desired final result 
\begin{align}
	\text{J}_\text{det}\scalebox{1.15}{\big(}t_0,x_\mu^\perp\!\!\:=\!\!\:0\scalebox{1.15}{\big)}  = \sqrt{\frac{m\omega}{\hbar}} \;\Bigg\lvert\!\Bigg\lvert \frac{\partial x_\text{crit}^{(t_0)}(t)}{\partial t_0}\Bigg\rvert\!\Bigg\rvert \, .
	\label{eq:JacobianFactor}
\end{align}
In the special case of time-translation symmetry, the $t_0$-derivative can directly be converted into a usual time derivative acting on $x\raisebox{-2.5pt}{\scalebox{0.7}{\text{crit}}}\!\!\!\!\!\!\!\,\raisebox{5pt}{\scalebox{0.65}{$(t_0)$}}\:\!(t)$.

\section{Auxiliary identity involving Hermite polynomials}
\label{sec:E_HermiteIdentity}

Let us briefly derive a useful relation involving Hermite polynomials, which will come in handy when dealing with the non-exponential contribution $f_\text{non-exp}$. We desire to evaluate an expression of the form
\begin{align}
	\mathcal{I}_n\Big(\alpha_1,\alpha_2;\beta_1,\beta_2,\beta_3\Big)\coloneqq \!\!\:\Bigg[H_n\!\!\!\;\left(\!\!\:\alpha_1+\frac{\partial}{\partial x}\right) H_n\!\!\!\;\left(\!\!\:\alpha_2+\frac{\partial}{\partial y}\right) \exp\!\Big(\beta_1 x^2 +\beta_2 y^2 + \beta_3 xy\Big)\Bigg]_{\substack{\scalebox{0.75}{$x\!\!\:\!\!\:=\!\!\:\!\!\: 0$} \\[0.05cm] \scalebox{0.75}{$y\!\!\:\!\!\:=\!\!\:\!\!\: 0$}}} \, ,
	\label{eq:6_Definition_Auxiliary_Function_HermitePolys}
\end{align}
where the Hermite polynomials act as operators on a Gaussian exponential. We can cast the problem into a much simpler form by employing a defining relation for Hermite polynomials, which can be represented as
\begin{align}
	H_n(x)= \left[\:\!\frac{\partial^n}{\partial v^n} \:\! \exp\!\Big(\!\!\!\:-v^2+2vx\Big)\right]_{\scalebox{0.75}{$v\!\!\:\!\!\:=\!\!\:\!\!\: 0$}} \, .
\end{align}
Notice that when inserting $\partial_x$ into this relation, the derivative inside the exponential acts as a shift operator due to $\exp\!\left(v \partial_x\right) f(x)= f\big(x+v\big)$. This allows us to simplify the former expression by interchanging the order of differentiation as well as readily evaluating $x=y=0$. We conclude
\begin{align}
	\mathcal{I}_n\scalebox{1.2}{\big(}\alpha_{1,2};\beta_{1,2,3}\scalebox{1.2}{\big)} &= \frac{\partial^{2n}}{\partial v^n\partial w^n} \Bigg[\exp\!\bigg(\!\!-v^2+2\alpha_1v +2v \,\frac{\partial}{\partial x}\bigg)  \label{eq:6_General_Result_Auxiliary_Function_HermitePolys} \\[-0.1cm] 
	&\qquad\qquad\:\! \times \exp\!\bigg(\!\!-w^2+2\alpha_2 w+2w\, \frac{\partial}{\partial y}\bigg) \exp\!\Big(\beta_1 x^2 +\beta_2 y^2 + \beta_3 xy\Big)\Bigg]_{\substack{$\scalebox{0.75}{$\;\! x\!\!\:=\!\!\: y \!\!\: =\!\!\: 0$} $ \\[0.02cm] $\scalebox{0.75}{$v\!\!\:=\!\!\: w \!\!\: =\!\!\: 0$} $} } \nonumber \\[0.1cm]
	&= \frac{\partial^{2n}}{\partial v^n\partial w^n}\,\Bigg\{\!\exp\!\Big(\!\!\!\:-v^2-w^2+2\alpha_1 v + 2\alpha_2 w\Big) \nonumber \\[-0.1cm] 
	&\qquad\qquad\:\times \exp\!\bigg[\beta_1\mathlarger{\big(}x+2v\mathlarger{\big)}^{\!\!\: 2} +\beta_2\mathlarger{\big(}y+2w\mathlarger{\big)}^{\!\!\: 2} + \beta_3\mathlarger{\big(}x+2v\mathlarger{\big)}\mathlarger{\big(}y+2w\mathlarger{\big)}\bigg]\!\!\:\Bigg\}_{\substack{$\scalebox{0.75}{$\;\! x\!\!\:=\!\!\: y \!\!\: =\!\!\: 0$} $\\[0.02cm] $\scalebox{0.75}{$v\!\!\:=\!\!\: w \!\!\: =\!\!\: 0$} $} } \nonumber\\[0.2cm]
	&= \frac{\partial^{2n}}{\partial v^n\partial w^n}\, \Bigg\{ \! \exp\!\bigg[2\alpha_1 v + 2\alpha_2 w+\mathlarger{\big(}4\beta_1-1\mathlarger{\big)}v^2+\mathlarger{\big(}4\beta_2-1\mathlarger{\big)}w^2+ 4\beta_3 v w\bigg]\!\!\:\Bigg\}_{\substack{\scalebox{0.75}{$\,v\!\!\:=\!\!\: 0$}\\[0.05cm] \scalebox{0.75}{$w \!\!\: =\!\!\: 0$} }} . \nonumber
\end{align}
By completing the square, the entailed expression can be simplified further. Defining the auxiliary constants
\begin{align}
	\begin{split}
	\sigma_1 &\coloneqq \frac{2 \alpha_2 \beta_3-\alpha_1 (4\beta_2-1) }{(4\beta_1 -1) (4\beta_2 -1)-4 \beta_3^2} \, ,\qquad\qquad \begin{matrix}
	\kappa_1 \coloneqq 4\beta_1-1\, , \\[0.15cm]
	\kappa_2 \coloneqq 4\beta_2-1\, ,
	\end{matrix} \\[0.1cm]
	\sigma_2 &\coloneqq \frac{2 \alpha_1 \beta_3-\alpha_2 (4\beta_1-1) }{(4\beta_1 -1) (4\beta_2 -1)-4 \beta_3^2} \, ,\qquad\qquad \begin{matrix}
		\kappa_3 \coloneqq 4\beta_3\, , \qquad\qquad\qquad\qquad\; \\[0.2cm]
		\varkappa \coloneqq \kappa_1 \sigma_1^2+\kappa_2 \sigma_2^2+ \kappa_3  \sigma_1\sigma_2 \, ,
	\end{matrix} 
	\end{split}
	\label{eq:Simplified_Parametrization_I_n}
\end{align}
one obtains the suggestive expression 
\begin{align}
	\mathcal{I}_n\Big(\alpha_1,\alpha_2;\beta_1,\beta_2,\beta_3\Big) &=e^{-\varkappa}\, \frac{\partial^{2n}}{\partial v^n\partial w^n}\, \Bigg\{ \! \exp\!\bigg[\kappa_1 v^2+\kappa_2 w^2+ \kappa_3 vw\bigg]\!\!\:\Bigg\}_{\substack{\scalebox{0.75}{$\,v\!\!\:=\!\!\: \sigma_1$}\\[0.05cm] \scalebox{0.75}{$w \!\!\: =\!\!\: \sigma_2$} }} \: .
	\label{eq:I_n_FinalForm}
\end{align}
The exponential prefactor $e^{-\varkappa}$ ensures the correct normalization $\mathcal{I}_0\big(\alpha_{1,2};\beta_{1,2,3}\big)=1$, thus we really only probe the polynomial emerging from differentiation. Even though it does not seem like a substantial improvement, the final expression~\eqref{eq:I_n_FinalForm} is much simpler to evaluate compared to the original representation~\eqref{eq:6_Definition_Auxiliary_Function_HermitePolys}, as the differential operator acting on the Gaussian exponential is, in terms of derivatives, a trivial monomial instead of a complicated polynomial with $\mathcal{O}(n^2)$ terms. Moreover, the attained representation~\eqref{eq:I_n_FinalForm} greatly simplifies in the relevant special cases, e.g. for $\alpha_1=\alpha_2=0$ and $\beta_{1}=\beta_2=\frac{1}{4}$. In that case, we obtain the important result
\begin{align}
	\mathcal{I}_n\bigg(0,0;\frac{1}{4},\frac{1}{4},\beta_3\bigg)&= H_n\!\!\!\:\left(\frac{\partial}{\partial x}\right) H_n\!\!\!\:\left(\frac{\partial}{\partial y}\right) \exp\!\bigg(\frac{x^2+y^2}{4} + \beta_3 xy\bigg) \nonumber \\[0.1cm] 
	&= \frac{\partial^{2n}}{\partial v^n\partial w^n}\, \bigg[\! \exp\!\Big(4\beta_3 vw\Big)\!\!\:\bigg]_{\substack{\scalebox{0.75}{$\,v\!\!\:=\!\!\: 0$}\\[0.05cm] \scalebox{0.75}{$w \!\!\: =\!\!\: 0$} }} = \big(4\beta_3\big)^{\!\!\: n}\:\! n! \: ,
	\label{eq:6_Special_Result_Auxiliary_Function_HermitePolys}
\end{align}
which in its original formulation is not easily spotted.\\[-0.25cm]

\section{Relevant aspects of Sturm--Liouville problems}
\label{sec:F_AspectsOfSLProblems}

In the following section we review some notions of regular Sturm--Liouville eigenvalue problems of the form 
\begin{align}
	O \;\! e_\mu(t)\coloneqq \frac{1}{w(t)} \,\Bigg\{\!-\frac{\text{d}}{\text{d}t} \bigg[p(t)\frac{\text{d}}{\text{d}t}\bigg] + q(t)  \Bigg\} \, e_{\mu}(t) = \lambda_\mu e_{\mu}(t) \, .
	\label{eq:A3_Sturm_Liouville_Eigenvalue_Problem}
\end{align}
The differential operator $O$ acts on real, continuously differentiable, square-integrable functions living on the time interval $[0,T]$, with $\mu\in \mathbb{N}^0$ and $w(t),p(t)$ constrained to be positive. The subsequently illustrated properties are well-known, with proofs provided by most textbooks on the subject matter, see e.g. the lucid treatment by Teschl~\cite{TeschlODETheory}. We define the operator $O$ such that both the eigenvalues $\lambda_\mu$ and eigenfunctions $e_\mu(t)$ are dimensionless quantities; thus a natural option is to choose $w(t)$, $p(t)^{-1}$ and $q(t)$ such that they all possess units of inverse time. The most general boundary conditions can be brought into the form
\begin{align}
	\underbrace{\begin{pmatrix}
			m_{11} & m_{12} \\
			m_{21} & m_{22}
	\end{pmatrix}}_{\displaystyle{\eqqcolon M}} \begin{pmatrix} \;\;\;\;\;\;\;\!\;\! e_{\mu}(0) \\ p(0)\;\!\dot{e}_{\mu}(0) \end{pmatrix} + \underbrace{\begin{pmatrix}
			n_{11} & n_{12} \\
			n_{21} & n_{22}
	\end{pmatrix}}_{\displaystyle{\eqqcolon N}} \begin{pmatrix} \;\;\;\;\;\;\;\:\!\;\! e_{\mu}(T) \\ p(T)\;\! \dot{e}_{\mu}(T) \end{pmatrix} = 0 \, ,
	\label{eq:A3_Bouncary_Conditions_Eigenvalue_Problem}
\end{align}
with $M$ and $N$ constituting real, dimensionless $2\!\!\:\times \!\!\:2\!\:$-matrices. In case the boundary conditions do not violate self-adjointness of the differential operator $O$, we know the eigenvalues $\lambda_\mu$ to be real, as well as the countable set of normalized eigenfunctions $e_\mu(t)$ to form an orthonormal basis of $L^2\big([0,T],w\big)$, i.e. 
\begin{align}
	\big\langle e_{\mu} , e_{\nu} \big\rangle_{\scalebox{0.8}{$L^2\big([0,T],w\big)$}} \coloneqq \int_0^T e_{\mu}(t) \:\! e_{\nu}(t)\:\! w(t) \,\text{d}t= \delta_{\mu\nu} \, .
	\label{eq:A3_Real_Scalar_Product_with_Weight}
\end{align}
It is furthermore known that the set of eigenvalues is discrete and unbounded, accumulating to $\infty$. Commonly encountered boundary conditions for Sturm--Liouville problems include:
\begin{table}[H]
	\centering
	\begin{tabular}{c|| c c}
		boundary condition & $M$ & $N$ \\ \hline\hline
		&  & \\[-0.3cm] 
		Dirichlet &  $\begin{pmatrix} 1 & 0 \\ 0 & 0\end{pmatrix}$ & $\begin{pmatrix} 0 & 0 \\ 1 & 0\end{pmatrix}$ \\
		&  & \\[-0.3cm] 
		Neumann &  $\begin{pmatrix} 0 & 1 \\ 0 & 0\end{pmatrix}$ & $\begin{pmatrix} 0 & 0 \\ 0 & 1\end{pmatrix}$ \\
		&  & \\[-0.3cm] 
		Robin &  $\begin{pmatrix} m_{11} & m_{12} \\ 0 & 0\end{pmatrix}$ & $\begin{pmatrix} 0 & 0 \\ n_{21} & n_{22} \end{pmatrix}$ \\
		&  & \\[-0.3cm] 
		(anti-)periodic &  $\begin{pmatrix} 1 & 0 \\ 0 & 1\end{pmatrix}$ & $\begin{pmatrix} \mp 1 & 0 \\ 0 & \mp 1\end{pmatrix}$ 
	\end{tabular}
\end{table}
\noindent
We will be especially interested in the functional determinant of the differential operator $O$ as well as its corresponding Green's function in the case of Robin boundary conditions.

\subsection{Determinants}
\label{sec:F1_Sturm_Liouville_Determinants}
The functional determinant of an operator $O$ is formally defined as the product of all its eigenvalues 
\begin{align}
	\det(O) = \text{det}_\zeta\scalebox{1.1}{\Bigg(}\!\!\:\frac{1}{w(t)} \,\Bigg\{\!-\frac{\text{d}}{\text{d}t} \bigg[p(t)\!\:\frac{\text{d}}{\text{d}t}\bigg] + q(t)  \Bigg\}\!\!\:\scalebox{1.1}{\Bigg)}\overset{\,\zeta}{\coloneqq} \prod_{\mu=0}^\infty \lambda_\mu \, ,
	\label{eq:A3_Definition_Determinant}
\end{align}
which however requires a proper regularization procedure to treat the tail of ever-growing eigenvalues, rendering a na\"{\i}ve definition ill-defined. There are two equivalent ways of regulating the emerging infinity, namely by either considering ratios of determinants or by applying $\zeta$-regularization~\cite{TakhtajanQMForMathematicians,DunneDeterminants}. We will formulate all relations in terms of $\zeta$-regularized determinants, being the predominant method to prove the subsequently given relations. In case the Sturm--Liouville problem contains an exact zero mode, the determinant vanishes identically. Thus, one instead defines the so-called primed determinant $\text{det}_\zeta'$ by 
\begin{align}
	\text{det}_\zeta'(O) \coloneqq \lim_{\lambda\rightarrow 0}\left[\frac{\det_\zeta\!\!\:\!\!\:\big(O+\lambda\big)}{\lambda}\right] =  \frac{\text{d}}{\text{d}\lambda}\bigg\vert_{\scalebox{0.8}{$\lambda\!\!\:\!\!\:=\!\!\:\!\!\:0$}} \text{det}_\zeta\big(O+\lambda\big) \, ,
	\label{eq:A3_Definition_Primed_Determinant}
\end{align} 
effectively subtracting the sole factor of zero while retaining the information about the remaining, non-trivial spectrum~\cite{TakhtajanQMForMathematicians}. \\ 

\noindent
Let us subsequently solely focus on the relevant case $w(t)=p(t)^{-1}=\omega$, which generally emerges when inspecting fluctuation operators $O_\text{crit}$ arising from the expansion of action functionals possessing a standard kinetic term, see equation~\eqref{eq:Expanded_Exponent}. Let us rescale $q(t)$ accordingly to spare us from carrying more factors of $\omega$ than necessary. With the weight function $w(t)$ given by a constant, the determinant computation solely boils down to applying knowledge about the behavior of two linearly independent solutions $\mathpzc{y}\raisebox{-3.25pt}{\scalebox{0.75}{$\lambda\!\!\:\!\!\:=\!\!\:\!\!\:0$}}\!\!\!\!\!\!\!\!\!\!\:\:\!\raisebox{4.75pt}{\scalebox{0.65}{$(1,2)$}}(t)$ to the ODE~\eqref{eq:A3_Sturm_Liouville_Eigenvalue_Problem} for vanishing $\lambda$. This method goes back to Gel'fand and Yaglom~\cite{GelfandFunctionalIntegration}, being improved upon to deal with arbitrary boundary conditions as well as emerging zero modes by Kirsten and McKane~\cite{KirstenMcKaneFuncDets1,KirstenMcKaneFuncDets2}. \\

\noindent
Stating the theorem in its full extent, one initially defines the two linearly independent solutions $\mathpzc{y}\raisebox{-3.25pt}{\scalebox{0.75}{$\lambda$}}\!\!\raisebox{4.75pt}{\scalebox{0.65}{$(1,2)$}}(t)$ of the ODE 
\begin{align}
	O\:\! \mathpzc{y}_{\scalebox{0.8}{$\lambda$}}^{\scalebox{0.8}{$(1,2)$}}(t)=\bigg\{\!\!\!\:-\!\!\:\frac{\text{d}^2}{\text{d}(\omega t)^2}+ q(t) \bigg\} \: \mathpzc{y}_{\scalebox{0.8}{$\lambda$}}^{\scalebox{0.8}{$(1,2)$}}(t) = \lambda \;\! \mathpzc{y}_{\scalebox{0.8}{$\lambda$}}^{\scalebox{0.8}{$(1,2)$}}(t)
	\label{eq:A3_ODE_Definition_y12}
\end{align}
for arbitrary fixed $\lambda\in \mathbb{R}$, obeying the boundary conditions
\begin{align}
	\begin{split}
		\mathpzc{y}_{\scalebox{0.8}{$\lambda$}}^{\scalebox{0.8}{$(1)$}}(0)&=1\, , \qquad\qquad \;\;\;\;\;\;\;\!\, \mathpzc{y}_{\scalebox{0.8}{$\lambda$}}^{\scalebox{0.8}{$(2)$}}(0)=0\, , \\
		\omega^{-1} \,\:\dot{\!\!\!\:\mathpzc{y}}_{\scalebox{0.8}{$\lambda$}}^{\scalebox{0.8}{$(1)$}}(0)&=0\, ,\qquad\qquad\,  \omega^{-1} \,\:\dot{\!\!\!\:\mathpzc{y}}_{\scalebox{0.8}{$\lambda$}}^{\scalebox{0.8}{$(2)$}}(0)=1 \, . 
	\end{split}
	\label{eq:A3_Boundary_Conditions_y12}
\end{align}
Introducing the matrix
\begin{align}
	\mathcal{Y}\coloneqq \lim_{\lambda \rightarrow 0}\!\begin{pmatrix}
		\;\;\;\;\;\;\;\!\: \mathpzc{y}_{\scalebox{0.8}{$\lambda$}}^{\scalebox{0.8}{$(1)$}}(T) & \;\;\;\;\;\;\;\;\;\!\,\, \mathpzc{y}_{\scalebox{0.8}{$\lambda$}}^{\scalebox{0.8}{$(2)$}}(T) \\[0.2cm]
		\,\omega^{-1} \,\:\dot{\!\!\!\:\mathpzc{y}}_{\scalebox{0.8}{$\lambda$}}^{\scalebox{0.8}{$(1)$}}(T) & \;\;\, \omega^{-1} \, \:\dot{\!\!\!\:\mathpzc{y}}_{\scalebox{0.8}{$\lambda$}}^{\scalebox{0.8}{$(2)$}}(T) 
	\end{pmatrix} \, ,
	\label{eq:A3_Definition_Y_Matrix}
\end{align}
the functional determinant is then found to be given by
\begin{align}
	\text{det}_\zeta(O) &=\hspace{-1.05cm} \underbrace{\frac{\text{det}_\zeta(O_\text{ref})}{\text{det}\big(M+N\mathcal{Y}_\text{ref}\big)}}_{\displaystyle{\text{proportionality constant } \mathcal{N}_\text{det}}} \hspace{-1.05cm} \,  \text{det}\big(M+N\mathcal{Y}\big) \, ,
	\label{eq:A3_Gelfand_Yaglom_Determinant_Ratio}
\end{align}
where the proportionality constant $\mathcal{N}_\text{det}$ is computed using a reference operator whose $\zeta$-regularized determinant is known~\cite{KirstenMcKaneFuncDets2}. As introduced in relation~\eqref{eq:A3_Bouncary_Conditions_Eigenvalue_Problem}, the dependence of the determinant on the boundary conditions of the eigenvalue problem is encoded in the $2\times 2$-matrices $M$ and $N$. In the presence of an exact zero mode $e_0(t)$, a similar relation can be obtained for the primed determinant~\cite{KirstenMcKaneFuncDets2}. Focusing on the relevant case of Robin boundary conditions and assuming $n_{22}\neq 0$, the acquired result simplifies to
\begin{align}
	\text{det}_\zeta'(O) &= \mathcal{N}_\text{det} \underbrace{\left[\int_0^T  \mathpzc{y}(t)^2 \;\!  \text{d}(\omega t)\right]}_{\displaystyle{=\big\lvert\!\big\lvert \;\!\mathpzc{y} \;\!\big\rvert\!\big\rvert^2}} \frac{n_{22}}{\mathpzc{y}(T)} \, ,
	\label{eq:Gelfand_Yaglom_Primed_Determinant}
\end{align}
where the auxiliary function $\mathpzc{y}(t)$ is defined by
\begin{align}
	\mathpzc{y}(t)=\lim_{\lambda\rightarrow 0} \Big[m_{11} \, \mathpzc{y}_{\scalebox{0.8}{$\lambda$}}^{\scalebox{0.8}{$(2)$}}(t) -m_{12} \, \mathpzc{y}_{\scalebox{0.8}{$\lambda$}}^{\scalebox{0.8}{$(1)$}}(t)\Big] = -m_{12}\, \big[e_0(0)\big]^{-1} e_0(t) \, .
	\label{eq:A3_Definition_Auxiliary_y}
\end{align}
In the last line we used that, for Robin boundary conditions, $\mathpzc{y}(t)$ is directly proportional to the zero mode $e_0(t)$, with the correct scaling factor obtained by utilizing relation~\eqref{eq:A3_Boundary_Conditions_y12}. \\ 

\noindent
Let us stress that the proportionality constant $\mathcal{N}_\text{det}$ appearing in relations~\eqref{eq:A3_Gelfand_Yaglom_Determinant_Ratio} and~\eqref{eq:Gelfand_Yaglom_Primed_Determinant} is identical for both ordinary and primed determinant. Since it depends non-trivially on the imposed boundary matrices $M$ and $N$, we are obliged to determine this proportionality constant up to the normalization factor $\mathcal{N}_\text{E}$ encountered in the definition of the path integral measure~\eqref{eq:EuclideanPathIntegralMeasure}. We will restrict our considerations to Robin boundary conditions, constituting the main case of interest. Obtaining the reference quantity $\mathcal{N}_\text{det}$ can be achieved by comparing the results of computing a purely Gaussian path integral both in the sequential and the composite approach. For this, it will prove useful to weight the free particle propagator $K_\text{E}^{(\text{free})}\big(x_0,x_T\:\!;T\big)$ using carefully chosen exponential weight functions. Since the free Euclidean propagator is fully known, it serves as the required reference quantity for our endeavor. Before evaluating the composite, endpoint-weighted path integral, we can simply proceed via a sequential computation. Given the exactly calculable quantity
\begin{align}
	F\Big(m_{11},m_{12};n_{21},n_{22}\Big) &\coloneqq \mathlarger{\mathlarger{\int}}_{\mathbb{R}^2} \exp\!\Bigg\{\!\!-\!\!\:\frac{m\omega}{2\hbar}\bigg[ \frac{n_{21}}{n_{22}}\;\! x_T^2-\frac{m_{11}}{m_{12}}\;\! x_0^2\bigg]\!\!\:\Bigg\} \, K_\text{E}^{(\text{free})}\big(x_0,x_T\:\!;T\big) \, \text{d}x_0 \:\!\text{d}x_T \nonumber \\
	&=  \sqrt{\frac{m}{2\pi\hbar T}} \mathlarger{\mathlarger{\int}}_{\mathbb{R}^2} \exp\!\Bigg\{\!\!-\!\!\:\frac{m\omega}{2\hbar}\bigg[ \frac{n_{21}}{n_{22}}\;\! x_T^2-\frac{m_{11}}{m_{12}}\;\! x_0^2\bigg]-\frac{m(x_T-x_0)^2}{2\hbar T}\Bigg\} \, \text{d}x_0 \:\!\text{d}x_T \nonumber \\
	&=\sqrt{\frac{2\pi  \hbar}{m\omega}} \,\bigg[1-\frac{m_{11} \:\!\omega\:\! T}{m_{12}}\bigg]^{-\frac{1}{2}} \bigg[\frac{m_{11}}{m_{11} \:\!\omega\:\! T-m_{12}}+\frac{n_{21}}{n_{22}}\bigg]^{-\frac{1}{2}} \, ,
	\label{eq:F_ResultSequential}
\end{align}
we can additionally represent $F$ as the composite path integral
\begin{align}
	\ldots=\mathlarger{\mathlarger{\int}}_{\scalebox{0.75}{$\mathcal{C}\big([0,T]\big)$}} \mathcal{D}\llbracket x \rrbracket \, \exp\!\Bigg\{\!\!-\!\!\:\frac{m\omega}{2\hbar}\bigg[ \frac{n_{21}}{n_{22}}\;\! x(T)^2-\frac{m_{11}}{m_{12}}\;\! x(0)^2+\omega^{-1}\int_0^T \dot{x}(t)^2 \, \text{d}t\bigg]\!\!\:\Bigg\} \, ,
\end{align}
which can be computed with the procedure illustrated in section~\ref{sec:3_2_GeneralEvaluationTactics}. Expanding around a critical trajectory $x_\text{crit}(t)$ as done in equation~\eqref{eq:Expanded_Exponent} reveals the necessary requirements
\begin{subequations}
	\begin{align}
		\ddot{x}_\text{crit}(t)&=0 & & \!\!\:\,\! \text{Euler--Lagrange equation} \, , \\[0.05cm]
		m_{12} \,\omega^{-1}\dot{x}_\text{crit}(0) + m_{11}\, x_\text{crit}(0) &=0 & & \text{left transversality condition} \, , \\[0.05cm]
		n_{22}\,\omega^{-1}\dot{x}_\text{crit}(T) + n_{21}\, x_\text{crit}(T) &=0 & & \text{right transversality condition}\, .
	\end{align}
\end{subequations}
For general $m_{11},m_{12},n_{21}$ and $n_{22}$, the only admissible solution is the trivial trajectory $x_\text{crit}(t)=0$.\footnote{In case of the parameters chosen such that several solutions exist, the spectrum of the fluctuation operator possesses an exact zero mode, a case we want to circumvent.} Additionally, the weight functions were chosen such that the eigenfunctions $e_\mu(t)$ are required to satisfy the desired Robin boundary conditions
\begin{align}
	\begin{pmatrix}
		m_{11} & m_{12} \\
		0 & 0
	\end{pmatrix} \begin{pmatrix} \;\;\;\;\;\;\;\!\;\! e_{\mu}(0) \\ \omega^{-1}\;\:\!\!\dot{e}_{\mu}(0) \end{pmatrix} + \begin{pmatrix}
		0 & 0 \\
		n_{21} & n_{22}
	\end{pmatrix} \begin{pmatrix} \;\;\;\;\;\;\;\:\!\;\! e_{\mu}(T) \\ \omega^{-1}\;\!\: \dot{e}_{\mu}(T) \end{pmatrix} = 0 
	\label{eq:RobinBoundaryConditions}
\end{align}
in order for the supplementary boundary terms~\eqref{eq:Robin_BoundaryConditions} to vanish. Employing the path integral measure~\eqref{eq:EuclideanPathIntegralMeasure} and computing the arising Gaussian integrals yields the final result 
\begin{align}
	F\Big(m_{11},m_{12};n_{21},n_{22}\Big)= \mathcal{N}_\text{E} \,\sqrt{\frac{\pi\hbar}{m\omega}}\; \text{det}_\zeta\big(O_\text{free}\big)^{-\frac{1}{2}} \, ,
	\label{eq:F_ResultComposite}
\end{align}
where it is understood that the determinant is taken with respect to the above Robin boundary conditions~\eqref{eq:RobinBoundaryConditions}. With $\text{det}_\zeta\big(O_\text{free}\big)$ fully known after equating~\eqref{eq:F_ResultSequential} and~\eqref{eq:F_ResultComposite}, the remaining quantity to be computed is $\text{det}\big(M+N\mathcal{Y}_\text{free}\big)$. With the operator being trivial, we can straightforwardly give the two required functions as $\smash{\mathpzc{y}\!\!\:\raisebox{-3.25pt}{\scalebox{0.75}{$\lambda\!\!\:\!\!\:=\!\!\:\!\!\:0$}}\!\!\!\!\!\!\!\!\raisebox{4.75pt}{\scalebox{0.65}{$(1)$}}\;\;(t)=1}$ and $\smash{\mathpzc{y}\!\!\:\raisebox{-3.25pt}{\scalebox{0.75}{$\lambda\!\!\:\!\!\:=\!\!\:\!\!\:0$}}\!\!\!\!\!\!\!\!\raisebox{4.75pt}{\scalebox{0.65}{$(2)$}}\;\;(t)=\omega t}$. Thus, we obtain
\begin{align}
	\text{det}\big(M+N\mathcal{Y}_\text{free}\big)&=  \text{det}\Bigg[\!\begin{pmatrix} m_{11} & m_{12} \\0 & 0 \end{pmatrix}+\begin{pmatrix} 0 & 0 \\ n_{21} & n_{22} \end{pmatrix} \begin{pmatrix} 1  & \omega T  \\ 0  & 1 \end{pmatrix} \! \Bigg] \nonumber \\[0.1cm]
	&= m_{11} \Big(n_{21}\;\!\omega\;\! T+n_{22}\Big) -m_{12} \;\! n_{21} \, ,
\end{align}
yielding the required determinant ratio 
\begin{align}
	\mathcal{N}_\text{det}&= \frac{\mathcal{N}_\text{E}\big.^{\!\!\!\!\: 2}}{2}\bigg[1-\frac{m_{11} \:\!\omega\:\! T}{m_{12}}\bigg] \bigg[\frac{m_{11}}{m_{11} \:\!\omega\:\! T-m_{12}}+\frac{n_{21}}{n_{22}}\bigg] \bigg[m_{11} \Big(n_{21}\;\!\omega\;\! T+n_{22}\Big) -m_{12} \;\! n_{21}\bigg]^{-1} \nonumber \\ 
	&= -\frac{\mathcal{N}_\text{E}\big.^{\!\!\!\!\: 2}}{2m_{12}\:\! n_{22}} \, .
	\label{eq:ProportionalityConstantDeterminants}
\end{align}
Given the found proportionality constant $\mathcal{N}_\text{det}$, we can greatly simplify expression~\eqref{eq:Gelfand_Yaglom_Primed_Determinant} for the primed determinant if expressed using the zero mode $e_0(t)$ of the differential operator. Employing relation~\eqref{eq:A3_Definition_Auxiliary_y}, we obtain the streamlined result
\begin{align}
	\text{det}_\zeta'(O) &=\frac{1}{2} \:\mathcal{N}_\text{E}\big.^{\!\!\!\!\: 2}\, \big\lvert\!\big\lvert \;\!e_0 \;\!\big\rvert\!\big\rvert^2 \:\! \Big[e_0(0) \, e_0(T)\Big]^{-1}  \, .
	\label{eq:StreamlinedPrimedDeterminant}
\end{align}
Notice that we can also insert scaled versions of the normalized mode $e_0$, which would not alter the given relation~\eqref{eq:StreamlinedPrimedDeterminant}.

\subsection{Green's function}
\label{sec:F2_GreensFunction}

Given the eigenfunctions $e_\mu(t)$ of the Sturm--Liouville problem~\eqref{eq:A3_Sturm_Liouville_Eigenvalue_Problem} pertaining to the boundary conditions~\eqref{eq:A3_Bouncary_Conditions_Eigenvalue_Problem}, we define the Green's function through the expression
\begin{align}
	G\big(t,t'\big)\coloneqq \sum_{\mu=0}^\infty \, \frac{e_\mu(t)\, e_\mu(t')}{\lambda_\mu} \, .
\end{align}
It satisfies the differential equation $O \:\! G(t,t')=\omega^{-1}\:\!\delta(t-t')$, where it is understood that the differential operator $O$ acts on the first argument. Once more assuming Robin boundary conditions as well as setting $w(t)=p(t)^{-1}=\omega$, the Green's function obeys the boundary constraints
\begin{subequations}
	\begin{align}
		m_{11} \, G\big(0,t'\big) + m_{12} \,\omega^{-1} \, \bigg[\frac{\partial \:\! G(t,t') }{\partial t}\bigg]_{\scalebox{0.8}{$t\!\!\:\!\!\:=\!\!\:\!\!\:0$}} &= 0 \, ,\\[0.05cm]
		n_{21} \, G\big(T,t'\big) + \: n_{22} \,\omega^{-1} \,\bigg[\frac{\partial \:\! G(t,t') }{\partial t}\bigg]_{\scalebox{0.8}{$t\!\!\:\!\!\:=\!\!\:\!\!\:T$}} \!\!\,&= 0  \, .
	\end{align}
	\label{eq:A3_Modified_Greens_Function_Boundary_Conditions}%
\end{subequations}
By definition, the Green's function is symmetric, satisfying $G(t,t')=G(t',t)$, thus automatically ensuring continuity at $t=t'$. In order for the $\delta$-distribution to emerge, one additionally requires a jump discontinuity in the derivative of $G$ at $t=t'$. This is most easily seen by integrating both sides over $\big(t'-\varepsilon,t'+\varepsilon\big)$ and subsequently taking the limit $\varepsilon\rightarrow 0^+$, revealing the additional condition 
\begin{align}
	\lim_{\varepsilon\rightarrow 0^+} \Bigg\{\frac{\partial \:\! G(t,t')}{\partial t} \bigg\vert_{\scalebox{0.8}{$t\!\!\:\!\!\:=\!\!\:\!\!\: t'\!\!\:\!\!\:-\!\!\:\!\!\:\varepsilon$}} - \frac{\partial \:\! G(t,t')}{\partial t} \bigg\vert_{\scalebox{0.8}{$t\!\!\:\!\!\:=\!\!\:\!\!\: t'\!\!\:\!\!\:+\!\!\:\!\!\:\varepsilon$}}\Bigg\} = \omega \, .
	\label{eq:A3_Modified_Greens_Function_Jump_Discontinuity}
\end{align} 
With the given constraints on the Green's function, we can fully determine it in terms of the formerly introduced solutions $\mathpzc{y}\raisebox{-3.25pt}{\scalebox{0.75}{$\lambda\!\!\:\!\!\:=\!\!\:\!\!\:0$}}\!\!\!\!\!\!\!\!\,\!\raisebox{4.75pt}{\scalebox{0.65}{$(1,2)$}}(t)$, see equations~\eqref{eq:A3_ODE_Definition_y12} and~\eqref{eq:A3_Boundary_Conditions_y12}. Employing the most general ansatz 
\begin{align}
	G\big(t,t'\big) = \left\{\begin{matrix}
		\alpha_1(t') \, \mathpzc{y}_{\scalebox{0.8}{$\lambda\!\!\:\!\!\:=\!\!\:\!\!\:0$}}^{\scalebox{0.8}{$(1)$}}(t) + \alpha_2(t') \, \mathpzc{y}_{\scalebox{0.8}{$\lambda\!\!\:\!\!\:=\!\!\:\!\!\:0$}}^{\scalebox{0.8}{$(2)$}}(t) && \text{for } t\leq t'  \\[0.3cm]
		\:\!\:\! \beta_1(t') \, \mathpzc{y}_{\scalebox{0.8}{$\lambda\!\!\:\!\!\:=\!\!\:\!\!\:0$}}^{\scalebox{0.8}{$(1)$}}(t) + \:\!\:\! \beta_2(t') \, \mathpzc{y}_{\scalebox{0.8}{$\lambda\!\!\:\!\!\:=\!\!\:\!\!\:0$}}^{\scalebox{0.8}{$(2)$}}(t) && \text{for } t'\leq t 
	\end{matrix} \right. \, ,
	\label{eq:A3_Greens_Function_General_Ansatz}
\end{align}
the boundary conditions~\eqref{eq:A3_Modified_Greens_Function_Boundary_Conditions} yield the two constraints
\begin{subequations}
\begin{align}
	0&=m_{11} \:\! \Big[\alpha_1(t') \, \mathpzc{y}_{\scalebox{0.8}{$\lambda\!\!\:\!\!\:=\!\!\:\!\!\:0$}}^{\scalebox{0.8}{$(1)$}}(0) + \alpha_2(t') \, \mathpzc{y}_{\scalebox{0.8}{$\lambda\!\!\:\!\!\:=\!\!\:\!\!\:0$}}^{\scalebox{0.8}{$(2)$}}(0)\Big] + m_{12} \,\omega^{-1} \, \Big[\alpha_1(t') \, \:\dot{\!\!\!\:\mathpzc{y}}_{\scalebox{0.8}{$\lambda\!\!\:\!\!\:=\!\!\:\!\!\:0$}}^{\scalebox{0.8}{$(1)$}}(0) + \alpha_2(t') \, \:\dot{\!\!\!\:\mathpzc{y}}_{\scalebox{0.8}{$\lambda\!\!\:\!\!\:=\!\!\:\!\!\:0$}}^{\scalebox{0.8}{$(2)$}}(0)\Big] \nonumber \\ 
	&= m_{11} \:\!\alpha_1(t') + m_{12} \,\alpha_2(t') \, , \\[0.2cm]
	0&=n_{21} \:\! \Big[\beta_1(t') \, \mathpzc{y}_{\scalebox{0.8}{$\lambda\!\!\:\!\!\:=\!\!\:\!\!\:0$}}^{\scalebox{0.8}{$(1)$}}(T) + \beta_2(t') \, \mathpzc{y}_{\scalebox{0.8}{$\lambda\!\!\:\!\!\:=\!\!\:\!\!\:0$}}^{\scalebox{0.8}{$(2)$}}(T)\Big] + n_{22} \,\omega^{-1} \, \Big[\beta_1(t') \, \:\dot{\!\!\!\:\mathpzc{y}}_{\scalebox{0.8}{$\lambda\!\!\:\!\!\:=\!\!\:\!\!\:0$}}^{\scalebox{0.8}{$(1)$}}(T) + \beta_2(t') \, \:\dot{\!\!\!\:\mathpzc{y}}_{\scalebox{0.8}{$\lambda\!\!\:\!\!\:=\!\!\:\!\!\:0$}}^{\scalebox{0.8}{$(2)$}}(T)\Big] \nonumber \\
	&=\hat{n}_{21}\:\! \beta_1(t')+ \hat{n}_{22}\:\! \beta_2(t') \, ,
\end{align}
\label{eq:GreensFunctionBoundaryRestrains}%
\end{subequations}
where we employed the former conditions~\eqref{eq:A3_Boundary_Conditions_y12} and defined the auxiliary constants
\begin{subequations}
	\begin{align}
		\hat{n}_{21}&\coloneqq n_{21} \, \mathpzc{y}_{\scalebox{0.8}{$\lambda\!\!\:\!\!\:=\!\!\:\!\!\:0$}}^{\scalebox{0.8}{$(1)$}}(T) + \,\!\;\! n_{22} \, \omega^{-1}\, \:\dot{\!\!\!\:\mathpzc{y}}_{\scalebox{0.8}{$\lambda\!\!\:\!\!\:=\!\!\:\!\!\:0$}}^{\scalebox{0.8}{$(1)$}}(T) \, ,\\
		\hat{n}_{22} &\coloneqq n_{21} \, \mathpzc{y}_{\scalebox{0.8}{$\lambda\!\!\:\!\!\:=\!\!\:\!\!\:0$}}^{\scalebox{0.8}{$(2)$}}(T) + n_{22} \, \omega^{-1} \, \:\dot{\!\!\!\:\mathpzc{y}}_{\scalebox{0.8}{$\lambda\!\!\:\!\!\:=\!\!\:\!\!\:0$}}^{\scalebox{0.8}{$(2)$}}(T) \, .
	\end{align}
	\label{eq:A3_Free_Greens_Function_Auxiliary_nhat}%
\end{subequations}
Both relations~\eqref{eq:GreensFunctionBoundaryRestrains} fix two of the four free functions. Symmetry as well as the jump discontinuity~\eqref{eq:A3_Modified_Greens_Function_Jump_Discontinuity} yield the remaining two conditions
\begin{subequations}
\begin{align}
	\beta_1(t) \, \mathpzc{y}_{\scalebox{0.8}{$\lambda\!\!\:\!\!\:=\!\!\:\!\!\:0$}}^{\scalebox{0.8}{$(1)$}}(t') +\, \beta_2(t) \; \mathpzc{y}_{\scalebox{0.8}{$\lambda\!\!\:\!\!\:=\!\!\:\!\!\:0$}}^{\scalebox{0.8}{$(2)$}}(t')&= \alpha_1(t') \; \mathpzc{y}_{\scalebox{0.8}{$\lambda\!\!\:\!\!\:=\!\!\:\!\!\:0$}}^{\scalebox{0.8}{$(1)$}}(t) \,+ \alpha_2(t') \, \mathpzc{y}_{\scalebox{0.8}{$\lambda\!\!\:\!\!\:=\!\!\:\!\!\:0$}}^{\scalebox{0.8}{$(2)$}}(t) \, , 
	\label{eq:SymmetryCondition} \\[0.2cm]
	\beta_1(t') \,  \:\dot{\!\!\!\:\mathpzc{y}}_{\scalebox{0.8}{$\lambda\!\!\:\!\!\:=\!\!\:\!\!\:0$}}^{\scalebox{0.8}{$(1)$}}(t') + \beta_2(t') \, \:\dot{\!\!\!\:\mathpzc{y}}_{\scalebox{0.8}{$\lambda\!\!\:\!\!\:=\!\!\:\!\!\:0$}}^{\scalebox{0.8}{$(2)$}}(t') &= \alpha_1(t') \, \:\dot{\!\!\!\:\mathpzc{y}}_{\scalebox{0.8}{$\lambda\!\!\:\!\!\:=\!\!\:\!\!\:0$}}^{\scalebox{0.8}{$(1)$}}(t') + \alpha_2(t') \, \:\dot{\!\!\!\:\mathpzc{y}}_{\scalebox{0.8}{$\lambda\!\!\:\!\!\:=\!\!\:\!\!\:0$}}^{\scalebox{0.8}{$(2)$}}(t') - \omega \, .
	\label{eq:JumpDiscontinuity}
\end{align}
\end{subequations}
Inserting the obtained ratios $\alpha_1/\alpha_2$ and $\beta_1/\beta_2$ into the symmetry relation~\eqref{eq:SymmetryCondition} and noticing that the equality has to hold for arbitrary $t,t'$, we find
\begin{align}
	\frac{1}{\beta_2(t)}\bigg[ \mathpzc{y}_{\scalebox{0.8}{$\lambda\!\!\:\!\!\:=\!\!\:\!\!\:0$}}^{\scalebox{0.8}{$(2)$}}(t) - \frac{m_{12}}{m_{11}}\, \mathpzc{y}_{\scalebox{0.8}{$\lambda\!\!\:\!\!\:=\!\!\:\!\!\:0$}}^{\scalebox{0.8}{$(1)$}}(t) \bigg] = \frac{1}{\alpha_2(t')} \bigg[ \mathpzc{y}_{\scalebox{0.8}{$\lambda\!\!\:\!\!\:=\!\!\:\!\!\:0$}}^{\scalebox{0.8}{$(2)$}}(t') - \frac{\hat{n}_{22}}{\hat{n}_{21}} \, \mathpzc{y}_{\scalebox{0.8}{$\lambda\!\!\:\!\!\:=\!\!\:\!\!\:0$}}^{\scalebox{0.8}{$(1)$}}(t') \bigg]=\Omega\, ,
	\label{eq:A3_Free_Greens_Function_Definition_Kappa}
\end{align}
with $\Omega \in \mathbb{R}$ being a to-be-determined constant. Employing equation~\eqref{eq:A3_Free_Greens_Function_Definition_Kappa}, all unknown functions $\alpha_{1,2}$ and $\beta_{1,2}$ are fully specified up to the constant $\Omega$, which is fixed by the remaining equation~\eqref{eq:JumpDiscontinuity}. Notice that the final restriction is valid for arbitrary $t'$ because the Wronskian of the two solutions $\mathpzc{y}\raisebox{-3.25pt}{\scalebox{0.75}{$\lambda\!\!\:\!\!\:=\!\!\:\!\!\:0$}}\!\!\!\!\!\!\!\!\,\!\raisebox{4.75pt}{\scalebox{0.65}{$(1,2)$}}(t)$ is unity. With $\Omega$ found to be
\begin{align}
	\Omega = \frac{m_{12}}{m_{11}} -\frac{\hat{n}_{22}}{\hat{n}_{21}} \, ,
\end{align}
we finally obtain the result 
\begin{equation}
	G\big(t,t'\big) \!\!\:= \bigg[\frac{m_{12}}{m_{11}} -\frac{\hat{n}_{22}}{\hat{n}_{21}} \bigg]^{-1} \bigg[ \mathpzc{y}_{\scalebox{0.8}{$\lambda\!\!\:\!\!\:=\!\!\:\!\!\:0$}}^{\scalebox{0.8}{$(2)$}}(t^+) - \frac{\hat{n}_{22}}{\hat{n}_{21}} \, \mathpzc{y}_{\scalebox{0.8}{$\lambda\!\!\:\!\!\:=\!\!\:\!\!\:0$}}^{\scalebox{0.8}{$(1)$}}(t^+) \bigg] \bigg[\mathpzc{y}_{\scalebox{0.8}{$\lambda\!\!\:\!\!\:=\!\!\:\!\!\:0$}}^{\scalebox{0.8}{$(2)$}}(t^-)-\frac{m_{12}}{m_{11}} \, \mathpzc{y}_{\scalebox{0.8}{$\lambda\!\!\:\!\!\:=\!\!\:\!\!\:0$}}^{\scalebox{0.8}{$(1)$}}(t^-)\bigg]\, ,
	\label{eq:A3_Free_Greens_Function_Result}
\end{equation}
with $t^+=\text{max}(t,t')$ and $t^-=\text{min}(t,t')$.

\subsection{Subtracted Green's function}
\label{sec:F3_Subtracted_Greens_Function}

The discussion for the subtracted Green's function
\begin{align}
	G^\perp\big(t,t'\big)\coloneqq \sum_{\mu=1}^\infty \, \frac{e_\mu(t)\, e_\mu(t')}{\lambda_\mu} 
\end{align}
is slightly more involved. While still required to satisfy the boundary conditions~\eqref{eq:A3_Modified_Greens_Function_Boundary_Conditions}, the jump discontinuity~\eqref{eq:A3_Modified_Greens_Function_Jump_Discontinuity}, as well as being symmetric under exchange of $t$ and $t'$, it additionally obeys the modified Green's function equation $O \:\! G^\perp(t,t')=\omega^{-1}\:\!\delta(t-t')-e_0(t)\, e_0(t')$. Again we denote the exact zero mode by $e_0(t)$, satisfying $O\:\! e_0(t)=0$ together with the appropriate Robin boundary conditions at $t=0,T$. Let us denote a second, linearly independent solution to the homogeneous ODE by $f_0(t)$, obeying $O f_0(t)=0$. We scale $f_0(t)$ accordingly such that both solutions possess a unit Wronskian 
\begin{align}
	W_{\!\!\:\omega}\!\!\:\big[e_0,f_0\big] = \omega^{-1} \Big[e_0(t) \, \dot{f}_0(t) - \dot{e}_0(t) \, f_0(t)\Big] = 1 \, .
	\label{eq:Wronskian_Property}
\end{align}
From this, we can straightforwardly infer a particular solution $f_\text{part}(t)$ to the differential equation $O f_\text{part}(t)=-e_0(t)$ in the form of
\begin{align}
	f_\text{part}(t) = f_0(t) \int_0^t e_0(\tau)^2 \,\text{d}(\omega \tau) - e_0(t) \int_0^t e_0(\tau)\, f_0(\tau) \,\text{d}(\omega \tau) \, .
\end{align}
With this particular solution at hand, the most general solution to the modified Green's function ODE can be given by 
\begin{align}
	G^\perp\big(t,t'\big) = f_\text{part}(t) \, e_0(t')+ \left\{\begin{matrix}
		\alpha_1(t') \, e_0(t) + \alpha_2(t') \, f_0(t) && \text{for } t\leq t' \, , \\[0.3cm]
		\:\!\:\! \beta_1(t') \, e_0(t) + \beta_2(t')\:\!\:\! f_0(t) && \text{for } t'\leq t \, .
	\end{matrix} \right.
	\label{eq:A3_Modified_Greens_Function_General_Ansatz}
\end{align}
With $e_0(t)$ obeying both Robin boundary conditions at $t=0,T$, $\alpha_1$ and $\beta_1$ drop out of the constraints~\eqref{eq:A3_Modified_Greens_Function_Boundary_Conditions}, leaving us with conditions involving only $\alpha_2$ and $\beta_2$. Employing $f_\text{part}(0)=\dot{f}_\text{part}(0)=0$, the boundary condition at $t=0$ requires $\alpha_2$ to vanish for all $t'$. In similar fashion, at the right boundary one uses $f_\text{part}(T) = f_0(T) - \big\langle e_0, f_0\big\rangle \, e_0(T)$ and $\dot{f}_\text{part}(T) = \dot{f}_0(T) - \big\langle e_0, f_0\big\rangle \, \dot{e}_0(T)$ to infer $\beta_2(t')=-e_0(t')$. Here we used that $e_0$ is normalized over the given time interval $[0,T]$ and $f_0$ does not satisfy the Robin boundary conditions at either of the two boundaries, otherwise $e_0$ and $f_0$ would be linearly dependent. The jump discontinuity~\eqref{eq:A3_Modified_Greens_Function_Jump_Discontinuity} again demands
\begin{align}
	\Big[\beta_1(t)-\alpha_1(t)\Big]\!\: \dot{e}_0(t) = \Big[\alpha_2(t)-\beta_2(t)\Big]\!\: \dot{f}_0(t) -\omega  \, ,
\end{align}
which then fixes $\beta_1(t)-\alpha_1(t)=f_0(t)$ after using the Wronskian~\eqref{eq:Wronskian_Property}. The remaining symmetry relation finally demands $\beta_1(t)=f_\text{part}(t)$, such that the full Green's function is found to be 
\begin{align}
	G^\perp\big(t,t'\big) = f_\text{part}(t) \, e_0(t')+f_\text{part}(t') \, e_0(t)- e_0(t^-)\, f_0(t^+)\, ,
	\label{eq:A3_Modified_Greens_Function_Result}
\end{align}
where we again use the convention $t^+=\text{max}(t,t')$ and $t^-=\text{min}(t,t')$. Notice that there is still a caveat entailed in the given result, as $f_0(t)$ is only fixed up to a multiple of $e_0(t)$, leaving the Wronskian invariant. However, such a shift will add an $e_0(t)\, e_0(t')$ contribution to the Green's function, thus we have to enforce the additional condition that the projection of $G^\perp$ on the zero mode vanishes, i.e.
\begin{align}
	\braket*{G^\perp\big(t,t'\big),e_0(t)}= \braket[\Big]{e_0,f_\text{part}-f_0} \, e_0(t') \overset{!}{=} 0 \, ,
\end{align}  
implicitly fixing the remaining freedom for $f_0(t)$. After integration by parts, the full condition on $f_0(t)$ can be brought into the convenient form 
\begin{align}
	\mathlarger{\mathlarger{\int}}_0^T e_0(t)\, f_0(t) \,\Bigg\{ \mathlarger{\int}_t^T \raisebox{-0.75pt}{$\scalebox{1.35}{\big[}$} e_0(t')\raisebox{-0.75pt}{$\scalebox{1.35}{\big]}$}^{\!\!\: 2} \,\text{d}(\omega t')\Bigg\} \, \text{d}(\omega t) \overset{!}{=} 0 \, .
	\label{eq:Restriction_SecondSolution}
\end{align}

\subsection{Homogeneous solutions}
\label{sec:Homogeneous_Solutions}

Assuming the first homogeneous solution $e_0$ to the differential operator $O$ to be known, we are interested in the behavior of the linearly independent solution $f_0$ required for the computation of the subtracted Green's function, see section~\ref{sec:F3_Subtracted_Greens_Function}. Typically, a variation of constants ansatz provides a second fundamental solution given by
\begin{align}
	f_0(t) = e_0(t) \mathlarger{\int}_0^t \raisebox{-0.75pt}{$\scalebox{1.35}{\big[}$} e_0(t')\raisebox{-0.75pt}{$\scalebox{1.35}{\big]}$}^{-2} \, \text{d} (\omega t') \, ,
	\label{eq:A3_Homogeneous_Solution_Na\"{\i}ve_Result}
\end{align}
however, in case the first solution entails zero crossings inside the interval $[0,T]$, this ansatz cannot be valid in the whole domain~\cite{TeschlODETheory}. Instead, Rofe-Beketov’s formula~\cite{RofeBeketovFormula} generates a solution not constrained by zeros of the first fundamental solution $e_0(t)$. Rewriting his attained formula conveniently for the present case yields the representation 
\begin{align}
 	\scalebox{0.99}{$\displaystyle{f_0(t) = e_0(t)\!\underbrace{ \mathlarger{\mathlarger{\mathlarger{\int}}}_{\scalebox{0.65}{$\!\:t_\text{low}$}}^{\scalebox{0.65}{$\!\:t$}} \bigg[1+\frac{\omega^{-2}\ddot{e}_0(t')}{e_0(t')}\bigg] \frac{\!\!\!\raisebox{-0.75pt}{$\scalebox{1.35}{\big[}$} e_0(t')\raisebox{-0.75pt}{$\scalebox{1.35}{\big]}$}^{\!\!\: 2}-\raisebox{-0.75pt}{$\scalebox{1.35}{\big[}$} \omega^{-1} \dot{e}_0(t')\raisebox{-0.75pt}{$\scalebox{1.35}{\big]}$}^{\!\!\: 2} }{\bigg\{\!\raisebox{-0.75pt}{$\scalebox{1.35}{\big[}$} e_0(t')\raisebox{-0.75pt}{$\scalebox{1.35}{\big]}$}^{\!\!\: 2}+\raisebox{-0.75pt}{$\scalebox{1.35}{\big[}$} \omega^{-1} \dot{e}_0(t')\raisebox{-0.75pt}{$\scalebox{1.35}{\big]}$}^{\!\!\: 2} \bigg\}^{\! 2}} \; \text{d}(\omega t')}_{\rule{0pt}{0.35cm}\displaystyle{\eqqcolon \Xi\big(t_\text{low},t\big)}} \: - \: \frac{\omega^{-1} \dot{e}_0(t)}{\raisebox{-0.75pt}{$\scalebox{1.35}{\big[}$} e_0(t)\raisebox{-0.75pt}{$\scalebox{1.35}{\big]}$}^{\!\!\: 2}+\raisebox{-0.75pt}{$\scalebox{1.35}{\big[}$} \omega^{-1} \dot{e}_0(t)\raisebox{-0.75pt}{$\scalebox{1.35}{\big]}$}^{\!\!\: 2}} \; .}$} \nonumber \\[-1.15cm]
 	 \vphantom{\bigg(}
 	\label{eq:A3_Homogeneous_Solution_Rofe_Beketov}
\end{align}
Note that the lower integration bound $t_\text{low}$ can be chosen freely, only resulting in an overall shift by some multiple of $e_0$. Differentiating the given representation and reinserting the definition of $f_0$, one obtains the simple expression $e_0(t)\,\dot{f}_0(t)= \dot{e}_0(t)\, f_0(t)+\omega$, showing that per construction the Wronskian of $e_0$ and $f_0$ is unity, such that the discussion in section~\ref{sec:F3_Subtracted_Greens_Function} can be applied. Differentiating once more shows that $f_0$ indeed solves the desired homogeneous ODE $O f_0(t)=0$. \\ 

\noindent
Let us now restrict to the relevant case of the zero mode being proportional to the (time-translated) bounce velocity $\dot{x}\raisebox{-2.5pt}{\scalebox{0.7}{\text{crit}}}\!\!\!\!\!\!\!\,\raisebox{5pt}{\scalebox{0.65}{$(t_0)$}}\:\!(t)$. Symmetry of the bounce motion enforces $e\raisebox{-2.5pt}{\scalebox{0.65}{0}}\!\!\!\,\raisebox{5pt}{\scalebox{0.65}{$(t_0)$}}\:\!\big(t_0-t\big)=-e\raisebox{-2.5pt}{\scalebox{0.65}{0}}\!\!\!\,\raisebox{5pt}{\scalebox{0.65}{$(t_0)$}}\:\!\big(t_0+t\big)$, such that it proves useful to choose $t_\text{low}=t_0$, for which the corresponding solution $f\!\!\:\raisebox{-3pt}{\scalebox{0.7}{0,\:\!\text{sym}}}\!\!\!\!\!\!\!\!\!\!\!\,\raisebox{5pt}{\scalebox{0.65}{$(t_0)$}}\;\;(t)$ is rendered symmetric. Defining the general solution 
\begin{align}
	f_0^{(t_0)}(t) = -2\eta \:\! e_0^{(t_0)}(t) +\hspace{-0.45cm}\underbrace{e_0^{(t_0)}(t) \: \Xi^{(t_0)}\big(t_0,t\big) - \frac{\omega^{-1} \dot{e}_0^{(t_0)}(t)}{\raisebox{-0.25pt}{$\scalebox{1.35}{\big[}$} e_0^{(t_0)}(t)\raisebox{-0.25pt}{$\scalebox{1.35}{\big]}$}^{\!\!\: 2}+\raisebox{-0.25pt}{$\scalebox{1.35}{\big[}$} \omega^{-1} \dot{e}^{(t_0)}_0(t)\raisebox{-0.25pt}{$\scalebox{1.35}{\big]}$}^{\!\!\: 2}}}_{\rule{0pt}{0.45cm}\displaystyle{\eqqcolon f_{0,\:\!\text{sym}}^{(t_0)}(t), \text{ satisfying } f_{0,\:\!\text{sym}}^{(t_0)}\big(t_0-t\big)=f_{0,\:\!\text{sym}}^{(t_0)}\big(t_0+t\big)}} \hspace{-0.45cm}
	\label{eq:General_Ansatz_Second_Solution}
\end{align}
with the constant $\eta$ parametrizing an arbitrary shift, we find the former restriction~\eqref{eq:Restriction_SecondSolution}  bestowed upon the special second solution given by 
\begin{align}
	\eta= \mathlarger{\mathlarger{\int}}_0^T e_0^{(t_0)}(t)\, f_{0,\:\!\text{sym}}^{(t_0)}(t) \,\Bigg\{ \mathlarger{\int}_t^T \raisebox{-0.25pt}{$\scalebox{1.35}{\big[}$} e_0^{(t_0)}(t')\raisebox{-0.25pt}{$\scalebox{1.35}{\big]}$}^{\!\!\: 2} \,\text{d}(\omega t')\Bigg\} \, \text{d}(\omega t) \, .
	\label{eq:Special_Shift_Eta_Constraint}
\end{align} 
Inserting the ansatz~\eqref{eq:General_Ansatz_Second_Solution} into the former result~\eqref{eq:A3_Modified_Greens_Function_Result} and evaluating the expression at the desired boundary values yields  
\begin{subequations}
	\begin{align}
		G_\text{bounce}^{\perp,(t_0)}\big(0,0\big) \;\!\;\!\;\!\;\! &= - \,\bigg\{f_{0,\:\!\text{sym}}^{(t_0)}(0)\;\!\;\!-2\eta \:\!e_0^{(t_0)}(0) \bigg\} \: e_0^{(t_0)}(0) \, , \\[0.05cm] 
		G_\text{bounce}^{\perp,(t_0)}\big(T,T\big) &=\phantom{+}\, \bigg\{ f_{0,\:\!\text{sym}}^{(t_0)}(T)+2 \raisebox{-0.2pt}{$\scalebox{1.2}{\Big[}$}\eta-\braket[\Big]{e_0^{(t_0)},f^{(t_0)}_{0,\:\!\text{sym}}}\raisebox{-0.2pt}{$\scalebox{1.2}{\Big]}$} e_0^{(t_0)}(T) \bigg\} \: e_0^{(t_0)}(T) \, ,\\[0.1cm]
		G_\text{bounce}^{\perp,(t_0)}\big(0,T\big) \;\!\;\! &=  \raisebox{-0.2pt}{$\scalebox{1.2}{\Big[}$}2\eta-\braket[\Big]{e_0^{(t_0)},f^{(t_0)}_{0,\:\!\text{sym}}}\raisebox{-0.2pt}{$\scalebox{1.2}{\Big]}$}\:\! e_0^{(t_0)}(0)\, e_0^{(t_0)}(T)\, .
	\end{align}
\end{subequations}
To find suitable approximations for large $T$, we are mandated to study the behavior of $f\!\!\:\raisebox{-3pt}{\scalebox{0.7}{0,\:\!\text{sym}}}\!\!\!\!\!\!\!\!\!\!\!\,\raisebox{5pt}{\scalebox{0.65}{$(t_0)$}}\;\;(t)$ far from $t_0$. For this, we require a convenient representation of the $\Xi$-integral, as the remaining expressions can be readily approximated. Employing the relations
\begin{subequations}
\begin{align}
	e_0^{(t_0)}(t)&= \raisebox{0.75pt}{$\Big\lvert\!\Big\lvert$} \dot{x}_\text{crit}^{(t_0)}\raisebox{0.75pt}{$\Big\rvert\!\Big\rvert$}^{-1} \, \dot{x}_\text{crit}^{(t_0)}(t)=\raisebox{0.75pt}{$\Big\lvert\!\Big\lvert$} \dot{x}_\text{crit}^{(t_0)}\raisebox{0.75pt}{$\Big\rvert\!\Big\rvert$}^{-1} \, \text{sign}\big(t_0-t\big) \,\omega \,x_\text{crit}^{(t_0)}(t)\, \bigg\{1+W\!\!\:\Big[x_\text{crit}^{(t_0)}(t)\Big]\!\!\:\bigg\}^{\!\frac{1}{2}} \, , \\[0.1cm]
	\dot{e}^{(t_0)}_0(t)&=\raisebox{0.75pt}{$\Big\lvert\!\Big\lvert$} \dot{x}_\text{crit}^{(t_0)}\raisebox{0.75pt}{$\Big\rvert\!\Big\rvert$}^{-1}\,\omega^2 \,\Bigg(x_\text{crit}^{(t_0)}(t)\, \bigg\{1+W\!\!\:\Big[x_\text{crit}^{(t_0)}(t)\Big]\!\!\:\bigg\}+\frac{1}{2}\, \Big[x_\text{crit}^{(t_0)}(t)\Big]^2\, W'\!\!\:\Big[x_\text{crit}^{(t_0)}(t)\Big]\Bigg) \, , \\[0.1cm]
	\frac{\ddot{e}^{(t_0)}_0(t)}{e_0^{(t_0)}(t)} &= \omega^2\, \Bigg\{1+W\!\!\:\Big[x_\text{crit}^{(t_0)}(t)\Big]+2\:\! x_\text{crit}^{(t_0)}(t)\, W'\!\!\:\Big[x_\text{crit}^{(t_0)}(t)\Big]+\frac{1}{2}\, \Big[x_\text{crit}^{(t_0)}(t)\Big]^2\, W''\!\!\:\Big[x_\text{crit}^{(t_0)}(t)\Big]\!\!\:\Bigg\} \, ,
\end{align} 
\end{subequations} 
we can rewrite the integral as
\begin{align}
	\Xi^{(t_0)}\big(t_0,t\big) &= -\frac{4}{\omega}\,\raisebox{0.75pt}{$\Big\lvert\!\Big\lvert$} \dot{x}_\text{crit}^{(t_0)}\raisebox{0.75pt}{$\Big\rvert\!\Big\rvert$}^{2} \mathlarger{\mathlarger{\mathlarger{\int}}}_{\scalebox{0.65}{$t_0$}}^{\scalebox{0.65}{$\!\:t$}} \; \text{d}t' \: \bigg[2+W(\xi)+2\:\!\xi \:\! W'(\xi)+\frac{\xi^2}{2} \:\! W''(\xi)\bigg] \, \xi^{-2} 
	\label{eq:Integral_Rewriting} \\[-0.2cm] 
	&\qquad\qquad\qquad\quad\!\!\!\!\;  \times  \frac{4\Big[1+W(\xi)\Big]\Big[W(\xi)+\xi \:\! W'(\xi)\Big]+ \Big[\xi \:\! W'(\xi)\Big]^2}{\bigg\{ 4\Big[1+W(\xi)\Big]\Big[2+W(\xi)+\xi \:\! W'(\xi)\Big]+ \Big[\xi \:\! W'(\xi)\Big]^2\bigg\}^{\! 2}}\left.\rule{0pt}{1.2cm}\right\vert_{\scalebox{0.9}{$\;\xi=x_\text{crit}^{(t_0)}(t')$}} \! . \nonumber
\end{align}
Changing the integration variable to be the bounce position, implicitly using invertibility of the bounce velocity for either $t<t_0$ or $t>t_0$, yields 
\begin{align}
	\Xi^{(t_0)}\big(t_0,t\big) &=  \frac{4\,\text{sign}(t_0-t)}{\omega^2}\,\raisebox{0.75pt}{$\Big\lvert\!\Big\lvert$} \dot{x}_\text{crit}^{(t_0)}\raisebox{0.75pt}{$\Big\rvert\!\Big\rvert$}^{2}\, \mathlarger{\mathlarger{\mathlarger{\int}}}_{\scalebox{0.65}{$\displaystyle{x_\text{crit}^{(t_0)}(t)}$}}^{\scalebox{0.65}{$\!\:x_\text{escape}$}} \; \text{d}\xi \;\, \xi^{-1} \Big[1+W(\xi)\Big]^{-\frac{1}{2}} \ldots \, ,
	\label{eq:Integral_Rewriting2}
\end{align}
where the dots abbreviate the remaining integrand as portrayed in~\eqref{eq:Integral_Rewriting}. This representation especially lends itself to be studied for times $t$ obeying $\big\lvert t-t_0\big\rvert \gg 1$, being far from the turning time $t_0$. In this case we formerly found $x\raisebox{-2.5pt}{\scalebox{0.7}{\text{crit}}}\!\!\!\!\!\!\!\,\raisebox{5pt}{\scalebox{0.65}{$(t_0)$}}\:\!(t)\sim \mathdutchbcal{A} e^{-\omega \lvert t-t_0\rvert} \ll 1$, see equation~\eqref{eq:ExactTimeRelations}, thus the integral should be expanded for small $x\raisebox{-2.5pt}{\scalebox{0.7}{\text{crit}}}\!\!\!\!\!\!\!\,\raisebox{5pt}{\scalebox{0.65}{$(t_0)$}}\:\!(t)$. One notices that the full integrand is singular at $\xi=0$, behaving like 
\begin{align}
	\text{integrand} \xrightarrow{\xi\rightarrow 0} \frac{W'(0)}{4 \xi^2}-\frac{3}{32 \xi}\, \Big[5 W'(0)^2-2 W''(0)\Big] + \mathcal{O}(1) \, .
	\label{eq:Integrand_RofeBeketov}
\end{align}
Thus, the largest contributions arise solely from the vicinity of $\xi=0$, such that we find the leading-order result 
\begin{align}
	\Xi^{(t_0)}\big(t_0,t\big) &\xrightarrow{\lvert t-t_0\rvert \gg 1}\, \text{sign}\big(t_0-t\big) \: \frac{W'(0) }{\omega^2}\,\raisebox{0.75pt}{$\Big\lvert\!\Big\lvert$} \dot{x}_\text{crit}^{(t_0)}\raisebox{0.75pt}{$\Big\rvert\!\Big\rvert$}^2\, \raisebox{-0.25pt}{$\scalebox{1.35}{\big[}$} x_\text{crit}^{(t_0)}(t)\raisebox{-0.25pt}{$\scalebox{1.35}{\big]}$}^{\!\!\: -1}  + \mathcal{O}\:\!\scalebox{1.1}{\Big\{}\!\log\!\!\!\:\raisebox{-0.25pt}{$\scalebox{1.35}{\big[}$} x_\text{crit}^{(t_0)}(t)\raisebox{-0.25pt}{$\scalebox{1.35}{\big]}$} \!\scalebox{1.1}{\Big\}} \, .
\end{align}
Performing an analogous expansion for the remaining term constituting $f\!\!\:\raisebox{-3pt}{\scalebox{0.7}{0,\:\!\text{sym}}}\!\!\!\!\!\!\!\!\!\!\!\,\raisebox{5pt}{\scalebox{0.65}{$(t_0)$}}\;\;(t)$ as seen in equation~\eqref{eq:General_Ansatz_Second_Solution}, we obtain
\begin{align}
	\frac{\omega^{-1} \dot{e}_0(t)}{\raisebox{-0.75pt}{$\scalebox{1.35}{\big[}$} e_0(t)\raisebox{-0.75pt}{$\scalebox{1.35}{\big]}$}^{\!\!\: 2}+\raisebox{-0.75pt}{$\scalebox{1.35}{\big[}$} \omega^{-1} \dot{e}_0(t)\raisebox{-0.75pt}{$\scalebox{1.35}{\big]}$}^{\!\!\: 2}} &= \frac{ 2\raisebox{0.75pt}{$\Big\lvert\!\Big\lvert$} \dot{x}_\text{crit}^{(t_0)}\raisebox{0.75pt}{$\Big\rvert\!\Big\rvert$}}{\omega \xi} \, \frac{2\Big[1+W(\xi)\Big] + \xi \:\! W'(\xi)}{4\Big[1+W(\xi)\Big]\Big[2+W(\xi)+\xi \:\! W'(\xi)\Big]+ \Big[\xi \:\! W'(\xi)\Big]^2} \left.\rule{0pt}{0.9cm}\right\vert_{\scalebox{0.9}{$\,\xi=x_\text{crit}^{(t_0)}(t)$}} \nonumber \\[0.15cm]
	&= \raisebox{0.75pt}{$\Big\lvert\!\Big\lvert$} \dot{x}_\text{crit}^{(t_0)}\raisebox{0.75pt}{$\Big\rvert\!\Big\rvert$} \: \Bigg\{\frac{1}{2\omega} \raisebox{-0.25pt}{$\scalebox{1.35}{\big[}$} x_\text{crit}^{(t_0)}(t)\raisebox{-0.25pt}{$\scalebox{1.35}{\big]}$}^{\!\!\: -1}-\frac{W'(0)}{4\omega}+ \mathcal{O}\raisebox{-0.25pt}{$\scalebox{1.35}{\big[}$} x_\text{crit}^{(t_0)}(t)\raisebox{-0.25pt}{$\scalebox{1.35}{\big]}$}\!\!\:\Bigg\} \, .
\end{align}
Collecting the results, far away from $t_0$ we find the asymptotic behavior
\begin{align}
	f_{0,\:\!\text{sym}}^{(t_0)}(t) = -\frac{1}{2\omega}\, \raisebox{0.75pt}{$\Big\lvert\!\Big\lvert$} \dot{x}_\text{crit}^{(t_0)}\raisebox{0.75pt}{$\Big\rvert\!\Big\rvert$} \: \bigg\{\! \raisebox{-0.25pt}{$\scalebox{1.35}{\big[}$} x_\text{crit}^{(t_0)}(t)\raisebox{-0.25pt}{$\scalebox{1.35}{\big]}$}^{\!\!\: -1}-\frac{5 W'(0)}{2}\!\!\:\bigg\} + \mathcal{O}\:\!\scalebox{1.2}{\Big\{}\!\log\!\!\!\:\raisebox{-0.25pt}{$\scalebox{1.35}{\big[}$} x_\text{crit}^{(t_0)}(t)\raisebox{-0.25pt}{$\scalebox{1.35}{\big]}$} x_\text{crit}^{(t_0)}(t)\scalebox{1.2}{\Big\}} \, ,
\end{align}
with the expansion being valid due to $x\raisebox{-2.5pt}{\scalebox{0.7}{\text{crit}}}\!\!\!\!\!\!\!\,\raisebox{5pt}{\scalebox{0.65}{$(t_0)$}}\:\!(t)\sim \mathdutchbcal{A} e^{-\omega \lvert t-t_0\rvert}\ll 1$. This especially holds for the endpoints $t=0,T$ in case we assume the turning point is to be reached well inside the time interval. Using the above result together with an expansion for $e\raisebox{-2.5pt}{\scalebox{0.65}{0}}\!\!\!\,\raisebox{5pt}{\scalebox{0.65}{$(t_0)$}}(t)$ leads to the important product 
\begin{align}
	e_0^{(t_0)}(t) \, f_{0,\:\!\text{sym}}^{(t_0)}(t) = \text{sign}\big(t-t_0\big)\, \Bigg(\!\:\frac{1}{2}-W'(0)\, x_\text{crit}^{(t_0)}(t) + \mathcal{O}\:\!\scalebox{1.3}{\Big\{}\!\!\!\;\raisebox{-0.25pt}{$\scalebox{1.35}{\big[}$} x_\text{crit}^{(t_0)}(t)\raisebox{-0.25pt}{$\scalebox{1.35}{\big]}$}^2 \log\!\!\!\:\raisebox{-0.25pt}{$\scalebox{1.35}{\big[}$} x_\text{crit}^{(t_0)}(t)\raisebox{-0.25pt}{$\scalebox{1.35}{\big]}$}\!\!\!\: \scalebox{1.3}{\Big\}}\Bigg) \, .
	\label{eq:Approximation_Product_e0_f0}
\end{align} 

\noindent
With this knowledge, we can easily approximate the projection $\braket{e_0,f_0}$. Using symmetry of the involved functions, we find
\begin{align}
	\braket[\Big]{e_0^{(t_0)},f^{(t_0)}_{0,\:\!\text{sym}}}= \mathlarger{\int}_0^T e_0^{(t_0)}(t)\, f^{(t_0)}_{0,\:\!\text{sym}}(t) \: \text{d}(\omega t) &= \mathlarger{\int}_{2t_0}^T \, e_0^{(t_0)}(t)\, f^{(t_0)}_{0,\:\!\text{sym}}(t) \: \text{d}(\omega t) \nonumber \\[0.1cm]
	&= \frac{\omega \big(T-2t_0\big)}{2} \:  \bigg\{1+\mathcal{O}\raisebox{-0.25pt}{$\scalebox{1.35}{\big[}$} x_\text{crit}^{(t_0)}(t)\raisebox{-0.25pt}{$\scalebox{1.35}{\big]}$}\!\bigg\} \, .
\end{align}
In the second line, given $t_0 \gg 1$, we used the leading-order approximation~\eqref{eq:Approximation_Product_e0_f0} for the product $e\raisebox{-2.5pt}{\scalebox{0.65}{0}}\!\!\!\,\raisebox{5pt}{\scalebox{0.65}{$(t_0)$}}(t) \, f\!\!\:\raisebox{-3pt}{\scalebox{0.7}{0,\:\!\text{sym}}}\!\!\!\!\!\!\!\!\!\!\!\,\raisebox{5pt}{\scalebox{0.65}{$(t_0)$}}\;\;(t)$. This is possible because the remaining integration domain $\big[2t_0,T\big]$ is far from the turning point of the bounce motion at $t=t_0$, thus the exponential behavior of both fundamental homogeneous solutions is already dominant. \\ 

\noindent
The sole remaining unknown is the shift $\eta$, given by relation~\eqref{eq:Special_Shift_Eta_Constraint}. It turns out to be excruciatingly difficult to obtain a solid approximation for which a quantitative error bound can be given, thus we resort to an order of magnitude estimate. Luckily such a rough approximation suffices for our current needs. We begin by simplifying the inner integral using
\begin{align}
	\mathlarger{\int}_t^T \raisebox{-0.25pt}{$\scalebox{1.35}{\big[}$} e_0^{(t_0)}(t')\raisebox{-0.25pt}{$\scalebox{1.35}{\big]}$}^{\!\!\: 2} \,\text{d}(\omega t') \approx \Theta\big(t_0-t\big) \, ,
\end{align}
knowing that for large times the bounce velocity is a sharply peaked function around $t=t_0$. The above substitution only introduces a substantial error inside a narrow $t$-region centered around $t_0$, in which the value of the integral transitions from approximately 0 to 1. With this, we find 
\begin{align}
	\eta\approx \mathlarger{\int}_0^{t_0} e_0^{(t_0)}(t)\, f_{0,\:\!\text{sym}}^{(t_0)}(t) \: \text{d}(\omega t) \approx -\frac{\omega t_0}{2}\, ,
	\label{eq:Special_Shift_Eta_Constraint_Simplified}
\end{align}
where we again used that the product $e\raisebox{-2.5pt}{\scalebox{0.65}{0}}\!\!\!\,\raisebox{5pt}{\scalebox{0.65}{$(t_0)$}}(t)\,f\!\!\:\raisebox{-3pt}{\scalebox{0.7}{0,\:\!\text{sym}}}\!\!\!\!\!\!\!\!\!\!\!\,\raisebox{5pt}{\scalebox{0.65}{$(t_0)$}}\;\;(t)$ over most of the time interval is practically $-\frac{1}{2}$, even though this former approximation~\eqref{eq:Approximation_Product_e0_f0} does not strictly apply to the narrow region around $t=t_0$. While we cannot rigorously give an estimate on the error, it should be sufficiently small since for large $T$ the regions in which the approximation breaks down becomes arbitrarily small. Additionally, the value of the functions is not expected to be extraordinarily large in those regions, such that the relative error should decrease with increasing $T$.\\

\noindent 
Collecting all former results, we obtain the final approximation
\begin{subequations}
	\begin{align}
		G_\text{bounce}^{\perp,(t_0)}\big(0,0\big) \;\!\;\!\;\!\;\! &= \frac{1}{2}-W'(0)\, x_\text{crit}^{(t_0)}(0) \;\;\!\! + \mathcal{O}\:\!\scalebox{1.3}{\Big\{}\!\!\!\;\raisebox{-0.25pt}{$\scalebox{1.35}{\big[}$} \:\! x_\text{crit}^{(t_0)}(0)\:\!\raisebox{-0.25pt}{$\scalebox{1.35}{\big]}$}^2 \;\! \log\!\!\!\:\raisebox{-0.25pt}{$\scalebox{1.35}{\big[}$} \:\! x_\text{crit}^{(t_0)}(0)\:\!\raisebox{-0.25pt}{$\scalebox{1.35}{\big]}$}\;\!\!\!\!\: \scalebox{1.3}{\Big\}}\, , \\[0.05cm] 
		G_\text{bounce}^{\perp,(t_0)}\big(T,T\big) &=\frac{1}{2}-W'(0)\, x_\text{crit}^{(t_0)}(T) + \mathcal{O}\:\!\scalebox{1.3}{\Big\{}\!\!\!\;\raisebox{-0.25pt}{$\scalebox{1.35}{\big[}$} x_\text{crit}^{(t_0)}(T)\raisebox{-0.25pt}{$\scalebox{1.35}{\big]}$}^2 \log\!\!\!\:\raisebox{-0.25pt}{$\scalebox{1.35}{\big[}$} x_\text{crit}^{(t_0)}(T)\raisebox{-0.25pt}{$\scalebox{1.35}{\big]}$}\!\!\!\: \scalebox{1.3}{\Big\}} \, ,\\[0.1cm]
		G_\text{bounce}^{\perp,(t_0)}\big(0,T\big) \;\!\;\! &= -\frac{\omega T}{2} \: \raisebox{0.75pt}{$\Big\lvert\!\Big\lvert$} \dot{x}_\text{crit}^{(t_0)}\raisebox{0.75pt}{$\Big\rvert\!\Big\rvert$}^{-2} \omega^2 x_\text{crit}^{(t_0)}(0) \, x_\text{crit}^{(t_0)}(T) + \text{subleading} \, .
	\end{align}
	\label{eq:Subtracted_Greens_Function_Bounce_Final}%
\end{subequations}
Beware that in the case $W'(0)=0$, one should also retain the first subleading contribution inside the integrand~\eqref{eq:Integrand_RofeBeketov} in order to capture the leading correction to the otherwise trivial result $G_\text{bounce}^{\perp,(t_0)}\big(0,0\big)=G_\text{bounce}^{\perp,(t_0)}\big(T,T\big)=\frac{1}{2}$.\\[0.25cm]

\bibliographystyle{CustomBibliography} 
\addcontentsline{toc}{section}{\protect\numberline{}References} 
\bibliography{FinalDraft}

\end{document}